%% file: draft1-24.tex
\documentstyle[12pt]{article}
\begin{document}
\hbadness=10000
\hbadness=10000
\begin{titlepage}
\nopagebreak
%\def\thefootnote{\fnsymbol{footnote}}
%%%%%%%%%%%%%%%%%%%%%%%%%% Preprint No. %%%%%%%%%%%
\begin{flushright}
%~\\
%~\\
%~\\
%\end{flushright}
{\normalsize
HIP-1998-04/TH\\
TUM-HEP-306/98\\
DPSU-98-98-1\\
Janurary, 1998}\\
\end{flushright}
%%%%%%%%%%%%%%%%%%%%%%%%%%%%%%%%%%%%%%%%%%%%%%%%%%%%
\vspace{0.7cm}
\begin{center}

{\large \bf  Bottom-up approach to $B$-parameter\\
in minimal supersymmetric standard model}

\vspace{1cm}

{\bf Yoshiharu Kawamura $^{a,b}$ 
\footnote[1]{e-mail: ykawamu@gipac.shinshu-u.ac.jp, Humboldt Fellow}}, 
{\bf Tatsuo Kobayashi $^{c,d}$
\footnote[2]{e-mail: kobayash@rock.helsinki.fi}}\\
and\\
{\bf Manabu Watanabe $^b$
\footnote[3]{e-mail: watanabe@azusa.shinshu-u.ac.jp}}

\vspace{0.7cm}
$^a$ Physik Department, Technische Universit\"at M\"unchen \\
   D-85748 Garching, Germany, \\
$^b$ Department of Physics, Shinshu University \\
   Matsumoto 390, Japan, \\
$^c$ Department of Physics, High Energy Physics Division\\
    University of Helsinki \\
and\\
$^d$ Helsinki Institute of Physics\\
    P.O. Box 9 (Siltavuorenpenger 20 C)\\
    FIN-00014 Helsinki, Finland \\

\end{center}
\vspace{0.7cm}

\nopagebreak
%%%%%%%%%%%%%%%%%%%%%%%%%%%%%

\begin{abstract}
We study $\mu$ and $B$-parameters in the
minimal supersymmetric standard model (MSSM)
based on the radiative electroweak symmetry breaking scenario
using $\lq\lq$bottom-up" approach and
show how useful our approach is to select a phenomenologically 
viable model beyond the MSSM
under the assumption that the underlying theory
is a string model or a gauge-Yukawa unified gauge model.
\end{abstract}
%PACS number(s):04.65.+e, 11.25.Mj, 12.60.Jv\\
%Keyword(s): mu-term, radiative electroweak breaking, 
%Soft SUSY breaking terms, string model, coupling reduction
\vfill
\end{titlepage}
\pagestyle{plain}
\newpage
\def\thefootnote{\fnsymbol{footnote}}

\section{Introduction}

Supersymmetric (SUSY) extension of the standard model 
is an attractive candidate 
beyond the standard model \cite{HK}.
Much effort has been devoted to study phenomenological aspects
of the minimal supersymmetric standard model (MSSM).
One of important issues in the MSSM is 
what is the origin of physical parameters such as soft SUSY 
breaking parameters and the $\mu$-parameter.
Their patterns and magnitudes are supposed to be related to 
the SUSY breaking mechanism in some underlying theory. 
For example, the pattern of soft SUSY breaking parameters 
depends on the structure of couplings between visible sector
and hidden sector, and their magnitudes 
are of order of the gravitino mass $m_{3/2}$ in the
hidden sector SUSY breaking scenario of supergravity (SUGRA) \cite{N}. 
It applies to the effective SUGRA derived 
from superstring theory.

There is another type of interesting scenario to fix the pattern
of physical parameters at some high energy scale.
It is coupling reduction theory, whose basic concept is to construct
renormalization group (RG) invariant relations including those 
between gauge and Yukawa couplings \cite{GYU}.
Thus, hereafter we call this type of model as gauge-Yukawa unification 
(GYU) model.
Recently it has been applied to reduction of dimensionful couplings, 
i.e., reduction of soft SUSY breaking parameters 
\cite{KMZ,KKK,KKMZ,KKK2}.
In this framework, soft SUSY breaking parameters are given as functions of 
a gaugino mass $M_{1/2}$ in the form of RG invariants.

In any cases, the $\mu$-parameter is in a special position.
The $\mu$-term as well as the $B$-term plays a role in electroweak 
symmetry breaking, but its magnitude has an arbitariness from theoretical 
viewpoint, because the $\mu$-parameter 
can, in principle, have a different origin from that of soft SUSY breaking 
parameters.
On the other hand, the condition that the electroweak symmetry 
is broken down at the weak scale
requires that magnitudes of the $\mu$-parameter and soft SUSY breaking
parameters should be around the weak scale.
Here we encounter the so-called $\mu$ problem \cite{mupro};
how is the $\mu$-parameter of $O(1)$ TeV generated?
Several interesting solutions have been proposed \cite{mu1,mupro,mu2,mu3,mu4}.
Also, recently the study of $\mu$-term has been carried out
from several viewpoints \cite{Higgs-st,LT,CLM,Imu}.
The exploration of the origin of $\mu$-term is one of important subjects
because it can give a key to high energy physics beyond the MSSM.
It would be possible to comprehend the high energy physics 
and the origin of $\mu$-term simultaneously using experimental data in future.
It would be troublesome to scan a whole range of the parameter space
on the MSSM.
Hence it is meaningful to investigate which type of $\mu$-term 
generation mechanism is 
favorable under some phenomenological requirement in advance.
In most cases, the $B$-parameter depends on the mechanism of $\mu$-term 
generation and there are certain relations of the $B$-parameter with 
other soft SUSY breaking parameters.\footnote{
In GYU model, the relation between the $B$-parameter and 
other soft SUSY breaking parameters is determined by the use of 
the coupling reduction
procedure independent of the $\mu$-term generation mechanism.}
Hence the study on $B$-parameter is indispensable to
quest the source of $\mu$-parameter and other soft SUSY breaking 
parameters.

If we specify the underlying theory at high energy scale, $M_X$,
we can check the reality of $\mu$ and $B$-term generation mechanism
comparing two kinds of formulae for $\mu$ and $B$-parameters at $M_X$
explained just below.
One set of formulae is derived from the phenomenological requirement that the 
electroweak symmetry is broken radiatively.
That is, using the minimization conditions of the Higgs potential, 
we can write down $\mu$ and $B$-parameters in terms of soft Higgs masses and 
$\tan \beta =\langle h_2 \rangle /\langle h_1 \rangle$, where 
$h_1$ and $h_2$ are neutral components of Higgs doublets.
After the introduction of radiative corrections,
$\mu$ and $B$-parameters at $M_X$ are written as functions of soft 
scalar masses $m_k$, $A$-parameters $A_{klm}$, 
gaugino masses $M_a$ and $\tan \beta$.
On the other hand, the formulae of SUSY breaking parameters
and $\mu$-parameter are written down in terms of 
parameters ($m_S$, $\langle F \rangle$, $\lambda$) in the underlying theory.
Here $m_S$ and $\langle F \rangle$ are some mass parameters (e.g., $m_{3/2}$
in SUGRA and $M_{1/2}$ in GYU-model) and some $F$-term condensations 
characteristic to the SUSY breaking, respectively. 
The parameters $\lambda$ are coupling constants related to $\mu$-term
generation mechanism.
Comparing with these two types of formulae for $\mu$ and $B$-parameters
at $M_X$,
we could find a constraint manifold of solutions in the 
$(m_S, \langle F \rangle, \tan \beta, \lambda)$-space for each of 
$\mu$ and $B$-term generation mechanisms.
As a degree of freedom of ($m_S$, $\langle F \rangle$, $\lambda$), in general,
is much less than the number of soft SUSY breaking parameters,
there are several relations among these soft parameters.
If the underlying theory leads to such specific relations,
the comparison with the formulae would become serious.
In this case, some generation mechanisms might
be ruled out in certain regions.
This ``bottom-up'' approach\footnote{
In \cite{Bup}, qualitative properties of the high-energy parameter space are
discussed using bottom-up approach, but the comparison with the prediction
of specific models, e.g. string model, has not been carried out.}
to study $\mu$ and $B$ is more 
powerful than the ``top-down'' approach where one $\mu$-term 
generation mechanism as well as other 
soft SUSY breaking parameters is fixed at a high energy scale and then 
the realization of electroweak symmetry breaking is investigated.
Actually the study by the bottom-up approach has been done 
in Ref. \cite{KKW} for some parameter regions of superstring models 
in the small $\tan \beta$ scenario.\footnote{
In \cite{KMVS}, the availability of $B$-term is discussed based on 
radiative symmetry breaking scenario using string model.}

In this paper, we develop it and show how useful our approach is 
to select a phenomenologically viable model beyond the MSSM.
We apply our strategy, the bottom-up approach, to the MSSM whose
underlying theory is superstring model or GYU-grand unified theory
(GUT).
The investigation is carried out for several parameter regions 
in both the small and large $\tan \beta$ scenarios.

This paper is organized as follows.
In the next section, we explain our strategy
to select a realistic $\mu$ and $B$-term generation mechanism.
We devide the parameter space of Yukawa couplings into the 
small and large $\tan \beta$ scenarios.
These two regions show different aspects for the $B$-parameter each other.
In section 3, we apply our method to the MSSM derived from a certain type 
of string model or a generic GYU-GUT.
Also comments are given for other cases.
Section 4 is devoted to conclusions and discussions.

\section{Bottom-up approach}

In this section, we give an outline of our strategy \cite{KKW}. 
The neutral components $h_1$ and $h_2$ of Higgs doublets
$H_1$ and $H_2$ have the following 
potential \cite{Higgs}\cite{RBS}:
\begin{eqnarray}
V(h_1,h_2) &=& m_1^2h_1^2+m_2^2h_2^2+(\mu Bh_1h_2+h.c.) 
\nonumber \\
 &~& + {1 \over 8}(g^2+g'^2)(h_1^2-h_2^2)^2 ,\\
m_1^2 &\equiv& m_{H_1}^2+\mu^2, \quad m_2^2~\equiv~m_{H_2}^2+\mu^2
\end{eqnarray}
where the values of all parameters correspond to those at the weak scale.
Through our analysis, we use the $Z$ boson mass $M_Z=91.2$ GeV
as the weak scale
and we neglect the threshold corrections due to the difference among
sparticle masses for simplicity.
The mass parameters $m_{H_1}^2$ and $m_{H_2}^2$ are 
soft SUSY breaking masses of Higgs scalars whose hypercharges are
$-1/2$ and $+1/2$, respectively, and $\mu$ is a SUSY Higgs mass parameter.
The condition for electroweak symmetry breaking is given by 
\begin{eqnarray}
m_1^2 m_2^2 < |\mu B|^2.
\label{SB}
\end{eqnarray}
The bounded from below condition along 
the $D$-flat direction requires 
\begin{eqnarray}
m_1^2+m_2^2 >2 |\mu B|.
\label{BFB}
\end{eqnarray}
Further the conditions that minimizing the potential are given by
\begin{eqnarray}
&~& m_1^2+m_2^2 = -{2 \mu B \over \sin 2 \beta}, 
\label{mini1}\\
&~& m_1^2-m_2^2 = -\cos 2 \beta (M_Z^2+m_1^2+m_2^2) 
\label{mini2}
\end{eqnarray}
where we have used the relation $M_Z^2 = {1 \over 4}(g^2+g'^2)v^2$
and $v^2 = \langle h_2 \rangle^2 + \langle h_1 \rangle^2$.

By the use of stationary conditions (\ref{mini1}) and (\ref{mini2}),
$\mu$ and $B$-parameters are expressed by 
\begin{eqnarray}
&~& |\mu| = {1 \over \sqrt{2}}
\left({m_{H_1}^2 - m_{H_2}^2 \over -\cos 2\beta} - m_{H_1}^2 - 
 m_{H_2}^2 -M_Z^2\right)^{1/2} ,
\label{mu}\\
&~& |B| = {\sin 2\beta \over 2|\mu|}
\left({m_{H_1}^2 - m_{H_2}^2 \over -\cos 2\beta} - M_Z^2 \right) .
\label{B}
\end{eqnarray}
At this stage, $|\mu|$ and $|B|$ are functions of $m_{H_i}^2$ and $\tan \beta$ 
at the weak scale.
Throughout this paper, we assume that there is no new CP violation source 
except that from the Yukawa coupling sector.
In this case, we can take $A$ and $B$-parameters real 
after the gaugino mass parameter is made real and positive
by the chiral rotation for the gaugino field.
We use this convention with $\mu B < 0$ made
by a suitable phase rotation for Higgs multiplets.

In order to probe underlying theory, we would like to have 
$\mu(M_X)$ and $B(M_X)$ as functions of other parameters at $M_X$.
Through our analysis, 
we take the gauge coupling unification scale
$M_X = 1.7 \times 10^{16}$GeV as an energy scale 
where boundary conditions are imposed.
Considering RG flows \cite{RBS},
%\footnote{
%In this paper, we discuss only scalar potential at the tree level
%and RG flows at one-loop level for soft parameters and $\mu$-parameter.
%It is straightforward to introduce the effects at a higher level
%to our analysis.}, 
we can obtain the $\mu$ and $B$ parameters 
at $M_X$ as follows,
\begin{eqnarray}
&~& \mu_{\pm}(M_X) = \pm c_\mu |\mu| 
\label{muMX}
\end{eqnarray}
and
\begin{eqnarray}
&~& B_{\pm}(M_X) = \mp |B| + \Delta B 
\label{BMX}
\end{eqnarray}
where $c_\mu$ and $\Delta B$ are renormalization factors from $M_Z$
to $M_X$.
Here and hereafter $B_{\pm}(M_X)$ and $\mu_{\pm}(M_X)$ denote the 
$B$ and $\mu$ parameters at $M_X$, which are consistent with
the radiative electroweak symmetry breaking scenario.
The radiative corrections $\Delta B$ depend on $M_a^{(0)}$ and $A_{klm}^{(0)}$ 
at $M_X$ as well as $\tan \beta$, 
while $c_\mu$ depends only on $\tan \beta$.
Here and hereafter the suffix $(0)$ of the parameter $a^{(0)}$ denotes 
a value of the parameter $a$ predicted from underlying theory at $M_X$.
Further here we have used the fact that Yukawa couplings are given as functions
of $\tan \beta$ when the fermion masses are fixed.
The top quark mass is given as $m_t(m_t)=f_t(m_t) v \sin \beta/\sqrt{2}$.
The pole mass of top quark $m_t^{pole}$ is related to the running mass 
by 
%\begin{eqnarray}
$m_t^{pole}=m_t(m_t)[1+{4\alpha_3(m_t)/ (3\pi )}+O(\alpha_3^2)]$. 
%\nonumber
%\end{eqnarray}
We use the values $m_t=175$GeV and $m_{\tau}=1.78$GeV as the pole masses
of top quark and tau lepton from the current experiments.
On the bottom quark mass, we impose the condition $f_b = f_\tau$ at $M_X$
on the Yukawa couplings of bottom quark and tau lepton
in the case with large $\tan \beta$. 
Because the experimental value of bottom quark still has some uncertainty
and large SUSY corrections can be demanded 
in the case with large $\tan \beta$ \cite{hall,COPW}.

The soft SUSY breaking scalar masses $m_{k}$ at $M_Z$
also receive radiative corrections such that
\begin{eqnarray}
&~&m_{k}^{2} = m_k^{(0)2} + \sum_a \xi_k^a M_a^{(0)2} + \Delta m_k^2 + S_k
\label{smass}
\end{eqnarray}
where $m_k^{(0)}$'s are soft scalar masses at $M_X$, the second term 
represents 
a renormalization effect due to gauginos of $U(1)_Y$, $SU(2)_L$ and $SU(3)_C$
whose masses at $M_X$ are given as $M_a^{(0)}$ ($a=1,2,3$), i.e., 
$\sum_a \xi_{H_1}^a = \sum_a \xi_{H_2}^a = 0.52$ for the universal 
gaugino mass $M_{1/2}^{(0)}$.
The third term $\Delta m_k^2$ is a renormalization factor 
due to Yukawa couplings and this is a function of $m_k^{(0)}$, $A_{klm}^{(0)}$
and $M_a^{(0)}$ as well as $\tan \beta$.
The fourth term $S_k$ is a tadpole contribution due to scalar fields 
with hypercharge interaction.
Further $m_{k}^{(0)}$, $A_{klm}^{(0)}$ and $M_a^{(0)}$, in general, 
are given as functions of parameters $m_S$ and
%(e.g., $m_{3/2}$ in SUGRA) and some $F$-term condensations 
$\langle F \rangle$ which appears in the underlying theory.
%characteristic to SUSY breaking.
Hence $\mu_{\pm}(M_X)$ and $B_{\pm}(M_X)$ are written as functions
of $m_S$, $\langle F \rangle$ and $\tan \beta$.

On the other hand, if we specify the underlying theory,
% and one $\mu$-term generation mechanism, 
we can write down formulae of the $\mu$ and $B$-parameters 
(they are denoted by $\mu^{(0)}$ and $B^{(0)}$) 
and other soft SUSY breaking parameters at $M_X$
using $m_S$, $\langle F \rangle$, $\tan \beta$
and other coupling constants $\lambda$
related to $\mu$-term generation mechanism.

Requiring that $B_{\pm}(M_X)$ and $\mu_{\pm}(M_X)$
should agree with $\mu^{(0)}$ and $B^{(0)}$,
we can find allowed parameter regions for 
($m_S$, $\langle F \rangle$, $\tan \beta$, $\lambda$)
leading to successful electroweak symmetry breaking
and
know which type of $\mu$-term generation mechanism is hopeful.
This is the outline of our approach 
to probe realistic $\mu$ and $B$-parameters.
Our approach can be more powerful in the presence of some non-trivial 
relations among soft SUSY breaking parameters at $M_X$
because the number of the free parameters ($m_S$, $\langle F \rangle$)
is reduced.
Such a restricted situation appears
in the MSSM based on a certain type of string model as well as in 
a generic GYU-GUT.
We will study such cases in the next section.

Next we discuss features of $\mu$ and $B$-parameters in the cases with 
small $\tan \beta$ and large $\tan \beta$, separately. 

(1) Small $\tan \beta$ case ($2 \leq \tan \beta \leq 10$)

In this case, the top Yukawa coupling $f_t$ is 
so strong compared with other Yukawa couplings
that the effects of other Yukawa couplings can be neglected.
Then we have analytical solutions 
for renormalization factors $c_{\mu}$ and $\Delta B$ at one-loop level.
For example, the solutions in the case with universal gaugino mass at $M_X$
are given by \cite{manual}\footnote{
See the second reference in \cite{Bup} for the analytic solutions
in the case with non-universal gaugino masses at $M_X$.}
\begin{eqnarray}
&~& c_\mu \equiv \left({\alpha_2(t_Z) \over \alpha^{(0)}}\right)^{3/2}
\left({\alpha_1(t_Z) \over \alpha^{(0)}}\right)^{1/22}
(1+6\alpha_t^{(0)} F(t_Z))^{1/4} ,
\label{cmu}\\
&~& \Delta B \equiv 3A_t^{(0)}{\alpha_t^{(0)} F(t_Z) \over
1 + 6\alpha_t^{(0)} F(t_Z)} 
\nonumber \\
&~& + M_{1/2}^{(0)}\left\{t_Z(3\alpha_2(t_Z)
+{3 \over 5}\alpha_1(t_Z)) - {3\alpha_t^{(0)}(t_Z F'(t_Z)-F(t_Z)) \over
1 + 6\alpha_t^{(0)} F(t_Z)}\right\}
\label{DeltaB}
\end{eqnarray}
where 
\begin{eqnarray}
&~&\alpha^{(0)} \equiv {g^{(0)2} \over 4\pi}, ~~
\alpha_t^{(0)} \equiv {f_t^{(0)2} \over 4\pi},\\
&~&F(t_Z) \equiv \int_0^{t_Z} 
\left({\alpha_3(t) \over \alpha^{(0)}}\right)^{16/9}
\left({\alpha_2(t) \over \alpha^{(0)}}\right)^{-3}
\left({\alpha_1(t) \over \alpha^{(0)}}\right)^{-13/99}dt ,\\
&~&t_Z = (4\pi)^{-1} \log {M_X^2 \over M_Z^2} .
\label{tZ}
\end{eqnarray}
Here $g^{(0)}$ and $f_t^{(0)}$ are the gauge coupling
and top Yukawa coupling at $M_X$, respectively.
Using analytical expressions for 
$\mu_{\pm}(M_X)$, $B_{\pm}(M_X)$ and $m_{H_i}^2$,
$\mu_{\pm}(M_X)$ and $B_{\pm}(M_X)$ are written down
as functions of ($m_{k}^{(0)2}$, $M_{a}^{(0)}$, $A_t^{(0)}$, $\tan\beta$).

(2)  Large $\tan \beta$ case ($20 \leq \tan \beta \leq 60$)
 
We calculate renormalization effects numerically in this case 
because no analytical solutions are known\footnote{
The approximate analytical solutions are obtained in the neighborhood
of quasi-fixed points of Yukawa couplings \cite{COPW}.}.
Here we discuss the large limit of $\tan \beta$.
In this limit, $\mu$ and $B$ parameters are approximately 
expressed by
\begin{eqnarray}
&~& |\mu|^2 \sim - m_{H_2}^2 - {1 \over 2}M_Z^2  ,
\label{mu-Lt}\\
&~& |B| \sim {1 \over \tan \beta}{m_{H_1}^2 - m_{H_2}^2 - M_Z^2
 \over |\mu|} .
\label{B-Lt}
\end{eqnarray}
~From the requirement that $|\mu|^2$ and $|B|$ should be positive-definite,
we have the following two conditions,
\begin{eqnarray}
&~&  m_{H_2}^2 < -{1 \over 2}M_Z^2  ,
\label{neq1}\\
&~&  m_{H_1}^2 > m_{H_2}^2 + M_Z^2  .
\label{neq2}
\end{eqnarray}
If the magnitude of $(m_{H_1}^2 - m_{H_2}^2 - M_Z^2)/|\mu|$ is of $O(m_S)$,
$|B|$ is of $O(m_S/\tan \beta)$, which is negligiblely small compared 
with its radiative corrections of $O(m_S)$.
%\footnote{
%The above relations (\ref{mu-Lt}) and (\ref{B-Lt}) are regarded as a kind of
%fine-tuning among physical parameters at the weak scale.
%They gave strong conditions for high energy theory  
%on the analysis based on RG equations at one-loop level 
%and one-loop effective potential \cite{Y-uni}.}
This statement is justified, in most cases,
by the comparison of the values of $|B(M_Z)|/M_{1/2}^{(0)}$ with
those of $\Delta B/M_{1/2}^{(0)}$ practically.
The values of $|B(M_Z)|/M_{1/2}^{(0)}$ are given in Fig. 1 
for the universal case 
with $m^{(0)2}_k=M^{(0)2}_{1/2}/3$ and 
$A_f^{(0)}=-M^{(0)}_{1/2}$.
(The implication of these conditions will be discussed later.)
%Here and hereafter we choose that $M_{1/2}^{(0)}$ is positive
%by suitable rephasing of gaugino.
The typical order of $|B(M_Z)|/M^{(0)}_{1/2}$ is $O(10^{-2})$ for
$\tan \beta \geq 30$ and, in particular, it is $O(10^{-3})$ for 
$\tan \beta \geq 50$.
Note that the values are suppressed, comparing with the order 
($O(1/\tan\beta)$) estimated in the general case, because 
$m_{H_1}^2 - m_{H_2}^2$ is proportional to $M_{1/2}^{(0)2}$ 
with a small coefficient in the universal case.
The values approach constants in the limit of
large gaugino mass because the dependency of $M_Z$ 
in $|B(M_Z)|/M_{1/2}^{(0)}$ becomes small.
The relatively small value of $M_{1/2}^{(0)}$ is not allowed  
from the condition (\ref{neq2}).
For example, $M^{(0)}_{1/2}$ should be bigger than about 180 GeV 
for $\tan \beta =50$.
\begin{center}
\input BMZ.tex

Fig. 1
\end{center}
On the other hand, the values of radiative corrections, 
$\Delta B/M_{1/2}^{(0)}$, are 
shown against $A^{(0)}/M^{(0)}_{1/2}$ in Fig. 2, 
where the solid and dotted lines correspond to the case with $\tan \beta =20$ 
and 50, respectively.
\begin{center}
\input BAM.tex

~\\
~\\
Fig. 2
\end{center}
The typical order of $\Delta B/M^{(0)}_{1/2}$ is $O(1)$
except for the case $M^{(0)}_{1/2} \sim A_f^{(0)}$.
Hence we can set $B(M_Z)$ as zero safely 
for the very large $\tan \beta$ and $M^{(0)}_{1/2} \neq A_f^{(0)}$.
This fact can lead to more powerful prediction in the very large $\tan \beta$ 
scenario than in the small $\tan \beta$ one 
because $B_{\pm}(M_X)$ is strongly
dependent of only $M^{(0)}_a$ and $A^{(0)}_{klm}$
in the large $\tan \beta$ scenario , 
while $m_k^{(0)}$'s are also indispensable to the determination of
$B_{\pm}(M_X)$ in the small $\tan \beta$ scenario.
For example, the underlying theory should predict $B^{(0)}$ which satisfies
\begin{equation}
-1.9 \leq {B^{(0)} \over M^{(0)}_{1/2}} \leq -0.7,
\label{regionB}
\end{equation}
for $20 \leq \tan \beta \leq 60$ and $A_f^{(0)}=-M^{(0)}_{1/2}$, 
where the value of $B(M_Z)/M_{1/2}^{(0)}$ is taken into account for 
$\tan \beta =20$.

\section{Applications}

As discussed in the previous section, the values $B_{\pm}(M_X)$ 
and $\mu_{\pm}(M_X)$ are functions of $m_k^{(0)}$, $M^{(0)}_a$ 
and $A^{(0)}_{klm}$ as well as $\tan \beta$.
%When these parameters are independent each other, our predictability 
%is not poweful.
In most cases, some underlying thoeries lead to certain specific relations 
among these parameters, e.g., string-inspired relations, 
RG-invariant relations or the relations from the no-scale SUGRA model.
In this case, our prediction power becomes strong.
In this section, we study the origin of 
$\mu$ and $B$-parameters based on the MSSM with such specific relations
from the phenomenological standpoint.

\subsection{Effective MSSM from string theory}
\subsubsection{Formulae from string theory}

At first we discuss a certain type of 4-dimensional string model 
where SUSY is broken by dilaton and/or moduli 
$F$-term condensations \cite{CCM,ST-soft,BIM}.
In this case, $M_{a}^{(0)}$ and $m_{k}^{(0)2}$ are given by 
\begin{eqnarray}
&~&M_a^{(0)}=\sqrt 3 |m_{3/2}| \sin \theta e^{-i\alpha_S} ,
\label{Mi(0)}\\
&~&m_{k}^{(0)2}=|m_{3/2}|^2 (1+n_k \cos^2 \theta ) + d_k |m_{3/2}|^2
\label{smass(0)}
\end{eqnarray}
where $\theta$ is a goldstino angle, $n_k$ a modular weight 
of the corresponding chiral matter field and $d_k |m_{3/2}|^2$ 
$D$-term contributions \cite{stringD}.
Here we assume that the vacuum energy vanishes.
The universality among gaugino masses originates from the fact that 
the gauge kinetic functions include only the dilaton field in
the same fashion at tree level.
%For SUSY-GUT, it is due to the grand unification of the SM gauge group.
In addition, the $A$-parameters are obtained as
\begin{eqnarray}
A_{klm}^{(0)} &=& - \sqrt{3} |m_{3/2}| \sin\theta e^{-i \alpha_S} 
- |m_{3/2}| \cos\theta
(3 + \sum_k n_k) e^{-i \alpha_T}
\label{A(0)}
\end{eqnarray}
where $\sum_k n_k$ is the sum of modular weights of matter 
fields in the corresponding Yukawa couplings $Y_{klm}$.
Here the phases $\alpha_S$ and $\alpha_T$ stem from 
phases of $F$-terms of $S$ and $T$.
If $Y_{klm}$ depends on $T$, there appears another contribution.
Here we assume that the top Yukawa coupling is independent of the moduli 
field.
(We will give a brief comment on the case where this assumption is relaxed in 
3.3.)
Using this assumption and the following relation from $T$-duality of 
lagrangian 
\begin{equation}
n_{Q_3}+n_{t}+n_{H_2}=-3,
\label{sumn}
\end{equation}
we have the following relations at $M_X$ \cite{BIMS}, 
\begin{eqnarray}
&~&A_{t}^{(0)} = - M_{1/2}^{(0)}  ,
\label{At(0)}\\
&~&m_{\Sigma (t)}^{(0)2}=m_{\tilde{Q}_3}^{(0)2} + m_{\tilde{t}}^{(0)2}
+m_{H_2}^{(0)2} =  M_{1/2}^{(0)2} . 
\label{sumsmass(0)}
\end{eqnarray}
In the case with large $\tan \beta$, we assume that the same type of 
relations as (\ref{sumn}), (\ref{At(0)}) and (\ref{sumsmass(0)}) 
hold for $A$-parameters 
and soft SUSY breaking masses related to the bottom and tau Yukawa couplings.
In this case, the radiative corrections to $B$ and 
$m_k^2$ can be written in terms of $M_{1/2}^{(0)}$ as well as $\tan \beta$.
On top of that, the initial values of $m_{H_i}^{(0)2}$ can be 
written in terms of $M_{1/2}^{(0)}$ and $\theta$.
Thus the 
values $\mu_{\pm}(M_X)$ and $B_{\pm}(M_X)$ can be written down only by 
the use of $M_{1/2}^{(0)}$, $\tan \beta$ and $\theta$.
In the very large $\tan \beta$ scenario, $B_{\pm}(M_X)$ can 
be treated as a function of $M_{1/2}^{(0)}$ and $\tan \beta$ because 
the radiative corrections are dominant.

Here we summarize several types of 
solutions for the $\mu$-problem \cite{BIM,Munoz} which we study. 

($\mu$-1) The $\mu$-parameter $\mu_Z^{(0)}$ of $O(m_{3/2})$ 
appears after SUSY breaking in the 
case where a K\"ahler potential includes a term such as 
$Z H_1 H_2$ \cite{mu2}.
%In this case we have 
%\begin{eqnarray}
%\mu_Z^{(0)} &=& m_{3/2} \langle Z \rangle 
%~ - \langle {F}_T \rangle \langle Z^T \rangle ,
%\label{mu1}\\
%B_Z^{(0)} &=& 2 m_{3/2} + \langle {F}_T \rangle \left(\partial^T \log\mu_Z
%~ -{n_{H_1}+n_{H_2} \over \langle T+T^* \rangle}\right)
%\nonumber\\
%&~& + {m_{3/2} \over \mu_Z}\langle {F}_T \rangle \langle Z^T \rangle .
%\label{B1}
%\end{eqnarray}
Hereafter we take $Z=1/(T+T^*)$ and then $\mu_Z^{(0)}$ and $B_Z^{(0)}$ 
are given by
\begin{eqnarray}
\mu_Z^{(0)} &=& |m_{3/2}| (e^{i\alpha_{3/2}}+e^{i\alpha_T}\cos\theta) ,
\label{mu1-ex}\\
B_Z^{(0)} &=& {|m_{3/2}| \over e^{i\alpha_{3/2}} 
+ e^{i\alpha_T}\cos\theta} \{2-
\cos\theta(e^{-i(\alpha_T-\alpha_{3/2})}
(1+n_{H_1}+n_{H_2})
\nonumber\\
&~& - e^{i(\alpha_T-\alpha_{3/2})}) - \cos^2\theta(2+n_{H_1}+n_{H_2})\} 
\label{B1-ex}
\end{eqnarray}
where the phase $\alpha_{3/2}$ comes from the phase of gravitino mass.
The above formula (\ref{mu1-ex}) and (\ref{B1-ex}) are obtained
from the orbifold models with a multi-moduli in the following way.
The $Z_{2n}$ and $Z_{2n} \times Z_M$ orbifold models \cite{Orb} 
have $U$-type of moduli fields corresponding to complex structures 
of orbifolds
and a mixing term in the K\"ahler potential as \cite{Umoduli}
\begin{eqnarray}
{1 \over (T_3 +T_3^*)(U_3 +U_3^*)}(H_1H_2+ h.c.).
\label{multi-H}
\end{eqnarray}
In this case, the Higgs fields $H_1$ and $H_2$ belong to 
the untwisted sector.
The $F$-terms of $S$, $T_i$ $(i=1,2,3)$ and $U_3$ 
are parametrized by $m_{3/2}$, $\theta$, 
$\Theta_i$ and $\Theta_3'$ following Refs.\cite{BIM,multiT,BIMS}.
In the case that $\Theta_3 = 1/\sqrt{3}$ and ${\Theta}'_3 = 0$,
the $B$ and $\mu$-parameters
reduce to $B_Z^{(0)}$ and $\mu_Z^{(0)}$, respectively.

($\mu$-2) The $\mu$-parameter $\mu_\lambda^{(0)}$ of $O(m_{3/2})$ 
appears after SUSY breaking in the 
case where a superpotential $W$ includes
a term such as $\lambda \tilde{W} H_1 H_2$ \cite{mu4}.
Here $\tilde{W}$ is a superpotential in the SUSY breaking sector.
In this case we have 
\begin{eqnarray}
\mu_\lambda^{(0)} &=& \lambda m_{3/2} ,
\label{mu2}\\
B_\lambda^{(0)} &=& |m_{3/2}|\{2e^{-i\alpha_{3/2}}-e^{-i\alpha_T}
\cos\theta(n_{H_1}+n_{H_2}
\nonumber\\
&~& - \langle T+T^* \rangle
\langle \partial^T \log \lambda \rangle)\} .
\label{B2}
\end{eqnarray}

($\mu$-3) The $\mu$-parameter can be generated
through some non-perturbative effects such as gaugino condensation 
\cite{mu3} and it generally depends on the VEVs of $S$ and $T$.
In this case we have 
\begin{eqnarray}
\mu_\mu^{(0)} &=& \mu_\mu(S,T) ,
\label{mu3}\\
B_\mu^{(0)} &=& |m_{3/2}| \{-e^{-i\alpha_{3/2}} - \sqrt{3} e^{-i\alpha_S}
\sin\theta(1- \langle S+S^* \rangle 
\langle \partial_S \log \mu_\mu \rangle)
\nonumber\\
&~& - e^{-i\alpha_T}\cos\theta (3+n_{H_1}+n_{H_2}- 
\langle T+T^* \rangle
\langle \partial^T \log \mu_\mu \rangle)\} .
\label{B3}
\end{eqnarray}

There can be an admixture of several $\mu$-term
generation mechanisms and, in this case,
$\mu$ and $B$ parameters are given by
\begin{eqnarray}
\mu_{\rm Mix}^{(0)} &=& \sum_p \mu_p^{(0)} ,
\label{mu-mix}\\
B_{\rm Mix}^{(0)} &=& \sum_p \mu_p^{(0)} B_p^{(0)} / \sum_q \mu_q^{(0)}
\label{B-mix}
\end{eqnarray}
where the indices $p$ and $q$ run over all $\mu$-term
generation mechanisms.

We plot $B^{(0)}/M_{1/2}^{(0)}$ for ($\mu$-1), ($\mu$-2)
and ($\mu$-3) in Fig. 3. 
Here we restrict ourselves to the case without CP phases, e.g., 
$e^{-i\alpha_S} = \pm 1$ for $\sin \theta = \pm 1$
and $e^{-i\alpha_{3/2}}=e^{-i\alpha_{T}}=1$.
(The other simple choice is $e^{-i\alpha_{3/2}}=e^{-i\alpha_{T}}=-1$
and the plots in this case are obtained under the reflection for 
the horizontal axis.)
Further we assume that the dependence of $S$ and $T$ is very small
in $\mu_{\lambda}^{(0)}$ and $\mu_{\mu}^{(0)}$, i.e.,
$\partial \lambda/\partial S, 
\partial \lambda/\partial T,
\partial \mu_{\mu}^{(0)}/\partial S, 
\partial \mu_{\mu}^{(0)}/\partial T \ll 1$.
\begin{center}
\input Bstring.tex

Fig. 3
\end{center}
In general, $B^{(0)}/M_{1/2}^{(0)}$ contains a small number of free
parameters compared with $\mu^{(0)}/M_{1/2}^{(0)}$ and so the analysis of
$B^{(0)}/M_{1/2}^{(0)}$ can be more predictable.

\subsubsection{Small $\tan \beta$ case}

The parameters $\mu_{\pm}(M_X)$ and $B_{\pm}(M_X)$ depend 
on $m_{k}^{(0)2}$ in the small $\tan \beta$ scenario and 
$m_{k}^{(0)2}$ are functions of $n_k$, $\theta$ and $d_k$.
First we discuss the simplest case, i.e., the case with the universal   
soft scalar mass $m^{(0)}$, which satisfies 
$m^{(0)2}=M_{1/2}^{(0)2}/3$.
Such universality can be realized in the dilaton-dominant SUSY breaking 
case or the case where all relevant matter fields have the same 
modular weight $n_k=-1$ in the absence of $D$-term contribution
to scalar masses.
Fig. 4 and 5 show $B_{\pm}(M_X)/M_{1/2}^{(0)}$
and $\mu_{\pm}(M_X)/M_{1/2}^{(0)}$ against $M_{1/2}^{(0)}$ for several 
values of $\tan \beta =2 \sim 10$.
Note that $S_k = 0$ in the universal case.
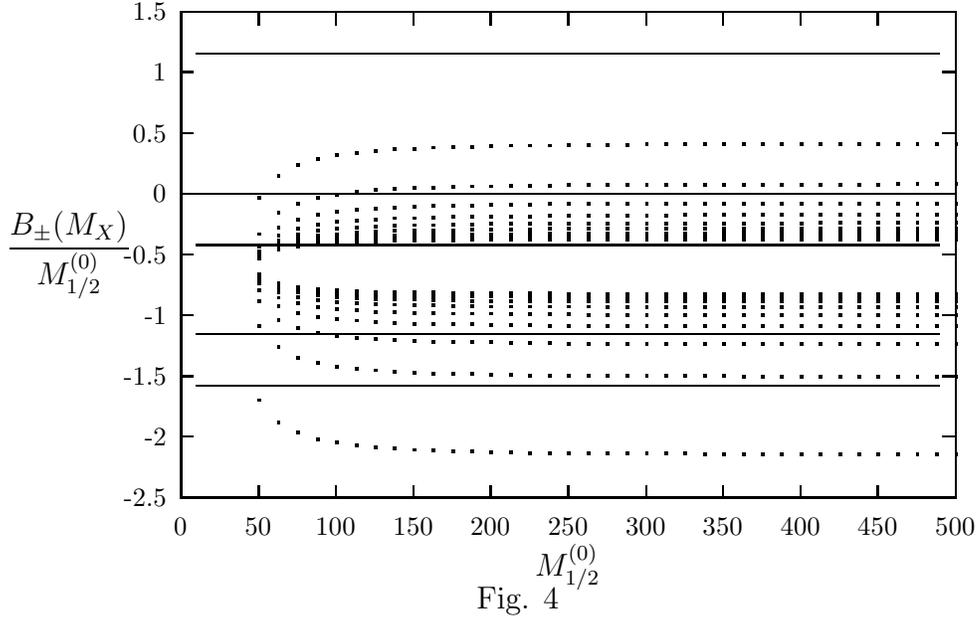
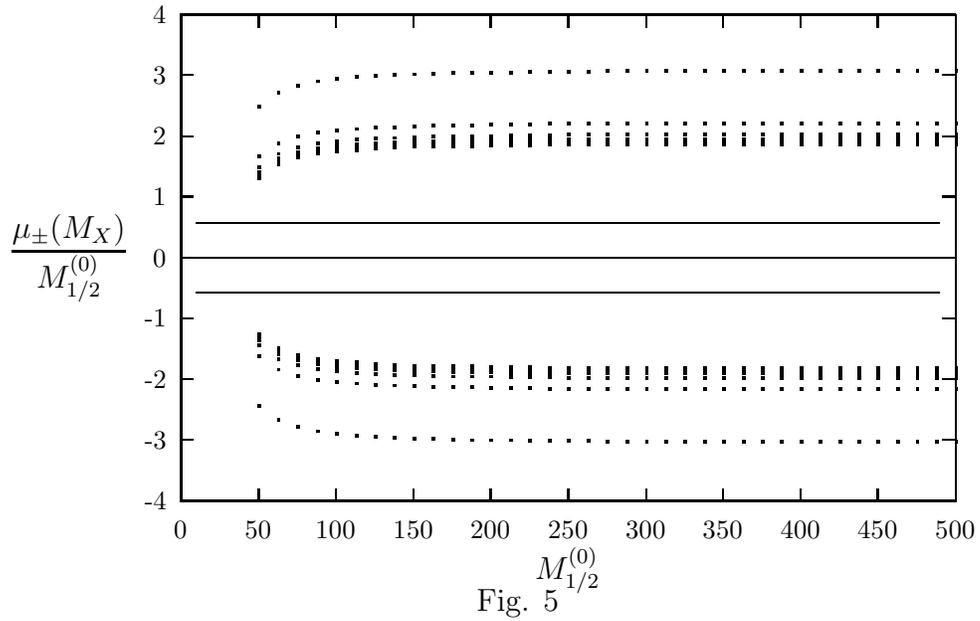
\begin{center}
\input tempfig1-2.tex

Fig. 4 
\end{center}
\begin{center}
\input tempfig2-2.tex

Fig. 5
\end{center}
In both figures, the outside (inside) curves correspond to the case with 
$\tan \beta =2~(10)$.
Only in the narrow region where the SUSY threshold is very close to $M_Z$, 
the dependence of $M_Z/M_{1/2}^{(0)}$ is not weak for
$B_{\pm}(M_X)/M_{1/2}^{(0)}$ and $\mu_{\pm}(M_X)/M_{1/2}^{(0)}$.
However, these values become stable against $M_{1/2}^{(0)}$ where 
$M_{1/2}^{(0)}$ is large enough compared with $M_Z$.

The ranges of $B_{-}(M_X)/M_{1/2}^{(0)}$ and $B_{+}(M_X)/M_{1/2}^{(0)}$
are $-0.5 \sim 0.4$ and $-2.2 \sim -0.7$ 
for $2 \leq \tan \beta \leq 10$.
The range of $|\mu_{\pm}(M_X)|/M_{1/2}^{(0)}$ is
$1.4 \sim 3.1$ for $2 \leq \tan \beta \leq 10$.
Here we use $M_{1/2}^{(0)} \geq 61$GeV as a lower bound of $M_{1/2}^{(0)}$, 
which is derived from the experimental bound on the gluino
mass $m_{\tilde{g}} \geq 154$ GeV \cite{gluino}.
The $B$ and $\mu$-parameters can be realistic if the values of
$B^{(0)}/M_{1/2}^{(0)}$ and $\mu^{(0)}/M_{1/2}^{(0)}$ hit the
ranges in Fig. 4 and Fig. 5, respectively.

Now let us compare values in Figs. 4 and 5 with values predicted
from each $\mu$-term generation mechanism.
First we consider the dilaton induced SUSY breaking case 
($\sin \theta = \pm 1$) for concreteness.

The first case ($\mu$-1) is not realistic
since there is no allowed region for $\mu$-parameter
as we see from Fig. 4 and $\mu_Z^{(0)}/M_{1/2}^{(0)} = \pm 1/\sqrt{3}$.
Next we discuss the case with the $D$-term contribution.
As discussed in Ref. \cite{stringD}, $D$-term contribution can survive
even in the limit of dilaton dominant SUSY breaking if string model
contains an anomalous $U(1)$ symmetry which is cancelled by the 
Green-Schwarz mechanism \cite{GS}.
On the other hand, $D$-term contributions related to anomaly-free
symmetries vanish at tree level in the limit of dilaton dominant 
SUSY breaking.
Further the tadpole contribution $S_k$ can also survive accompanied with
$D$-term contributions.
It is shown that there are no solutions
which satisfy $\mu_{\pm}(M_X)=\mu_Z^{(0)}$ and $B_{\pm}(M_X)=B_Z^{(0)}$
for $2 \leq \tan\beta \leq 10$ even
in the presence of $D$-term contribution for $\mu_Z^{(0)} B_Z^{(0)} = 2/3
M_{1/2}^{(0)2}$.
On the other hand, we can find solutions after the introduction of 
$D$-term contribution for $\mu_Z^{(0)} B_Z^{(0)} = -2/3 M_{1/2}^{(0)2}$.
Note that the condition $\mu_Z^{(0)} B_Z^{(0)} = -2/3 M_{1/2}^{(0)2}$ 
is derived by
the phase rotation for Higgs multiplets and this case 
should be also examined based on our convention with $\mu B < 0$ at $M_Z$.

In the second case ($\mu$-2), the $\mu$-parameter includes 
an unknown parameter $\lambda$ and so
what we can do is to estimate a favorable value of $\lambda$
using the condition $\mu_{\pm}(M_X)=\mu_{\lambda}^{(0)}$.
There is a solution which satisfies 
$B_{\pm}(M_X)/M_{1/2}^{(0)} = -2/\sqrt{3}$ at $\tan \beta \sim 5$.
For $B_{\pm}(M_X)/M_{1/2}^{(0)} = 2/\sqrt{3}$,
there is a solution at $\tan\beta = 1.3$, but it is not realistic since 
the top Yukawa coupling blows up below $M_X$.
If we assume the existence of $D$-term contribution, 
there appears a region 
which satisfies the condition $B_{\pm}(M_X)/M_{1/2}^{(0)} = 2/\sqrt{3}$
with a positive value of $S_{H_1}$.
\footnote{
The $D$-term contributions to Higgs masses can be absorbed into the tadpole 
contribution by the redefinition of $S_{H_i}$.
Here and hereafter we suppose that such a redefinition has been carried out.}
For example, the region with $\tan\beta \sim 2$ and 
$S_{H_1} \sim 2.3 M_{1/2}^{(0)2}$ is allowed.
Hence the radiative breaking scenario can be realized in the models 
with $B_{\lambda}^{(0)}/M_{1/2}^{(0)}=-2/\sqrt{3}$ 
and/or a contribution of $S_{H_1}$ under a suitable value of $\lambda$.

In the third case ($\mu$-3), $B_{\mu}^{(0)}/M_{1/2}^{(0)}$ takes the 
real values $-(\sqrt{3} \pm 1)/\sqrt{3}$ and $(\sqrt{3} \pm 1)/\sqrt{3}$.
We have a solution for $B_{\mu}^{(0)}/M_{1/2}^{(0)} = (-\sqrt{3}-1)/\sqrt{3}$
even in the absence of $D$-term and tadpole contributions.
The favorable value is $\tan\beta \sim 3$.
Similarly, there appears a solution for 
$B_{\mu}^{(0)}/M_{1/2}^{(0)} = (-\sqrt{3}+1)/\sqrt{3}$
at $\tan \beta \sim 10$.
% even without $D$-term contribution.
%The introduction of $D$-term and tadpole contributions yields 
%$\tan\beta > 2.8$.
At present, though we treat $\mu$ as a free parameter as well as
in the second case, we can select a $\mu$-term generation mechanism, when
its origin is specified through some non-perturbative effect, 
by using the allowed region given in Fig. 5. 

We consider the effect of overall moduli $F$-term condensation.
The first case ($\mu$-1) is not realistic without $D$-term contributions.
Because the inequality $|\mu_Z^{(0)}/B_Z^{(0)}| = |1+\cos \theta|/2 \leq 1$
derived from Eqs. (\ref{mu1-ex}) and (\ref{B1-ex}) is incompatible with
the values $|\mu_{\pm}(M_X)/B_{\pm}(M_X)|$ as we see from Figs. 4 and 5.
On the other hand, both of $B^{(0)}_\lambda$ and $B^{(0)}_\mu$ have 
solutions of $B^{(0)}_{\lambda(\mu)}=B_{\pm}(M_X)$
in the wider regions of $\sin \theta$ as well as $\tan \beta$
as we see from Figs. 3 and 4.

Up to now, we have discussed the case with the universal soft scalar masses
except the $D$-term contribution.
Similarly we can study cases with non-universal soft scalar masses.
Note that radiative corrections do not change from the universal case
up to the tadpole and $D$-term contributions 
as far as the sum rule (\ref{sumsmass(0)}) holds on.
%Only the initial condition of $m_{Hi}^{(0)}$ is altered.
Here we consider the case with $n_{H_1}=-2$, $n_{H_2}=-1$ and $S_{H_i}=0$.
Fig. 6 shows the values of $B_{\pm}(M_X)/M_{1/2}^{(0)}$ against 
$\cos \theta$ in the limit $M_Z/M_{1/2}^{(0)} \to 0$.
Further Fig. 7 shows predicted values of $B^{(0)}/M_{1/2}^{(0)}$ 
from string models with $n_{H_1}=-2$ and $n_{H_2}=-1$
with a choice $e^{-i\alpha_{3/2}} = e^{-i\alpha_T} = 1$.
We can get the allowed values for $\cos\theta$ and $\tan\beta$ comparing with 
the values in Figs. 6 and 7 for each of $\mu$-term generation mechanisms.
\begin{center}
\input fignp.tex

Fig. 6
\end{center}
\begin{center}
\input Bstring22.tex

Fig. 7
\end{center}
The difference between the values in the case with
$n_{H_1}=-1$ and $n_{H_1}=-2$ increases as $\cos^2\theta$ increases.
Hence the similar conclusion holds for small values of $\cos \theta$ 
on the reality of the radiative breaking scenario
as the case with $n_{H_1}=-1$.

\subsubsection{Large $\tan \beta$ case}

First let us discuss the conditions (\ref{neq1}) and (\ref{neq2}).
If there exists positive sizable $S_{H_1}$, the above conditions
are fulfilled easily.
Even in the universal case, there is allowed region for the large
$M_{1/2}^{(0)2}$.
Because the radiative correction to the difference 
$m_{H_1}^2-m_{H_2}^2$ is protortional to $M_{1/2}^{(0)2}$ and 
its coefficient is always positive for $\tan \beta \leq 60$, although 
such coefficient becomes small for a very large value of $\tan \beta$.
For example, we have $0.2 \times M_{1/2}^{(0)2}$ for $\tan \beta = 50$
and so it leads to a constraint $M_{1/2}^{(0)} \geq 180$GeV as described
before.

Second let us analyze the $B$-parameter based on Fig. 8 which shows
$\Delta B/M_{1/2}^{(0)}$ against $\tan\beta$ in the case with 
$A_f^{(0)}=-M_{1/2}^{(0)}$.
\begin{center}
\input tanbb.tex

Fig. 8 
\end{center}
As mentioned in Eq.(\ref{regionB}), the values $B^{(0)}/M_{1/2}^{(0)}$ 
should be within the region $[-1.9,-0.7]$ in the large $\tan \beta$ 
scenario.

For example, string models with the dilaton dominant SUSY breaking
predict $B_Z^{(0)}/M_{1/2}^{(0)}=B_\lambda^{(0)}/M_{1/2}^{(0)}
\sim \pm 1.15$ and $B_\mu^{(0)}/M_{1/2}^{(0)} \sim \pm 1.58, \pm 0.42$. 
%for $\sin \theta = 1, -1$, respectively.
Thus, the first and second types of 
$\mu$-term generation mechanisms cannot realize for 
$B_\lambda^{(0)}/M_{1/2}^{(0)} \sim 1.15$ and
the third one can realize
for $B_\mu^{(0)}/M_{1/2}^{(0)} \sim -1.58$.
Though there is a solution for
$B_Z^{(0)}/M_{1/2}^{(0)} \sim -1.15$ around $\tan \beta \sim 30$, 
it is not realistic because we have no solution which satisfies
the condition ${|\mu_{\pm}(M_X)| / M_{1/2}^{(0)}} \sim 0.58$
as we see from Fig. 9. 
Fig. 9 shows ${|\mu_{\pm}(M_X)| / M_{1/2}^{(0)}}$ against $M_{1/2}^{(0)}$
for several values of $\tan \beta = 20 \sim 60$
where we take the universal case, $m_k^{(0)2}=M_{1/2}^{(0)2}/3$. 
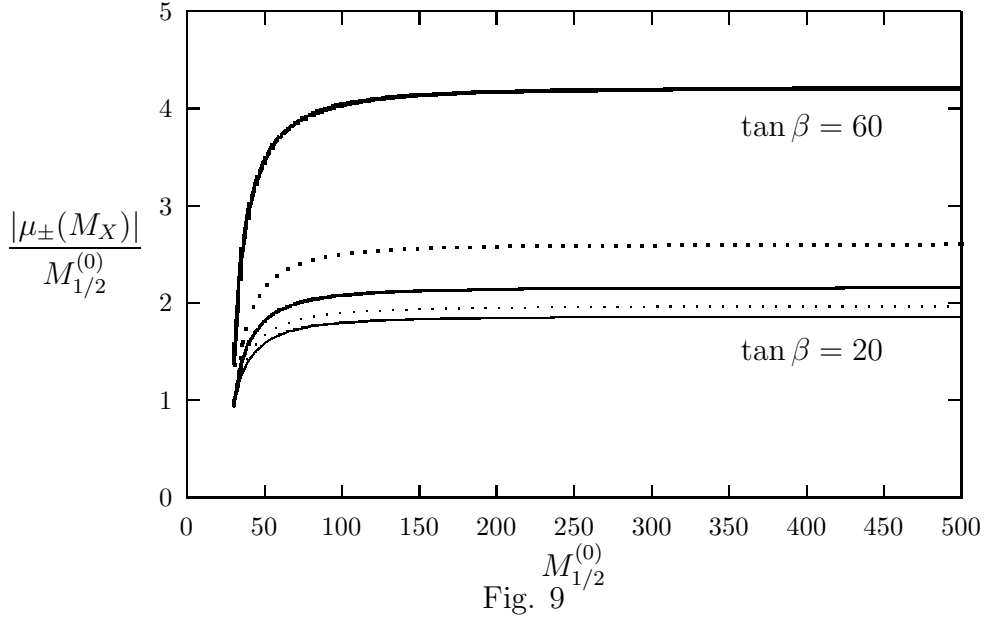
\begin{center}
\input muMZ.tex

Fig. 9 
\end{center}
For the second or third ones, we can estimate a favorable value for an
unknown parameter in $\mu^{(0)}$ using Fig. 9.
The above result holds on except for the case where $B(M_Z)$ is not 
neglected accidentally.

We study the region of $B$-parameters after incorporating with the effect
of overall moduli $F$-term condensation.
The first type of $\mu$-term generation mechanism does not realize
without $D$-term contributions since 
$|\mu_Z^{(0)}/B_Z^{(0)}| < |\mu_{\pm}(M_X)/B_{\pm}(M_X)|$.
For the second one, the region such that $|\sin \theta| > 0.9$ is allowed
for $B_{\lambda}^{(0)}$.
For the third one,
Fig. 10 shows ($\tan \beta$, $\sin \theta$)
which satisfies the condition 
\begin{eqnarray}
&~&{\Delta B \over M_{1/2}^{(0)}} = {-1-\sqrt{3}\sin \theta - \cos \theta
\over \sqrt{3}\sin \theta} .
\label{BZsol}
\end{eqnarray}
\begin{center}
\input tanbsin2.tex

Fig. 10
\end{center}
In addition to the solution shown in Fig. 10, we always have a 
solution, $\sin \theta  \rightarrow 0$ for Eq.(\ref{BZsol}).
In this limit, the parameters $A^{(0)}$ and $M_{1/2}^{(0)}$ 
become very small.
Furthermore, the radiative correction $\Delta B$ also becomes small.
To obtain precise results in this limit, we need to consider 
one-loop threshold corrections and $B(M_Z)$.

\subsection{Gauge-Yukawa unification model as theory beyond MSSM}

Recently the coupling reduction technique has been applied to
the reduction of soft SUSY breaking parameters \cite{KMZ,KKK,KKMZ,KKK2}.
Then it is found that the relations (\ref{At(0)}) and (\ref{sumsmass(0)}) 
are RG-invariant.\footnote{
In Ref. \cite{JJ}, it was found the relations 
(\ref{At(0)}) and (\ref{sumsmass(0)})  with the universal soft mass 
$m^{(0)2}_k=M^{(0)2}_{1/2}/3$ are two-loop RG invariant in finite models.}
Thus, results in the previous subsection are applicable to 
GYU-GUTs.

In addition, the 
following relation \cite{KKK},
\begin{eqnarray}
&~&B^{(0)} M^{(0)}_{1/2} = - m_{H_1} ^{(0)2}- m_{H_2}^{(0)2} 
\label{B(0)}
\end{eqnarray}
is also RG-invariant and so the following condition should be satisfied,
\begin{eqnarray}
&~&{B_{\pm}(M_X) \over M_{1/2}^{(0)}} = - h_{H_1} - h_{H_2} ,
\label{GYU-B}
\end{eqnarray}
where $h_{H_i}=m_{H_i}^{(0)2}/M_{1/2}^{(0)2}$.

In the universal case,
% where $m_{0} ^{(0)2}=M^{(0)2}_{1/2}/3$, 
we have $B^{(0)}/M^{(0)}_{1/2}=-2/3$ \cite{JJ}.
Figs. 4 and 6 show that it is difficult to realize the electroweak breaking 
scenario in the universal case with the small $\tan \beta$.
Similarly it is not realized in the very large 
$\tan \beta$ scenario
% for values of $20 \leq \tan \beta \leq 60$
as we see from Fig. 2 and 8.
It is known that there is a solution 
which satisfies $B_{\pm}(M_{X})/M^{(0)}_{1/2}=-2/3$
for $\tan \beta \sim 18$ \cite{JJR}.

Next we discuss a general case.
We can estimate the allowed values for $h_{H_1}$ and $h_{H_2}$ using
Eq. (\ref{GYU-B}) and Figs. 4 and 8.
For example, $h_{H_1} + h_{H_2}$ should be in $[0.7, 1.9]$ for the large 
$\tan\beta$, e.g. $h_{H_1}=h_{H_2}=0.8$ for the degenerate case 
with $\tan \beta =50$.
For the small $\tan\beta$, we study the degenerate case, i.e.,
$h \equiv h_{H_1} = h_{H_2}$ which includes the universal case as a special
one with $h = 1/3$, and a more general case with non-degenerate soft Higgs 
masses.
Fig. 11 shows the allowed regions for ($\tan \beta$, $h$)
in the limit $M_Z/M_{1/2}^{(0)} \to 0$ for the degenerate case.
\begin{center}
\input Sub1.tex

Fig. 11
\end{center}
For a more general case,
solutions satisfying the relation (\ref{GYU-B}) 
are given in Fig. 12 in the limit $M_Z/M_{1/2}^{(0)} \to 0$.
\begin{center}
\input Sub2.tex

Fig. 12
\end{center}
We can calculate the explicit values of $h_{H_1}$ and $h_{H_2}$
in the non-finite case if we specify a model of GYU-GUT.
It is not obvious whether the values shown in Figs. 11 and 12
can be realized in the explicit model, although in 
finite models these remain free parameters \cite{KKMZ}.
Actually there is a restriction for the values of 
$h_{H_1}$ and $h_{H_2}$ in a certain situation.
In such a case, we can show that it is, in general, difficult 
to realize radiative electroweak breaking scenario with large $\tan \beta$ 
in asymptotically free GYU-GUTs under some assumptions 
as we will discuss in the appendix.

\subsection{Other relations between $M_{1/2}^{(0)}$ and $A_f^{(0)}$}

Up to now we have discussed the cases with $A_f^{(0)}=-M_{1/2}^{(0)}$.
In this subsection, comments are given for the case with
$A_f^{(0)} \neq -M^{(0)}_{1/2}$.
There exist several factors to deviate from the relation
$A_f^{(0)}=-M_{1/2}^{(0)}$, e.g., the existence of the modular weights
of matter fields with $\sum n_k \neq -3$, the moduli-dependence of the
Yukawa couplings in the string model.
The Yukawa couplings are moduli-independent if the top Yukawa coupling is 
realized as
a renormalizable coupling in the untwisted sector.
It is known that the quark doublet $Q_3$ and up-type
singlet quark $t$ in the third family and the Higgs field $H_2$
belong to the untwisted sector in semi-realistic models \cite{CM-FIQS}.
(See also Ref. \cite{dienes}.)  
In other case, the Yukawa couplings include moduli as a volume factor and
an exponential factor which are model-dependent.

First we study the violation of the relations (\ref{sumn}) and (\ref{At(0)})
as a trial to make the first or second $\mu$-term generation mechanism 
in string model realistic from the viewpoint of $B$-parameter
in the large $\tan \beta$ scenario.
%That is, we treat $A^{(0)}$ and $M_{1/2}^{(0)}$ as independent parameters,
For example, the $A$-parameter is required as 
$A^{(0)}=2.7 \times M_{1/2}^{(0)}$ for $\tan \beta =50$ in order to
obtain $\Delta B/M_{1/2}^{(0)}$=1.15 as we infer from Fig. 2.
% which correspond the prediction from the first and second $\mu$-term 
%generation mechanisms in the dilaton-dominat SUSY breaking case.
It is impossible to get such large $A$-parameter 
in the dilaton-dominant SUSY breaking unless the Yukawa couplings have no
sizable moduli dependence.
Further such a large absolute values of $A$-parameters 
seems to be unfavorable due to the occurence of dangerous charge and/or 
color breaking (CCB) \cite{CCB}.
There exists a constraint on the $(M_{1/2}^{(0)},A^{(0)})$-plane, 
i.e., the $(\theta,\Sigma n_k,\tan \beta)$ plane,
to satisfy the condition $B_{\pm}(M_X) = B^{(0)}$.
%Fig. 8 shows such relations between $\theta$ and $\Sigma n_k$ for 
%$\tan \beta =50$ for the first $\mu$-term generation mechanism.
%\begin{center}
%\input ABZ.tex
%
%Fig. 8
%\end{center}
Modular weights of quark and lepton fields are constrained 
in explicit string models \cite{IL,KKO}.
Thus it is not obvious whether values of $\Sigma n_k$ 
on the constraint plane
%shown in Fig. 8
can be really obtained in explicit string models or not.
%For the second type of the $\mu$-term generation mechanism, 
%similar discussions can be done.

Second we discuss the extreme case with $A_f^{(0)}=0$.
Such an initial condition is realized in the no-scale model \cite{noscale} 
and at the same time the no-scale model predicts that soft scalar 
masses also vanish, $m_k^{(0)}=0$.
The values of $B_{\pm}(M_X)/M_{1/2}^{(0)}$ in 
this case, $A_f^{(0)}=m_k^{(0)}=0$, are plotted in Fig. 13 
in the small $\tan \beta$ scenario.
\begin{center}
\input newfig3.tex 

Fig. 13
\end{center}
Here we take a large gaugino mass limit compared with $M_Z$.
We see that the relation $B^{(0)}=0$ can be realized around 
$\tan \beta \sim 9$.
On the other hand, the condition $B^{(0)}=0$ cannot be realized 
in the large $\tan \beta$ scenario
because 
$\Delta B/M^{(0)}_{1/2} = -0.9 \sim -0.3$ for $\tan \beta = 20 \sim 60$
and $A_f^{(0)}=0$.

In the case with $0<|A^{(0)}|< \sqrt{3} M^{(0)}_{1/2}$,
the value of $\Delta B/M^{(0)}_{1/2}$ allows a wider range 
[$-2$, 0.5] than the universal case, in the large $\tan \beta$ scenario
as shown in Fig. 2.
%The negative values of $B^{(0)}/M^{(0)}_{1/2}$ are favorable.

% can be discussed in a smilar way. For such cases 
%we can estimate roughtly $B_{\pm}(M_X)$ 
%in both of the small and large $\tan \beta$ scenarios from 
%refering the above-mentioned
%discussions for $A^{(0)}=-M^{(0)}_{1/2}$ and $A^{(0)}=0$, i.e., Figs. 2,
%5 and 11.
%For example, we have $\Delta B/M^{(0)}_{1/2}= -0.9 \sim -1.6$ for 
%$A^{(0)}=0 \sim -M^{(0)}_{1/2}$ 
%and $\Delta B/M^{(0)}_{1/2}= -0.9 \sim -0.1$ 
%for $A^{(0)}=0 \sim M^{(0)}_{1/2}$ with $\tan \beta =50$.
%Thus Figs. 1, 4 and 10 would be useful for generic case.
%The case with large absolute values of $A$-parameters such that
%$|A^{(0)}| > \sqrt{3} |M^{(0)}_{1/2}|$ seems to be unfavorable 
%because it could lead to dangerous charge and/or 
%color breaking \cite{CCB}.

\section{Conclusions and Discussions}

Using the bottom-up approach, we have studied the $\mu$ and 
$B$ parameters in the framework of the MSSM under the assumption
that the underlying theory is a certain type of superstring model
or GYU model.
These models predict specific relations 
among soft SUSY breaking parameters, i.e., string-inspired relations and
RG invariant relations, which make our predictability strong.
% or the relations derived from the no-scale model.
Two types of formulae for $\mu$ and $B$-parameters at 
$M_X$ can be written in terms of a few number of independent parameters 
($m_S$, $\langle F \rangle$, $\tan \beta$).
%such as soft scalar masses, the gaugino mass and $\tan \beta$.
One set is ($\mu_{\pm}(M_X)$, $B_{\pm}(M_X)$), which is 
derived from the conditions of the realization of
radiative electroweak symmetry breaking.
(See Eqs. (\ref{muMX}) and (\ref{BMX}).) 
The other set is ($\mu^{(0)}$, $B^{(0)}$), which is obtained from 
the underlying theory.
We have examined which type of $\mu$ and $B$-term
generation mechanism is hopeful from the requirement
that the values from two types of formulae should agree with.

In the MSSM inspired by a certain type of string model,
the $\mu$-term generated by the non-renormalizable term such as $\lambda 
\tilde{W} H_1 H_2$, some non-perturbative effects or both cases,
 i.e., ($\mu$-2), ($\mu$-3) or both cases, is hopeful even in the dilaton 
dominant SUSY breaking
without $D$-term contribution.
We have discussed effects of the moduli $F$-term condensation 
and $D$-term contribution to soft scalar masses.
The first mechanism is impossible to realize the radiative
scenario without $D$-term contributions even after the introduction of
over-all moduli $F$-term contribution.
For the second and third ones, we have solutions of 
$B_{\lambda(\mu)}^{(0)} = B_{\pm}(M_X)$ in the wider regions of 
$\sin \theta$ and $\tan \beta$ than the dilaton dominant SUSY breaking case.
For $\mu$-parameter, we can select a $\mu$-term generation mechanism, when
its origin is specified through some non-perturbative effect, 
by using the allowed region for $\mu_{\pm}^{(0)}/M_{1/2}^{(0)}$
plotted in Figs. 5 and 9. 

In the MSSM based on GYU-GUTs with the reduced $B$-parameter 
(See Eq.({\ref{B(0)})), the small or intermediate
$\tan \beta$ scenario is favorable to realize the radiative
electroweak symmetry breaking.
There is a solution for Eqs. (\ref{mini1}) and (\ref{mini2}) 
at $\tan \beta \sim 18$ in the universal case, i.e.,
$m_i^{(0)2}=M_{1/2}^{(0)2}/3$ and $A_f^{(0)}=-M_{1/2}^{(0)}$.
We have studied the case with the non-universal masses
only for a small $\tan \beta$ region because
it is difficult to realize in the large $\tan \beta$ scenario
in the framework of asymptotically free GYU-GUTs under some assumptions.

We have also discussed $B$-parameter under the initial condition, 
i.e., $A_f^{(0)}=m_k^{(0)}=0$ predicted by the no-scale model. 
The relation $B^{(0)}=0$ can be realized around 
$\tan \beta \sim 9$.

Our bottom-up approach to select a realistic $\mu$ and $B$-term generation 
mechanism is so generic and powerful that we can apply it to
the models where the formulae of soft SUSY breaking terms are derived,
the case with an improvement of approximation and more complex situations.
For example, the improvement by the incorporation
of 1-loop effective potential \cite{1loop},
the case with large moduli-dominant threshold corrections for gaugino masses,
other assignments of modular weight for matter fields,
the case where the coupling constants in $\mu$ and $B$ parameters
depend on $S$ and/or $T_i$ fields,
the modular dominant SUSY breaking case \cite{mmd} and 
$\mu$ and $B$-term generation mechanism by one-loop effects 
in SUGRA \cite{CLM}.

\section*{Acknowledgments}  
This work was partially supported by the European Commission TMR programs
ERBFMRX-CT96-0045 and CT96-0090.
We would like to thank J. Kubo, H.P. Nilles and M. Olechowski 
for useful discussions and suggestions.

\section*{Appendix} 

In this appendix, we show that it is impossible to realize radiative 
electroweak breaking scenario
with large $\tan \beta$ in the framework of asymptotically free
GYU-GUTs under some assumptions.

First the formula of soft SUSY breaking scalar masses is given by
\begin{eqnarray}
m_k^{(0)2} &=& {2 C_2(R_k) - \sum_f y_f N_k^f \over -b} M_{1/2}^{(0)2} 
\equiv h_k M_{1/2}^{(0)2}
\label{mk-formula}
\end{eqnarray}
up to $D$-term contribution due to extra gauge symmetry breaking.
Here $C_2(R_k)$ is the quadratic Casimir invariants of representation $R_k$,
$y_f$ the coefficient in the GYU conditions, 
i.e., $\alpha_f^{(0)} = y_f \alpha^{(0)}$,
$N_k^f$ the number of independent diagrams contributed to the wave function
renormalization due to the Yukawa coupling $\alpha_f^{(0)}$ and
$b$ the coefficient of $\beta$ function of the gauge coupling.
Note that $y_f$ should be positive-definite by definition.
We can show the model-independent sum rule, 
$m_{\Sigma(f)}^{(0)2} = M_{1/2}^{(0)2}$,
by using the following relation derived from the GYU conditions
\begin{eqnarray}
\sum_k \sum_f y_f N_k^f = 2 \sum_k C_2(R_k) + b  .
\end{eqnarray}

Next let us list our basic assumptions in GYU-GUT based on $SU(5)$ or
$SO(10)$ gauge group.
\begin{enumerate}
\item The theory is asymptotically free.
The magnitude of Yukawa coupling related to the bottom quark 
is comparable to that related to the top quark.

\item The squarks in the third generation, $\tilde{q}_3$, $\tilde{t}$
and $\tilde{b}$, and Higgs fields $H_i$ belong to 
the following representations in $SU(5)$ and $SO(10)$-GUT, respectively,
% and they do not appear as mixing states of several fields,
\begin{eqnarray}
&~&\tilde{q}_3, \tilde{t}  \in {\bf 10} ,~~~ \tilde{b}  \in {\bf 5}^* ,~~~
H_1 \in {\bf 5}^*,~~~ H_2 \in {\bf 5} ~~~\mbox{under} ~~SU(5) ,\\
%\end{eqnarray}
%and
%\begin{eqnarray}
&~&\tilde{q}_3, \tilde{t}, \tilde{b}  \in {\bf 16} ,~~~H_1 , H_2 
\in {\bf 10} ~~~\mbox{under} ~~SO(10) . 
\end{eqnarray}

\item The multiplets including $\tilde{q}_3$, $\tilde{t}$ and $\tilde{b}$
have no sizable Yukawa couplings other than those among Higgs multiplets
including $H_i$.
However, through sizable Yukawa interactions, the Higgs multiplets
including $H_i$ can couple to other multiplets, which trigger 
the breakdown of GUT symmetry, in order that extra Higgs fields 
acquire heavy masses.
%(There appears no extra light Higgs fields except for two $SU(2)_L$ 
%doublet in the low-energy spectrum.)
\end{enumerate}

The sum rule, $m_{\Sigma(f)}^{(0)2} = M_{1/2}^{(0)2}$,
is equivalent to the following relations,
\begin{eqnarray}
&~& h_{H_1} + h_{\tilde{q}_3} + h_{\tilde{b}} = 1 ,
\label{sumh-1}\\
&~& h_{H_2} + h_{\tilde{q}_3} + h_{\tilde{t}} = 1 .
\label{sumh-2}
\end{eqnarray}

Let us consider GYU-$SU(5)$-GUT.
The coefficients $h_k$ are given by
\begin{eqnarray}
&~& h_{\tilde{q_3}} = h_{\tilde{t}} 
= {1 \over -b}({36 \over 5} - 3 y_t - 2 y_b),
\label{su5-1}\\
&~& h_{\tilde{b}} 
= {1 \over -b}({24 \over 5}  - 4 y_b),
\label{su5-2}\\
&~& h_{H_1} 
= {1 \over -b}({24 \over 5}  - 4 y_b - {\sum_f}' y_f N_{H_1}^f),
\label{su5-3}\\
&~& h_{H_2} 
= {1 \over -b}({24 \over 5}  - 3 y_t - {\sum_f}' y_f N_{H_2}^f) 
\label{su5-4}
\end{eqnarray}
where ${\sum}'$ means the omission of the contribution of the top and
bottom Yukawa couplings.
%Using the above Eqs. (\ref{su5-1}) - (\ref{su5-2}) and
%the first assumption which means $b < 0$, we can derive the following 
%inequalities
%\begin{eqnarray}
%&~&  h_{H_1} < h_{\tilde{b}} ,
%\label{h1-ineq}\\
%&~&  h_{H_1} + 2 h_{H_2} < h_{\tilde{q_3}} + h_{\tilde{t}} .
%\label{h2-ineq}
%\end{eqnarray}
The following inequality is derived
\begin{eqnarray}
&~&  h_{H_1} +  h_{H_2} < {7 \over 13} 
\label{h12-ineq}
\end{eqnarray}
from the relations (\ref{sumh-1}) -- (\ref{su5-4}) and $b < 0$.
Hence we get the inequality $B^{(0)}/M_{1/2}^{(0)} > -7/13$.
% related to $B$-parameter
%\begin{eqnarray}
%{B^{(0)} \over M_{1/2}^{(0)}} > -{7 \over 13} .
%\label{B-ineq-su5} 
%\end{eqnarray}
It is incompatible to the result
$-1.9 \leq B(M_X)/M_{1/2}^{(0)} \leq -0.7$ 
from the radiative electroweak breaking scenario with large $\tan \beta$.

In the same way, it is shown that $B^{(0)}/M_{1/2}^{(0)} > -2/3$
in GYU-$SO(10)$-GUT with the reduced $B$-parameter.
%The inequality $h_{H_1} +  h_{H_2} < {2 \over 3}$ 
%which means the incompatibility to the radiative breaking scenario
%with large $\tan \beta$, 
The same argument applies to a more generic asymptotically free GUT 
with the following features.

(1) Every field (the third generation squarks $Q_{3j}$ and Higgs fields
$H_i$) does not realize as a mixing state but belongs to a single
representation.
Let us denote them $R_{Q_{3j}}$ and $R_{H_i}$, respectively,
bearing the case where some fields belong to the same representation
in mind.

(2) The dimensions of $R_{Q_{3j}}$ equal to or are bigger 
than those of $R_{H_i}$.

(3) Every multiplet including $Q_{3j}$ has
only one sizable Yukawa coupling, which couples to $H_1$ and/or $H_2$.

The second feature leads to
\begin{eqnarray}
C_2(R_{H_i}) \leq C_2(R_{Q_{3j}})  .
\label{C2-ineq}
\end{eqnarray}
The third feature leads to the following inequality
\begin{eqnarray}
\sum_f y_f N_{H_i}^f > \sum_f y_f N_{Q_{3j}}^f  .
\label{yN-ineq}
\end{eqnarray}
Using (\ref{C2-ineq}) and (\ref{yN-ineq}), we can get the inequality
$h_{H_i} < h_{Q_{3j}}$.
Further considering the sum rules (\ref{sumh-1}) and (\ref{sumh-2}),
we find that both of $h_{H_i}$'s are less than 1/3, which means
$B^{(0)}/M_{1/2}^{(0)} > -2/3$.
The presence of $D$-term contribution does not change the above
conclusion as far as the quantum numbers of $H_1$ for broken diagonal
generators take the opposite values of those of $H_2$.

\newpage

\section*{Figure Captions}

\renewcommand{\labelenumi}{Figure~\arabic{enumi}}

\begin{enumerate}

\item Values of $|B(M_Z)|/M_{1/2}^{(0)}$ against $M_{1/2}^{(0)}$ 
in the universal case with $m^{(0)2}_i=M^{(0)2}_{1/2}/3$ and 
$A_f^{(0)}=-M^{(0)}_{1/2}$.

\item Values of $\Delta B/M_{1/2}^{(0)}$ against $A_f^{(0)}/M_{1/2}^{(0)}$
in the case with non-universal parameters.
The solid and dotted curves correspond to the cases with 
$\tan \beta = 20$ and $60$, respectively.

\item Values of $B^{(0)}/M_{1/2}^{(0)}$ against $\sin \theta$ 
for $\mu$-term generation mechanisms ($\mu$-1), ($\mu$-2) and ($\mu$-3)
with a choice $e^{-i\alpha_{3/2}} = e^{-i\alpha_T} = 1$, i.e.,
$B_Z^{(0)}/M_{1/2}^{(0)} = 2 /(\sqrt{3} \sin \theta)$, 
$B_\lambda^{(0)}/M_{1/2}^{(0)} = (2 + 2 \cos \theta)/(\sqrt{3} \sin \theta)$ 
and $B_\mu^{(0)}/M_{1/2}^{(0)} 
= -(1+\sqrt{3}\sin\theta+\cos\theta)/(\sqrt{3} \sin \theta)$.

\item Values of $B_{\pm}(M_X)/M_{1/2}^{(0)}$ against $M_{1/2}^{(0)}$ 
in the universal case with $m^{(0)2}_i=M^{(0)2}_{1/2}/3$ and 
$A_t^{(0)}=-M^{(0)}_{1/2}$.
The solid lines represent the values, 
$B^{(0)}/M_{1/2}^{(0)} = \pm 2/\sqrt{3}$ and 
$(-1\mp\sqrt{3})/\sqrt{3}$, predicted from ($\mu$-1) -- ($\mu$-3) 
in the dilaton dominant SUSY breaking case.
(The lines represent $B^{(0)}/M_{1/2}^{(0)} = (1\pm\sqrt{3})/\sqrt{3}$
are omitted.)

\item Values of $\mu_{\pm}(M_X)/M_{1/2}^{(0)}$ against $M_{1/2}^{(0)}$ 
in the universal case with $m^{(0)2}_i=M^{(0)2}_{1/2}/3$ and 
$A_t^{(0)}=-M^{(0)}_{1/2}$.
The solid line represents the value,
$\mu^{(0)}/M_{1/2}^{(0)}$ $ = \pm 1/\sqrt{3}$,
predicted from ($\mu$-1).

%\item Values of $B_{\pm}(M_X)/M_{1/2}^{(0)}$ against $\tan \beta$ 
%in the limit $M_Z/M_{1/2}^{(0)} \to 0$
%in the universal case in the dilaton dominant SUSY breaking case 
%with $\sin \theta=1$.

\item Values of $B_{\pm}(M_X)/M_{1/2}^{(0)}$ against 
$\cos \theta$ in the case with $n_{H_1}=-2$ and $n_{H_2}=-1$
in the limit $M_Z/M_{1/2}^{(0)} \to 0$.
The dotted curves correspond to the case for $\tan \beta =2$, 
while curves with closed (open) circles correspond to the case 
for $\tan \beta = 5$ (10).

\item Values of $B^{(0)}/M_{1/2}^{(0)}$ against $\cos \theta$ 
in the case with $n_{H_1}=-2$ and $n_{H_2}=-1$
for $\mu$-term generation mechanisms ($\mu$-1), ($\mu$-2) and ($\mu$-3)
with a choice $e^{-i\alpha_{3/2}} = e^{-i\alpha_T} = 1$.

\item Values of radiative corrections, $\Delta B/M_{1/2}^{(0)}$, 
against $M_{1/2}^{(0)}$ in the case with $A_f^{(0)}=-M^{(0)}_{1/2}$.

\item Values of $|\mu_{\pm}(M_X)|/M_{1/2}^{(0)}$ against $M_{1/2}^{(0)}$
for several values of $\tan \beta = 20, 30, 40, 50$ and 60
where we take the universal case. 

\item Values ($\tan \beta$, $\sin \theta$) which satisfy
the condition $\Delta B = B_{\mu}^{(0)}$.

\item Values ($\tan \beta$, $h$) which satisfy
the condition $B_{\pm}(M_X)/M_{1/2}^{(0)} = -2h$
in the limit $M_Z/M_{1/2}^{(0)} \to 0$.

\item Values ($h_{H_{1}}$, $h_{H_{2}}$) which satisfy
the condition $B_{\pm}(M_X)/M_{1/2}^{(0)} = -h_{H_{1}}-h_{H_{2}}$
in the limit $M_Z/M_{1/2}^{(0)} \to 0$.
The dotted curves correspond to the case with $\tan \beta =2$, 
while curves with closed (open) circles correspond to the case 
with $\tan \beta = 5$ (10).

\item Values of $B_{\pm}(M_X)/M_{1/2}^{(0)}$ in 
the case with $A_f^{(0)}=m_k^{(0)}=0$
and in the limit $M_Z/M_{1/2}^{(0)} \to 0$.

\end{enumerate}

\end{document}

%% file: BMZ.tex
% GNUPLOT: LaTeX picture
\setlength{\unitlength}{0.240900pt}
\ifx\plotpoint\undefined\newsavebox{\plotpoint}\fi
\sbox{\plotpoint}{\rule[-0.200pt]{0.400pt}{0.400pt}}%
\begin{picture}(1500,900)(0,0)
\font\gnuplot=cmr10 at 10pt
\gnuplot
\sbox{\plotpoint}{\rule[-0.200pt]{0.400pt}{0.400pt}}%
\put(220.0,113.0){\rule[-0.200pt]{292.934pt}{0.400pt}}
\put(220.0,113.0){\rule[-0.200pt]{4.818pt}{0.400pt}}
\put(198,113){\makebox(0,0)[r]{0}}
\put(1416.0,113.0){\rule[-0.200pt]{4.818pt}{0.400pt}}
\put(220.0,189.0){\rule[-0.200pt]{4.818pt}{0.400pt}}
\put(198,189){\makebox(0,0)[r]{0.01}}
\put(1416.0,189.0){\rule[-0.200pt]{4.818pt}{0.400pt}}
\put(220.0,266.0){\rule[-0.200pt]{4.818pt}{0.400pt}}
\put(198,266){\makebox(0,0)[r]{0.02}}
\put(1416.0,266.0){\rule[-0.200pt]{4.818pt}{0.400pt}}
\put(220.0,342.0){\rule[-0.200pt]{4.818pt}{0.400pt}}
\put(198,342){\makebox(0,0)[r]{0.03}}
\put(1416.0,342.0){\rule[-0.200pt]{4.818pt}{0.400pt}}
\put(220.0,419.0){\rule[-0.200pt]{4.818pt}{0.400pt}}
\put(198,419){\makebox(0,0)[r]{0.04}}
\put(1416.0,419.0){\rule[-0.200pt]{4.818pt}{0.400pt}}
\put(220.0,495.0){\rule[-0.200pt]{4.818pt}{0.400pt}}
\put(198,495){\makebox(0,0)[r]{0.05}}
\put(1416.0,495.0){\rule[-0.200pt]{4.818pt}{0.400pt}}
\put(220.0,571.0){\rule[-0.200pt]{4.818pt}{0.400pt}}
\put(198,571){\makebox(0,0)[r]{0.06}}
\put(1416.0,571.0){\rule[-0.200pt]{4.818pt}{0.400pt}}
\put(220.0,648.0){\rule[-0.200pt]{4.818pt}{0.400pt}}
\put(198,648){\makebox(0,0)[r]{0.07}}
\put(1416.0,648.0){\rule[-0.200pt]{4.818pt}{0.400pt}}
\put(220.0,724.0){\rule[-0.200pt]{4.818pt}{0.400pt}}
\put(198,724){\makebox(0,0)[r]{0.08}}
\put(1416.0,724.0){\rule[-0.200pt]{4.818pt}{0.400pt}}
\put(220.0,801.0){\rule[-0.200pt]{4.818pt}{0.400pt}}
\put(198,801){\makebox(0,0)[r]{0.09}}
\put(1416.0,801.0){\rule[-0.200pt]{4.818pt}{0.400pt}}
\put(220.0,877.0){\rule[-0.200pt]{4.818pt}{0.400pt}}
\put(198,877){\makebox(0,0)[r]{0.1}}
\put(1416.0,877.0){\rule[-0.200pt]{4.818pt}{0.400pt}}
\put(220.0,113.0){\rule[-0.200pt]{0.400pt}{4.818pt}}
\put(220,68){\makebox(0,0){100}}
\put(220.0,857.0){\rule[-0.200pt]{0.400pt}{4.818pt}}
\put(372.0,113.0){\rule[-0.200pt]{0.400pt}{4.818pt}}
\put(372,68){\makebox(0,0){150}}
\put(372.0,857.0){\rule[-0.200pt]{0.400pt}{4.818pt}}
\put(524.0,113.0){\rule[-0.200pt]{0.400pt}{4.818pt}}
\put(524,68){\makebox(0,0){200}}
\put(524.0,857.0){\rule[-0.200pt]{0.400pt}{4.818pt}}
\put(676.0,113.0){\rule[-0.200pt]{0.400pt}{4.818pt}}
\put(676,68){\makebox(0,0){250}}
\put(676.0,857.0){\rule[-0.200pt]{0.400pt}{4.818pt}}
\put(828.0,113.0){\rule[-0.200pt]{0.400pt}{4.818pt}}
\put(828,68){\makebox(0,0){300}}
\put(828.0,857.0){\rule[-0.200pt]{0.400pt}{4.818pt}}
\put(980.0,113.0){\rule[-0.200pt]{0.400pt}{4.818pt}}
\put(980,68){\makebox(0,0){350}}
\put(980.0,857.0){\rule[-0.200pt]{0.400pt}{4.818pt}}
\put(1132.0,113.0){\rule[-0.200pt]{0.400pt}{4.818pt}}
\put(1132,68){\makebox(0,0){400}}
\put(1132.0,857.0){\rule[-0.200pt]{0.400pt}{4.818pt}}
\put(1284.0,113.0){\rule[-0.200pt]{0.400pt}{4.818pt}}
\put(1284,68){\makebox(0,0){450}}
\put(1284.0,857.0){\rule[-0.200pt]{0.400pt}{4.818pt}}
\put(1436.0,113.0){\rule[-0.200pt]{0.400pt}{4.818pt}}
\put(1436,68){\makebox(0,0){500}}
\put(1436.0,857.0){\rule[-0.200pt]{0.400pt}{4.818pt}}
\put(220.0,113.0){\rule[-0.200pt]{292.934pt}{0.400pt}}
\put(1436.0,113.0){\rule[-0.200pt]{0.400pt}{184.048pt}}
\put(220.0,877.0){\rule[-0.200pt]{292.934pt}{0.400pt}}
\put(45,495){\makebox(0,0)
{$\displaystyle{\frac{|B(M_Z)|}{M_{1/2}^{(0)}}}~~~$}}
\put(828,23){\makebox(0,0){\shortstack{\\ \\ \\ $M_{1/2}^{(0)}$}}}
\put(1284,648){\makebox(0,0)[r]{$\tan \beta =20$}}
\put(1284,342){\makebox(0,0)[r]{$\tan \beta =30$}}
\put(1284,228){\makebox(0,0)[r]{$\tan \beta =40$}}
\put(1284,151){\makebox(0,0)[r]{$\tan \beta =50$}}
\put(220.0,113.0){\rule[-0.200pt]{0.400pt}{184.048pt}}
\put(220,556){\usebox{\plotpoint}}
\multiput(220.00,556.58)(0.496,0.492){21}{\rule{0.500pt}{0.119pt}}
\multiput(220.00,555.17)(10.962,12.000){2}{\rule{0.250pt}{0.400pt}}
\multiput(232.00,568.58)(0.652,0.491){17}{\rule{0.620pt}{0.118pt}}
\multiput(232.00,567.17)(11.713,10.000){2}{\rule{0.310pt}{0.400pt}}
\multiput(245.00,578.59)(0.669,0.489){15}{\rule{0.633pt}{0.118pt}}
\multiput(245.00,577.17)(10.685,9.000){2}{\rule{0.317pt}{0.400pt}}
\multiput(257.00,587.59)(0.669,0.489){15}{\rule{0.633pt}{0.118pt}}
\multiput(257.00,586.17)(10.685,9.000){2}{\rule{0.317pt}{0.400pt}}
\multiput(269.00,596.59)(0.874,0.485){11}{\rule{0.786pt}{0.117pt}}
\multiput(269.00,595.17)(10.369,7.000){2}{\rule{0.393pt}{0.400pt}}
\multiput(281.00,603.59)(0.950,0.485){11}{\rule{0.843pt}{0.117pt}}
\multiput(281.00,602.17)(11.251,7.000){2}{\rule{0.421pt}{0.400pt}}
\multiput(294.00,610.59)(1.033,0.482){9}{\rule{0.900pt}{0.116pt}}
\multiput(294.00,609.17)(10.132,6.000){2}{\rule{0.450pt}{0.400pt}}
\multiput(306.00,616.59)(1.267,0.477){7}{\rule{1.060pt}{0.115pt}}
\multiput(306.00,615.17)(9.800,5.000){2}{\rule{0.530pt}{0.400pt}}
\multiput(318.00,621.59)(1.378,0.477){7}{\rule{1.140pt}{0.115pt}}
\multiput(318.00,620.17)(10.634,5.000){2}{\rule{0.570pt}{0.400pt}}
\multiput(331.00,626.60)(1.651,0.468){5}{\rule{1.300pt}{0.113pt}}
\multiput(331.00,625.17)(9.302,4.000){2}{\rule{0.650pt}{0.400pt}}
\multiput(343.00,630.60)(1.651,0.468){5}{\rule{1.300pt}{0.113pt}}
\multiput(343.00,629.17)(9.302,4.000){2}{\rule{0.650pt}{0.400pt}}
\multiput(355.00,634.60)(1.651,0.468){5}{\rule{1.300pt}{0.113pt}}
\multiput(355.00,633.17)(9.302,4.000){2}{\rule{0.650pt}{0.400pt}}
\multiput(367.00,638.60)(1.797,0.468){5}{\rule{1.400pt}{0.113pt}}
\multiput(367.00,637.17)(10.094,4.000){2}{\rule{0.700pt}{0.400pt}}
\multiput(380.00,642.61)(2.472,0.447){3}{\rule{1.700pt}{0.108pt}}
\multiput(380.00,641.17)(8.472,3.000){2}{\rule{0.850pt}{0.400pt}}
\multiput(392.00,645.61)(2.472,0.447){3}{\rule{1.700pt}{0.108pt}}
\multiput(392.00,644.17)(8.472,3.000){2}{\rule{0.850pt}{0.400pt}}
\put(404,648.17){\rule{2.700pt}{0.400pt}}
\multiput(404.00,647.17)(7.396,2.000){2}{\rule{1.350pt}{0.400pt}}
\multiput(417.00,650.61)(2.472,0.447){3}{\rule{1.700pt}{0.108pt}}
\multiput(417.00,649.17)(8.472,3.000){2}{\rule{0.850pt}{0.400pt}}
\put(429,653.17){\rule{2.500pt}{0.400pt}}
\multiput(429.00,652.17)(6.811,2.000){2}{\rule{1.250pt}{0.400pt}}
\put(441,655.17){\rule{2.500pt}{0.400pt}}
\multiput(441.00,654.17)(6.811,2.000){2}{\rule{1.250pt}{0.400pt}}
\put(453,657.17){\rule{2.700pt}{0.400pt}}
\multiput(453.00,656.17)(7.396,2.000){2}{\rule{1.350pt}{0.400pt}}
\put(466,659.17){\rule{2.500pt}{0.400pt}}
\multiput(466.00,658.17)(6.811,2.000){2}{\rule{1.250pt}{0.400pt}}
\put(478,661.17){\rule{2.500pt}{0.400pt}}
\multiput(478.00,660.17)(6.811,2.000){2}{\rule{1.250pt}{0.400pt}}
\put(490,663.17){\rule{2.700pt}{0.400pt}}
\multiput(490.00,662.17)(7.396,2.000){2}{\rule{1.350pt}{0.400pt}}
\put(503,664.67){\rule{2.891pt}{0.400pt}}
\multiput(503.00,664.17)(6.000,1.000){2}{\rule{1.445pt}{0.400pt}}
\put(515,666.17){\rule{2.500pt}{0.400pt}}
\multiput(515.00,665.17)(6.811,2.000){2}{\rule{1.250pt}{0.400pt}}
\put(527,667.67){\rule{2.891pt}{0.400pt}}
\multiput(527.00,667.17)(6.000,1.000){2}{\rule{1.445pt}{0.400pt}}
\put(539,668.67){\rule{3.132pt}{0.400pt}}
\multiput(539.00,668.17)(6.500,1.000){2}{\rule{1.566pt}{0.400pt}}
\put(552,670.17){\rule{2.500pt}{0.400pt}}
\multiput(552.00,669.17)(6.811,2.000){2}{\rule{1.250pt}{0.400pt}}
\put(564,671.67){\rule{2.891pt}{0.400pt}}
\multiput(564.00,671.17)(6.000,1.000){2}{\rule{1.445pt}{0.400pt}}
\put(576,672.67){\rule{2.891pt}{0.400pt}}
\multiput(576.00,672.17)(6.000,1.000){2}{\rule{1.445pt}{0.400pt}}
\put(588,673.67){\rule{3.132pt}{0.400pt}}
\multiput(588.00,673.17)(6.500,1.000){2}{\rule{1.566pt}{0.400pt}}
\put(601,674.67){\rule{2.891pt}{0.400pt}}
\multiput(601.00,674.17)(6.000,1.000){2}{\rule{1.445pt}{0.400pt}}
\put(613,675.67){\rule{2.891pt}{0.400pt}}
\multiput(613.00,675.17)(6.000,1.000){2}{\rule{1.445pt}{0.400pt}}
\put(625,676.67){\rule{3.132pt}{0.400pt}}
\multiput(625.00,676.17)(6.500,1.000){2}{\rule{1.566pt}{0.400pt}}
\put(638,677.67){\rule{2.891pt}{0.400pt}}
\multiput(638.00,677.17)(6.000,1.000){2}{\rule{1.445pt}{0.400pt}}
\put(662,678.67){\rule{2.891pt}{0.400pt}}
\multiput(662.00,678.17)(6.000,1.000){2}{\rule{1.445pt}{0.400pt}}
\put(674,679.67){\rule{3.132pt}{0.400pt}}
\multiput(674.00,679.17)(6.500,1.000){2}{\rule{1.566pt}{0.400pt}}
\put(650.0,679.0){\rule[-0.200pt]{2.891pt}{0.400pt}}
\put(699,680.67){\rule{2.891pt}{0.400pt}}
\multiput(699.00,680.17)(6.000,1.000){2}{\rule{1.445pt}{0.400pt}}
\put(711,681.67){\rule{3.132pt}{0.400pt}}
\multiput(711.00,681.17)(6.500,1.000){2}{\rule{1.566pt}{0.400pt}}
\put(687.0,681.0){\rule[-0.200pt]{2.891pt}{0.400pt}}
\put(736,682.67){\rule{2.891pt}{0.400pt}}
\multiput(736.00,682.17)(6.000,1.000){2}{\rule{1.445pt}{0.400pt}}
\put(724.0,683.0){\rule[-0.200pt]{2.891pt}{0.400pt}}
\put(760,683.67){\rule{3.132pt}{0.400pt}}
\multiput(760.00,683.17)(6.500,1.000){2}{\rule{1.566pt}{0.400pt}}
\put(748.0,684.0){\rule[-0.200pt]{2.891pt}{0.400pt}}
\put(785,684.67){\rule{2.891pt}{0.400pt}}
\multiput(785.00,684.17)(6.000,1.000){2}{\rule{1.445pt}{0.400pt}}
\put(773.0,685.0){\rule[-0.200pt]{2.891pt}{0.400pt}}
\put(810,685.67){\rule{2.891pt}{0.400pt}}
\multiput(810.00,685.17)(6.000,1.000){2}{\rule{1.445pt}{0.400pt}}
\put(797.0,686.0){\rule[-0.200pt]{3.132pt}{0.400pt}}
\put(834,686.67){\rule{2.891pt}{0.400pt}}
\multiput(834.00,686.17)(6.000,1.000){2}{\rule{1.445pt}{0.400pt}}
\put(822.0,687.0){\rule[-0.200pt]{2.891pt}{0.400pt}}
\put(871,687.67){\rule{2.891pt}{0.400pt}}
\multiput(871.00,687.17)(6.000,1.000){2}{\rule{1.445pt}{0.400pt}}
\put(846.0,688.0){\rule[-0.200pt]{6.022pt}{0.400pt}}
\put(908,688.67){\rule{2.891pt}{0.400pt}}
\multiput(908.00,688.17)(6.000,1.000){2}{\rule{1.445pt}{0.400pt}}
\put(883.0,689.0){\rule[-0.200pt]{6.022pt}{0.400pt}}
\put(945,689.67){\rule{2.891pt}{0.400pt}}
\multiput(945.00,689.17)(6.000,1.000){2}{\rule{1.445pt}{0.400pt}}
\put(920.0,690.0){\rule[-0.200pt]{6.022pt}{0.400pt}}
\put(994,690.67){\rule{2.891pt}{0.400pt}}
\multiput(994.00,690.17)(6.000,1.000){2}{\rule{1.445pt}{0.400pt}}
\put(957.0,691.0){\rule[-0.200pt]{8.913pt}{0.400pt}}
\put(1043,691.67){\rule{2.891pt}{0.400pt}}
\multiput(1043.00,691.17)(6.000,1.000){2}{\rule{1.445pt}{0.400pt}}
\put(1006.0,692.0){\rule[-0.200pt]{8.913pt}{0.400pt}}
\put(1104,692.67){\rule{3.132pt}{0.400pt}}
\multiput(1104.00,692.17)(6.500,1.000){2}{\rule{1.566pt}{0.400pt}}
\put(1055.0,693.0){\rule[-0.200pt]{11.804pt}{0.400pt}}
\put(1178,693.67){\rule{2.891pt}{0.400pt}}
\multiput(1178.00,693.17)(6.000,1.000){2}{\rule{1.445pt}{0.400pt}}
\put(1117.0,694.0){\rule[-0.200pt]{14.695pt}{0.400pt}}
\put(1264,694.67){\rule{2.891pt}{0.400pt}}
\multiput(1264.00,694.17)(6.000,1.000){2}{\rule{1.445pt}{0.400pt}}
\put(1190.0,695.0){\rule[-0.200pt]{17.827pt}{0.400pt}}
\put(1375,695.67){\rule{2.891pt}{0.400pt}}
\multiput(1375.00,695.17)(6.000,1.000){2}{\rule{1.445pt}{0.400pt}}
\put(1276.0,696.0){\rule[-0.200pt]{23.849pt}{0.400pt}}
\put(1387.0,697.0){\rule[-0.200pt]{11.804pt}{0.400pt}}
\put(220,195){\usebox{\plotpoint}}
\put(220.00,195.00){\usebox{\plotpoint}}
\put(236.90,207.02){\usebox{\plotpoint}}
\put(254.71,217.67){\usebox{\plotpoint}}
\multiput(257,219)(18.564,9.282){0}{\usebox{\plotpoint}}
\put(273.20,227.10){\usebox{\plotpoint}}
\put(292.23,235.32){\usebox{\plotpoint}}
\multiput(294,236)(19.159,7.983){0}{\usebox{\plotpoint}}
\put(311.56,242.85){\usebox{\plotpoint}}
\multiput(318,245)(19.838,6.104){0}{\usebox{\plotpoint}}
\put(331.35,249.09){\usebox{\plotpoint}}
\put(351.49,254.12){\usebox{\plotpoint}}
\multiput(355,255)(20.136,5.034){0}{\usebox{\plotpoint}}
\put(371.65,259.07){\usebox{\plotpoint}}
\multiput(380,261)(20.473,3.412){0}{\usebox{\plotpoint}}
\put(392.02,263.00){\usebox{\plotpoint}}
\put(412.30,267.28){\usebox{\plotpoint}}
\multiput(417,268)(20.473,3.412){0}{\usebox{\plotpoint}}
\put(432.83,270.32){\usebox{\plotpoint}}
\multiput(441,271)(20.473,3.412){0}{\usebox{\plotpoint}}
\put(453.38,273.06){\usebox{\plotpoint}}
\put(473.96,275.66){\usebox{\plotpoint}}
\multiput(478,276)(20.684,1.724){0}{\usebox{\plotpoint}}
\put(494.61,277.71){\usebox{\plotpoint}}
\multiput(503,279)(20.684,1.724){0}{\usebox{\plotpoint}}
\put(515.22,280.02){\usebox{\plotpoint}}
\put(535.91,281.74){\usebox{\plotpoint}}
\multiput(539,282)(20.694,1.592){0}{\usebox{\plotpoint}}
\put(556.60,283.38){\usebox{\plotpoint}}
\multiput(564,284)(20.684,1.724){0}{\usebox{\plotpoint}}
\put(577.28,285.11){\usebox{\plotpoint}}
\put(598.00,286.00){\usebox{\plotpoint}}
\multiput(601,286)(20.684,1.724){0}{\usebox{\plotpoint}}
\put(618.69,287.47){\usebox{\plotpoint}}
\multiput(625,288)(20.694,1.592){0}{\usebox{\plotpoint}}
\put(639.39,289.00){\usebox{\plotpoint}}
\put(660.11,289.84){\usebox{\plotpoint}}
\multiput(662,290)(20.756,0.000){0}{\usebox{\plotpoint}}
\put(680.84,290.53){\usebox{\plotpoint}}
\multiput(687,291)(20.756,0.000){0}{\usebox{\plotpoint}}
\put(701.57,291.21){\usebox{\plotpoint}}
\put(722.29,292.00){\usebox{\plotpoint}}
\multiput(724,292)(20.684,1.724){0}{\usebox{\plotpoint}}
\put(743.00,293.00){\usebox{\plotpoint}}
\multiput(748,293)(20.684,1.724){0}{\usebox{\plotpoint}}
\put(763.72,294.00){\usebox{\plotpoint}}
\put(784.47,294.00){\usebox{\plotpoint}}
\multiput(785,294)(20.684,1.724){0}{\usebox{\plotpoint}}
\put(805.19,295.00){\usebox{\plotpoint}}
\multiput(810,295)(20.684,1.724){0}{\usebox{\plotpoint}}
\put(825.90,296.00){\usebox{\plotpoint}}
\multiput(834,296)(20.756,0.000){0}{\usebox{\plotpoint}}
\put(846.66,296.00){\usebox{\plotpoint}}
\put(867.38,296.70){\usebox{\plotpoint}}
\multiput(871,297)(20.756,0.000){0}{\usebox{\plotpoint}}
\put(888.12,297.00){\usebox{\plotpoint}}
\multiput(896,297)(20.684,1.724){0}{\usebox{\plotpoint}}
\put(908.84,298.00){\usebox{\plotpoint}}
\put(929.59,298.00){\usebox{\plotpoint}}
\multiput(932,298)(20.756,0.000){0}{\usebox{\plotpoint}}
\put(950.35,298.00){\usebox{\plotpoint}}
\multiput(957,298)(20.684,1.724){0}{\usebox{\plotpoint}}
\put(971.06,299.00){\usebox{\plotpoint}}
\put(991.82,299.00){\usebox{\plotpoint}}
\multiput(994,299)(20.756,0.000){0}{\usebox{\plotpoint}}
\put(1012.57,299.00){\usebox{\plotpoint}}
\multiput(1018,299)(20.694,1.592){0}{\usebox{\plotpoint}}
\put(1033.29,300.00){\usebox{\plotpoint}}
\put(1054.05,300.00){\usebox{\plotpoint}}
\multiput(1055,300)(20.756,0.000){0}{\usebox{\plotpoint}}
\put(1074.80,300.00){\usebox{\plotpoint}}
\multiput(1080,300)(20.756,0.000){0}{\usebox{\plotpoint}}
\put(1095.55,300.30){\usebox{\plotpoint}}
\put(1116.27,301.00){\usebox{\plotpoint}}
\multiput(1117,301)(20.756,0.000){0}{\usebox{\plotpoint}}
\put(1137.03,301.00){\usebox{\plotpoint}}
\multiput(1141,301)(20.756,0.000){0}{\usebox{\plotpoint}}
\put(1157.78,301.00){\usebox{\plotpoint}}
\multiput(1166,301)(20.756,0.000){0}{\usebox{\plotpoint}}
\put(1178.54,301.00){\usebox{\plotpoint}}
\put(1199.27,301.71){\usebox{\plotpoint}}
\multiput(1203,302)(20.756,0.000){0}{\usebox{\plotpoint}}
\put(1220.01,302.00){\usebox{\plotpoint}}
\multiput(1227,302)(20.756,0.000){0}{\usebox{\plotpoint}}
\put(1240.77,302.00){\usebox{\plotpoint}}
\put(1261.52,302.00){\usebox{\plotpoint}}
\multiput(1264,302)(20.756,0.000){0}{\usebox{\plotpoint}}
\put(1282.28,302.00){\usebox{\plotpoint}}
\multiput(1289,302)(20.756,0.000){0}{\usebox{\plotpoint}}
\put(1303.03,302.17){\usebox{\plotpoint}}
\put(1323.75,303.00){\usebox{\plotpoint}}
\multiput(1325,303)(20.756,0.000){0}{\usebox{\plotpoint}}
\put(1344.50,303.00){\usebox{\plotpoint}}
\multiput(1350,303)(20.756,0.000){0}{\usebox{\plotpoint}}
\put(1365.26,303.00){\usebox{\plotpoint}}
\put(1386.01,303.00){\usebox{\plotpoint}}
\multiput(1387,303)(20.756,0.000){0}{\usebox{\plotpoint}}
\put(1406.77,303.00){\usebox{\plotpoint}}
\multiput(1411,303)(20.756,0.000){0}{\usebox{\plotpoint}}
\put(1427.52,303.00){\usebox{\plotpoint}}
\put(1436,303){\usebox{\plotpoint}}
\sbox{\plotpoint}{\rule[-0.400pt]{0.800pt}{0.800pt}}%
\multiput(221.00,114.40)(0.825,0.526){7}{\rule{1.457pt}{0.127pt}}
\multiput(221.00,111.34)(7.976,7.000){2}{\rule{0.729pt}{0.800pt}}
\multiput(232.00,121.39)(1.244,0.536){5}{\rule{1.933pt}{0.129pt}}
\multiput(232.00,118.34)(8.987,6.000){2}{\rule{0.967pt}{0.800pt}}
\multiput(245.00,127.39)(1.132,0.536){5}{\rule{1.800pt}{0.129pt}}
\multiput(245.00,124.34)(8.264,6.000){2}{\rule{0.900pt}{0.800pt}}
\multiput(257.00,133.38)(1.600,0.560){3}{\rule{2.120pt}{0.135pt}}
\multiput(257.00,130.34)(7.600,5.000){2}{\rule{1.060pt}{0.800pt}}
\multiput(269.00,138.38)(1.600,0.560){3}{\rule{2.120pt}{0.135pt}}
\multiput(269.00,135.34)(7.600,5.000){2}{\rule{1.060pt}{0.800pt}}
\put(281,142.34){\rule{2.800pt}{0.800pt}}
\multiput(281.00,140.34)(7.188,4.000){2}{\rule{1.400pt}{0.800pt}}
\put(294,145.84){\rule{2.891pt}{0.800pt}}
\multiput(294.00,144.34)(6.000,3.000){2}{\rule{1.445pt}{0.800pt}}
\put(306,149.34){\rule{2.600pt}{0.800pt}}
\multiput(306.00,147.34)(6.604,4.000){2}{\rule{1.300pt}{0.800pt}}
\put(318,152.84){\rule{3.132pt}{0.800pt}}
\multiput(318.00,151.34)(6.500,3.000){2}{\rule{1.566pt}{0.800pt}}
\put(331,155.84){\rule{2.891pt}{0.800pt}}
\multiput(331.00,154.34)(6.000,3.000){2}{\rule{1.445pt}{0.800pt}}
\put(343,158.34){\rule{2.891pt}{0.800pt}}
\multiput(343.00,157.34)(6.000,2.000){2}{\rule{1.445pt}{0.800pt}}
\put(355,160.34){\rule{2.891pt}{0.800pt}}
\multiput(355.00,159.34)(6.000,2.000){2}{\rule{1.445pt}{0.800pt}}
\put(367,162.84){\rule{3.132pt}{0.800pt}}
\multiput(367.00,161.34)(6.500,3.000){2}{\rule{1.566pt}{0.800pt}}
\put(380,165.34){\rule{2.891pt}{0.800pt}}
\multiput(380.00,164.34)(6.000,2.000){2}{\rule{1.445pt}{0.800pt}}
\put(392,166.84){\rule{2.891pt}{0.800pt}}
\multiput(392.00,166.34)(6.000,1.000){2}{\rule{1.445pt}{0.800pt}}
\put(404,168.34){\rule{3.132pt}{0.800pt}}
\multiput(404.00,167.34)(6.500,2.000){2}{\rule{1.566pt}{0.800pt}}
\put(417,170.34){\rule{2.891pt}{0.800pt}}
\multiput(417.00,169.34)(6.000,2.000){2}{\rule{1.445pt}{0.800pt}}
\put(429,171.84){\rule{2.891pt}{0.800pt}}
\multiput(429.00,171.34)(6.000,1.000){2}{\rule{1.445pt}{0.800pt}}
\put(441,172.84){\rule{2.891pt}{0.800pt}}
\multiput(441.00,172.34)(6.000,1.000){2}{\rule{1.445pt}{0.800pt}}
\put(453,174.34){\rule{3.132pt}{0.800pt}}
\multiput(453.00,173.34)(6.500,2.000){2}{\rule{1.566pt}{0.800pt}}
\put(466,175.84){\rule{2.891pt}{0.800pt}}
\multiput(466.00,175.34)(6.000,1.000){2}{\rule{1.445pt}{0.800pt}}
\put(478,176.84){\rule{2.891pt}{0.800pt}}
\multiput(478.00,176.34)(6.000,1.000){2}{\rule{1.445pt}{0.800pt}}
\put(490,177.84){\rule{3.132pt}{0.800pt}}
\multiput(490.00,177.34)(6.500,1.000){2}{\rule{1.566pt}{0.800pt}}
\put(503,178.84){\rule{2.891pt}{0.800pt}}
\multiput(503.00,178.34)(6.000,1.000){2}{\rule{1.445pt}{0.800pt}}
\put(515,179.84){\rule{2.891pt}{0.800pt}}
\multiput(515.00,179.34)(6.000,1.000){2}{\rule{1.445pt}{0.800pt}}
\put(527,180.84){\rule{2.891pt}{0.800pt}}
\multiput(527.00,180.34)(6.000,1.000){2}{\rule{1.445pt}{0.800pt}}
\put(552,181.84){\rule{2.891pt}{0.800pt}}
\multiput(552.00,181.34)(6.000,1.000){2}{\rule{1.445pt}{0.800pt}}
\put(564,182.84){\rule{2.891pt}{0.800pt}}
\multiput(564.00,182.34)(6.000,1.000){2}{\rule{1.445pt}{0.800pt}}
\put(539.0,183.0){\rule[-0.400pt]{3.132pt}{0.800pt}}
\put(588,183.84){\rule{3.132pt}{0.800pt}}
\multiput(588.00,183.34)(6.500,1.000){2}{\rule{1.566pt}{0.800pt}}
\put(601,184.84){\rule{2.891pt}{0.800pt}}
\multiput(601.00,184.34)(6.000,1.000){2}{\rule{1.445pt}{0.800pt}}
\put(576.0,185.0){\rule[-0.400pt]{2.891pt}{0.800pt}}
\put(625,185.84){\rule{3.132pt}{0.800pt}}
\multiput(625.00,185.34)(6.500,1.000){2}{\rule{1.566pt}{0.800pt}}
\put(613.0,187.0){\rule[-0.400pt]{2.891pt}{0.800pt}}
\put(650,186.84){\rule{2.891pt}{0.800pt}}
\multiput(650.00,186.34)(6.000,1.000){2}{\rule{1.445pt}{0.800pt}}
\put(638.0,188.0){\rule[-0.400pt]{2.891pt}{0.800pt}}
\put(674,187.84){\rule{3.132pt}{0.800pt}}
\multiput(674.00,187.34)(6.500,1.000){2}{\rule{1.566pt}{0.800pt}}
\put(662.0,189.0){\rule[-0.400pt]{2.891pt}{0.800pt}}
\put(711,188.84){\rule{3.132pt}{0.800pt}}
\multiput(711.00,188.34)(6.500,1.000){2}{\rule{1.566pt}{0.800pt}}
\put(687.0,190.0){\rule[-0.400pt]{5.782pt}{0.800pt}}
\put(736,189.84){\rule{2.891pt}{0.800pt}}
\multiput(736.00,189.34)(6.000,1.000){2}{\rule{1.445pt}{0.800pt}}
\put(724.0,191.0){\rule[-0.400pt]{2.891pt}{0.800pt}}
\put(773,190.84){\rule{2.891pt}{0.800pt}}
\multiput(773.00,190.34)(6.000,1.000){2}{\rule{1.445pt}{0.800pt}}
\put(748.0,192.0){\rule[-0.400pt]{6.022pt}{0.800pt}}
\put(822,191.84){\rule{2.891pt}{0.800pt}}
\multiput(822.00,191.34)(6.000,1.000){2}{\rule{1.445pt}{0.800pt}}
\put(785.0,193.0){\rule[-0.400pt]{8.913pt}{0.800pt}}
\put(871,192.84){\rule{2.891pt}{0.800pt}}
\multiput(871.00,192.34)(6.000,1.000){2}{\rule{1.445pt}{0.800pt}}
\put(834.0,194.0){\rule[-0.400pt]{8.913pt}{0.800pt}}
\put(932,193.84){\rule{3.132pt}{0.800pt}}
\multiput(932.00,193.34)(6.500,1.000){2}{\rule{1.566pt}{0.800pt}}
\put(883.0,195.0){\rule[-0.400pt]{11.804pt}{0.800pt}}
\put(1006,194.84){\rule{2.891pt}{0.800pt}}
\multiput(1006.00,194.34)(6.000,1.000){2}{\rule{1.445pt}{0.800pt}}
\put(945.0,196.0){\rule[-0.400pt]{14.695pt}{0.800pt}}
\put(1104,195.84){\rule{3.132pt}{0.800pt}}
\multiput(1104.00,195.34)(6.500,1.000){2}{\rule{1.566pt}{0.800pt}}
\put(1018.0,197.0){\rule[-0.400pt]{20.717pt}{0.800pt}}
\put(1239,196.84){\rule{3.132pt}{0.800pt}}
\multiput(1239.00,196.34)(6.500,1.000){2}{\rule{1.566pt}{0.800pt}}
\put(1117.0,198.0){\rule[-0.400pt]{29.390pt}{0.800pt}}
\put(1411,197.84){\rule{3.132pt}{0.800pt}}
\multiput(1411.00,197.34)(6.500,1.000){2}{\rule{1.566pt}{0.800pt}}
\put(1252.0,199.0){\rule[-0.400pt]{38.303pt}{0.800pt}}
\put(1424.0,200.0){\rule[-0.400pt]{2.891pt}{0.800pt}}
\sbox{\plotpoint}{\rule[-0.500pt]{1.000pt}{1.000pt}}%
\put(486.00,113.00){\usebox{\plotpoint}}
\multiput(490,113)(20.694,1.592){0}{\usebox{\plotpoint}}
\put(506.70,114.31){\usebox{\plotpoint}}
\multiput(515,115)(20.684,1.724){0}{\usebox{\plotpoint}}
\put(527.39,116.00){\usebox{\plotpoint}}
\put(548.12,116.70){\usebox{\plotpoint}}
\multiput(552,117)(20.684,1.724){0}{\usebox{\plotpoint}}
\put(568.82,118.00){\usebox{\plotpoint}}
\multiput(576,118)(20.684,1.724){0}{\usebox{\plotpoint}}
\put(589.53,119.12){\usebox{\plotpoint}}
\put(610.25,120.00){\usebox{\plotpoint}}
\multiput(613,120)(20.684,1.724){0}{\usebox{\plotpoint}}
\put(630.97,121.00){\usebox{\plotpoint}}
\multiput(638,121)(20.756,0.000){0}{\usebox{\plotpoint}}
\put(651.71,121.14){\usebox{\plotpoint}}
\put(672.43,122.00){\usebox{\plotpoint}}
\multiput(674,122)(20.694,1.592){0}{\usebox{\plotpoint}}
\put(693.15,123.00){\usebox{\plotpoint}}
\multiput(699,123)(20.756,0.000){0}{\usebox{\plotpoint}}
\put(713.90,123.22){\usebox{\plotpoint}}
\put(734.62,124.00){\usebox{\plotpoint}}
\multiput(736,124)(20.756,0.000){0}{\usebox{\plotpoint}}
\put(755.35,124.61){\usebox{\plotpoint}}
\multiput(760,125)(20.756,0.000){0}{\usebox{\plotpoint}}
\put(776.09,125.00){\usebox{\plotpoint}}
\put(796.85,125.00){\usebox{\plotpoint}}
\multiput(797,125)(20.694,1.592){0}{\usebox{\plotpoint}}
\put(817.57,126.00){\usebox{\plotpoint}}
\multiput(822,126)(20.756,0.000){0}{\usebox{\plotpoint}}
\put(838.32,126.00){\usebox{\plotpoint}}
\multiput(846,126)(20.756,0.000){0}{\usebox{\plotpoint}}
\put(859.08,126.01){\usebox{\plotpoint}}
\put(879.79,127.00){\usebox{\plotpoint}}
\multiput(883,127)(20.756,0.000){0}{\usebox{\plotpoint}}
\put(900.55,127.00){\usebox{\plotpoint}}
\multiput(908,127)(20.756,0.000){0}{\usebox{\plotpoint}}
\put(921.30,127.00){\usebox{\plotpoint}}
\put(942.03,127.77){\usebox{\plotpoint}}
\multiput(945,128)(20.756,0.000){0}{\usebox{\plotpoint}}
\put(962.77,128.00){\usebox{\plotpoint}}
\multiput(969,128)(20.756,0.000){0}{\usebox{\plotpoint}}
\put(983.53,128.00){\usebox{\plotpoint}}
\put(1004.29,128.00){\usebox{\plotpoint}}
\multiput(1006,128)(20.756,0.000){0}{\usebox{\plotpoint}}
\put(1025.02,128.54){\usebox{\plotpoint}}
\multiput(1031,129)(20.756,0.000){0}{\usebox{\plotpoint}}
\put(1045.76,129.00){\usebox{\plotpoint}}
\put(1066.51,129.00){\usebox{\plotpoint}}
\multiput(1068,129)(20.756,0.000){0}{\usebox{\plotpoint}}
\put(1087.27,129.00){\usebox{\plotpoint}}
\multiput(1092,129)(20.756,0.000){0}{\usebox{\plotpoint}}
\put(1108.03,129.00){\usebox{\plotpoint}}
\put(1128.78,129.00){\usebox{\plotpoint}}
\multiput(1129,129)(20.684,1.724){0}{\usebox{\plotpoint}}
\put(1149.49,130.00){\usebox{\plotpoint}}
\multiput(1153,130)(20.756,0.000){0}{\usebox{\plotpoint}}
\put(1170.25,130.00){\usebox{\plotpoint}}
\multiput(1178,130)(20.756,0.000){0}{\usebox{\plotpoint}}
\put(1191.01,130.00){\usebox{\plotpoint}}
\put(1211.76,130.00){\usebox{\plotpoint}}
\multiput(1215,130)(20.756,0.000){0}{\usebox{\plotpoint}}
\put(1232.52,130.00){\usebox{\plotpoint}}
\multiput(1239,130)(20.756,0.000){0}{\usebox{\plotpoint}}
\put(1253.27,130.00){\usebox{\plotpoint}}
\put(1274.03,130.00){\usebox{\plotpoint}}
\multiput(1276,130)(20.756,0.000){0}{\usebox{\plotpoint}}
\put(1294.78,130.00){\usebox{\plotpoint}}
\multiput(1301,130)(20.684,1.724){0}{\usebox{\plotpoint}}
\put(1315.50,131.00){\usebox{\plotpoint}}
\put(1336.25,131.00){\usebox{\plotpoint}}
\multiput(1338,131)(20.756,0.000){0}{\usebox{\plotpoint}}
\put(1357.01,131.00){\usebox{\plotpoint}}
\multiput(1362,131)(20.756,0.000){0}{\usebox{\plotpoint}}
\put(1377.76,131.00){\usebox{\plotpoint}}
\put(1398.52,131.00){\usebox{\plotpoint}}
\multiput(1399,131)(20.756,0.000){0}{\usebox{\plotpoint}}
\put(1419.27,131.00){\usebox{\plotpoint}}
\multiput(1424,131)(20.756,0.000){0}{\usebox{\plotpoint}}
\put(1436,131){\usebox{\plotpoint}}
\end{picture}

%% file: BAM.tex
% GNUPLOT: LaTeX picture
\setlength{\unitlength}{0.240900pt}
\ifx\plotpoint\undefined\newsavebox{\plotpoint}\fi
\sbox{\plotpoint}{\rule[-0.200pt]{0.400pt}{0.400pt}}%
\begin{picture}(1500,900)(0,0)
\font\gnuplot=cmr10 at 10pt
\gnuplot
\sbox{\plotpoint}{\rule[-0.200pt]{0.400pt}{0.400pt}}%
\put(220.0,622.0){\rule[-0.200pt]{292.934pt}{0.400pt}}
\put(828.0,113.0){\rule[-0.200pt]{0.400pt}{184.048pt}}
\put(220.0,113.0){\rule[-0.200pt]{4.818pt}{0.400pt}}
\put(198,113){\makebox(0,0)[r]{-4}}
\put(1416.0,113.0){\rule[-0.200pt]{4.818pt}{0.400pt}}
\put(220.0,240.0){\rule[-0.200pt]{4.818pt}{0.400pt}}
\put(198,240){\makebox(0,0)[r]{-3}}
\put(1416.0,240.0){\rule[-0.200pt]{4.818pt}{0.400pt}}
\put(220.0,368.0){\rule[-0.200pt]{4.818pt}{0.400pt}}
\put(198,368){\makebox(0,0)[r]{-2}}
\put(1416.0,368.0){\rule[-0.200pt]{4.818pt}{0.400pt}}
\put(220.0,495.0){\rule[-0.200pt]{4.818pt}{0.400pt}}
\put(198,495){\makebox(0,0)[r]{-1}}
\put(1416.0,495.0){\rule[-0.200pt]{4.818pt}{0.400pt}}
\put(220.0,622.0){\rule[-0.200pt]{4.818pt}{0.400pt}}
\put(198,622){\makebox(0,0)[r]{0}}
\put(1416.0,622.0){\rule[-0.200pt]{4.818pt}{0.400pt}}
\put(220.0,750.0){\rule[-0.200pt]{4.818pt}{0.400pt}}
\put(198,750){\makebox(0,0)[r]{1}}
\put(1416.0,750.0){\rule[-0.200pt]{4.818pt}{0.400pt}}
\put(220.0,877.0){\rule[-0.200pt]{4.818pt}{0.400pt}}
\put(198,877){\makebox(0,0)[r]{2}}
\put(1416.0,877.0){\rule[-0.200pt]{4.818pt}{0.400pt}}
\put(220.0,113.0){\rule[-0.200pt]{0.400pt}{4.818pt}}
\put(220,68){\makebox(0,0){-3}}
\put(220.0,857.0){\rule[-0.200pt]{0.400pt}{4.818pt}}
\put(423.0,113.0){\rule[-0.200pt]{0.400pt}{4.818pt}}
\put(423,68){\makebox(0,0){-2}}
\put(423.0,857.0){\rule[-0.200pt]{0.400pt}{4.818pt}}
\put(625.0,113.0){\rule[-0.200pt]{0.400pt}{4.818pt}}
\put(625,68){\makebox(0,0){-1}}
\put(625.0,857.0){\rule[-0.200pt]{0.400pt}{4.818pt}}
\put(828.0,113.0){\rule[-0.200pt]{0.400pt}{4.818pt}}
\put(828,68){\makebox(0,0){0}}
\put(828.0,857.0){\rule[-0.200pt]{0.400pt}{4.818pt}}
\put(1031.0,113.0){\rule[-0.200pt]{0.400pt}{4.818pt}}
\put(1031,68){\makebox(0,0){1}}
\put(1031.0,857.0){\rule[-0.200pt]{0.400pt}{4.818pt}}
\put(1233.0,113.0){\rule[-0.200pt]{0.400pt}{4.818pt}}
\put(1233,68){\makebox(0,0){2}}
\put(1233.0,857.0){\rule[-0.200pt]{0.400pt}{4.818pt}}
\put(1436.0,113.0){\rule[-0.200pt]{0.400pt}{4.818pt}}
\put(1436,68){\makebox(0,0){3}}
\put(1436.0,857.0){\rule[-0.200pt]{0.400pt}{4.818pt}}
\put(220.0,113.0){\rule[-0.200pt]{292.934pt}{0.400pt}}
\put(1436.0,113.0){\rule[-0.200pt]{0.400pt}{184.048pt}}
\put(220.0,877.0){\rule[-0.200pt]{292.934pt}{0.400pt}}
\put(45,495){\makebox(0,0)
{$\displaystyle{\frac{\Delta B}{M_{1/2}^{(0)}}}$}}
\put(828,23){\makebox(0,0)
{\shortstack{\\ \\ \\ \\ \\ \\ \\ $\displaystyle{\frac{A_f^{(0)}}{M_{1/2}^{(0)}}}$}}}
\put(220.0,113.0){\rule[-0.200pt]{0.400pt}{184.048pt}}
\sbox{\plotpoint}{\rule[-0.400pt]{0.800pt}{0.800pt}}%
\put(220,407){\usebox{\plotpoint}}
\put(220,406.84){\rule{2.891pt}{0.800pt}}
\multiput(220.00,405.34)(6.000,3.000){2}{\rule{1.445pt}{0.800pt}}
\put(232,410.34){\rule{2.800pt}{0.800pt}}
\multiput(232.00,408.34)(7.188,4.000){2}{\rule{1.400pt}{0.800pt}}
\put(245,413.84){\rule{2.891pt}{0.800pt}}
\multiput(245.00,412.34)(6.000,3.000){2}{\rule{1.445pt}{0.800pt}}
\put(257,417.34){\rule{2.600pt}{0.800pt}}
\multiput(257.00,415.34)(6.604,4.000){2}{\rule{1.300pt}{0.800pt}}
\put(269,420.84){\rule{2.891pt}{0.800pt}}
\multiput(269.00,419.34)(6.000,3.000){2}{\rule{1.445pt}{0.800pt}}
\put(281,423.84){\rule{3.132pt}{0.800pt}}
\multiput(281.00,422.34)(6.500,3.000){2}{\rule{1.566pt}{0.800pt}}
\put(294,427.34){\rule{2.600pt}{0.800pt}}
\multiput(294.00,425.34)(6.604,4.000){2}{\rule{1.300pt}{0.800pt}}
\put(306,430.84){\rule{2.891pt}{0.800pt}}
\multiput(306.00,429.34)(6.000,3.000){2}{\rule{1.445pt}{0.800pt}}
\put(318,434.34){\rule{2.800pt}{0.800pt}}
\multiput(318.00,432.34)(7.188,4.000){2}{\rule{1.400pt}{0.800pt}}
\put(331,437.84){\rule{2.891pt}{0.800pt}}
\multiput(331.00,436.34)(6.000,3.000){2}{\rule{1.445pt}{0.800pt}}
\put(343,441.34){\rule{2.600pt}{0.800pt}}
\multiput(343.00,439.34)(6.604,4.000){2}{\rule{1.300pt}{0.800pt}}
\put(355,444.84){\rule{2.891pt}{0.800pt}}
\multiput(355.00,443.34)(6.000,3.000){2}{\rule{1.445pt}{0.800pt}}
\put(367,448.34){\rule{2.800pt}{0.800pt}}
\multiput(367.00,446.34)(7.188,4.000){2}{\rule{1.400pt}{0.800pt}}
\put(380,451.84){\rule{2.891pt}{0.800pt}}
\multiput(380.00,450.34)(6.000,3.000){2}{\rule{1.445pt}{0.800pt}}
\put(392,455.34){\rule{2.600pt}{0.800pt}}
\multiput(392.00,453.34)(6.604,4.000){2}{\rule{1.300pt}{0.800pt}}
\put(404,458.84){\rule{3.132pt}{0.800pt}}
\multiput(404.00,457.34)(6.500,3.000){2}{\rule{1.566pt}{0.800pt}}
\put(417,461.84){\rule{2.891pt}{0.800pt}}
\multiput(417.00,460.34)(6.000,3.000){2}{\rule{1.445pt}{0.800pt}}
\put(429,465.34){\rule{2.600pt}{0.800pt}}
\multiput(429.00,463.34)(6.604,4.000){2}{\rule{1.300pt}{0.800pt}}
\put(441,468.84){\rule{2.891pt}{0.800pt}}
\multiput(441.00,467.34)(6.000,3.000){2}{\rule{1.445pt}{0.800pt}}
\put(453,472.34){\rule{2.800pt}{0.800pt}}
\multiput(453.00,470.34)(7.188,4.000){2}{\rule{1.400pt}{0.800pt}}
\put(466,475.84){\rule{2.891pt}{0.800pt}}
\multiput(466.00,474.34)(6.000,3.000){2}{\rule{1.445pt}{0.800pt}}
\put(478,479.34){\rule{2.600pt}{0.800pt}}
\multiput(478.00,477.34)(6.604,4.000){2}{\rule{1.300pt}{0.800pt}}
\put(490,482.84){\rule{3.132pt}{0.800pt}}
\multiput(490.00,481.34)(6.500,3.000){2}{\rule{1.566pt}{0.800pt}}
\put(503,486.34){\rule{2.600pt}{0.800pt}}
\multiput(503.00,484.34)(6.604,4.000){2}{\rule{1.300pt}{0.800pt}}
\put(515,489.84){\rule{2.891pt}{0.800pt}}
\multiput(515.00,488.34)(6.000,3.000){2}{\rule{1.445pt}{0.800pt}}
\put(527,493.34){\rule{2.600pt}{0.800pt}}
\multiput(527.00,491.34)(6.604,4.000){2}{\rule{1.300pt}{0.800pt}}
\put(539,496.84){\rule{3.132pt}{0.800pt}}
\multiput(539.00,495.34)(6.500,3.000){2}{\rule{1.566pt}{0.800pt}}
\put(552,499.84){\rule{2.891pt}{0.800pt}}
\multiput(552.00,498.34)(6.000,3.000){2}{\rule{1.445pt}{0.800pt}}
\put(564,503.34){\rule{2.600pt}{0.800pt}}
\multiput(564.00,501.34)(6.604,4.000){2}{\rule{1.300pt}{0.800pt}}
\put(576,506.84){\rule{2.891pt}{0.800pt}}
\multiput(576.00,505.34)(6.000,3.000){2}{\rule{1.445pt}{0.800pt}}
\put(588,510.34){\rule{2.800pt}{0.800pt}}
\multiput(588.00,508.34)(7.188,4.000){2}{\rule{1.400pt}{0.800pt}}
\put(601,513.84){\rule{2.891pt}{0.800pt}}
\multiput(601.00,512.34)(6.000,3.000){2}{\rule{1.445pt}{0.800pt}}
\put(613,517.34){\rule{2.600pt}{0.800pt}}
\multiput(613.00,515.34)(6.604,4.000){2}{\rule{1.300pt}{0.800pt}}
\put(625,520.84){\rule{3.132pt}{0.800pt}}
\multiput(625.00,519.34)(6.500,3.000){2}{\rule{1.566pt}{0.800pt}}
\put(638,524.34){\rule{2.600pt}{0.800pt}}
\multiput(638.00,522.34)(6.604,4.000){2}{\rule{1.300pt}{0.800pt}}
\put(650,527.84){\rule{2.891pt}{0.800pt}}
\multiput(650.00,526.34)(6.000,3.000){2}{\rule{1.445pt}{0.800pt}}
\put(662,531.34){\rule{2.600pt}{0.800pt}}
\multiput(662.00,529.34)(6.604,4.000){2}{\rule{1.300pt}{0.800pt}}
\put(674,534.84){\rule{3.132pt}{0.800pt}}
\multiput(674.00,533.34)(6.500,3.000){2}{\rule{1.566pt}{0.800pt}}
\put(687,537.84){\rule{2.891pt}{0.800pt}}
\multiput(687.00,536.34)(6.000,3.000){2}{\rule{1.445pt}{0.800pt}}
\put(699,541.34){\rule{2.600pt}{0.800pt}}
\multiput(699.00,539.34)(6.604,4.000){2}{\rule{1.300pt}{0.800pt}}
\put(711,544.84){\rule{3.132pt}{0.800pt}}
\multiput(711.00,543.34)(6.500,3.000){2}{\rule{1.566pt}{0.800pt}}
\put(724,548.34){\rule{2.600pt}{0.800pt}}
\multiput(724.00,546.34)(6.604,4.000){2}{\rule{1.300pt}{0.800pt}}
\put(736,551.84){\rule{2.891pt}{0.800pt}}
\multiput(736.00,550.34)(6.000,3.000){2}{\rule{1.445pt}{0.800pt}}
\put(748,555.34){\rule{2.600pt}{0.800pt}}
\multiput(748.00,553.34)(6.604,4.000){2}{\rule{1.300pt}{0.800pt}}
\put(760,558.84){\rule{3.132pt}{0.800pt}}
\multiput(760.00,557.34)(6.500,3.000){2}{\rule{1.566pt}{0.800pt}}
\put(773,562.34){\rule{2.600pt}{0.800pt}}
\multiput(773.00,560.34)(6.604,4.000){2}{\rule{1.300pt}{0.800pt}}
\put(785,565.84){\rule{2.891pt}{0.800pt}}
\multiput(785.00,564.34)(6.000,3.000){2}{\rule{1.445pt}{0.800pt}}
\put(797,569.34){\rule{2.800pt}{0.800pt}}
\multiput(797.00,567.34)(7.188,4.000){2}{\rule{1.400pt}{0.800pt}}
\put(810,572.84){\rule{2.891pt}{0.800pt}}
\multiput(810.00,571.34)(6.000,3.000){2}{\rule{1.445pt}{0.800pt}}
\put(822,576.34){\rule{2.600pt}{0.800pt}}
\multiput(822.00,574.34)(6.604,4.000){2}{\rule{1.300pt}{0.800pt}}
\put(834,579.84){\rule{2.891pt}{0.800pt}}
\multiput(834.00,578.34)(6.000,3.000){2}{\rule{1.445pt}{0.800pt}}
\put(846,582.84){\rule{3.132pt}{0.800pt}}
\multiput(846.00,581.34)(6.500,3.000){2}{\rule{1.566pt}{0.800pt}}
\put(859,586.34){\rule{2.600pt}{0.800pt}}
\multiput(859.00,584.34)(6.604,4.000){2}{\rule{1.300pt}{0.800pt}}
\put(871,589.84){\rule{2.891pt}{0.800pt}}
\multiput(871.00,588.34)(6.000,3.000){2}{\rule{1.445pt}{0.800pt}}
\put(883,593.34){\rule{2.800pt}{0.800pt}}
\multiput(883.00,591.34)(7.188,4.000){2}{\rule{1.400pt}{0.800pt}}
\put(896,596.84){\rule{2.891pt}{0.800pt}}
\multiput(896.00,595.34)(6.000,3.000){2}{\rule{1.445pt}{0.800pt}}
\put(908,600.34){\rule{2.600pt}{0.800pt}}
\multiput(908.00,598.34)(6.604,4.000){2}{\rule{1.300pt}{0.800pt}}
\put(920,603.84){\rule{2.891pt}{0.800pt}}
\multiput(920.00,602.34)(6.000,3.000){2}{\rule{1.445pt}{0.800pt}}
\put(932,607.34){\rule{2.800pt}{0.800pt}}
\multiput(932.00,605.34)(7.188,4.000){2}{\rule{1.400pt}{0.800pt}}
\put(945,610.84){\rule{2.891pt}{0.800pt}}
\multiput(945.00,609.34)(6.000,3.000){2}{\rule{1.445pt}{0.800pt}}
\put(957,614.34){\rule{2.600pt}{0.800pt}}
\multiput(957.00,612.34)(6.604,4.000){2}{\rule{1.300pt}{0.800pt}}
\put(969,617.84){\rule{3.132pt}{0.800pt}}
\multiput(969.00,616.34)(6.500,3.000){2}{\rule{1.566pt}{0.800pt}}
\put(982,620.84){\rule{2.891pt}{0.800pt}}
\multiput(982.00,619.34)(6.000,3.000){2}{\rule{1.445pt}{0.800pt}}
\put(994,624.34){\rule{2.600pt}{0.800pt}}
\multiput(994.00,622.34)(6.604,4.000){2}{\rule{1.300pt}{0.800pt}}
\put(1006,627.84){\rule{2.891pt}{0.800pt}}
\multiput(1006.00,626.34)(6.000,3.000){2}{\rule{1.445pt}{0.800pt}}
\put(1018,631.34){\rule{2.800pt}{0.800pt}}
\multiput(1018.00,629.34)(7.188,4.000){2}{\rule{1.400pt}{0.800pt}}
\put(1031,634.84){\rule{2.891pt}{0.800pt}}
\multiput(1031.00,633.34)(6.000,3.000){2}{\rule{1.445pt}{0.800pt}}
\put(1043,638.34){\rule{2.600pt}{0.800pt}}
\multiput(1043.00,636.34)(6.604,4.000){2}{\rule{1.300pt}{0.800pt}}
\put(1055,641.84){\rule{3.132pt}{0.800pt}}
\multiput(1055.00,640.34)(6.500,3.000){2}{\rule{1.566pt}{0.800pt}}
\put(1068,645.34){\rule{2.600pt}{0.800pt}}
\multiput(1068.00,643.34)(6.604,4.000){2}{\rule{1.300pt}{0.800pt}}
\put(1080,648.84){\rule{2.891pt}{0.800pt}}
\multiput(1080.00,647.34)(6.000,3.000){2}{\rule{1.445pt}{0.800pt}}
\put(1092,652.34){\rule{2.600pt}{0.800pt}}
\multiput(1092.00,650.34)(6.604,4.000){2}{\rule{1.300pt}{0.800pt}}
\put(1104,655.84){\rule{3.132pt}{0.800pt}}
\multiput(1104.00,654.34)(6.500,3.000){2}{\rule{1.566pt}{0.800pt}}
\put(1117,658.84){\rule{2.891pt}{0.800pt}}
\multiput(1117.00,657.34)(6.000,3.000){2}{\rule{1.445pt}{0.800pt}}
\put(1129,662.34){\rule{2.600pt}{0.800pt}}
\multiput(1129.00,660.34)(6.604,4.000){2}{\rule{1.300pt}{0.800pt}}
\put(1141,665.84){\rule{2.891pt}{0.800pt}}
\multiput(1141.00,664.34)(6.000,3.000){2}{\rule{1.445pt}{0.800pt}}
\put(1153,669.34){\rule{2.800pt}{0.800pt}}
\multiput(1153.00,667.34)(7.188,4.000){2}{\rule{1.400pt}{0.800pt}}
\put(1166,672.84){\rule{2.891pt}{0.800pt}}
\multiput(1166.00,671.34)(6.000,3.000){2}{\rule{1.445pt}{0.800pt}}
\put(1178,676.34){\rule{2.600pt}{0.800pt}}
\multiput(1178.00,674.34)(6.604,4.000){2}{\rule{1.300pt}{0.800pt}}
\put(1190,679.84){\rule{3.132pt}{0.800pt}}
\multiput(1190.00,678.34)(6.500,3.000){2}{\rule{1.566pt}{0.800pt}}
\put(1203,683.34){\rule{2.600pt}{0.800pt}}
\multiput(1203.00,681.34)(6.604,4.000){2}{\rule{1.300pt}{0.800pt}}
\put(1215,686.84){\rule{2.891pt}{0.800pt}}
\multiput(1215.00,685.34)(6.000,3.000){2}{\rule{1.445pt}{0.800pt}}
\put(1227,690.34){\rule{2.600pt}{0.800pt}}
\multiput(1227.00,688.34)(6.604,4.000){2}{\rule{1.300pt}{0.800pt}}
\put(1239,693.84){\rule{3.132pt}{0.800pt}}
\multiput(1239.00,692.34)(6.500,3.000){2}{\rule{1.566pt}{0.800pt}}
\put(1252,697.34){\rule{2.600pt}{0.800pt}}
\multiput(1252.00,695.34)(6.604,4.000){2}{\rule{1.300pt}{0.800pt}}
\put(1264,700.84){\rule{2.891pt}{0.800pt}}
\multiput(1264.00,699.34)(6.000,3.000){2}{\rule{1.445pt}{0.800pt}}
\put(1276,703.84){\rule{3.132pt}{0.800pt}}
\multiput(1276.00,702.34)(6.500,3.000){2}{\rule{1.566pt}{0.800pt}}
\put(1289,707.34){\rule{2.600pt}{0.800pt}}
\multiput(1289.00,705.34)(6.604,4.000){2}{\rule{1.300pt}{0.800pt}}
\put(1301,710.84){\rule{2.891pt}{0.800pt}}
\multiput(1301.00,709.34)(6.000,3.000){2}{\rule{1.445pt}{0.800pt}}
\put(1313,714.34){\rule{2.600pt}{0.800pt}}
\multiput(1313.00,712.34)(6.604,4.000){2}{\rule{1.300pt}{0.800pt}}
\put(1325,717.84){\rule{3.132pt}{0.800pt}}
\multiput(1325.00,716.34)(6.500,3.000){2}{\rule{1.566pt}{0.800pt}}
\put(1338,721.34){\rule{2.600pt}{0.800pt}}
\multiput(1338.00,719.34)(6.604,4.000){2}{\rule{1.300pt}{0.800pt}}
\put(1350,724.84){\rule{2.891pt}{0.800pt}}
\multiput(1350.00,723.34)(6.000,3.000){2}{\rule{1.445pt}{0.800pt}}
\put(1362,728.34){\rule{2.800pt}{0.800pt}}
\multiput(1362.00,726.34)(7.188,4.000){2}{\rule{1.400pt}{0.800pt}}
\put(1375,731.84){\rule{2.891pt}{0.800pt}}
\multiput(1375.00,730.34)(6.000,3.000){2}{\rule{1.445pt}{0.800pt}}
\put(1387,735.34){\rule{2.600pt}{0.800pt}}
\multiput(1387.00,733.34)(6.604,4.000){2}{\rule{1.300pt}{0.800pt}}
\put(1399,738.84){\rule{2.891pt}{0.800pt}}
\multiput(1399.00,737.34)(6.000,3.000){2}{\rule{1.445pt}{0.800pt}}
\put(1411,741.84){\rule{3.132pt}{0.800pt}}
\multiput(1411.00,740.34)(6.500,3.000){2}{\rule{1.566pt}{0.800pt}}
\put(1424,745.34){\rule{2.600pt}{0.800pt}}
\multiput(1424.00,743.34)(6.604,4.000){2}{\rule{1.300pt}{0.800pt}}
\sbox{\plotpoint}{\rule[-0.500pt]{1.000pt}{1.000pt}}%
\put(220,218){\usebox{\plotpoint}}
\put(220.00,218.00){\usebox{\plotpoint}}
\put(238.85,226.63){\usebox{\plotpoint}}
\multiput(245,229)(18.564,9.282){0}{\usebox{\plotpoint}}
\put(257.67,235.34){\usebox{\plotpoint}}
\put(276.23,244.62){\usebox{\plotpoint}}
\multiput(281,247)(18.845,8.698){0}{\usebox{\plotpoint}}
\put(294.99,253.50){\usebox{\plotpoint}}
\put(313.56,262.78){\usebox{\plotpoint}}
\multiput(318,265)(19.372,7.451){0}{\usebox{\plotpoint}}
\put(332.66,270.83){\usebox{\plotpoint}}
\put(351.23,280.11){\usebox{\plotpoint}}
\multiput(355,282)(18.564,9.282){0}{\usebox{\plotpoint}}
\put(369.83,289.31){\usebox{\plotpoint}}
\put(388.55,298.28){\usebox{\plotpoint}}
\multiput(392,300)(18.564,9.282){0}{\usebox{\plotpoint}}
\put(407.25,307.25){\usebox{\plotpoint}}
\put(426.22,315.61){\usebox{\plotpoint}}
\multiput(429,317)(18.564,9.282){0}{\usebox{\plotpoint}}
\put(444.79,324.89){\usebox{\plotpoint}}
\put(463.51,333.85){\usebox{\plotpoint}}
\multiput(466,335)(18.564,9.282){0}{\usebox{\plotpoint}}
\put(482.11,343.05){\usebox{\plotpoint}}
\put(501.14,351.28){\usebox{\plotpoint}}
\multiput(503,352)(18.564,9.282){0}{\usebox{\plotpoint}}
\put(519.78,360.39){\usebox{\plotpoint}}
\put(538.34,369.67){\usebox{\plotpoint}}
\multiput(539,370)(18.845,8.698){0}{\usebox{\plotpoint}}
\put(557.10,378.55){\usebox{\plotpoint}}
\put(575.66,387.83){\usebox{\plotpoint}}
\multiput(576,388)(18.564,9.282){0}{\usebox{\plotpoint}}
\put(594.50,396.50){\usebox{\plotpoint}}
\multiput(601,399)(18.564,9.282){0}{\usebox{\plotpoint}}
\put(613.34,405.17){\usebox{\plotpoint}}
\put(632.00,414.23){\usebox{\plotpoint}}
\multiput(638,417)(18.564,9.282){0}{\usebox{\plotpoint}}
\put(650.66,423.33){\usebox{\plotpoint}}
\put(669.22,432.61){\usebox{\plotpoint}}
\multiput(674,435)(19.372,7.451){0}{\usebox{\plotpoint}}
\put(688.33,440.66){\usebox{\plotpoint}}
\put(706.89,449.95){\usebox{\plotpoint}}
\multiput(711,452)(18.845,8.698){0}{\usebox{\plotpoint}}
\put(725.65,458.83){\usebox{\plotpoint}}
\put(744.21,468.11){\usebox{\plotpoint}}
\multiput(748,470)(18.564,9.282){0}{\usebox{\plotpoint}}
\put(762.90,477.12){\usebox{\plotpoint}}
\put(781.89,485.44){\usebox{\plotpoint}}
\multiput(785,487)(18.564,9.282){0}{\usebox{\plotpoint}}
\put(800.50,494.62){\usebox{\plotpoint}}
\put(819.21,503.60){\usebox{\plotpoint}}
\multiput(822,505)(18.564,9.282){0}{\usebox{\plotpoint}}
\put(837.77,512.89){\usebox{\plotpoint}}
\put(856.79,521.15){\usebox{\plotpoint}}
\multiput(859,522)(18.564,9.282){0}{\usebox{\plotpoint}}
\put(875.44,530.22){\usebox{\plotpoint}}
\put(894.17,539.16){\usebox{\plotpoint}}
\multiput(896,540)(18.564,9.282){0}{\usebox{\plotpoint}}
\put(912.76,548.38){\usebox{\plotpoint}}
\put(931.33,557.66){\usebox{\plotpoint}}
\multiput(932,558)(18.845,8.698){0}{\usebox{\plotpoint}}
\put(950.25,566.19){\usebox{\plotpoint}}
\multiput(957,569)(18.564,9.282){0}{\usebox{\plotpoint}}
\put(969.02,575.01){\usebox{\plotpoint}}
\put(987.78,583.89){\usebox{\plotpoint}}
\multiput(994,587)(18.564,9.282){0}{\usebox{\plotpoint}}
\put(1006.35,593.17){\usebox{\plotpoint}}
\put(1025.02,602.24){\usebox{\plotpoint}}
\multiput(1031,605)(19.159,7.983){0}{\usebox{\plotpoint}}
\put(1044.04,610.52){\usebox{\plotpoint}}
\put(1062.72,619.56){\usebox{\plotpoint}}
\multiput(1068,622)(18.564,9.282){0}{\usebox{\plotpoint}}
\put(1081.36,628.68){\usebox{\plotpoint}}
\put(1099.93,637.96){\usebox{\plotpoint}}
\multiput(1104,640)(18.845,8.698){0}{\usebox{\plotpoint}}
\put(1118.74,646.72){\usebox{\plotpoint}}
\put(1137.62,655.31){\usebox{\plotpoint}}
\multiput(1141,657)(18.564,9.282){0}{\usebox{\plotpoint}}
\put(1156.23,664.49){\usebox{\plotpoint}}
\put(1174.94,673.47){\usebox{\plotpoint}}
\multiput(1178,675)(18.564,9.282){0}{\usebox{\plotpoint}}
\put(1193.56,682.64){\usebox{\plotpoint}}
\put(1212.56,690.98){\usebox{\plotpoint}}
\multiput(1215,692)(18.564,9.282){0}{\usebox{\plotpoint}}
\put(1231.20,700.10){\usebox{\plotpoint}}
\put(1249.93,709.05){\usebox{\plotpoint}}
\multiput(1252,710)(18.564,9.282){0}{\usebox{\plotpoint}}
\put(1268.53,718.26){\usebox{\plotpoint}}
\put(1287.26,727.20){\usebox{\plotpoint}}
\multiput(1289,728)(18.564,9.282){0}{\usebox{\plotpoint}}
\put(1306.00,736.08){\usebox{\plotpoint}}
\put(1324.79,744.89){\usebox{\plotpoint}}
\multiput(1325,745)(18.845,8.698){0}{\usebox{\plotpoint}}
\put(1343.54,753.77){\usebox{\plotpoint}}
\multiput(1350,757)(18.564,9.282){0}{\usebox{\plotpoint}}
\put(1362.11,763.05){\usebox{\plotpoint}}
\put(1380.87,771.93){\usebox{\plotpoint}}
\multiput(1387,775)(19.159,7.983){0}{\usebox{\plotpoint}}
\put(1399.80,780.40){\usebox{\plotpoint}}
\put(1418.48,789.45){\usebox{\plotpoint}}
\multiput(1424,792)(18.564,9.282){0}{\usebox{\plotpoint}}
\put(1436,798){\usebox{\plotpoint}}
\end{picture}

%% file: Bstring.tex
% GNUPLOT: LaTeX picture
\setlength{\unitlength}{0.240900pt}
\ifx\plotpoint\undefined\newsavebox{\plotpoint}\fi
\sbox{\plotpoint}{\rule[-0.200pt]{0.400pt}{0.400pt}}%
\begin{picture}(1500,900)(0,0)
\font\gnuplot=cmr10 at 10pt
\gnuplot
\sbox{\plotpoint}{\rule[-0.200pt]{0.400pt}{0.400pt}}%
\put(220.0,495.0){\rule[-0.200pt]{292.934pt}{0.400pt}}
\put(828.0,113.0){\rule[-0.200pt]{0.400pt}{184.048pt}}
\put(220.0,113.0){\rule[-0.200pt]{4.818pt}{0.400pt}}
\put(198,113){\makebox(0,0)[r]{-3}}
\put(1416.0,113.0){\rule[-0.200pt]{4.818pt}{0.400pt}}
\put(220.0,240.0){\rule[-0.200pt]{4.818pt}{0.400pt}}
\put(198,240){\makebox(0,0)[r]{-2}}
\put(1416.0,240.0){\rule[-0.200pt]{4.818pt}{0.400pt}}
\put(220.0,368.0){\rule[-0.200pt]{4.818pt}{0.400pt}}
\put(198,368){\makebox(0,0)[r]{-1}}
\put(1416.0,368.0){\rule[-0.200pt]{4.818pt}{0.400pt}}
\put(220.0,495.0){\rule[-0.200pt]{4.818pt}{0.400pt}}
\put(198,495){\makebox(0,0)[r]{0}}
\put(1416.0,495.0){\rule[-0.200pt]{4.818pt}{0.400pt}}
\put(220.0,622.0){\rule[-0.200pt]{4.818pt}{0.400pt}}
\put(198,622){\makebox(0,0)[r]{1}}
\put(1416.0,622.0){\rule[-0.200pt]{4.818pt}{0.400pt}}
\put(220.0,750.0){\rule[-0.200pt]{4.818pt}{0.400pt}}
\put(198,750){\makebox(0,0)[r]{2}}
\put(1416.0,750.0){\rule[-0.200pt]{4.818pt}{0.400pt}}
\put(220.0,877.0){\rule[-0.200pt]{4.818pt}{0.400pt}}
\put(198,877){\makebox(0,0)[r]{3}}
\put(1416.0,877.0){\rule[-0.200pt]{4.818pt}{0.400pt}}
\put(220.0,113.0){\rule[-0.200pt]{0.400pt}{4.818pt}}
\put(220,68){\makebox(0,0){-1}}
\put(220.0,857.0){\rule[-0.200pt]{0.400pt}{4.818pt}}
\put(342.0,113.0){\rule[-0.200pt]{0.400pt}{4.818pt}}
\put(342,68){\makebox(0,0){-0.8}}
\put(342.0,857.0){\rule[-0.200pt]{0.400pt}{4.818pt}}
\put(463.0,113.0){\rule[-0.200pt]{0.400pt}{4.818pt}}
\put(463,68){\makebox(0,0){-0.6}}
\put(463.0,857.0){\rule[-0.200pt]{0.400pt}{4.818pt}}
\put(585.0,113.0){\rule[-0.200pt]{0.400pt}{4.818pt}}
\put(585,68){\makebox(0,0){-0.4}}
\put(585.0,857.0){\rule[-0.200pt]{0.400pt}{4.818pt}}
\put(706.0,113.0){\rule[-0.200pt]{0.400pt}{4.818pt}}
\put(706,68){\makebox(0,0){-0.2}}
\put(706.0,857.0){\rule[-0.200pt]{0.400pt}{4.818pt}}
\put(828.0,113.0){\rule[-0.200pt]{0.400pt}{4.818pt}}
\put(828,68){\makebox(0,0){0}}
\put(828.0,857.0){\rule[-0.200pt]{0.400pt}{4.818pt}}
\put(950.0,113.0){\rule[-0.200pt]{0.400pt}{4.818pt}}
\put(950,68){\makebox(0,0){0.2}}
\put(950.0,857.0){\rule[-0.200pt]{0.400pt}{4.818pt}}
\put(1071.0,113.0){\rule[-0.200pt]{0.400pt}{4.818pt}}
\put(1071,68){\makebox(0,0){0.4}}
\put(1071.0,857.0){\rule[-0.200pt]{0.400pt}{4.818pt}}
\put(1193.0,113.0){\rule[-0.200pt]{0.400pt}{4.818pt}}
\put(1193,68){\makebox(0,0){0.6}}
\put(1193.0,857.0){\rule[-0.200pt]{0.400pt}{4.818pt}}
\put(1314.0,113.0){\rule[-0.200pt]{0.400pt}{4.818pt}}
\put(1314,68){\makebox(0,0){0.8}}
\put(1314.0,857.0){\rule[-0.200pt]{0.400pt}{4.818pt}}
\put(1436.0,113.0){\rule[-0.200pt]{0.400pt}{4.818pt}}
\put(1436,68){\makebox(0,0){1}}
\put(1436.0,857.0){\rule[-0.200pt]{0.400pt}{4.818pt}}
\put(220.0,113.0){\rule[-0.200pt]{292.934pt}{0.400pt}}
\put(1436.0,113.0){\rule[-0.200pt]{0.400pt}{184.048pt}}
\put(220.0,877.0){\rule[-0.200pt]{292.934pt}{0.400pt}}
\put(45,495){\makebox(0,0)
{$\displaystyle{\frac{B^{(0)}}{M_{1/2}^{(0)}}}$}}
\put(828,23){\makebox(0,0){$\sin \theta$}}
\put(1071,813){\makebox(0,0)[r]{$B_Z^{(0)}$}}
\put(646,177){\makebox(0,0)[r]{$B_Z^{(0)}$}}
\put(1314,520){\makebox(0,0)[r]{$B_\lambda^{(0)}$ }}
\put(585,813){\makebox(0,0)[r]{$B_\mu^{(0)}$ }}
\put(220.0,113.0){\rule[-0.200pt]{0.400pt}{184.048pt}}
\sbox{\plotpoint}{\rule[-0.400pt]{0.800pt}{0.800pt}}%
\put(220,348){\usebox{\plotpoint}}
\put(220,344.84){\rule{2.891pt}{0.800pt}}
\multiput(220.00,346.34)(6.000,-3.000){2}{\rule{1.445pt}{0.800pt}}
\put(232,341.84){\rule{3.132pt}{0.800pt}}
\multiput(232.00,343.34)(6.500,-3.000){2}{\rule{1.566pt}{0.800pt}}
\put(245,338.34){\rule{2.600pt}{0.800pt}}
\multiput(245.00,340.34)(6.604,-4.000){2}{\rule{1.300pt}{0.800pt}}
\put(257,334.84){\rule{2.891pt}{0.800pt}}
\multiput(257.00,336.34)(6.000,-3.000){2}{\rule{1.445pt}{0.800pt}}
\put(269,331.34){\rule{2.600pt}{0.800pt}}
\multiput(269.00,333.34)(6.604,-4.000){2}{\rule{1.300pt}{0.800pt}}
\put(281,327.84){\rule{3.132pt}{0.800pt}}
\multiput(281.00,329.34)(6.500,-3.000){2}{\rule{1.566pt}{0.800pt}}
\put(294,324.34){\rule{2.600pt}{0.800pt}}
\multiput(294.00,326.34)(6.604,-4.000){2}{\rule{1.300pt}{0.800pt}}
\put(306,320.34){\rule{2.600pt}{0.800pt}}
\multiput(306.00,322.34)(6.604,-4.000){2}{\rule{1.300pt}{0.800pt}}
\multiput(318.00,318.06)(1.768,-0.560){3}{\rule{2.280pt}{0.135pt}}
\multiput(318.00,318.34)(8.268,-5.000){2}{\rule{1.140pt}{0.800pt}}
\put(331,311.34){\rule{2.600pt}{0.800pt}}
\multiput(331.00,313.34)(6.604,-4.000){2}{\rule{1.300pt}{0.800pt}}
\multiput(343.00,309.06)(1.600,-0.560){3}{\rule{2.120pt}{0.135pt}}
\multiput(343.00,309.34)(7.600,-5.000){2}{\rule{1.060pt}{0.800pt}}
\multiput(355.00,304.06)(1.600,-0.560){3}{\rule{2.120pt}{0.135pt}}
\multiput(355.00,304.34)(7.600,-5.000){2}{\rule{1.060pt}{0.800pt}}
\multiput(367.00,299.06)(1.768,-0.560){3}{\rule{2.280pt}{0.135pt}}
\multiput(367.00,299.34)(8.268,-5.000){2}{\rule{1.140pt}{0.800pt}}
\multiput(380.00,294.07)(1.132,-0.536){5}{\rule{1.800pt}{0.129pt}}
\multiput(380.00,294.34)(8.264,-6.000){2}{\rule{0.900pt}{0.800pt}}
\multiput(392.00,288.07)(1.132,-0.536){5}{\rule{1.800pt}{0.129pt}}
\multiput(392.00,288.34)(8.264,-6.000){2}{\rule{0.900pt}{0.800pt}}
\multiput(404.00,282.07)(1.244,-0.536){5}{\rule{1.933pt}{0.129pt}}
\multiput(404.00,282.34)(8.987,-6.000){2}{\rule{0.967pt}{0.800pt}}
\multiput(417.00,276.08)(0.913,-0.526){7}{\rule{1.571pt}{0.127pt}}
\multiput(417.00,276.34)(8.738,-7.000){2}{\rule{0.786pt}{0.800pt}}
\multiput(429.00,269.08)(0.913,-0.526){7}{\rule{1.571pt}{0.127pt}}
\multiput(429.00,269.34)(8.738,-7.000){2}{\rule{0.786pt}{0.800pt}}
\multiput(441.00,262.08)(0.774,-0.520){9}{\rule{1.400pt}{0.125pt}}
\multiput(441.00,262.34)(9.094,-8.000){2}{\rule{0.700pt}{0.800pt}}
\multiput(453.00,254.08)(0.847,-0.520){9}{\rule{1.500pt}{0.125pt}}
\multiput(453.00,254.34)(9.887,-8.000){2}{\rule{0.750pt}{0.800pt}}
\multiput(466.00,246.08)(0.774,-0.520){9}{\rule{1.400pt}{0.125pt}}
\multiput(466.00,246.34)(9.094,-8.000){2}{\rule{0.700pt}{0.800pt}}
\multiput(478.00,238.08)(0.599,-0.514){13}{\rule{1.160pt}{0.124pt}}
\multiput(478.00,238.34)(9.592,-10.000){2}{\rule{0.580pt}{0.800pt}}
\multiput(490.00,228.08)(0.654,-0.514){13}{\rule{1.240pt}{0.124pt}}
\multiput(490.00,228.34)(10.426,-10.000){2}{\rule{0.620pt}{0.800pt}}
\multiput(503.00,218.08)(0.599,-0.514){13}{\rule{1.160pt}{0.124pt}}
\multiput(503.00,218.34)(9.592,-10.000){2}{\rule{0.580pt}{0.800pt}}
\multiput(515.00,208.08)(0.491,-0.511){17}{\rule{1.000pt}{0.123pt}}
\multiput(515.00,208.34)(9.924,-12.000){2}{\rule{0.500pt}{0.800pt}}
\multiput(528.41,193.57)(0.511,-0.536){17}{\rule{0.123pt}{1.067pt}}
\multiput(525.34,195.79)(12.000,-10.786){2}{\rule{0.800pt}{0.533pt}}
\multiput(539.00,183.08)(0.492,-0.509){19}{\rule{1.000pt}{0.123pt}}
\multiput(539.00,183.34)(10.924,-13.000){2}{\rule{0.500pt}{0.800pt}}
\multiput(553.41,166.74)(0.511,-0.671){17}{\rule{0.123pt}{1.267pt}}
\multiput(550.34,169.37)(12.000,-13.371){2}{\rule{0.800pt}{0.633pt}}
\multiput(565.41,150.74)(0.511,-0.671){17}{\rule{0.123pt}{1.267pt}}
\multiput(562.34,153.37)(12.000,-13.371){2}{\rule{0.800pt}{0.633pt}}
\multiput(577.41,134.19)(0.511,-0.762){17}{\rule{0.123pt}{1.400pt}}
\multiput(574.34,137.09)(12.000,-15.094){2}{\rule{0.800pt}{0.700pt}}
\multiput(589.39,116.19)(0.536,-0.797){5}{\rule{0.129pt}{1.400pt}}
\multiput(586.34,119.09)(6.000,-6.094){2}{\rule{0.800pt}{0.700pt}}
\multiput(1063.39,871.19)(0.536,-0.797){5}{\rule{0.129pt}{1.400pt}}
\multiput(1060.34,874.09)(6.000,-6.094){2}{\rule{0.800pt}{0.700pt}}
\multiput(1069.41,862.19)(0.511,-0.762){17}{\rule{0.123pt}{1.400pt}}
\multiput(1066.34,865.09)(12.000,-15.094){2}{\rule{0.800pt}{0.700pt}}
\multiput(1081.41,844.74)(0.511,-0.671){17}{\rule{0.123pt}{1.267pt}}
\multiput(1078.34,847.37)(12.000,-13.371){2}{\rule{0.800pt}{0.633pt}}
\multiput(1093.41,828.74)(0.511,-0.671){17}{\rule{0.123pt}{1.267pt}}
\multiput(1090.34,831.37)(12.000,-13.371){2}{\rule{0.800pt}{0.633pt}}
\multiput(1104.00,816.08)(0.492,-0.509){19}{\rule{1.000pt}{0.123pt}}
\multiput(1104.00,816.34)(10.924,-13.000){2}{\rule{0.500pt}{0.800pt}}
\multiput(1118.41,800.57)(0.511,-0.536){17}{\rule{0.123pt}{1.067pt}}
\multiput(1115.34,802.79)(12.000,-10.786){2}{\rule{0.800pt}{0.533pt}}
\multiput(1129.00,790.08)(0.491,-0.511){17}{\rule{1.000pt}{0.123pt}}
\multiput(1129.00,790.34)(9.924,-12.000){2}{\rule{0.500pt}{0.800pt}}
\multiput(1141.00,778.08)(0.599,-0.514){13}{\rule{1.160pt}{0.124pt}}
\multiput(1141.00,778.34)(9.592,-10.000){2}{\rule{0.580pt}{0.800pt}}
\multiput(1153.00,768.08)(0.654,-0.514){13}{\rule{1.240pt}{0.124pt}}
\multiput(1153.00,768.34)(10.426,-10.000){2}{\rule{0.620pt}{0.800pt}}
\multiput(1166.00,758.08)(0.599,-0.514){13}{\rule{1.160pt}{0.124pt}}
\multiput(1166.00,758.34)(9.592,-10.000){2}{\rule{0.580pt}{0.800pt}}
\multiput(1178.00,748.08)(0.774,-0.520){9}{\rule{1.400pt}{0.125pt}}
\multiput(1178.00,748.34)(9.094,-8.000){2}{\rule{0.700pt}{0.800pt}}
\multiput(1190.00,740.08)(0.847,-0.520){9}{\rule{1.500pt}{0.125pt}}
\multiput(1190.00,740.34)(9.887,-8.000){2}{\rule{0.750pt}{0.800pt}}
\multiput(1203.00,732.08)(0.774,-0.520){9}{\rule{1.400pt}{0.125pt}}
\multiput(1203.00,732.34)(9.094,-8.000){2}{\rule{0.700pt}{0.800pt}}
\multiput(1215.00,724.08)(0.913,-0.526){7}{\rule{1.571pt}{0.127pt}}
\multiput(1215.00,724.34)(8.738,-7.000){2}{\rule{0.786pt}{0.800pt}}
\multiput(1227.00,717.08)(0.913,-0.526){7}{\rule{1.571pt}{0.127pt}}
\multiput(1227.00,717.34)(8.738,-7.000){2}{\rule{0.786pt}{0.800pt}}
\multiput(1239.00,710.07)(1.244,-0.536){5}{\rule{1.933pt}{0.129pt}}
\multiput(1239.00,710.34)(8.987,-6.000){2}{\rule{0.967pt}{0.800pt}}
\multiput(1252.00,704.07)(1.132,-0.536){5}{\rule{1.800pt}{0.129pt}}
\multiput(1252.00,704.34)(8.264,-6.000){2}{\rule{0.900pt}{0.800pt}}
\multiput(1264.00,698.07)(1.132,-0.536){5}{\rule{1.800pt}{0.129pt}}
\multiput(1264.00,698.34)(8.264,-6.000){2}{\rule{0.900pt}{0.800pt}}
\multiput(1276.00,692.06)(1.768,-0.560){3}{\rule{2.280pt}{0.135pt}}
\multiput(1276.00,692.34)(8.268,-5.000){2}{\rule{1.140pt}{0.800pt}}
\multiput(1289.00,687.06)(1.600,-0.560){3}{\rule{2.120pt}{0.135pt}}
\multiput(1289.00,687.34)(7.600,-5.000){2}{\rule{1.060pt}{0.800pt}}
\multiput(1301.00,682.06)(1.600,-0.560){3}{\rule{2.120pt}{0.135pt}}
\multiput(1301.00,682.34)(7.600,-5.000){2}{\rule{1.060pt}{0.800pt}}
\put(1313,675.34){\rule{2.600pt}{0.800pt}}
\multiput(1313.00,677.34)(6.604,-4.000){2}{\rule{1.300pt}{0.800pt}}
\multiput(1325.00,673.06)(1.768,-0.560){3}{\rule{2.280pt}{0.135pt}}
\multiput(1325.00,673.34)(8.268,-5.000){2}{\rule{1.140pt}{0.800pt}}
\put(1338,666.34){\rule{2.600pt}{0.800pt}}
\multiput(1338.00,668.34)(6.604,-4.000){2}{\rule{1.300pt}{0.800pt}}
\put(1350,662.34){\rule{2.600pt}{0.800pt}}
\multiput(1350.00,664.34)(6.604,-4.000){2}{\rule{1.300pt}{0.800pt}}
\put(1362,658.84){\rule{3.132pt}{0.800pt}}
\multiput(1362.00,660.34)(6.500,-3.000){2}{\rule{1.566pt}{0.800pt}}
\put(1375,655.34){\rule{2.600pt}{0.800pt}}
\multiput(1375.00,657.34)(6.604,-4.000){2}{\rule{1.300pt}{0.800pt}}
\put(1387,651.84){\rule{2.891pt}{0.800pt}}
\multiput(1387.00,653.34)(6.000,-3.000){2}{\rule{1.445pt}{0.800pt}}
\put(1399,648.34){\rule{2.600pt}{0.800pt}}
\multiput(1399.00,650.34)(6.604,-4.000){2}{\rule{1.300pt}{0.800pt}}
\put(1411,644.84){\rule{3.132pt}{0.800pt}}
\multiput(1411.00,646.34)(6.500,-3.000){2}{\rule{1.566pt}{0.800pt}}
\put(1424,641.84){\rule{2.891pt}{0.800pt}}
\multiput(1424.00,643.34)(6.000,-3.000){2}{\rule{1.445pt}{0.800pt}}
\sbox{\plotpoint}{\rule[-0.500pt]{1.000pt}{1.000pt}}%
\put(220,348){\usebox{\plotpoint}}
\multiput(220,348)(7.093,-19.506){2}{\usebox{\plotpoint}}
\put(236.03,310.04){\usebox{\plotpoint}}
\put(249.25,294.04){\usebox{\plotpoint}}
\put(263.01,278.49){\usebox{\plotpoint}}
\put(277.43,263.57){\usebox{\plotpoint}}
\put(292.54,249.35){\usebox{\plotpoint}}
\multiput(294,248)(14.676,-14.676){0}{\usebox{\plotpoint}}
\put(307.27,234.73){\usebox{\plotpoint}}
\put(322.10,220.22){\usebox{\plotpoint}}
\put(337.11,205.89){\usebox{\plotpoint}}
\put(351.43,190.87){\usebox{\plotpoint}}
\put(365.51,175.62){\usebox{\plotpoint}}
\multiput(367,174)(14.676,-14.676){0}{\usebox{\plotpoint}}
\put(380.11,160.87){\usebox{\plotpoint}}
\put(393.62,145.11){\usebox{\plotpoint}}
\put(407.15,129.37){\usebox{\plotpoint}}
\put(419.94,113.10){\usebox{\plotpoint}}
\put(420,113){\usebox{\plotpoint}}
\put(1236.00,877.00){\usebox{\plotpoint}}
\put(1248.77,860.72){\usebox{\plotpoint}}
\put(1262.30,844.98){\usebox{\plotpoint}}
\put(1275.81,829.22){\usebox{\plotpoint}}
\multiput(1276,829)(14.676,-14.676){0}{\usebox{\plotpoint}}
\put(1290.41,814.47){\usebox{\plotpoint}}
\put(1304.49,799.22){\usebox{\plotpoint}}
\put(1318.80,784.20){\usebox{\plotpoint}}
\put(1333.81,769.87){\usebox{\plotpoint}}
\put(1348.65,755.35){\usebox{\plotpoint}}
\multiput(1350,754)(14.676,-14.676){0}{\usebox{\plotpoint}}
\put(1363.37,740.73){\usebox{\plotpoint}}
\put(1378.49,726.51){\usebox{\plotpoint}}
\put(1392.91,711.59){\usebox{\plotpoint}}
\put(1406.67,696.06){\usebox{\plotpoint}}
\put(1419.89,680.06){\usebox{\plotpoint}}
\multiput(1424,675)(7.093,-19.506){2}{\usebox{\plotpoint}}
\put(1436,642){\usebox{\plotpoint}}
\put(220,348){\usebox{\plotpoint}}
\multiput(220,348)(8.430,18.967){2}{\usebox{\plotpoint}}
\put(241.48,382.29){\usebox{\plotpoint}}
\multiput(245,385)(17.928,10.458){0}{\usebox{\plotpoint}}
\put(259.17,393.09){\usebox{\plotpoint}}
\put(278.01,401.76){\usebox{\plotpoint}}
\multiput(281,403)(19.372,7.451){0}{\usebox{\plotpoint}}
\put(297.41,409.14){\usebox{\plotpoint}}
\put(317.35,414.84){\usebox{\plotpoint}}
\multiput(318,415)(19.838,6.104){0}{\usebox{\plotpoint}}
\put(337.29,420.57){\usebox{\plotpoint}}
\multiput(343,422)(20.136,5.034){0}{\usebox{\plotpoint}}
\put(357.43,425.61){\usebox{\plotpoint}}
\put(377.76,429.66){\usebox{\plotpoint}}
\multiput(380,430)(20.136,5.034){0}{\usebox{\plotpoint}}
\put(398.04,434.01){\usebox{\plotpoint}}
\multiput(404,435)(20.224,4.667){0}{\usebox{\plotpoint}}
\put(418.35,438.22){\usebox{\plotpoint}}
\put(438.82,441.64){\usebox{\plotpoint}}
\multiput(441,442)(20.473,3.412){0}{\usebox{\plotpoint}}
\put(459.31,444.97){\usebox{\plotpoint}}
\multiput(466,446)(20.473,3.412){0}{\usebox{\plotpoint}}
\put(479.79,448.30){\usebox{\plotpoint}}
\put(500.29,451.58){\usebox{\plotpoint}}
\multiput(503,452)(20.473,3.412){0}{\usebox{\plotpoint}}
\put(520.77,454.96){\usebox{\plotpoint}}
\multiput(527,456)(20.473,3.412){0}{\usebox{\plotpoint}}
\put(541.24,458.35){\usebox{\plotpoint}}
\put(561.84,460.82){\usebox{\plotpoint}}
\multiput(564,461)(20.473,3.412){0}{\usebox{\plotpoint}}
\put(582.33,464.06){\usebox{\plotpoint}}
\multiput(588,465)(20.694,1.592){0}{\usebox{\plotpoint}}
\put(602.95,466.32){\usebox{\plotpoint}}
\put(623.42,469.74){\usebox{\plotpoint}}
\multiput(625,470)(20.694,1.592){0}{\usebox{\plotpoint}}
\put(644.03,472.01){\usebox{\plotpoint}}
\multiput(650,473)(20.473,3.412){0}{\usebox{\plotpoint}}
\put(664.53,475.21){\usebox{\plotpoint}}
\put(685.12,477.71){\usebox{\plotpoint}}
\multiput(687,478)(20.684,1.724){0}{\usebox{\plotpoint}}
\put(705.72,480.12){\usebox{\plotpoint}}
\multiput(711,481)(20.694,1.592){0}{\usebox{\plotpoint}}
\put(726.33,482.39){\usebox{\plotpoint}}
\put(746.92,484.91){\usebox{\plotpoint}}
\multiput(748,485)(20.473,3.412){0}{\usebox{\plotpoint}}
\put(767.48,487.58){\usebox{\plotpoint}}
\multiput(773,488)(20.473,3.412){0}{\usebox{\plotpoint}}
\put(788.05,490.25){\usebox{\plotpoint}}
\put(808.63,492.79){\usebox{\plotpoint}}
\multiput(810,493)(20.684,1.724){0}{\usebox{\plotpoint}}
\put(829.23,495.21){\usebox{\plotpoint}}
\multiput(834,496)(20.684,1.724){0}{\usebox{\plotpoint}}
\put(849.83,497.59){\usebox{\plotpoint}}
\put(870.44,499.95){\usebox{\plotpoint}}
\multiput(871,500)(20.473,3.412){0}{\usebox{\plotpoint}}
\put(891.01,502.62){\usebox{\plotpoint}}
\multiput(896,503)(20.473,3.412){0}{\usebox{\plotpoint}}
\put(911.57,505.30){\usebox{\plotpoint}}
\multiput(920,506)(20.473,3.412){0}{\usebox{\plotpoint}}
\put(932.13,508.01){\usebox{\plotpoint}}
\put(952.74,510.29){\usebox{\plotpoint}}
\multiput(957,511)(20.684,1.724){0}{\usebox{\plotpoint}}
\put(973.34,512.67){\usebox{\plotpoint}}
\put(993.96,515.00){\usebox{\plotpoint}}
\multiput(994,515)(20.473,3.412){0}{\usebox{\plotpoint}}
\put(1014.43,518.41){\usebox{\plotpoint}}
\multiput(1018,519)(20.694,1.592){0}{\usebox{\plotpoint}}
\put(1035.04,520.67){\usebox{\plotpoint}}
\multiput(1043,522)(20.473,3.412){0}{\usebox{\plotpoint}}
\put(1055.52,524.04){\usebox{\plotpoint}}
\put(1076.13,526.35){\usebox{\plotpoint}}
\multiput(1080,527)(20.473,3.412){0}{\usebox{\plotpoint}}
\put(1096.65,529.39){\usebox{\plotpoint}}
\multiput(1104,530)(20.514,3.156){0}{\usebox{\plotpoint}}
\put(1117.22,532.04){\usebox{\plotpoint}}
\put(1137.70,535.45){\usebox{\plotpoint}}
\multiput(1141,536)(20.473,3.412){0}{\usebox{\plotpoint}}
\put(1158.18,538.80){\usebox{\plotpoint}}
\multiput(1166,540)(20.473,3.412){0}{\usebox{\plotpoint}}
\put(1178.67,542.11){\usebox{\plotpoint}}
\put(1199.16,545.41){\usebox{\plotpoint}}
\multiput(1203,546)(20.473,3.412){0}{\usebox{\plotpoint}}
\put(1219.64,548.77){\usebox{\plotpoint}}
\multiput(1227,550)(20.473,3.412){0}{\usebox{\plotpoint}}
\put(1240.10,552.25){\usebox{\plotpoint}}
\put(1260.43,556.40){\usebox{\plotpoint}}
\multiput(1264,557)(20.136,5.034){0}{\usebox{\plotpoint}}
\put(1280.71,560.72){\usebox{\plotpoint}}
\put(1301.00,565.00){\usebox{\plotpoint}}
\multiput(1301,565)(20.136,5.034){0}{\usebox{\plotpoint}}
\put(1321.13,570.03){\usebox{\plotpoint}}
\multiput(1325,571)(19.838,6.104){0}{\usebox{\plotpoint}}
\put(1341.07,575.77){\usebox{\plotpoint}}
\put(1360.96,581.65){\usebox{\plotpoint}}
\multiput(1362,582)(19.372,7.451){0}{\usebox{\plotpoint}}
\put(1380.29,589.20){\usebox{\plotpoint}}
\multiput(1387,592)(18.564,9.282){0}{\usebox{\plotpoint}}
\put(1399.06,598.04){\usebox{\plotpoint}}
\put(1416.50,609.23){\usebox{\plotpoint}}
\put(1428.58,625.31){\usebox{\plotpoint}}
\put(1436,642){\usebox{\plotpoint}}
\sbox{\plotpoint}{\rule[-0.600pt]{1.200pt}{1.200pt}}%
\put(220,441){\usebox{\plotpoint}}
\multiput(222.24,434.36)(0.501,-0.489){14}{\rule{0.121pt}{1.600pt}}
\multiput(217.51,437.68)(12.000,-9.679){2}{\rule{1.200pt}{0.800pt}}
\put(232,423.01){\rule{3.132pt}{1.200pt}}
\multiput(232.00,425.51)(6.500,-5.000){2}{\rule{1.566pt}{1.200pt}}
\put(245,418.51){\rule{2.891pt}{1.200pt}}
\multiput(245.00,420.51)(6.000,-4.000){2}{\rule{1.445pt}{1.200pt}}
\put(257,415.01){\rule{2.891pt}{1.200pt}}
\multiput(257.00,416.51)(6.000,-3.000){2}{\rule{1.445pt}{1.200pt}}
\put(269,412.51){\rule{2.891pt}{1.200pt}}
\multiput(269.00,413.51)(6.000,-2.000){2}{\rule{1.445pt}{1.200pt}}
\put(281,410.01){\rule{3.132pt}{1.200pt}}
\multiput(281.00,411.51)(6.500,-3.000){2}{\rule{1.566pt}{1.200pt}}
\put(294,407.51){\rule{2.891pt}{1.200pt}}
\multiput(294.00,408.51)(6.000,-2.000){2}{\rule{1.445pt}{1.200pt}}
\put(306,406.01){\rule{2.891pt}{1.200pt}}
\multiput(306.00,406.51)(6.000,-1.000){2}{\rule{1.445pt}{1.200pt}}
\put(318,404.51){\rule{3.132pt}{1.200pt}}
\multiput(318.00,405.51)(6.500,-2.000){2}{\rule{1.566pt}{1.200pt}}
\put(331,402.51){\rule{2.891pt}{1.200pt}}
\multiput(331.00,403.51)(6.000,-2.000){2}{\rule{1.445pt}{1.200pt}}
\put(343,401.01){\rule{2.891pt}{1.200pt}}
\multiput(343.00,401.51)(6.000,-1.000){2}{\rule{1.445pt}{1.200pt}}
\put(355,399.51){\rule{2.891pt}{1.200pt}}
\multiput(355.00,400.51)(6.000,-2.000){2}{\rule{1.445pt}{1.200pt}}
\put(367,398.01){\rule{3.132pt}{1.200pt}}
\multiput(367.00,398.51)(6.500,-1.000){2}{\rule{1.566pt}{1.200pt}}
\put(380,397.01){\rule{2.891pt}{1.200pt}}
\multiput(380.00,397.51)(6.000,-1.000){2}{\rule{1.445pt}{1.200pt}}
\put(392,396.01){\rule{2.891pt}{1.200pt}}
\multiput(392.00,396.51)(6.000,-1.000){2}{\rule{1.445pt}{1.200pt}}
\put(404,394.51){\rule{3.132pt}{1.200pt}}
\multiput(404.00,395.51)(6.500,-2.000){2}{\rule{1.566pt}{1.200pt}}
\put(417,393.01){\rule{2.891pt}{1.200pt}}
\multiput(417.00,393.51)(6.000,-1.000){2}{\rule{1.445pt}{1.200pt}}
\put(429,392.01){\rule{2.891pt}{1.200pt}}
\multiput(429.00,392.51)(6.000,-1.000){2}{\rule{1.445pt}{1.200pt}}
\put(441,391.01){\rule{2.891pt}{1.200pt}}
\multiput(441.00,391.51)(6.000,-1.000){2}{\rule{1.445pt}{1.200pt}}
\put(453,390.01){\rule{3.132pt}{1.200pt}}
\multiput(453.00,390.51)(6.500,-1.000){2}{\rule{1.566pt}{1.200pt}}
\put(466,389.01){\rule{2.891pt}{1.200pt}}
\multiput(466.00,389.51)(6.000,-1.000){2}{\rule{1.445pt}{1.200pt}}
\put(478,388.01){\rule{2.891pt}{1.200pt}}
\multiput(478.00,388.51)(6.000,-1.000){2}{\rule{1.445pt}{1.200pt}}
\put(490,387.01){\rule{3.132pt}{1.200pt}}
\multiput(490.00,387.51)(6.500,-1.000){2}{\rule{1.566pt}{1.200pt}}
\put(503,386.01){\rule{2.891pt}{1.200pt}}
\multiput(503.00,386.51)(6.000,-1.000){2}{\rule{1.445pt}{1.200pt}}
\put(515,385.01){\rule{2.891pt}{1.200pt}}
\multiput(515.00,385.51)(6.000,-1.000){2}{\rule{1.445pt}{1.200pt}}
\put(527,384.01){\rule{2.891pt}{1.200pt}}
\multiput(527.00,384.51)(6.000,-1.000){2}{\rule{1.445pt}{1.200pt}}
\put(539,383.01){\rule{3.132pt}{1.200pt}}
\multiput(539.00,383.51)(6.500,-1.000){2}{\rule{1.566pt}{1.200pt}}
\put(552,382.01){\rule{2.891pt}{1.200pt}}
\multiput(552.00,382.51)(6.000,-1.000){2}{\rule{1.445pt}{1.200pt}}
\put(576,381.01){\rule{2.891pt}{1.200pt}}
\multiput(576.00,381.51)(6.000,-1.000){2}{\rule{1.445pt}{1.200pt}}
\put(588,380.01){\rule{3.132pt}{1.200pt}}
\multiput(588.00,380.51)(6.500,-1.000){2}{\rule{1.566pt}{1.200pt}}
\put(601,379.01){\rule{2.891pt}{1.200pt}}
\multiput(601.00,379.51)(6.000,-1.000){2}{\rule{1.445pt}{1.200pt}}
\put(613,378.01){\rule{2.891pt}{1.200pt}}
\multiput(613.00,378.51)(6.000,-1.000){2}{\rule{1.445pt}{1.200pt}}
\put(625,377.01){\rule{3.132pt}{1.200pt}}
\multiput(625.00,377.51)(6.500,-1.000){2}{\rule{1.566pt}{1.200pt}}
\put(564.0,384.0){\rule[-0.600pt]{2.891pt}{1.200pt}}
\put(650,376.01){\rule{2.891pt}{1.200pt}}
\multiput(650.00,376.51)(6.000,-1.000){2}{\rule{1.445pt}{1.200pt}}
\put(662,375.01){\rule{2.891pt}{1.200pt}}
\multiput(662.00,375.51)(6.000,-1.000){2}{\rule{1.445pt}{1.200pt}}
\put(674,374.01){\rule{3.132pt}{1.200pt}}
\multiput(674.00,374.51)(6.500,-1.000){2}{\rule{1.566pt}{1.200pt}}
\put(638.0,379.0){\rule[-0.600pt]{2.891pt}{1.200pt}}
\put(699,373.01){\rule{2.891pt}{1.200pt}}
\multiput(699.00,373.51)(6.000,-1.000){2}{\rule{1.445pt}{1.200pt}}
\put(711,372.01){\rule{3.132pt}{1.200pt}}
\multiput(711.00,372.51)(6.500,-1.000){2}{\rule{1.566pt}{1.200pt}}
\put(724,371.01){\rule{2.891pt}{1.200pt}}
\multiput(724.00,371.51)(6.000,-1.000){2}{\rule{1.445pt}{1.200pt}}
\put(687.0,376.0){\rule[-0.600pt]{2.891pt}{1.200pt}}
\put(748,370.01){\rule{2.891pt}{1.200pt}}
\multiput(748.00,370.51)(6.000,-1.000){2}{\rule{1.445pt}{1.200pt}}
\put(760,369.01){\rule{3.132pt}{1.200pt}}
\multiput(760.00,369.51)(6.500,-1.000){2}{\rule{1.566pt}{1.200pt}}
\put(773,368.01){\rule{2.891pt}{1.200pt}}
\multiput(773.00,368.51)(6.000,-1.000){2}{\rule{1.445pt}{1.200pt}}
\put(736.0,373.0){\rule[-0.600pt]{2.891pt}{1.200pt}}
\put(797,367.01){\rule{3.132pt}{1.200pt}}
\multiput(797.00,367.51)(6.500,-1.000){2}{\rule{1.566pt}{1.200pt}}
\put(810,366.01){\rule{2.891pt}{1.200pt}}
\multiput(810.00,366.51)(6.000,-1.000){2}{\rule{1.445pt}{1.200pt}}
\put(822,365.01){\rule{2.891pt}{1.200pt}}
\multiput(822.00,365.51)(6.000,-1.000){2}{\rule{1.445pt}{1.200pt}}
\put(785.0,370.0){\rule[-0.600pt]{2.891pt}{1.200pt}}
\put(846,364.01){\rule{3.132pt}{1.200pt}}
\multiput(846.00,364.51)(6.500,-1.000){2}{\rule{1.566pt}{1.200pt}}
\put(859,363.01){\rule{2.891pt}{1.200pt}}
\multiput(859.00,363.51)(6.000,-1.000){2}{\rule{1.445pt}{1.200pt}}
\put(871,362.01){\rule{2.891pt}{1.200pt}}
\multiput(871.00,362.51)(6.000,-1.000){2}{\rule{1.445pt}{1.200pt}}
\put(834.0,367.0){\rule[-0.600pt]{2.891pt}{1.200pt}}
\put(896,361.01){\rule{2.891pt}{1.200pt}}
\multiput(896.00,361.51)(6.000,-1.000){2}{\rule{1.445pt}{1.200pt}}
\put(908,360.01){\rule{2.891pt}{1.200pt}}
\multiput(908.00,360.51)(6.000,-1.000){2}{\rule{1.445pt}{1.200pt}}
\put(920,359.01){\rule{2.891pt}{1.200pt}}
\multiput(920.00,359.51)(6.000,-1.000){2}{\rule{1.445pt}{1.200pt}}
\put(883.0,364.0){\rule[-0.600pt]{3.132pt}{1.200pt}}
\put(945,358.01){\rule{2.891pt}{1.200pt}}
\multiput(945.00,358.51)(6.000,-1.000){2}{\rule{1.445pt}{1.200pt}}
\put(957,357.01){\rule{2.891pt}{1.200pt}}
\multiput(957.00,357.51)(6.000,-1.000){2}{\rule{1.445pt}{1.200pt}}
\put(969,356.01){\rule{3.132pt}{1.200pt}}
\multiput(969.00,356.51)(6.500,-1.000){2}{\rule{1.566pt}{1.200pt}}
\put(982,355.01){\rule{2.891pt}{1.200pt}}
\multiput(982.00,355.51)(6.000,-1.000){2}{\rule{1.445pt}{1.200pt}}
\put(932.0,361.0){\rule[-0.600pt]{3.132pt}{1.200pt}}
\put(1006,354.01){\rule{2.891pt}{1.200pt}}
\multiput(1006.00,354.51)(6.000,-1.000){2}{\rule{1.445pt}{1.200pt}}
\put(1018,353.01){\rule{3.132pt}{1.200pt}}
\multiput(1018.00,353.51)(6.500,-1.000){2}{\rule{1.566pt}{1.200pt}}
\put(1031,352.01){\rule{2.891pt}{1.200pt}}
\multiput(1031.00,352.51)(6.000,-1.000){2}{\rule{1.445pt}{1.200pt}}
\put(1043,351.01){\rule{2.891pt}{1.200pt}}
\multiput(1043.00,351.51)(6.000,-1.000){2}{\rule{1.445pt}{1.200pt}}
\put(994.0,357.0){\rule[-0.600pt]{2.891pt}{1.200pt}}
\put(1068,350.01){\rule{2.891pt}{1.200pt}}
\multiput(1068.00,350.51)(6.000,-1.000){2}{\rule{1.445pt}{1.200pt}}
\put(1080,349.01){\rule{2.891pt}{1.200pt}}
\multiput(1080.00,349.51)(6.000,-1.000){2}{\rule{1.445pt}{1.200pt}}
\put(1092,348.01){\rule{2.891pt}{1.200pt}}
\multiput(1092.00,348.51)(6.000,-1.000){2}{\rule{1.445pt}{1.200pt}}
\put(1104,347.01){\rule{3.132pt}{1.200pt}}
\multiput(1104.00,347.51)(6.500,-1.000){2}{\rule{1.566pt}{1.200pt}}
\put(1117,346.01){\rule{2.891pt}{1.200pt}}
\multiput(1117.00,346.51)(6.000,-1.000){2}{\rule{1.445pt}{1.200pt}}
\put(1129,345.01){\rule{2.891pt}{1.200pt}}
\multiput(1129.00,345.51)(6.000,-1.000){2}{\rule{1.445pt}{1.200pt}}
\put(1141,344.01){\rule{2.891pt}{1.200pt}}
\multiput(1141.00,344.51)(6.000,-1.000){2}{\rule{1.445pt}{1.200pt}}
\put(1153,343.01){\rule{3.132pt}{1.200pt}}
\multiput(1153.00,343.51)(6.500,-1.000){2}{\rule{1.566pt}{1.200pt}}
\put(1166,342.01){\rule{2.891pt}{1.200pt}}
\multiput(1166.00,342.51)(6.000,-1.000){2}{\rule{1.445pt}{1.200pt}}
\put(1178,341.01){\rule{2.891pt}{1.200pt}}
\multiput(1178.00,341.51)(6.000,-1.000){2}{\rule{1.445pt}{1.200pt}}
\put(1190,340.01){\rule{3.132pt}{1.200pt}}
\multiput(1190.00,340.51)(6.500,-1.000){2}{\rule{1.566pt}{1.200pt}}
\put(1203,339.01){\rule{2.891pt}{1.200pt}}
\multiput(1203.00,339.51)(6.000,-1.000){2}{\rule{1.445pt}{1.200pt}}
\put(1215,338.01){\rule{2.891pt}{1.200pt}}
\multiput(1215.00,338.51)(6.000,-1.000){2}{\rule{1.445pt}{1.200pt}}
\put(1227,337.01){\rule{2.891pt}{1.200pt}}
\multiput(1227.00,337.51)(6.000,-1.000){2}{\rule{1.445pt}{1.200pt}}
\put(1239,336.01){\rule{3.132pt}{1.200pt}}
\multiput(1239.00,336.51)(6.500,-1.000){2}{\rule{1.566pt}{1.200pt}}
\put(1252,335.01){\rule{2.891pt}{1.200pt}}
\multiput(1252.00,335.51)(6.000,-1.000){2}{\rule{1.445pt}{1.200pt}}
\put(1264,333.51){\rule{2.891pt}{1.200pt}}
\multiput(1264.00,334.51)(6.000,-2.000){2}{\rule{1.445pt}{1.200pt}}
\put(1276,332.01){\rule{3.132pt}{1.200pt}}
\multiput(1276.00,332.51)(6.500,-1.000){2}{\rule{1.566pt}{1.200pt}}
\put(1289,331.01){\rule{2.891pt}{1.200pt}}
\multiput(1289.00,331.51)(6.000,-1.000){2}{\rule{1.445pt}{1.200pt}}
\put(1301,329.51){\rule{2.891pt}{1.200pt}}
\multiput(1301.00,330.51)(6.000,-2.000){2}{\rule{1.445pt}{1.200pt}}
\put(1313,327.51){\rule{2.891pt}{1.200pt}}
\multiput(1313.00,328.51)(6.000,-2.000){2}{\rule{1.445pt}{1.200pt}}
\put(1325,326.01){\rule{3.132pt}{1.200pt}}
\multiput(1325.00,326.51)(6.500,-1.000){2}{\rule{1.566pt}{1.200pt}}
\put(1338,324.51){\rule{2.891pt}{1.200pt}}
\multiput(1338.00,325.51)(6.000,-2.000){2}{\rule{1.445pt}{1.200pt}}
\put(1350,322.51){\rule{2.891pt}{1.200pt}}
\multiput(1350.00,323.51)(6.000,-2.000){2}{\rule{1.445pt}{1.200pt}}
\put(1362,320.51){\rule{3.132pt}{1.200pt}}
\multiput(1362.00,321.51)(6.500,-2.000){2}{\rule{1.566pt}{1.200pt}}
\put(1375,318.01){\rule{2.891pt}{1.200pt}}
\multiput(1375.00,319.51)(6.000,-3.000){2}{\rule{1.445pt}{1.200pt}}
\put(1387,315.01){\rule{2.891pt}{1.200pt}}
\multiput(1387.00,316.51)(6.000,-3.000){2}{\rule{1.445pt}{1.200pt}}
\put(1399,312.01){\rule{2.891pt}{1.200pt}}
\multiput(1399.00,313.51)(6.000,-3.000){2}{\rule{1.445pt}{1.200pt}}
\put(1411,308.01){\rule{3.132pt}{1.200pt}}
\multiput(1411.00,310.51)(6.500,-5.000){2}{\rule{1.566pt}{1.200pt}}
\multiput(1426.24,300.94)(0.501,-0.534){14}{\rule{0.121pt}{1.700pt}}
\multiput(1421.51,304.47)(12.000,-10.472){2}{\rule{1.200pt}{0.850pt}}
\put(1055.0,353.0){\rule[-0.600pt]{3.132pt}{1.200pt}}
\put(220,441){\usebox{\plotpoint}}
\multiput(222.24,441.00)(0.501,0.669){14}{\rule{0.121pt}{2.000pt}}
\multiput(217.51,441.00)(12.000,12.849){2}{\rule{1.200pt}{1.000pt}}
\multiput(232.00,460.24)(0.732,0.503){6}{\rule{2.250pt}{0.121pt}}
\multiput(232.00,455.51)(8.330,8.000){2}{\rule{1.125pt}{1.200pt}}
\multiput(245.00,468.24)(0.738,0.505){4}{\rule{2.357pt}{0.122pt}}
\multiput(245.00,463.51)(7.108,7.000){2}{\rule{1.179pt}{1.200pt}}
\multiput(257.00,475.24)(0.792,0.509){2}{\rule{2.700pt}{0.123pt}}
\multiput(257.00,470.51)(6.396,6.000){2}{\rule{1.350pt}{1.200pt}}
\multiput(269.00,481.24)(0.792,0.509){2}{\rule{2.700pt}{0.123pt}}
\multiput(269.00,476.51)(6.396,6.000){2}{\rule{1.350pt}{1.200pt}}
\multiput(281.00,487.24)(0.962,0.509){2}{\rule{2.900pt}{0.123pt}}
\multiput(281.00,482.51)(6.981,6.000){2}{\rule{1.450pt}{1.200pt}}
\multiput(294.00,493.24)(0.792,0.509){2}{\rule{2.700pt}{0.123pt}}
\multiput(294.00,488.51)(6.396,6.000){2}{\rule{1.350pt}{1.200pt}}
\multiput(306.00,499.24)(0.792,0.509){2}{\rule{2.700pt}{0.123pt}}
\multiput(306.00,494.51)(6.396,6.000){2}{\rule{1.350pt}{1.200pt}}
\multiput(318.00,505.24)(0.962,0.509){2}{\rule{2.900pt}{0.123pt}}
\multiput(318.00,500.51)(6.981,6.000){2}{\rule{1.450pt}{1.200pt}}
\multiput(331.00,511.24)(0.792,0.509){2}{\rule{2.700pt}{0.123pt}}
\multiput(331.00,506.51)(6.396,6.000){2}{\rule{1.350pt}{1.200pt}}
\multiput(343.00,517.24)(0.738,0.505){4}{\rule{2.357pt}{0.122pt}}
\multiput(343.00,512.51)(7.108,7.000){2}{\rule{1.179pt}{1.200pt}}
\multiput(355.00,524.24)(0.792,0.509){2}{\rule{2.700pt}{0.123pt}}
\multiput(355.00,519.51)(6.396,6.000){2}{\rule{1.350pt}{1.200pt}}
\multiput(367.00,530.24)(0.835,0.505){4}{\rule{2.529pt}{0.122pt}}
\multiput(367.00,525.51)(7.752,7.000){2}{\rule{1.264pt}{1.200pt}}
\multiput(380.00,537.24)(0.738,0.505){4}{\rule{2.357pt}{0.122pt}}
\multiput(380.00,532.51)(7.108,7.000){2}{\rule{1.179pt}{1.200pt}}
\multiput(392.00,544.24)(0.738,0.505){4}{\rule{2.357pt}{0.122pt}}
\multiput(392.00,539.51)(7.108,7.000){2}{\rule{1.179pt}{1.200pt}}
\multiput(404.00,551.24)(0.835,0.505){4}{\rule{2.529pt}{0.122pt}}
\multiput(404.00,546.51)(7.752,7.000){2}{\rule{1.264pt}{1.200pt}}
\multiput(417.00,558.24)(0.657,0.503){6}{\rule{2.100pt}{0.121pt}}
\multiput(417.00,553.51)(7.641,8.000){2}{\rule{1.050pt}{1.200pt}}
\multiput(429.00,566.24)(0.657,0.503){6}{\rule{2.100pt}{0.121pt}}
\multiput(429.00,561.51)(7.641,8.000){2}{\rule{1.050pt}{1.200pt}}
\multiput(441.00,574.24)(0.588,0.502){8}{\rule{1.900pt}{0.121pt}}
\multiput(441.00,569.51)(8.056,9.000){2}{\rule{0.950pt}{1.200pt}}
\multiput(453.00,583.24)(0.651,0.502){8}{\rule{2.033pt}{0.121pt}}
\multiput(453.00,578.51)(8.780,9.000){2}{\rule{1.017pt}{1.200pt}}
\multiput(466.00,592.24)(0.531,0.502){10}{\rule{1.740pt}{0.121pt}}
\multiput(466.00,587.51)(8.389,10.000){2}{\rule{0.870pt}{1.200pt}}
\multiput(478.00,602.24)(0.531,0.502){10}{\rule{1.740pt}{0.121pt}}
\multiput(478.00,597.51)(8.389,10.000){2}{\rule{0.870pt}{1.200pt}}
\multiput(490.00,612.24)(0.533,0.502){12}{\rule{1.718pt}{0.121pt}}
\multiput(490.00,607.51)(9.434,11.000){2}{\rule{0.859pt}{1.200pt}}
\multiput(503.00,623.24)(0.444,0.501){14}{\rule{1.500pt}{0.121pt}}
\multiput(503.00,618.51)(8.887,12.000){2}{\rule{0.750pt}{1.200pt}}
\multiput(515.00,635.24)(0.444,0.501){14}{\rule{1.500pt}{0.121pt}}
\multiput(515.00,630.51)(8.887,12.000){2}{\rule{0.750pt}{1.200pt}}
\multiput(529.24,645.00)(0.501,0.534){14}{\rule{0.121pt}{1.700pt}}
\multiput(524.51,645.00)(12.000,10.472){2}{\rule{1.200pt}{0.850pt}}
\multiput(541.24,659.00)(0.501,0.493){16}{\rule{0.121pt}{1.592pt}}
\multiput(536.51,659.00)(13.000,10.695){2}{\rule{1.200pt}{0.796pt}}
\multiput(554.24,673.00)(0.501,0.624){14}{\rule{0.121pt}{1.900pt}}
\multiput(549.51,673.00)(12.000,12.056){2}{\rule{1.200pt}{0.950pt}}
\multiput(566.24,689.00)(0.501,0.714){14}{\rule{0.121pt}{2.100pt}}
\multiput(561.51,689.00)(12.000,13.641){2}{\rule{1.200pt}{1.050pt}}
\multiput(578.24,707.00)(0.501,0.759){14}{\rule{0.121pt}{2.200pt}}
\multiput(573.51,707.00)(12.000,14.434){2}{\rule{1.200pt}{1.100pt}}
\multiput(590.24,726.00)(0.501,0.781){16}{\rule{0.121pt}{2.238pt}}
\multiput(585.51,726.00)(13.000,16.354){2}{\rule{1.200pt}{1.119pt}}
\multiput(603.24,747.00)(0.501,0.939){14}{\rule{0.121pt}{2.600pt}}
\multiput(598.51,747.00)(12.000,17.604){2}{\rule{1.200pt}{1.300pt}}
\multiput(615.24,770.00)(0.501,1.074){14}{\rule{0.121pt}{2.900pt}}
\multiput(610.51,770.00)(12.000,19.981){2}{\rule{1.200pt}{1.450pt}}
\multiput(627.24,796.00)(0.501,1.109){16}{\rule{0.121pt}{2.977pt}}
\multiput(622.51,796.00)(13.000,22.821){2}{\rule{1.200pt}{1.488pt}}
\multiput(640.24,825.00)(0.501,1.434){14}{\rule{0.121pt}{3.700pt}}
\multiput(635.51,825.00)(12.000,26.320){2}{\rule{1.200pt}{1.850pt}}
\multiput(652.24,859.00)(0.509,1.811){2}{\rule{0.123pt}{3.900pt}}
\multiput(647.51,859.00)(6.000,9.905){2}{\rule{1.200pt}{1.950pt}}
\put(1152,111.01){\rule{0.241pt}{1.200pt}}
\multiput(1152.00,110.51)(0.500,1.000){2}{\rule{0.120pt}{1.200pt}}
\multiput(1153.00,116.24)(0.533,0.502){12}{\rule{1.718pt}{0.121pt}}
\multiput(1153.00,111.51)(9.434,11.000){2}{\rule{0.859pt}{1.200pt}}
\multiput(1166.00,127.24)(0.484,0.502){12}{\rule{1.609pt}{0.121pt}}
\multiput(1166.00,122.51)(8.660,11.000){2}{\rule{0.805pt}{1.200pt}}
\multiput(1178.00,138.24)(0.588,0.502){8}{\rule{1.900pt}{0.121pt}}
\multiput(1178.00,133.51)(8.056,9.000){2}{\rule{0.950pt}{1.200pt}}
\multiput(1190.00,147.24)(0.651,0.502){8}{\rule{2.033pt}{0.121pt}}
\multiput(1190.00,142.51)(8.780,9.000){2}{\rule{1.017pt}{1.200pt}}
\multiput(1203.00,156.24)(0.588,0.502){8}{\rule{1.900pt}{0.121pt}}
\multiput(1203.00,151.51)(8.056,9.000){2}{\rule{0.950pt}{1.200pt}}
\multiput(1215.00,165.24)(0.657,0.503){6}{\rule{2.100pt}{0.121pt}}
\multiput(1215.00,160.51)(7.641,8.000){2}{\rule{1.050pt}{1.200pt}}
\multiput(1227.00,173.24)(0.657,0.503){6}{\rule{2.100pt}{0.121pt}}
\multiput(1227.00,168.51)(7.641,8.000){2}{\rule{1.050pt}{1.200pt}}
\multiput(1239.00,181.24)(0.732,0.503){6}{\rule{2.250pt}{0.121pt}}
\multiput(1239.00,176.51)(8.330,8.000){2}{\rule{1.125pt}{1.200pt}}
\multiput(1252.00,189.24)(0.738,0.505){4}{\rule{2.357pt}{0.122pt}}
\multiput(1252.00,184.51)(7.108,7.000){2}{\rule{1.179pt}{1.200pt}}
\multiput(1264.00,196.24)(0.738,0.505){4}{\rule{2.357pt}{0.122pt}}
\multiput(1264.00,191.51)(7.108,7.000){2}{\rule{1.179pt}{1.200pt}}
\multiput(1276.00,203.24)(0.962,0.509){2}{\rule{2.900pt}{0.123pt}}
\multiput(1276.00,198.51)(6.981,6.000){2}{\rule{1.450pt}{1.200pt}}
\multiput(1289.00,209.24)(0.738,0.505){4}{\rule{2.357pt}{0.122pt}}
\multiput(1289.00,204.51)(7.108,7.000){2}{\rule{1.179pt}{1.200pt}}
\multiput(1301.00,216.24)(0.792,0.509){2}{\rule{2.700pt}{0.123pt}}
\multiput(1301.00,211.51)(6.396,6.000){2}{\rule{1.350pt}{1.200pt}}
\multiput(1313.00,222.24)(0.792,0.509){2}{\rule{2.700pt}{0.123pt}}
\multiput(1313.00,217.51)(6.396,6.000){2}{\rule{1.350pt}{1.200pt}}
\multiput(1325.00,228.24)(0.962,0.509){2}{\rule{2.900pt}{0.123pt}}
\multiput(1325.00,223.51)(6.981,6.000){2}{\rule{1.450pt}{1.200pt}}
\multiput(1338.00,234.24)(0.792,0.509){2}{\rule{2.700pt}{0.123pt}}
\multiput(1338.00,229.51)(6.396,6.000){2}{\rule{1.350pt}{1.200pt}}
\multiput(1350.00,240.24)(0.792,0.509){2}{\rule{2.700pt}{0.123pt}}
\multiput(1350.00,235.51)(6.396,6.000){2}{\rule{1.350pt}{1.200pt}}
\multiput(1362.00,246.24)(0.962,0.509){2}{\rule{2.900pt}{0.123pt}}
\multiput(1362.00,241.51)(6.981,6.000){2}{\rule{1.450pt}{1.200pt}}
\multiput(1375.00,252.24)(0.792,0.509){2}{\rule{2.700pt}{0.123pt}}
\multiput(1375.00,247.51)(6.396,6.000){2}{\rule{1.350pt}{1.200pt}}
\multiput(1387.00,258.24)(0.738,0.505){4}{\rule{2.357pt}{0.122pt}}
\multiput(1387.00,253.51)(7.108,7.000){2}{\rule{1.179pt}{1.200pt}}
\multiput(1399.00,265.24)(0.792,0.509){2}{\rule{2.700pt}{0.123pt}}
\multiput(1399.00,260.51)(6.396,6.000){2}{\rule{1.350pt}{1.200pt}}
\multiput(1411.00,271.24)(0.651,0.502){8}{\rule{2.033pt}{0.121pt}}
\multiput(1411.00,266.51)(8.780,9.000){2}{\rule{1.017pt}{1.200pt}}
\multiput(1426.24,278.00)(0.501,0.624){14}{\rule{0.121pt}{1.900pt}}
\multiput(1421.51,278.00)(12.000,12.056){2}{\rule{1.200pt}{0.950pt}}
\end{picture}

%% file: tempfig1-2.tex
% GNUPLOT: LaTeX picture
\setlength{\unitlength}{0.240900pt}
\ifx\plotpoint\undefined\newsavebox{\plotpoint}\fi
\begin{picture}(1500,900)(0,0)
\font\gnuplot=cmr10 at 10pt
\gnuplot
\sbox{\plotpoint}{\rule[-0.200pt]{0.400pt}{0.400pt}}%
\put(220.0,591.0){\rule[-0.200pt]{292.934pt}{0.400pt}}
\put(220.0,113.0){\rule[-0.200pt]{0.400pt}{184.048pt}}
\put(220.0,113.0){\rule[-0.200pt]{4.818pt}{0.400pt}}
\put(198,113){\makebox(0,0)[r]{-2.5}}
\put(1416.0,113.0){\rule[-0.200pt]{4.818pt}{0.400pt}}
\put(220.0,209.0){\rule[-0.200pt]{4.818pt}{0.400pt}}
\put(198,209){\makebox(0,0)[r]{-2}}
\put(1416.0,209.0){\rule[-0.200pt]{4.818pt}{0.400pt}}
\put(220.0,304.0){\rule[-0.200pt]{4.818pt}{0.400pt}}
\put(198,304){\makebox(0,0)[r]{-1.5}}
\put(1416.0,304.0){\rule[-0.200pt]{4.818pt}{0.400pt}}
\put(220.0,400.0){\rule[-0.200pt]{4.818pt}{0.400pt}}
\put(198,400){\makebox(0,0)[r]{-1}}
\put(1416.0,400.0){\rule[-0.200pt]{4.818pt}{0.400pt}}
\put(220.0,495.0){\rule[-0.200pt]{4.818pt}{0.400pt}}
\put(198,495){\makebox(0,0)[r]{-0.5}}
\put(1416.0,495.0){\rule[-0.200pt]{4.818pt}{0.400pt}}
\put(220.0,591.0){\rule[-0.200pt]{4.818pt}{0.400pt}}
\put(198,591){\makebox(0,0)[r]{0}}
\put(1416.0,591.0){\rule[-0.200pt]{4.818pt}{0.400pt}}
\put(220.0,686.0){\rule[-0.200pt]{4.818pt}{0.400pt}}
\put(198,686){\makebox(0,0)[r]{0.5}}
\put(1416.0,686.0){\rule[-0.200pt]{4.818pt}{0.400pt}}
\put(220.0,782.0){\rule[-0.200pt]{4.818pt}{0.400pt}}
\put(198,782){\makebox(0,0)[r]{1}}
\put(1416.0,782.0){\rule[-0.200pt]{4.818pt}{0.400pt}}
\put(220.0,877.0){\rule[-0.200pt]{4.818pt}{0.400pt}}
\put(198,877){\makebox(0,0)[r]{1.5}}
\put(1416.0,877.0){\rule[-0.200pt]{4.818pt}{0.400pt}}
\put(220.0,113.0){\rule[-0.200pt]{0.400pt}{4.818pt}}
\put(220,68){\makebox(0,0){0}}
\put(220.0,857.0){\rule[-0.200pt]{0.400pt}{4.818pt}}
\put(342.0,113.0){\rule[-0.200pt]{0.400pt}{4.818pt}}
\put(342,68){\makebox(0,0){50}}
\put(342.0,857.0){\rule[-0.200pt]{0.400pt}{4.818pt}}
\put(463.0,113.0){\rule[-0.200pt]{0.400pt}{4.818pt}}
\put(463,68){\makebox(0,0){100}}
\put(463.0,857.0){\rule[-0.200pt]{0.400pt}{4.818pt}}
\put(585.0,113.0){\rule[-0.200pt]{0.400pt}{4.818pt}}
\put(585,68){\makebox(0,0){150}}
\put(585.0,857.0){\rule[-0.200pt]{0.400pt}{4.818pt}}
\put(706.0,113.0){\rule[-0.200pt]{0.400pt}{4.818pt}}
\put(706,68){\makebox(0,0){200}}
\put(706.0,857.0){\rule[-0.200pt]{0.400pt}{4.818pt}}
\put(828.0,113.0){\rule[-0.200pt]{0.400pt}{4.818pt}}
\put(828,68){\makebox(0,0){250}}
\put(828.0,857.0){\rule[-0.200pt]{0.400pt}{4.818pt}}
\put(950.0,113.0){\rule[-0.200pt]{0.400pt}{4.818pt}}
\put(950,68){\makebox(0,0){300}}
\put(950.0,857.0){\rule[-0.200pt]{0.400pt}{4.818pt}}
\put(1071.0,113.0){\rule[-0.200pt]{0.400pt}{4.818pt}}
\put(1071,68){\makebox(0,0){350}}
\put(1071.0,857.0){\rule[-0.200pt]{0.400pt}{4.818pt}}
\put(1193.0,113.0){\rule[-0.200pt]{0.400pt}{4.818pt}}
\put(1193,68){\makebox(0,0){400}}
\put(1193.0,857.0){\rule[-0.200pt]{0.400pt}{4.818pt}}
\put(1314.0,113.0){\rule[-0.200pt]{0.400pt}{4.818pt}}
\put(1314,68){\makebox(0,0){450}}
\put(1314.0,857.0){\rule[-0.200pt]{0.400pt}{4.818pt}}
\put(1436.0,113.0){\rule[-0.200pt]{0.400pt}{4.818pt}}
\put(1436,68){\makebox(0,0){500}}
\put(1436.0,857.0){\rule[-0.200pt]{0.400pt}{4.818pt}}
\put(220.0,113.0){\rule[-0.200pt]{292.934pt}{0.400pt}}
\put(1436.0,113.0){\rule[-0.200pt]{0.400pt}{184.048pt}}
\put(220.0,877.0){\rule[-0.200pt]{292.934pt}{0.400pt}}
\put(45,495)
{\makebox(0,0){$\displaystyle{\frac{B_{\pm}(M_X)}{M_{1/2}^{(0)}}}$}}
\put(828,23){\makebox(0,0){\shortstack{\\ \\ \\ $M_{1/2}^{(0)}$}}}
\put(220.0,113.0){\rule[-0.200pt]{0.400pt}{184.048pt}}
\put(342,582){\rule{1pt}{1pt}}
\put(372,616){\rule{1pt}{1pt}}
\put(402,633){\rule{1pt}{1pt}}
\put(433,643){\rule{1pt}{1pt}}
\put(463,649){\rule{1pt}{1pt}}
\put(494,653){\rule{1pt}{1pt}}
\put(524,656){\rule{1pt}{1pt}}
\put(554,658){\rule{1pt}{1pt}}
\put(585,659){\rule{1pt}{1pt}}
\put(615,661){\rule{1pt}{1pt}}
\put(646,662){\rule{1pt}{1pt}}
\put(676,662){\rule{1pt}{1pt}}
\put(706,663){\rule{1pt}{1pt}}
\put(737,664){\rule{1pt}{1pt}}
\put(767,664){\rule{1pt}{1pt}}
\put(798,664){\rule{1pt}{1pt}}
\put(828,665){\rule{1pt}{1pt}}
\put(858,665){\rule{1pt}{1pt}}
\put(889,665){\rule{1pt}{1pt}}
\put(919,665){\rule{1pt}{1pt}}
\put(950,666){\rule{1pt}{1pt}}
\put(980,666){\rule{1pt}{1pt}}
\put(1010,666){\rule{1pt}{1pt}}
\put(1041,666){\rule{1pt}{1pt}}
\put(1071,666){\rule{1pt}{1pt}}
\put(1102,666){\rule{1pt}{1pt}}
\put(1132,666){\rule{1pt}{1pt}}
\put(1162,666){\rule{1pt}{1pt}}
\put(1193,667){\rule{1pt}{1pt}}
\put(1223,667){\rule{1pt}{1pt}}
\put(1254,667){\rule{1pt}{1pt}}
\put(1284,667){\rule{1pt}{1pt}}
\put(1314,667){\rule{1pt}{1pt}}
\put(1345,667){\rule{1pt}{1pt}}
\put(1375,667){\rule{1pt}{1pt}}
\put(1406,667){\rule{1pt}{1pt}}
\put(1436,667){\rule{1pt}{1pt}}
\put(342,525){\rule{1pt}{1pt}}
\put(372,558){\rule{1pt}{1pt}}
\put(402,573){\rule{1pt}{1pt}}
\put(433,582){\rule{1pt}{1pt}}
\put(463,587){\rule{1pt}{1pt}}
\put(494,591){\rule{1pt}{1pt}}
\put(524,594){\rule{1pt}{1pt}}
\put(554,596){\rule{1pt}{1pt}}
\put(585,597){\rule{1pt}{1pt}}
\put(615,598){\rule{1pt}{1pt}}
\put(646,599){\rule{1pt}{1pt}}
\put(676,600){\rule{1pt}{1pt}}
\put(706,600){\rule{1pt}{1pt}}
\put(737,601){\rule{1pt}{1pt}}
\put(767,601){\rule{1pt}{1pt}}
\put(798,601){\rule{1pt}{1pt}}
\put(828,602){\rule{1pt}{1pt}}
\put(858,602){\rule{1pt}{1pt}}
\put(889,602){\rule{1pt}{1pt}}
\put(919,602){\rule{1pt}{1pt}}
\put(950,602){\rule{1pt}{1pt}}
\put(980,603){\rule{1pt}{1pt}}
\put(1010,603){\rule{1pt}{1pt}}
\put(1041,603){\rule{1pt}{1pt}}
\put(1071,603){\rule{1pt}{1pt}}
\put(1102,603){\rule{1pt}{1pt}}
\put(1132,603){\rule{1pt}{1pt}}
\put(1162,603){\rule{1pt}{1pt}}
\put(1193,603){\rule{1pt}{1pt}}
\put(1223,603){\rule{1pt}{1pt}}
\put(1254,603){\rule{1pt}{1pt}}
\put(1284,603){\rule{1pt}{1pt}}
\put(1314,603){\rule{1pt}{1pt}}
\put(1345,604){\rule{1pt}{1pt}}
\put(1375,604){\rule{1pt}{1pt}}
\put(1406,604){\rule{1pt}{1pt}}
\put(1436,604){\rule{1pt}{1pt}}
\put(342,506){\rule{1pt}{1pt}}
\put(372,535){\rule{1pt}{1pt}}
\put(402,548){\rule{1pt}{1pt}}
\put(433,555){\rule{1pt}{1pt}}
\put(463,560){\rule{1pt}{1pt}}
\put(494,563){\rule{1pt}{1pt}}
\put(524,565){\rule{1pt}{1pt}}
\put(554,567){\rule{1pt}{1pt}}
\put(585,568){\rule{1pt}{1pt}}
\put(615,569){\rule{1pt}{1pt}}
\put(646,570){\rule{1pt}{1pt}}
\put(676,570){\rule{1pt}{1pt}}
\put(706,571){\rule{1pt}{1pt}}
\put(737,571){\rule{1pt}{1pt}}
\put(767,571){\rule{1pt}{1pt}}
\put(798,572){\rule{1pt}{1pt}}
\put(828,572){\rule{1pt}{1pt}}
\put(858,572){\rule{1pt}{1pt}}
\put(889,572){\rule{1pt}{1pt}}
\put(919,572){\rule{1pt}{1pt}}
\put(950,572){\rule{1pt}{1pt}}
\put(980,573){\rule{1pt}{1pt}}
\put(1010,573){\rule{1pt}{1pt}}
\put(1041,573){\rule{1pt}{1pt}}
\put(1071,573){\rule{1pt}{1pt}}
\put(1102,573){\rule{1pt}{1pt}}
\put(1132,573){\rule{1pt}{1pt}}
\put(1162,573){\rule{1pt}{1pt}}
\put(1193,573){\rule{1pt}{1pt}}
\put(1223,573){\rule{1pt}{1pt}}
\put(1254,573){\rule{1pt}{1pt}}
\put(1284,573){\rule{1pt}{1pt}}
\put(1314,573){\rule{1pt}{1pt}}
\put(1345,573){\rule{1pt}{1pt}}
\put(1375,573){\rule{1pt}{1pt}}
\put(1406,573){\rule{1pt}{1pt}}
\put(1436,573){\rule{1pt}{1pt}}
\put(342,498){\rule{1pt}{1pt}}
\put(372,523){\rule{1pt}{1pt}}
\put(402,534){\rule{1pt}{1pt}}
\put(433,540){\rule{1pt}{1pt}}
\put(463,544){\rule{1pt}{1pt}}
\put(494,546){\rule{1pt}{1pt}}
\put(524,548){\rule{1pt}{1pt}}
\put(554,550){\rule{1pt}{1pt}}
\put(585,550){\rule{1pt}{1pt}}
\put(615,551){\rule{1pt}{1pt}}
\put(646,552){\rule{1pt}{1pt}}
\put(676,552){\rule{1pt}{1pt}}
\put(706,553){\rule{1pt}{1pt}}
\put(737,553){\rule{1pt}{1pt}}
\put(767,553){\rule{1pt}{1pt}}
\put(798,554){\rule{1pt}{1pt}}
\put(828,554){\rule{1pt}{1pt}}
\put(858,554){\rule{1pt}{1pt}}
\put(889,554){\rule{1pt}{1pt}}
\put(919,554){\rule{1pt}{1pt}}
\put(950,554){\rule{1pt}{1pt}}
\put(980,554){\rule{1pt}{1pt}}
\put(1010,555){\rule{1pt}{1pt}}
\put(1041,555){\rule{1pt}{1pt}}
\put(1071,555){\rule{1pt}{1pt}}
\put(1102,555){\rule{1pt}{1pt}}
\put(1132,555){\rule{1pt}{1pt}}
\put(1162,555){\rule{1pt}{1pt}}
\put(1193,555){\rule{1pt}{1pt}}
\put(1223,555){\rule{1pt}{1pt}}
\put(1254,555){\rule{1pt}{1pt}}
\put(1284,555){\rule{1pt}{1pt}}
\put(1314,555){\rule{1pt}{1pt}}
\put(1345,555){\rule{1pt}{1pt}}
\put(1375,555){\rule{1pt}{1pt}}
\put(1406,555){\rule{1pt}{1pt}}
\put(1436,555){\rule{1pt}{1pt}}
\put(342,493){\rule{1pt}{1pt}}
\put(372,515){\rule{1pt}{1pt}}
\put(402,524){\rule{1pt}{1pt}}
\put(433,530){\rule{1pt}{1pt}}
\put(463,533){\rule{1pt}{1pt}}
\put(494,535){\rule{1pt}{1pt}}
\put(524,537){\rule{1pt}{1pt}}
\put(554,538){\rule{1pt}{1pt}}
\put(585,539){\rule{1pt}{1pt}}
\put(615,539){\rule{1pt}{1pt}}
\put(646,540){\rule{1pt}{1pt}}
\put(676,540){\rule{1pt}{1pt}}
\put(706,541){\rule{1pt}{1pt}}
\put(737,541){\rule{1pt}{1pt}}
\put(767,541){\rule{1pt}{1pt}}
\put(798,541){\rule{1pt}{1pt}}
\put(828,542){\rule{1pt}{1pt}}
\put(858,542){\rule{1pt}{1pt}}
\put(889,542){\rule{1pt}{1pt}}
\put(919,542){\rule{1pt}{1pt}}
\put(950,542){\rule{1pt}{1pt}}
\put(980,542){\rule{1pt}{1pt}}
\put(1010,542){\rule{1pt}{1pt}}
\put(1041,542){\rule{1pt}{1pt}}
\put(1071,542){\rule{1pt}{1pt}}
\put(1102,542){\rule{1pt}{1pt}}
\put(1132,542){\rule{1pt}{1pt}}
\put(1162,542){\rule{1pt}{1pt}}
\put(1193,542){\rule{1pt}{1pt}}
\put(1223,543){\rule{1pt}{1pt}}
\put(1254,543){\rule{1pt}{1pt}}
\put(1284,543){\rule{1pt}{1pt}}
\put(1314,543){\rule{1pt}{1pt}}
\put(1345,543){\rule{1pt}{1pt}}
\put(1375,543){\rule{1pt}{1pt}}
\put(1406,543){\rule{1pt}{1pt}}
\put(1436,543){\rule{1pt}{1pt}}
\put(342,491){\rule{1pt}{1pt}}
\put(372,509){\rule{1pt}{1pt}}
\put(402,518){\rule{1pt}{1pt}}
\put(433,522){\rule{1pt}{1pt}}
\put(463,525){\rule{1pt}{1pt}}
\put(494,527){\rule{1pt}{1pt}}
\put(524,529){\rule{1pt}{1pt}}
\put(554,530){\rule{1pt}{1pt}}
\put(585,530){\rule{1pt}{1pt}}
\put(615,531){\rule{1pt}{1pt}}
\put(646,531){\rule{1pt}{1pt}}
\put(676,532){\rule{1pt}{1pt}}
\put(706,532){\rule{1pt}{1pt}}
\put(737,532){\rule{1pt}{1pt}}
\put(767,532){\rule{1pt}{1pt}}
\put(798,533){\rule{1pt}{1pt}}
\put(828,533){\rule{1pt}{1pt}}
\put(858,533){\rule{1pt}{1pt}}
\put(889,533){\rule{1pt}{1pt}}
\put(919,533){\rule{1pt}{1pt}}
\put(950,533){\rule{1pt}{1pt}}
\put(980,533){\rule{1pt}{1pt}}
\put(1010,533){\rule{1pt}{1pt}}
\put(1041,533){\rule{1pt}{1pt}}
\put(1071,533){\rule{1pt}{1pt}}
\put(1102,533){\rule{1pt}{1pt}}
\put(1132,533){\rule{1pt}{1pt}}
\put(1162,533){\rule{1pt}{1pt}}
\put(1193,534){\rule{1pt}{1pt}}
\put(1223,534){\rule{1pt}{1pt}}
\put(1254,534){\rule{1pt}{1pt}}
\put(1284,534){\rule{1pt}{1pt}}
\put(1314,534){\rule{1pt}{1pt}}
\put(1345,534){\rule{1pt}{1pt}}
\put(1375,534){\rule{1pt}{1pt}}
\put(1406,534){\rule{1pt}{1pt}}
\put(1436,534){\rule{1pt}{1pt}}
\put(342,489){\rule{1pt}{1pt}}
\put(372,505){\rule{1pt}{1pt}}
\put(402,513){\rule{1pt}{1pt}}
\put(433,517){\rule{1pt}{1pt}}
\put(463,519){\rule{1pt}{1pt}}
\put(494,521){\rule{1pt}{1pt}}
\put(524,522){\rule{1pt}{1pt}}
\put(554,523){\rule{1pt}{1pt}}
\put(585,524){\rule{1pt}{1pt}}
\put(615,524){\rule{1pt}{1pt}}
\put(646,525){\rule{1pt}{1pt}}
\put(676,525){\rule{1pt}{1pt}}
\put(706,525){\rule{1pt}{1pt}}
\put(737,525){\rule{1pt}{1pt}}
\put(767,526){\rule{1pt}{1pt}}
\put(798,526){\rule{1pt}{1pt}}
\put(828,526){\rule{1pt}{1pt}}
\put(858,526){\rule{1pt}{1pt}}
\put(889,526){\rule{1pt}{1pt}}
\put(919,526){\rule{1pt}{1pt}}
\put(950,526){\rule{1pt}{1pt}}
\put(980,526){\rule{1pt}{1pt}}
\put(1010,526){\rule{1pt}{1pt}}
\put(1041,526){\rule{1pt}{1pt}}
\put(1071,527){\rule{1pt}{1pt}}
\put(1102,527){\rule{1pt}{1pt}}
\put(1132,527){\rule{1pt}{1pt}}
\put(1162,527){\rule{1pt}{1pt}}
\put(1193,527){\rule{1pt}{1pt}}
\put(1223,527){\rule{1pt}{1pt}}
\put(1254,527){\rule{1pt}{1pt}}
\put(1284,527){\rule{1pt}{1pt}}
\put(1314,527){\rule{1pt}{1pt}}
\put(1345,527){\rule{1pt}{1pt}}
\put(1375,527){\rule{1pt}{1pt}}
\put(1406,527){\rule{1pt}{1pt}}
\put(1436,527){\rule{1pt}{1pt}}
\put(342,487){\rule{1pt}{1pt}}
\put(372,502){\rule{1pt}{1pt}}
\put(402,509){\rule{1pt}{1pt}}
\put(433,513){\rule{1pt}{1pt}}
\put(463,515){\rule{1pt}{1pt}}
\put(494,516){\rule{1pt}{1pt}}
\put(524,517){\rule{1pt}{1pt}}
\put(554,518){\rule{1pt}{1pt}}
\put(585,519){\rule{1pt}{1pt}}
\put(615,519){\rule{1pt}{1pt}}
\put(646,520){\rule{1pt}{1pt}}
\put(676,520){\rule{1pt}{1pt}}
\put(706,520){\rule{1pt}{1pt}}
\put(737,520){\rule{1pt}{1pt}}
\put(767,520){\rule{1pt}{1pt}}
\put(798,521){\rule{1pt}{1pt}}
\put(828,521){\rule{1pt}{1pt}}
\put(858,521){\rule{1pt}{1pt}}
\put(889,521){\rule{1pt}{1pt}}
\put(919,521){\rule{1pt}{1pt}}
\put(950,521){\rule{1pt}{1pt}}
\put(980,521){\rule{1pt}{1pt}}
\put(1010,521){\rule{1pt}{1pt}}
\put(1041,521){\rule{1pt}{1pt}}
\put(1071,521){\rule{1pt}{1pt}}
\put(1102,521){\rule{1pt}{1pt}}
\put(1132,521){\rule{1pt}{1pt}}
\put(1162,521){\rule{1pt}{1pt}}
\put(1193,521){\rule{1pt}{1pt}}
\put(1223,521){\rule{1pt}{1pt}}
\put(1254,521){\rule{1pt}{1pt}}
\put(1284,521){\rule{1pt}{1pt}}
\put(1314,521){\rule{1pt}{1pt}}
\put(1345,521){\rule{1pt}{1pt}}
\put(1375,521){\rule{1pt}{1pt}}
\put(1406,521){\rule{1pt}{1pt}}
\put(1436,521){\rule{1pt}{1pt}}
\put(342,486){\rule{1pt}{1pt}}
\put(372,500){\rule{1pt}{1pt}}
\put(402,506){\rule{1pt}{1pt}}
\put(433,509){\rule{1pt}{1pt}}
\put(463,511){\rule{1pt}{1pt}}
\put(494,512){\rule{1pt}{1pt}}
\put(524,513){\rule{1pt}{1pt}}
\put(554,514){\rule{1pt}{1pt}}
\put(585,515){\rule{1pt}{1pt}}
\put(615,515){\rule{1pt}{1pt}}
\put(646,515){\rule{1pt}{1pt}}
\put(676,516){\rule{1pt}{1pt}}
\put(706,516){\rule{1pt}{1pt}}
\put(737,516){\rule{1pt}{1pt}}
\put(767,516){\rule{1pt}{1pt}}
\put(798,516){\rule{1pt}{1pt}}
\put(828,516){\rule{1pt}{1pt}}
\put(858,516){\rule{1pt}{1pt}}
\put(889,517){\rule{1pt}{1pt}}
\put(919,517){\rule{1pt}{1pt}}
\put(950,517){\rule{1pt}{1pt}}
\put(980,517){\rule{1pt}{1pt}}
\put(1010,517){\rule{1pt}{1pt}}
\put(1041,517){\rule{1pt}{1pt}}
\put(1071,517){\rule{1pt}{1pt}}
\put(1102,517){\rule{1pt}{1pt}}
\put(1132,517){\rule{1pt}{1pt}}
\put(1162,517){\rule{1pt}{1pt}}
\put(1193,517){\rule{1pt}{1pt}}
\put(1223,517){\rule{1pt}{1pt}}
\put(1254,517){\rule{1pt}{1pt}}
\put(1284,517){\rule{1pt}{1pt}}
\put(1314,517){\rule{1pt}{1pt}}
\put(1345,517){\rule{1pt}{1pt}}
\put(1375,517){\rule{1pt}{1pt}}
\put(1406,517){\rule{1pt}{1pt}}
\put(1436,517){\rule{1pt}{1pt}}
\put(342,264){\rule{1pt}{1pt}}
\put(372,229){\rule{1pt}{1pt}}
\put(402,213){\rule{1pt}{1pt}}
\put(433,203){\rule{1pt}{1pt}}
\put(463,197){\rule{1pt}{1pt}}
\put(494,193){\rule{1pt}{1pt}}
\put(524,190){\rule{1pt}{1pt}}
\put(554,188){\rule{1pt}{1pt}}
\put(585,186){\rule{1pt}{1pt}}
\put(615,185){\rule{1pt}{1pt}}
\put(646,184){\rule{1pt}{1pt}}
\put(676,183){\rule{1pt}{1pt}}
\put(706,182){\rule{1pt}{1pt}}
\put(737,182){\rule{1pt}{1pt}}
\put(767,181){\rule{1pt}{1pt}}
\put(798,181){\rule{1pt}{1pt}}
\put(828,181){\rule{1pt}{1pt}}
\put(858,180){\rule{1pt}{1pt}}
\put(889,180){\rule{1pt}{1pt}}
\put(919,180){\rule{1pt}{1pt}}
\put(950,180){\rule{1pt}{1pt}}
\put(980,180){\rule{1pt}{1pt}}
\put(1010,180){\rule{1pt}{1pt}}
\put(1041,179){\rule{1pt}{1pt}}
\put(1071,179){\rule{1pt}{1pt}}
\put(1102,179){\rule{1pt}{1pt}}
\put(1132,179){\rule{1pt}{1pt}}
\put(1162,179){\rule{1pt}{1pt}}
\put(1193,179){\rule{1pt}{1pt}}
\put(1223,179){\rule{1pt}{1pt}}
\put(1254,179){\rule{1pt}{1pt}}
\put(1284,179){\rule{1pt}{1pt}}
\put(1314,179){\rule{1pt}{1pt}}
\put(1345,179){\rule{1pt}{1pt}}
\put(1375,179){\rule{1pt}{1pt}}
\put(1406,179){\rule{1pt}{1pt}}
\put(1436,179){\rule{1pt}{1pt}}
\put(342,380){\rule{1pt}{1pt}}
\put(372,347){\rule{1pt}{1pt}}
\put(402,331){\rule{1pt}{1pt}}
\put(433,322){\rule{1pt}{1pt}}
\put(463,317){\rule{1pt}{1pt}}
\put(494,313){\rule{1pt}{1pt}}
\put(524,311){\rule{1pt}{1pt}}
\put(554,309){\rule{1pt}{1pt}}
\put(585,307){\rule{1pt}{1pt}}
\put(615,306){\rule{1pt}{1pt}}
\put(646,306){\rule{1pt}{1pt}}
\put(676,305){\rule{1pt}{1pt}}
\put(706,304){\rule{1pt}{1pt}}
\put(737,304){\rule{1pt}{1pt}}
\put(767,303){\rule{1pt}{1pt}}
\put(798,303){\rule{1pt}{1pt}}
\put(828,303){\rule{1pt}{1pt}}
\put(858,303){\rule{1pt}{1pt}}
\put(889,302){\rule{1pt}{1pt}}
\put(919,302){\rule{1pt}{1pt}}
\put(950,302){\rule{1pt}{1pt}}
\put(980,302){\rule{1pt}{1pt}}
\put(1010,302){\rule{1pt}{1pt}}
\put(1041,302){\rule{1pt}{1pt}}
\put(1071,302){\rule{1pt}{1pt}}
\put(1102,301){\rule{1pt}{1pt}}
\put(1132,301){\rule{1pt}{1pt}}
\put(1162,301){\rule{1pt}{1pt}}
\put(1193,301){\rule{1pt}{1pt}}
\put(1223,301){\rule{1pt}{1pt}}
\put(1254,301){\rule{1pt}{1pt}}
\put(1284,301){\rule{1pt}{1pt}}
\put(1314,301){\rule{1pt}{1pt}}
\put(1345,301){\rule{1pt}{1pt}}
\put(1375,301){\rule{1pt}{1pt}}
\put(1406,301){\rule{1pt}{1pt}}
\put(1436,301){\rule{1pt}{1pt}}
\put(342,419){\rule{1pt}{1pt}}
\put(372,390){\rule{1pt}{1pt}}
\put(402,377){\rule{1pt}{1pt}}
\put(433,370){\rule{1pt}{1pt}}
\put(463,365){\rule{1pt}{1pt}}
\put(494,362){\rule{1pt}{1pt}}
\put(524,360){\rule{1pt}{1pt}}
\put(554,359){\rule{1pt}{1pt}}
\put(585,357){\rule{1pt}{1pt}}
\put(615,356){\rule{1pt}{1pt}}
\put(646,356){\rule{1pt}{1pt}}
\put(676,355){\rule{1pt}{1pt}}
\put(706,355){\rule{1pt}{1pt}}
\put(737,354){\rule{1pt}{1pt}}
\put(767,354){\rule{1pt}{1pt}}
\put(798,354){\rule{1pt}{1pt}}
\put(828,353){\rule{1pt}{1pt}}
\put(858,353){\rule{1pt}{1pt}}
\put(889,353){\rule{1pt}{1pt}}
\put(919,353){\rule{1pt}{1pt}}
\put(950,353){\rule{1pt}{1pt}}
\put(980,353){\rule{1pt}{1pt}}
\put(1010,353){\rule{1pt}{1pt}}
\put(1041,352){\rule{1pt}{1pt}}
\put(1071,352){\rule{1pt}{1pt}}
\put(1102,352){\rule{1pt}{1pt}}
\put(1132,352){\rule{1pt}{1pt}}
\put(1162,352){\rule{1pt}{1pt}}
\put(1193,352){\rule{1pt}{1pt}}
\put(1223,352){\rule{1pt}{1pt}}
\put(1254,352){\rule{1pt}{1pt}}
\put(1284,352){\rule{1pt}{1pt}}
\put(1314,352){\rule{1pt}{1pt}}
\put(1345,352){\rule{1pt}{1pt}}
\put(1375,352){\rule{1pt}{1pt}}
\put(1406,352){\rule{1pt}{1pt}}
\put(1436,352){\rule{1pt}{1pt}}
\put(342,437){\rule{1pt}{1pt}}
\put(372,412){\rule{1pt}{1pt}}
\put(402,401){\rule{1pt}{1pt}}
\put(433,395){\rule{1pt}{1pt}}
\put(463,391){\rule{1pt}{1pt}}
\put(494,389){\rule{1pt}{1pt}}
\put(524,387){\rule{1pt}{1pt}}
\put(554,385){\rule{1pt}{1pt}}
\put(585,384){\rule{1pt}{1pt}}
\put(615,384){\rule{1pt}{1pt}}
\put(646,383){\rule{1pt}{1pt}}
\put(676,383){\rule{1pt}{1pt}}
\put(706,382){\rule{1pt}{1pt}}
\put(737,382){\rule{1pt}{1pt}}
\put(767,382){\rule{1pt}{1pt}}
\put(798,381){\rule{1pt}{1pt}}
\put(828,381){\rule{1pt}{1pt}}
\put(858,381){\rule{1pt}{1pt}}
\put(889,381){\rule{1pt}{1pt}}
\put(919,381){\rule{1pt}{1pt}}
\put(950,381){\rule{1pt}{1pt}}
\put(980,380){\rule{1pt}{1pt}}
\put(1010,380){\rule{1pt}{1pt}}
\put(1041,380){\rule{1pt}{1pt}}
\put(1071,380){\rule{1pt}{1pt}}
\put(1102,380){\rule{1pt}{1pt}}
\put(1132,380){\rule{1pt}{1pt}}
\put(1162,380){\rule{1pt}{1pt}}
\put(1193,380){\rule{1pt}{1pt}}
\put(1223,380){\rule{1pt}{1pt}}
\put(1254,380){\rule{1pt}{1pt}}
\put(1284,380){\rule{1pt}{1pt}}
\put(1314,380){\rule{1pt}{1pt}}
\put(1345,380){\rule{1pt}{1pt}}
\put(1375,380){\rule{1pt}{1pt}}
\put(1406,380){\rule{1pt}{1pt}}
\put(1436,380){\rule{1pt}{1pt}}
\put(342,447){\rule{1pt}{1pt}}
\put(372,425){\rule{1pt}{1pt}}
\put(402,416){\rule{1pt}{1pt}}
\put(433,410){\rule{1pt}{1pt}}
\put(463,407){\rule{1pt}{1pt}}
\put(494,405){\rule{1pt}{1pt}}
\put(524,403){\rule{1pt}{1pt}}
\put(554,402){\rule{1pt}{1pt}}
\put(585,401){\rule{1pt}{1pt}}
\put(615,401){\rule{1pt}{1pt}}
\put(646,400){\rule{1pt}{1pt}}
\put(676,400){\rule{1pt}{1pt}}
\put(706,400){\rule{1pt}{1pt}}
\put(737,399){\rule{1pt}{1pt}}
\put(767,399){\rule{1pt}{1pt}}
\put(798,399){\rule{1pt}{1pt}}
\put(828,399){\rule{1pt}{1pt}}
\put(858,398){\rule{1pt}{1pt}}
\put(889,398){\rule{1pt}{1pt}}
\put(919,398){\rule{1pt}{1pt}}
\put(950,398){\rule{1pt}{1pt}}
\put(980,398){\rule{1pt}{1pt}}
\put(1010,398){\rule{1pt}{1pt}}
\put(1041,398){\rule{1pt}{1pt}}
\put(1071,398){\rule{1pt}{1pt}}
\put(1102,398){\rule{1pt}{1pt}}
\put(1132,398){\rule{1pt}{1pt}}
\put(1162,398){\rule{1pt}{1pt}}
\put(1193,398){\rule{1pt}{1pt}}
\put(1223,398){\rule{1pt}{1pt}}
\put(1254,398){\rule{1pt}{1pt}}
\put(1284,398){\rule{1pt}{1pt}}
\put(1314,398){\rule{1pt}{1pt}}
\put(1345,398){\rule{1pt}{1pt}}
\put(1375,398){\rule{1pt}{1pt}}
\put(1406,397){\rule{1pt}{1pt}}
\put(1436,397){\rule{1pt}{1pt}}
\put(342,453){\rule{1pt}{1pt}}
\put(372,434){\rule{1pt}{1pt}}
\put(402,426){\rule{1pt}{1pt}}
\put(433,421){\rule{1pt}{1pt}}
\put(463,418){\rule{1pt}{1pt}}
\put(494,416){\rule{1pt}{1pt}}
\put(524,415){\rule{1pt}{1pt}}
\put(554,414){\rule{1pt}{1pt}}
\put(585,413){\rule{1pt}{1pt}}
\put(615,413){\rule{1pt}{1pt}}
\put(646,412){\rule{1pt}{1pt}}
\put(676,412){\rule{1pt}{1pt}}
\put(706,411){\rule{1pt}{1pt}}
\put(737,411){\rule{1pt}{1pt}}
\put(767,411){\rule{1pt}{1pt}}
\put(798,411){\rule{1pt}{1pt}}
\put(828,411){\rule{1pt}{1pt}}
\put(858,411){\rule{1pt}{1pt}}
\put(889,410){\rule{1pt}{1pt}}
\put(919,410){\rule{1pt}{1pt}}
\put(950,410){\rule{1pt}{1pt}}
\put(980,410){\rule{1pt}{1pt}}
\put(1010,410){\rule{1pt}{1pt}}
\put(1041,410){\rule{1pt}{1pt}}
\put(1071,410){\rule{1pt}{1pt}}
\put(1102,410){\rule{1pt}{1pt}}
\put(1132,410){\rule{1pt}{1pt}}
\put(1162,410){\rule{1pt}{1pt}}
\put(1193,410){\rule{1pt}{1pt}}
\put(1223,410){\rule{1pt}{1pt}}
\put(1254,410){\rule{1pt}{1pt}}
\put(1284,410){\rule{1pt}{1pt}}
\put(1314,410){\rule{1pt}{1pt}}
\put(1345,410){\rule{1pt}{1pt}}
\put(1375,410){\rule{1pt}{1pt}}
\put(1406,410){\rule{1pt}{1pt}}
\put(1436,410){\rule{1pt}{1pt}}
\put(342,457){\rule{1pt}{1pt}}
\put(372,440){\rule{1pt}{1pt}}
\put(402,433){\rule{1pt}{1pt}}
\put(433,429){\rule{1pt}{1pt}}
\put(463,426){\rule{1pt}{1pt}}
\put(494,424){\rule{1pt}{1pt}}
\put(524,423){\rule{1pt}{1pt}}
\put(554,422){\rule{1pt}{1pt}}
\put(585,422){\rule{1pt}{1pt}}
\put(615,421){\rule{1pt}{1pt}}
\put(646,421){\rule{1pt}{1pt}}
\put(676,420){\rule{1pt}{1pt}}
\put(706,420){\rule{1pt}{1pt}}
\put(737,420){\rule{1pt}{1pt}}
\put(767,420){\rule{1pt}{1pt}}
\put(798,420){\rule{1pt}{1pt}}
\put(828,419){\rule{1pt}{1pt}}
\put(858,419){\rule{1pt}{1pt}}
\put(889,419){\rule{1pt}{1pt}}
\put(919,419){\rule{1pt}{1pt}}
\put(950,419){\rule{1pt}{1pt}}
\put(980,419){\rule{1pt}{1pt}}
\put(1010,419){\rule{1pt}{1pt}}
\put(1041,419){\rule{1pt}{1pt}}
\put(1071,419){\rule{1pt}{1pt}}
\put(1102,419){\rule{1pt}{1pt}}
\put(1132,419){\rule{1pt}{1pt}}
\put(1162,419){\rule{1pt}{1pt}}
\put(1193,419){\rule{1pt}{1pt}}
\put(1223,419){\rule{1pt}{1pt}}
\put(1254,419){\rule{1pt}{1pt}}
\put(1284,419){\rule{1pt}{1pt}}
\put(1314,419){\rule{1pt}{1pt}}
\put(1345,419){\rule{1pt}{1pt}}
\put(1375,419){\rule{1pt}{1pt}}
\put(1406,419){\rule{1pt}{1pt}}
\put(1436,419){\rule{1pt}{1pt}}
\put(342,453){\rule{1pt}{1pt}}
\put(372,434){\rule{1pt}{1pt}}
\put(402,426){\rule{1pt}{1pt}}
\put(433,421){\rule{1pt}{1pt}}
\put(463,418){\rule{1pt}{1pt}}
\put(494,416){\rule{1pt}{1pt}}
\put(524,415){\rule{1pt}{1pt}}
\put(554,414){\rule{1pt}{1pt}}
\put(585,413){\rule{1pt}{1pt}}
\put(615,413){\rule{1pt}{1pt}}
\put(646,412){\rule{1pt}{1pt}}
\put(676,412){\rule{1pt}{1pt}}
\put(706,411){\rule{1pt}{1pt}}
\put(737,411){\rule{1pt}{1pt}}
\put(767,411){\rule{1pt}{1pt}}
\put(798,411){\rule{1pt}{1pt}}
\put(828,411){\rule{1pt}{1pt}}
\put(858,411){\rule{1pt}{1pt}}
\put(889,410){\rule{1pt}{1pt}}
\put(919,410){\rule{1pt}{1pt}}
\put(950,410){\rule{1pt}{1pt}}
\put(980,410){\rule{1pt}{1pt}}
\put(1010,410){\rule{1pt}{1pt}}
\put(1041,410){\rule{1pt}{1pt}}
\put(1071,410){\rule{1pt}{1pt}}
\put(1102,410){\rule{1pt}{1pt}}
\put(1132,410){\rule{1pt}{1pt}}
\put(1162,410){\rule{1pt}{1pt}}
\put(1193,410){\rule{1pt}{1pt}}
\put(1223,410){\rule{1pt}{1pt}}
\put(1254,410){\rule{1pt}{1pt}}
\put(1284,410){\rule{1pt}{1pt}}
\put(1314,410){\rule{1pt}{1pt}}
\put(1345,410){\rule{1pt}{1pt}}
\put(1375,410){\rule{1pt}{1pt}}
\put(1406,410){\rule{1pt}{1pt}}
\put(1436,410){\rule{1pt}{1pt}}
\put(342,457){\rule{1pt}{1pt}}
\put(372,440){\rule{1pt}{1pt}}
\put(402,433){\rule{1pt}{1pt}}
\put(433,429){\rule{1pt}{1pt}}
\put(463,426){\rule{1pt}{1pt}}
\put(494,424){\rule{1pt}{1pt}}
\put(524,423){\rule{1pt}{1pt}}
\put(554,422){\rule{1pt}{1pt}}
\put(585,422){\rule{1pt}{1pt}}
\put(615,421){\rule{1pt}{1pt}}
\put(646,421){\rule{1pt}{1pt}}
\put(676,420){\rule{1pt}{1pt}}
\put(706,420){\rule{1pt}{1pt}}
\put(737,420){\rule{1pt}{1pt}}
\put(767,420){\rule{1pt}{1pt}}
\put(798,420){\rule{1pt}{1pt}}
\put(828,419){\rule{1pt}{1pt}}
\put(858,419){\rule{1pt}{1pt}}
\put(889,419){\rule{1pt}{1pt}}
\put(919,419){\rule{1pt}{1pt}}
\put(950,419){\rule{1pt}{1pt}}
\put(980,419){\rule{1pt}{1pt}}
\put(1010,419){\rule{1pt}{1pt}}
\put(1041,419){\rule{1pt}{1pt}}
\put(1071,419){\rule{1pt}{1pt}}
\put(1102,419){\rule{1pt}{1pt}}
\put(1132,419){\rule{1pt}{1pt}}
\put(1162,419){\rule{1pt}{1pt}}
\put(1193,419){\rule{1pt}{1pt}}
\put(1223,419){\rule{1pt}{1pt}}
\put(1254,419){\rule{1pt}{1pt}}
\put(1284,419){\rule{1pt}{1pt}}
\put(1314,419){\rule{1pt}{1pt}}
\put(1345,419){\rule{1pt}{1pt}}
\put(1375,419){\rule{1pt}{1pt}}
\put(1406,419){\rule{1pt}{1pt}}
\put(1436,419){\rule{1pt}{1pt}}
\put(342,460){\rule{1pt}{1pt}}
\put(372,445){\rule{1pt}{1pt}}
\put(402,438){\rule{1pt}{1pt}}
\put(433,434){\rule{1pt}{1pt}}
\put(463,432){\rule{1pt}{1pt}}
\put(494,431){\rule{1pt}{1pt}}
\put(524,429){\rule{1pt}{1pt}}
\put(554,429){\rule{1pt}{1pt}}
\put(585,428){\rule{1pt}{1pt}}
\put(615,428){\rule{1pt}{1pt}}
\put(646,427){\rule{1pt}{1pt}}
\put(676,427){\rule{1pt}{1pt}}
\put(706,427){\rule{1pt}{1pt}}
\put(737,427){\rule{1pt}{1pt}}
\put(767,426){\rule{1pt}{1pt}}
\put(798,426){\rule{1pt}{1pt}}
\put(828,426){\rule{1pt}{1pt}}
\put(858,426){\rule{1pt}{1pt}}
\put(889,426){\rule{1pt}{1pt}}
\put(919,426){\rule{1pt}{1pt}}
\put(950,426){\rule{1pt}{1pt}}
\put(980,426){\rule{1pt}{1pt}}
\put(1010,426){\rule{1pt}{1pt}}
\put(1041,426){\rule{1pt}{1pt}}
\put(1071,426){\rule{1pt}{1pt}}
\put(1102,426){\rule{1pt}{1pt}}
\put(1132,426){\rule{1pt}{1pt}}
\put(1162,426){\rule{1pt}{1pt}}
\put(1193,426){\rule{1pt}{1pt}}
\put(1223,425){\rule{1pt}{1pt}}
\put(1254,425){\rule{1pt}{1pt}}
\put(1284,425){\rule{1pt}{1pt}}
\put(1314,425){\rule{1pt}{1pt}}
\put(1345,425){\rule{1pt}{1pt}}
\put(1375,425){\rule{1pt}{1pt}}
\put(1406,425){\rule{1pt}{1pt}}
\put(1436,425){\rule{1pt}{1pt}}
\put(342,462){\rule{1pt}{1pt}}
\put(372,448){\rule{1pt}{1pt}}
\put(402,442){\rule{1pt}{1pt}}
\put(433,439){\rule{1pt}{1pt}}
\put(463,437){\rule{1pt}{1pt}}
\put(494,435){\rule{1pt}{1pt}}
\put(524,434){\rule{1pt}{1pt}}
\put(554,434){\rule{1pt}{1pt}}
\put(585,433){\rule{1pt}{1pt}}
\put(615,433){\rule{1pt}{1pt}}
\put(646,432){\rule{1pt}{1pt}}
\put(676,432){\rule{1pt}{1pt}}
\put(706,432){\rule{1pt}{1pt}}
\put(737,432){\rule{1pt}{1pt}}
\put(767,432){\rule{1pt}{1pt}}
\put(798,432){\rule{1pt}{1pt}}
\put(828,431){\rule{1pt}{1pt}}
\put(858,431){\rule{1pt}{1pt}}
\put(889,431){\rule{1pt}{1pt}}
\put(919,431){\rule{1pt}{1pt}}
\put(950,431){\rule{1pt}{1pt}}
\put(980,431){\rule{1pt}{1pt}}
\put(1010,431){\rule{1pt}{1pt}}
\put(1041,431){\rule{1pt}{1pt}}
\put(1071,431){\rule{1pt}{1pt}}
\put(1102,431){\rule{1pt}{1pt}}
\put(1132,431){\rule{1pt}{1pt}}
\put(1162,431){\rule{1pt}{1pt}}
\put(1193,431){\rule{1pt}{1pt}}
\put(1223,431){\rule{1pt}{1pt}}
\put(1254,431){\rule{1pt}{1pt}}
\put(1284,431){\rule{1pt}{1pt}}
\put(1314,431){\rule{1pt}{1pt}}
\put(1345,431){\rule{1pt}{1pt}}
\put(1375,431){\rule{1pt}{1pt}}
\put(1406,431){\rule{1pt}{1pt}}
\put(1436,431){\rule{1pt}{1pt}}
\put(244,510){\usebox{\plotpoint}}
\put(244.0,510.0){\rule[-0.200pt]{281.371pt}{0.400pt}}
\put(244,289){\usebox{\plotpoint}}
\put(244.0,289.0){\rule[-0.200pt]{281.371pt}{0.400pt}}
\put(244,811){\usebox{\plotpoint}}
\put(244.0,811.0){\rule[-0.200pt]{281.371pt}{0.400pt}}
\put(244,370){\usebox{\plotpoint}}
\put(244.0,370.0){\rule[-0.200pt]{281.371pt}{0.400pt}}
\end{picture}

%% file: tempfig2-2.tex
% GNUPLOT: LaTeX picture
\setlength{\unitlength}{0.240900pt}
\ifx\plotpoint\undefined\newsavebox{\plotpoint}\fi
\begin{picture}(1500,900)(0,0)
\font\gnuplot=cmr10 at 10pt
\gnuplot
\sbox{\plotpoint}{\rule[-0.200pt]{0.400pt}{0.400pt}}%
\put(220.0,495.0){\rule[-0.200pt]{292.934pt}{0.400pt}}
\put(220.0,113.0){\rule[-0.200pt]{0.400pt}{184.048pt}}
\put(220.0,113.0){\rule[-0.200pt]{4.818pt}{0.400pt}}
\put(198,113){\makebox(0,0)[r]{-4}}
\put(1416.0,113.0){\rule[-0.200pt]{4.818pt}{0.400pt}}
\put(220.0,209.0){\rule[-0.200pt]{4.818pt}{0.400pt}}
\put(198,209){\makebox(0,0)[r]{-3}}
\put(1416.0,209.0){\rule[-0.200pt]{4.818pt}{0.400pt}}
\put(220.0,304.0){\rule[-0.200pt]{4.818pt}{0.400pt}}
\put(198,304){\makebox(0,0)[r]{-2}}
\put(1416.0,304.0){\rule[-0.200pt]{4.818pt}{0.400pt}}
\put(220.0,400.0){\rule[-0.200pt]{4.818pt}{0.400pt}}
\put(198,400){\makebox(0,0)[r]{-1}}
\put(1416.0,400.0){\rule[-0.200pt]{4.818pt}{0.400pt}}
\put(220.0,495.0){\rule[-0.200pt]{4.818pt}{0.400pt}}
\put(198,495){\makebox(0,0)[r]{0}}
\put(1416.0,495.0){\rule[-0.200pt]{4.818pt}{0.400pt}}
\put(220.0,591.0){\rule[-0.200pt]{4.818pt}{0.400pt}}
\put(198,591){\makebox(0,0)[r]{1}}
\put(1416.0,591.0){\rule[-0.200pt]{4.818pt}{0.400pt}}
\put(220.0,686.0){\rule[-0.200pt]{4.818pt}{0.400pt}}
\put(198,686){\makebox(0,0)[r]{2}}
\put(1416.0,686.0){\rule[-0.200pt]{4.818pt}{0.400pt}}
\put(220.0,782.0){\rule[-0.200pt]{4.818pt}{0.400pt}}
\put(198,782){\makebox(0,0)[r]{3}}
\put(1416.0,782.0){\rule[-0.200pt]{4.818pt}{0.400pt}}
\put(220.0,877.0){\rule[-0.200pt]{4.818pt}{0.400pt}}
\put(198,877){\makebox(0,0)[r]{4}}
\put(1416.0,877.0){\rule[-0.200pt]{4.818pt}{0.400pt}}
\put(220.0,113.0){\rule[-0.200pt]{0.400pt}{4.818pt}}
\put(220,68){\makebox(0,0){0}}
\put(220.0,857.0){\rule[-0.200pt]{0.400pt}{4.818pt}}
\put(342.0,113.0){\rule[-0.200pt]{0.400pt}{4.818pt}}
\put(342,68){\makebox(0,0){50}}
\put(342.0,857.0){\rule[-0.200pt]{0.400pt}{4.818pt}}
\put(463.0,113.0){\rule[-0.200pt]{0.400pt}{4.818pt}}
\put(463,68){\makebox(0,0){100}}
\put(463.0,857.0){\rule[-0.200pt]{0.400pt}{4.818pt}}
\put(585.0,113.0){\rule[-0.200pt]{0.400pt}{4.818pt}}
\put(585,68){\makebox(0,0){150}}
\put(585.0,857.0){\rule[-0.200pt]{0.400pt}{4.818pt}}
\put(706.0,113.0){\rule[-0.200pt]{0.400pt}{4.818pt}}
\put(706,68){\makebox(0,0){200}}
\put(706.0,857.0){\rule[-0.200pt]{0.400pt}{4.818pt}}
\put(828.0,113.0){\rule[-0.200pt]{0.400pt}{4.818pt}}
\put(828,68){\makebox(0,0){250}}
\put(828.0,857.0){\rule[-0.200pt]{0.400pt}{4.818pt}}
\put(950.0,113.0){\rule[-0.200pt]{0.400pt}{4.818pt}}
\put(950,68){\makebox(0,0){300}}
\put(950.0,857.0){\rule[-0.200pt]{0.400pt}{4.818pt}}
\put(1071.0,113.0){\rule[-0.200pt]{0.400pt}{4.818pt}}
\put(1071,68){\makebox(0,0){350}}
\put(1071.0,857.0){\rule[-0.200pt]{0.400pt}{4.818pt}}
\put(1193.0,113.0){\rule[-0.200pt]{0.400pt}{4.818pt}}
\put(1193,68){\makebox(0,0){400}}
\put(1193.0,857.0){\rule[-0.200pt]{0.400pt}{4.818pt}}
\put(1314.0,113.0){\rule[-0.200pt]{0.400pt}{4.818pt}}
\put(1314,68){\makebox(0,0){450}}
\put(1314.0,857.0){\rule[-0.200pt]{0.400pt}{4.818pt}}
\put(1436.0,113.0){\rule[-0.200pt]{0.400pt}{4.818pt}}
\put(1436,68){\makebox(0,0){500}}
\put(1436.0,857.0){\rule[-0.200pt]{0.400pt}{4.818pt}}
\put(220.0,113.0){\rule[-0.200pt]{292.934pt}{0.400pt}}
\put(1436.0,113.0){\rule[-0.200pt]{0.400pt}{184.048pt}}
\put(220.0,877.0){\rule[-0.200pt]{292.934pt}{0.400pt}}
\put(45,495)
{\makebox(0,0){$\displaystyle{\frac{\mu_{\pm}(M_X)}{M_{1/2}^{(0)}}}$}}
\put(828,23){\makebox(0,0){\shortstack{\\ \\ \\ $M_{1/2}^{(0)}$}}}
\put(220.0,113.0){\rule[-0.200pt]{0.400pt}{184.048pt}}
\put(342,730){\rule{1pt}{1pt}}
\put(372,752){\rule{1pt}{1pt}}
\put(402,763){\rule{1pt}{1pt}}
\put(433,770){\rule{1pt}{1pt}}
\put(463,774){\rule{1pt}{1pt}}
\put(494,777){\rule{1pt}{1pt}}
\put(524,779){\rule{1pt}{1pt}}
\put(554,780){\rule{1pt}{1pt}}
\put(585,781){\rule{1pt}{1pt}}
\put(615,782){\rule{1pt}{1pt}}
\put(646,783){\rule{1pt}{1pt}}
\put(676,784){\rule{1pt}{1pt}}
\put(706,784){\rule{1pt}{1pt}}
\put(737,784){\rule{1pt}{1pt}}
\put(767,785){\rule{1pt}{1pt}}
\put(798,785){\rule{1pt}{1pt}}
\put(828,785){\rule{1pt}{1pt}}
\put(858,785){\rule{1pt}{1pt}}
\put(889,786){\rule{1pt}{1pt}}
\put(919,786){\rule{1pt}{1pt}}
\put(950,786){\rule{1pt}{1pt}}
\put(980,786){\rule{1pt}{1pt}}
\put(1010,786){\rule{1pt}{1pt}}
\put(1041,786){\rule{1pt}{1pt}}
\put(1071,786){\rule{1pt}{1pt}}
\put(1102,786){\rule{1pt}{1pt}}
\put(1132,786){\rule{1pt}{1pt}}
\put(1162,786){\rule{1pt}{1pt}}
\put(1193,786){\rule{1pt}{1pt}}
\put(1223,787){\rule{1pt}{1pt}}
\put(1254,787){\rule{1pt}{1pt}}
\put(1284,787){\rule{1pt}{1pt}}
\put(1314,787){\rule{1pt}{1pt}}
\put(1345,787){\rule{1pt}{1pt}}
\put(1375,787){\rule{1pt}{1pt}}
\put(1406,787){\rule{1pt}{1pt}}
\put(1436,787){\rule{1pt}{1pt}}
\put(342,652){\rule{1pt}{1pt}}
\put(372,673){\rule{1pt}{1pt}}
\put(402,683){\rule{1pt}{1pt}}
\put(433,689){\rule{1pt}{1pt}}
\put(463,693){\rule{1pt}{1pt}}
\put(494,695){\rule{1pt}{1pt}}
\put(524,697){\rule{1pt}{1pt}}
\put(554,698){\rule{1pt}{1pt}}
\put(585,699){\rule{1pt}{1pt}}
\put(615,700){\rule{1pt}{1pt}}
\put(646,701){\rule{1pt}{1pt}}
\put(676,701){\rule{1pt}{1pt}}
\put(706,702){\rule{1pt}{1pt}}
\put(737,702){\rule{1pt}{1pt}}
\put(767,702){\rule{1pt}{1pt}}
\put(798,703){\rule{1pt}{1pt}}
\put(828,703){\rule{1pt}{1pt}}
\put(858,703){\rule{1pt}{1pt}}
\put(889,703){\rule{1pt}{1pt}}
\put(919,703){\rule{1pt}{1pt}}
\put(950,703){\rule{1pt}{1pt}}
\put(980,703){\rule{1pt}{1pt}}
\put(1010,703){\rule{1pt}{1pt}}
\put(1041,704){\rule{1pt}{1pt}}
\put(1071,704){\rule{1pt}{1pt}}
\put(1102,704){\rule{1pt}{1pt}}
\put(1132,704){\rule{1pt}{1pt}}
\put(1162,704){\rule{1pt}{1pt}}
\put(1193,704){\rule{1pt}{1pt}}
\put(1223,704){\rule{1pt}{1pt}}
\put(1254,704){\rule{1pt}{1pt}}
\put(1284,704){\rule{1pt}{1pt}}
\put(1314,704){\rule{1pt}{1pt}}
\put(1345,704){\rule{1pt}{1pt}}
\put(1375,704){\rule{1pt}{1pt}}
\put(1406,704){\rule{1pt}{1pt}}
\put(1436,704){\rule{1pt}{1pt}}
\put(342,635){\rule{1pt}{1pt}}
\put(372,656){\rule{1pt}{1pt}}
\put(402,666){\rule{1pt}{1pt}}
\put(433,672){\rule{1pt}{1pt}}
\put(463,676){\rule{1pt}{1pt}}
\put(494,678){\rule{1pt}{1pt}}
\put(524,680){\rule{1pt}{1pt}}
\put(554,681){\rule{1pt}{1pt}}
\put(585,682){\rule{1pt}{1pt}}
\put(615,683){\rule{1pt}{1pt}}
\put(646,684){\rule{1pt}{1pt}}
\put(676,684){\rule{1pt}{1pt}}
\put(706,684){\rule{1pt}{1pt}}
\put(737,685){\rule{1pt}{1pt}}
\put(767,685){\rule{1pt}{1pt}}
\put(798,685){\rule{1pt}{1pt}}
\put(828,686){\rule{1pt}{1pt}}
\put(858,686){\rule{1pt}{1pt}}
\put(889,686){\rule{1pt}{1pt}}
\put(919,686){\rule{1pt}{1pt}}
\put(950,686){\rule{1pt}{1pt}}
\put(980,686){\rule{1pt}{1pt}}
\put(1010,686){\rule{1pt}{1pt}}
\put(1041,686){\rule{1pt}{1pt}}
\put(1071,686){\rule{1pt}{1pt}}
\put(1102,686){\rule{1pt}{1pt}}
\put(1132,687){\rule{1pt}{1pt}}
\put(1162,687){\rule{1pt}{1pt}}
\put(1193,687){\rule{1pt}{1pt}}
\put(1223,687){\rule{1pt}{1pt}}
\put(1254,687){\rule{1pt}{1pt}}
\put(1284,687){\rule{1pt}{1pt}}
\put(1314,687){\rule{1pt}{1pt}}
\put(1345,687){\rule{1pt}{1pt}}
\put(1375,687){\rule{1pt}{1pt}}
\put(1406,687){\rule{1pt}{1pt}}
\put(1436,687){\rule{1pt}{1pt}}
\put(342,627){\rule{1pt}{1pt}}
\put(372,649){\rule{1pt}{1pt}}
\put(402,659){\rule{1pt}{1pt}}
\put(433,665){\rule{1pt}{1pt}}
\put(463,669){\rule{1pt}{1pt}}
\put(494,671){\rule{1pt}{1pt}}
\put(524,673){\rule{1pt}{1pt}}
\put(554,674){\rule{1pt}{1pt}}
\put(585,675){\rule{1pt}{1pt}}
\put(615,676){\rule{1pt}{1pt}}
\put(646,677){\rule{1pt}{1pt}}
\put(676,677){\rule{1pt}{1pt}}
\put(706,677){\rule{1pt}{1pt}}
\put(737,678){\rule{1pt}{1pt}}
\put(767,678){\rule{1pt}{1pt}}
\put(798,678){\rule{1pt}{1pt}}
\put(828,678){\rule{1pt}{1pt}}
\put(858,679){\rule{1pt}{1pt}}
\put(889,679){\rule{1pt}{1pt}}
\put(919,679){\rule{1pt}{1pt}}
\put(950,679){\rule{1pt}{1pt}}
\put(980,679){\rule{1pt}{1pt}}
\put(1010,679){\rule{1pt}{1pt}}
\put(1041,679){\rule{1pt}{1pt}}
\put(1071,679){\rule{1pt}{1pt}}
\put(1102,679){\rule{1pt}{1pt}}
\put(1132,679){\rule{1pt}{1pt}}
\put(1162,680){\rule{1pt}{1pt}}
\put(1193,680){\rule{1pt}{1pt}}
\put(1223,680){\rule{1pt}{1pt}}
\put(1254,680){\rule{1pt}{1pt}}
\put(1284,680){\rule{1pt}{1pt}}
\put(1314,680){\rule{1pt}{1pt}}
\put(1345,680){\rule{1pt}{1pt}}
\put(1375,680){\rule{1pt}{1pt}}
\put(1406,680){\rule{1pt}{1pt}}
\put(1436,680){\rule{1pt}{1pt}}
\put(342,624){\rule{1pt}{1pt}}
\put(372,645){\rule{1pt}{1pt}}
\put(402,655){\rule{1pt}{1pt}}
\put(433,661){\rule{1pt}{1pt}}
\put(463,665){\rule{1pt}{1pt}}
\put(494,667){\rule{1pt}{1pt}}
\put(524,669){\rule{1pt}{1pt}}
\put(554,671){\rule{1pt}{1pt}}
\put(585,672){\rule{1pt}{1pt}}
\put(615,672){\rule{1pt}{1pt}}
\put(646,673){\rule{1pt}{1pt}}
\put(676,673){\rule{1pt}{1pt}}
\put(706,674){\rule{1pt}{1pt}}
\put(737,674){\rule{1pt}{1pt}}
\put(767,674){\rule{1pt}{1pt}}
\put(798,675){\rule{1pt}{1pt}}
\put(828,675){\rule{1pt}{1pt}}
\put(858,675){\rule{1pt}{1pt}}
\put(889,675){\rule{1pt}{1pt}}
\put(919,675){\rule{1pt}{1pt}}
\put(950,675){\rule{1pt}{1pt}}
\put(980,676){\rule{1pt}{1pt}}
\put(1010,676){\rule{1pt}{1pt}}
\put(1041,676){\rule{1pt}{1pt}}
\put(1071,676){\rule{1pt}{1pt}}
\put(1102,676){\rule{1pt}{1pt}}
\put(1132,676){\rule{1pt}{1pt}}
\put(1162,676){\rule{1pt}{1pt}}
\put(1193,676){\rule{1pt}{1pt}}
\put(1223,676){\rule{1pt}{1pt}}
\put(1254,676){\rule{1pt}{1pt}}
\put(1284,676){\rule{1pt}{1pt}}
\put(1314,676){\rule{1pt}{1pt}}
\put(1345,676){\rule{1pt}{1pt}}
\put(1375,676){\rule{1pt}{1pt}}
\put(1406,676){\rule{1pt}{1pt}}
\put(1436,676){\rule{1pt}{1pt}}
\put(342,621){\rule{1pt}{1pt}}
\put(372,643){\rule{1pt}{1pt}}
\put(402,653){\rule{1pt}{1pt}}
\put(433,659){\rule{1pt}{1pt}}
\put(463,663){\rule{1pt}{1pt}}
\put(494,665){\rule{1pt}{1pt}}
\put(524,667){\rule{1pt}{1pt}}
\put(554,668){\rule{1pt}{1pt}}
\put(585,669){\rule{1pt}{1pt}}
\put(615,670){\rule{1pt}{1pt}}
\put(646,671){\rule{1pt}{1pt}}
\put(676,671){\rule{1pt}{1pt}}
\put(706,672){\rule{1pt}{1pt}}
\put(737,672){\rule{1pt}{1pt}}
\put(767,672){\rule{1pt}{1pt}}
\put(798,673){\rule{1pt}{1pt}}
\put(828,673){\rule{1pt}{1pt}}
\put(858,673){\rule{1pt}{1pt}}
\put(889,673){\rule{1pt}{1pt}}
\put(919,673){\rule{1pt}{1pt}}
\put(950,673){\rule{1pt}{1pt}}
\put(980,673){\rule{1pt}{1pt}}
\put(1010,673){\rule{1pt}{1pt}}
\put(1041,674){\rule{1pt}{1pt}}
\put(1071,674){\rule{1pt}{1pt}}
\put(1102,674){\rule{1pt}{1pt}}
\put(1132,674){\rule{1pt}{1pt}}
\put(1162,674){\rule{1pt}{1pt}}
\put(1193,674){\rule{1pt}{1pt}}
\put(1223,674){\rule{1pt}{1pt}}
\put(1254,674){\rule{1pt}{1pt}}
\put(1284,674){\rule{1pt}{1pt}}
\put(1314,674){\rule{1pt}{1pt}}
\put(1345,674){\rule{1pt}{1pt}}
\put(1375,674){\rule{1pt}{1pt}}
\put(1406,674){\rule{1pt}{1pt}}
\put(1436,674){\rule{1pt}{1pt}}
\put(342,620){\rule{1pt}{1pt}}
\put(372,641){\rule{1pt}{1pt}}
\put(402,652){\rule{1pt}{1pt}}
\put(433,658){\rule{1pt}{1pt}}
\put(463,661){\rule{1pt}{1pt}}
\put(494,664){\rule{1pt}{1pt}}
\put(524,666){\rule{1pt}{1pt}}
\put(554,667){\rule{1pt}{1pt}}
\put(585,668){\rule{1pt}{1pt}}
\put(615,669){\rule{1pt}{1pt}}
\put(646,669){\rule{1pt}{1pt}}
\put(676,670){\rule{1pt}{1pt}}
\put(706,670){\rule{1pt}{1pt}}
\put(737,671){\rule{1pt}{1pt}}
\put(767,671){\rule{1pt}{1pt}}
\put(798,671){\rule{1pt}{1pt}}
\put(828,671){\rule{1pt}{1pt}}
\put(858,672){\rule{1pt}{1pt}}
\put(889,672){\rule{1pt}{1pt}}
\put(919,672){\rule{1pt}{1pt}}
\put(950,672){\rule{1pt}{1pt}}
\put(980,672){\rule{1pt}{1pt}}
\put(1010,672){\rule{1pt}{1pt}}
\put(1041,672){\rule{1pt}{1pt}}
\put(1071,672){\rule{1pt}{1pt}}
\put(1102,672){\rule{1pt}{1pt}}
\put(1132,672){\rule{1pt}{1pt}}
\put(1162,672){\rule{1pt}{1pt}}
\put(1193,672){\rule{1pt}{1pt}}
\put(1223,673){\rule{1pt}{1pt}}
\put(1254,673){\rule{1pt}{1pt}}
\put(1284,673){\rule{1pt}{1pt}}
\put(1314,673){\rule{1pt}{1pt}}
\put(1345,673){\rule{1pt}{1pt}}
\put(1375,673){\rule{1pt}{1pt}}
\put(1406,673){\rule{1pt}{1pt}}
\put(1436,673){\rule{1pt}{1pt}}
\put(342,619){\rule{1pt}{1pt}}
\put(372,640){\rule{1pt}{1pt}}
\put(402,651){\rule{1pt}{1pt}}
\put(433,657){\rule{1pt}{1pt}}
\put(463,661){\rule{1pt}{1pt}}
\put(494,663){\rule{1pt}{1pt}}
\put(524,665){\rule{1pt}{1pt}}
\put(554,666){\rule{1pt}{1pt}}
\put(585,667){\rule{1pt}{1pt}}
\put(615,668){\rule{1pt}{1pt}}
\put(646,669){\rule{1pt}{1pt}}
\put(676,669){\rule{1pt}{1pt}}
\put(706,669){\rule{1pt}{1pt}}
\put(737,670){\rule{1pt}{1pt}}
\put(767,670){\rule{1pt}{1pt}}
\put(798,670){\rule{1pt}{1pt}}
\put(828,670){\rule{1pt}{1pt}}
\put(858,671){\rule{1pt}{1pt}}
\put(889,671){\rule{1pt}{1pt}}
\put(919,671){\rule{1pt}{1pt}}
\put(950,671){\rule{1pt}{1pt}}
\put(980,671){\rule{1pt}{1pt}}
\put(1010,671){\rule{1pt}{1pt}}
\put(1041,671){\rule{1pt}{1pt}}
\put(1071,671){\rule{1pt}{1pt}}
\put(1102,671){\rule{1pt}{1pt}}
\put(1132,671){\rule{1pt}{1pt}}
\put(1162,672){\rule{1pt}{1pt}}
\put(1193,672){\rule{1pt}{1pt}}
\put(1223,672){\rule{1pt}{1pt}}
\put(1254,672){\rule{1pt}{1pt}}
\put(1284,672){\rule{1pt}{1pt}}
\put(1314,672){\rule{1pt}{1pt}}
\put(1345,672){\rule{1pt}{1pt}}
\put(1375,672){\rule{1pt}{1pt}}
\put(1406,672){\rule{1pt}{1pt}}
\put(1436,672){\rule{1pt}{1pt}}
\put(342,618){\rule{1pt}{1pt}}
\put(372,640){\rule{1pt}{1pt}}
\put(402,650){\rule{1pt}{1pt}}
\put(433,656){\rule{1pt}{1pt}}
\put(463,660){\rule{1pt}{1pt}}
\put(494,662){\rule{1pt}{1pt}}
\put(524,664){\rule{1pt}{1pt}}
\put(554,666){\rule{1pt}{1pt}}
\put(585,667){\rule{1pt}{1pt}}
\put(615,667){\rule{1pt}{1pt}}
\put(646,668){\rule{1pt}{1pt}}
\put(676,668){\rule{1pt}{1pt}}
\put(706,669){\rule{1pt}{1pt}}
\put(737,669){\rule{1pt}{1pt}}
\put(767,669){\rule{1pt}{1pt}}
\put(798,670){\rule{1pt}{1pt}}
\put(828,670){\rule{1pt}{1pt}}
\put(858,670){\rule{1pt}{1pt}}
\put(889,670){\rule{1pt}{1pt}}
\put(919,670){\rule{1pt}{1pt}}
\put(950,670){\rule{1pt}{1pt}}
\put(980,670){\rule{1pt}{1pt}}
\put(1010,671){\rule{1pt}{1pt}}
\put(1041,671){\rule{1pt}{1pt}}
\put(1071,671){\rule{1pt}{1pt}}
\put(1102,671){\rule{1pt}{1pt}}
\put(1132,671){\rule{1pt}{1pt}}
\put(1162,671){\rule{1pt}{1pt}}
\put(1193,671){\rule{1pt}{1pt}}
\put(1223,671){\rule{1pt}{1pt}}
\put(1254,671){\rule{1pt}{1pt}}
\put(1284,671){\rule{1pt}{1pt}}
\put(1314,671){\rule{1pt}{1pt}}
\put(1345,671){\rule{1pt}{1pt}}
\put(1375,671){\rule{1pt}{1pt}}
\put(1406,671){\rule{1pt}{1pt}}
\put(1436,671){\rule{1pt}{1pt}}
\put(342,260){\rule{1pt}{1pt}}
\put(372,238){\rule{1pt}{1pt}}
\put(402,227){\rule{1pt}{1pt}}
\put(433,220){\rule{1pt}{1pt}}
\put(463,216){\rule{1pt}{1pt}}
\put(494,213){\rule{1pt}{1pt}}
\put(524,211){\rule{1pt}{1pt}}
\put(554,210){\rule{1pt}{1pt}}
\put(585,209){\rule{1pt}{1pt}}
\put(615,208){\rule{1pt}{1pt}}
\put(646,207){\rule{1pt}{1pt}}
\put(676,206){\rule{1pt}{1pt}}
\put(706,206){\rule{1pt}{1pt}}
\put(737,206){\rule{1pt}{1pt}}
\put(767,205){\rule{1pt}{1pt}}
\put(798,205){\rule{1pt}{1pt}}
\put(828,205){\rule{1pt}{1pt}}
\put(858,205){\rule{1pt}{1pt}}
\put(889,204){\rule{1pt}{1pt}}
\put(919,204){\rule{1pt}{1pt}}
\put(950,204){\rule{1pt}{1pt}}
\put(980,204){\rule{1pt}{1pt}}
\put(1010,204){\rule{1pt}{1pt}}
\put(1041,204){\rule{1pt}{1pt}}
\put(1071,204){\rule{1pt}{1pt}}
\put(1102,204){\rule{1pt}{1pt}}
\put(1132,204){\rule{1pt}{1pt}}
\put(1162,204){\rule{1pt}{1pt}}
\put(1193,204){\rule{1pt}{1pt}}
\put(1223,203){\rule{1pt}{1pt}}
\put(1254,203){\rule{1pt}{1pt}}
\put(1284,203){\rule{1pt}{1pt}}
\put(1314,203){\rule{1pt}{1pt}}
\put(1345,203){\rule{1pt}{1pt}}
\put(1375,203){\rule{1pt}{1pt}}
\put(1406,203){\rule{1pt}{1pt}}
\put(1436,203){\rule{1pt}{1pt}}
\put(342,338){\rule{1pt}{1pt}}
\put(372,317){\rule{1pt}{1pt}}
\put(402,307){\rule{1pt}{1pt}}
\put(433,301){\rule{1pt}{1pt}}
\put(463,297){\rule{1pt}{1pt}}
\put(494,295){\rule{1pt}{1pt}}
\put(524,293){\rule{1pt}{1pt}}
\put(554,292){\rule{1pt}{1pt}}
\put(585,291){\rule{1pt}{1pt}}
\put(615,290){\rule{1pt}{1pt}}
\put(646,289){\rule{1pt}{1pt}}
\put(676,289){\rule{1pt}{1pt}}
\put(706,288){\rule{1pt}{1pt}}
\put(737,288){\rule{1pt}{1pt}}
\put(767,288){\rule{1pt}{1pt}}
\put(798,287){\rule{1pt}{1pt}}
\put(828,287){\rule{1pt}{1pt}}
\put(858,287){\rule{1pt}{1pt}}
\put(889,287){\rule{1pt}{1pt}}
\put(919,287){\rule{1pt}{1pt}}
\put(950,287){\rule{1pt}{1pt}}
\put(980,287){\rule{1pt}{1pt}}
\put(1010,287){\rule{1pt}{1pt}}
\put(1041,286){\rule{1pt}{1pt}}
\put(1071,286){\rule{1pt}{1pt}}
\put(1102,286){\rule{1pt}{1pt}}
\put(1132,286){\rule{1pt}{1pt}}
\put(1162,286){\rule{1pt}{1pt}}
\put(1193,286){\rule{1pt}{1pt}}
\put(1223,286){\rule{1pt}{1pt}}
\put(1254,286){\rule{1pt}{1pt}}
\put(1284,286){\rule{1pt}{1pt}}
\put(1314,286){\rule{1pt}{1pt}}
\put(1345,286){\rule{1pt}{1pt}}
\put(1375,286){\rule{1pt}{1pt}}
\put(1406,286){\rule{1pt}{1pt}}
\put(1436,286){\rule{1pt}{1pt}}
\put(342,355){\rule{1pt}{1pt}}
\put(372,334){\rule{1pt}{1pt}}
\put(402,324){\rule{1pt}{1pt}}
\put(433,318){\rule{1pt}{1pt}}
\put(463,314){\rule{1pt}{1pt}}
\put(494,312){\rule{1pt}{1pt}}
\put(524,310){\rule{1pt}{1pt}}
\put(554,309){\rule{1pt}{1pt}}
\put(585,308){\rule{1pt}{1pt}}
\put(615,307){\rule{1pt}{1pt}}
\put(646,306){\rule{1pt}{1pt}}
\put(676,306){\rule{1pt}{1pt}}
\put(706,306){\rule{1pt}{1pt}}
\put(737,305){\rule{1pt}{1pt}}
\put(767,305){\rule{1pt}{1pt}}
\put(798,305){\rule{1pt}{1pt}}
\put(828,304){\rule{1pt}{1pt}}
\put(858,304){\rule{1pt}{1pt}}
\put(889,304){\rule{1pt}{1pt}}
\put(919,304){\rule{1pt}{1pt}}
\put(950,304){\rule{1pt}{1pt}}
\put(980,304){\rule{1pt}{1pt}}
\put(1010,304){\rule{1pt}{1pt}}
\put(1041,304){\rule{1pt}{1pt}}
\put(1071,304){\rule{1pt}{1pt}}
\put(1102,304){\rule{1pt}{1pt}}
\put(1132,303){\rule{1pt}{1pt}}
\put(1162,303){\rule{1pt}{1pt}}
\put(1193,303){\rule{1pt}{1pt}}
\put(1223,303){\rule{1pt}{1pt}}
\put(1254,303){\rule{1pt}{1pt}}
\put(1284,303){\rule{1pt}{1pt}}
\put(1314,303){\rule{1pt}{1pt}}
\put(1345,303){\rule{1pt}{1pt}}
\put(1375,303){\rule{1pt}{1pt}}
\put(1406,303){\rule{1pt}{1pt}}
\put(1436,303){\rule{1pt}{1pt}}
\put(342,363){\rule{1pt}{1pt}}
\put(372,341){\rule{1pt}{1pt}}
\put(402,331){\rule{1pt}{1pt}}
\put(433,325){\rule{1pt}{1pt}}
\put(463,321){\rule{1pt}{1pt}}
\put(494,319){\rule{1pt}{1pt}}
\put(524,317){\rule{1pt}{1pt}}
\put(554,316){\rule{1pt}{1pt}}
\put(585,315){\rule{1pt}{1pt}}
\put(615,314){\rule{1pt}{1pt}}
\put(646,313){\rule{1pt}{1pt}}
\put(676,313){\rule{1pt}{1pt}}
\put(706,313){\rule{1pt}{1pt}}
\put(737,312){\rule{1pt}{1pt}}
\put(767,312){\rule{1pt}{1pt}}
\put(798,312){\rule{1pt}{1pt}}
\put(828,312){\rule{1pt}{1pt}}
\put(858,311){\rule{1pt}{1pt}}
\put(889,311){\rule{1pt}{1pt}}
\put(919,311){\rule{1pt}{1pt}}
\put(950,311){\rule{1pt}{1pt}}
\put(980,311){\rule{1pt}{1pt}}
\put(1010,311){\rule{1pt}{1pt}}
\put(1041,311){\rule{1pt}{1pt}}
\put(1071,311){\rule{1pt}{1pt}}
\put(1102,311){\rule{1pt}{1pt}}
\put(1132,311){\rule{1pt}{1pt}}
\put(1162,310){\rule{1pt}{1pt}}
\put(1193,310){\rule{1pt}{1pt}}
\put(1223,310){\rule{1pt}{1pt}}
\put(1254,310){\rule{1pt}{1pt}}
\put(1284,310){\rule{1pt}{1pt}}
\put(1314,310){\rule{1pt}{1pt}}
\put(1345,310){\rule{1pt}{1pt}}
\put(1375,310){\rule{1pt}{1pt}}
\put(1406,310){\rule{1pt}{1pt}}
\put(1436,310){\rule{1pt}{1pt}}
\put(342,366){\rule{1pt}{1pt}}
\put(372,345){\rule{1pt}{1pt}}
\put(402,335){\rule{1pt}{1pt}}
\put(433,329){\rule{1pt}{1pt}}
\put(463,325){\rule{1pt}{1pt}}
\put(494,323){\rule{1pt}{1pt}}
\put(524,321){\rule{1pt}{1pt}}
\put(554,319){\rule{1pt}{1pt}}
\put(585,318){\rule{1pt}{1pt}}
\put(615,318){\rule{1pt}{1pt}}
\put(646,317){\rule{1pt}{1pt}}
\put(676,317){\rule{1pt}{1pt}}
\put(706,316){\rule{1pt}{1pt}}
\put(737,316){\rule{1pt}{1pt}}
\put(767,316){\rule{1pt}{1pt}}
\put(798,315){\rule{1pt}{1pt}}
\put(828,315){\rule{1pt}{1pt}}
\put(858,315){\rule{1pt}{1pt}}
\put(889,315){\rule{1pt}{1pt}}
\put(919,315){\rule{1pt}{1pt}}
\put(950,315){\rule{1pt}{1pt}}
\put(980,314){\rule{1pt}{1pt}}
\put(1010,314){\rule{1pt}{1pt}}
\put(1041,314){\rule{1pt}{1pt}}
\put(1071,314){\rule{1pt}{1pt}}
\put(1102,314){\rule{1pt}{1pt}}
\put(1132,314){\rule{1pt}{1pt}}
\put(1162,314){\rule{1pt}{1pt}}
\put(1193,314){\rule{1pt}{1pt}}
\put(1223,314){\rule{1pt}{1pt}}
\put(1254,314){\rule{1pt}{1pt}}
\put(1284,314){\rule{1pt}{1pt}}
\put(1314,314){\rule{1pt}{1pt}}
\put(1345,314){\rule{1pt}{1pt}}
\put(1375,314){\rule{1pt}{1pt}}
\put(1406,314){\rule{1pt}{1pt}}
\put(1436,314){\rule{1pt}{1pt}}
\put(342,369){\rule{1pt}{1pt}}
\put(372,347){\rule{1pt}{1pt}}
\put(402,337){\rule{1pt}{1pt}}
\put(433,331){\rule{1pt}{1pt}}
\put(463,327){\rule{1pt}{1pt}}
\put(494,325){\rule{1pt}{1pt}}
\put(524,323){\rule{1pt}{1pt}}
\put(554,322){\rule{1pt}{1pt}}
\put(585,321){\rule{1pt}{1pt}}
\put(615,320){\rule{1pt}{1pt}}
\put(646,319){\rule{1pt}{1pt}}
\put(676,319){\rule{1pt}{1pt}}
\put(706,318){\rule{1pt}{1pt}}
\put(737,318){\rule{1pt}{1pt}}
\put(767,318){\rule{1pt}{1pt}}
\put(798,317){\rule{1pt}{1pt}}
\put(828,317){\rule{1pt}{1pt}}
\put(858,317){\rule{1pt}{1pt}}
\put(889,317){\rule{1pt}{1pt}}
\put(919,317){\rule{1pt}{1pt}}
\put(950,317){\rule{1pt}{1pt}}
\put(980,317){\rule{1pt}{1pt}}
\put(1010,317){\rule{1pt}{1pt}}
\put(1041,316){\rule{1pt}{1pt}}
\put(1071,316){\rule{1pt}{1pt}}
\put(1102,316){\rule{1pt}{1pt}}
\put(1132,316){\rule{1pt}{1pt}}
\put(1162,316){\rule{1pt}{1pt}}
\put(1193,316){\rule{1pt}{1pt}}
\put(1223,316){\rule{1pt}{1pt}}
\put(1254,316){\rule{1pt}{1pt}}
\put(1284,316){\rule{1pt}{1pt}}
\put(1314,316){\rule{1pt}{1pt}}
\put(1345,316){\rule{1pt}{1pt}}
\put(1375,316){\rule{1pt}{1pt}}
\put(1406,316){\rule{1pt}{1pt}}
\put(1436,316){\rule{1pt}{1pt}}
\put(342,370){\rule{1pt}{1pt}}
\put(372,349){\rule{1pt}{1pt}}
\put(402,338){\rule{1pt}{1pt}}
\put(433,332){\rule{1pt}{1pt}}
\put(463,329){\rule{1pt}{1pt}}
\put(494,326){\rule{1pt}{1pt}}
\put(524,324){\rule{1pt}{1pt}}
\put(554,323){\rule{1pt}{1pt}}
\put(585,322){\rule{1pt}{1pt}}
\put(615,321){\rule{1pt}{1pt}}
\put(646,321){\rule{1pt}{1pt}}
\put(676,320){\rule{1pt}{1pt}}
\put(706,320){\rule{1pt}{1pt}}
\put(737,319){\rule{1pt}{1pt}}
\put(767,319){\rule{1pt}{1pt}}
\put(798,319){\rule{1pt}{1pt}}
\put(828,319){\rule{1pt}{1pt}}
\put(858,318){\rule{1pt}{1pt}}
\put(889,318){\rule{1pt}{1pt}}
\put(919,318){\rule{1pt}{1pt}}
\put(950,318){\rule{1pt}{1pt}}
\put(980,318){\rule{1pt}{1pt}}
\put(1010,318){\rule{1pt}{1pt}}
\put(1041,318){\rule{1pt}{1pt}}
\put(1071,318){\rule{1pt}{1pt}}
\put(1102,318){\rule{1pt}{1pt}}
\put(1132,318){\rule{1pt}{1pt}}
\put(1162,318){\rule{1pt}{1pt}}
\put(1193,318){\rule{1pt}{1pt}}
\put(1223,317){\rule{1pt}{1pt}}
\put(1254,317){\rule{1pt}{1pt}}
\put(1284,317){\rule{1pt}{1pt}}
\put(1314,317){\rule{1pt}{1pt}}
\put(1345,317){\rule{1pt}{1pt}}
\put(1375,317){\rule{1pt}{1pt}}
\put(1406,317){\rule{1pt}{1pt}}
\put(1436,317){\rule{1pt}{1pt}}
\put(342,371){\rule{1pt}{1pt}}
\put(372,350){\rule{1pt}{1pt}}
\put(402,339){\rule{1pt}{1pt}}
\put(433,333){\rule{1pt}{1pt}}
\put(463,329){\rule{1pt}{1pt}}
\put(494,327){\rule{1pt}{1pt}}
\put(524,325){\rule{1pt}{1pt}}
\put(554,324){\rule{1pt}{1pt}}
\put(585,323){\rule{1pt}{1pt}}
\put(615,322){\rule{1pt}{1pt}}
\put(646,321){\rule{1pt}{1pt}}
\put(676,321){\rule{1pt}{1pt}}
\put(706,321){\rule{1pt}{1pt}}
\put(737,320){\rule{1pt}{1pt}}
\put(767,320){\rule{1pt}{1pt}}
\put(798,320){\rule{1pt}{1pt}}
\put(828,320){\rule{1pt}{1pt}}
\put(858,319){\rule{1pt}{1pt}}
\put(889,319){\rule{1pt}{1pt}}
\put(919,319){\rule{1pt}{1pt}}
\put(950,319){\rule{1pt}{1pt}}
\put(980,319){\rule{1pt}{1pt}}
\put(1010,319){\rule{1pt}{1pt}}
\put(1041,319){\rule{1pt}{1pt}}
\put(1071,319){\rule{1pt}{1pt}}
\put(1102,319){\rule{1pt}{1pt}}
\put(1132,319){\rule{1pt}{1pt}}
\put(1162,318){\rule{1pt}{1pt}}
\put(1193,318){\rule{1pt}{1pt}}
\put(1223,318){\rule{1pt}{1pt}}
\put(1254,318){\rule{1pt}{1pt}}
\put(1284,318){\rule{1pt}{1pt}}
\put(1314,318){\rule{1pt}{1pt}}
\put(1345,318){\rule{1pt}{1pt}}
\put(1375,318){\rule{1pt}{1pt}}
\put(1406,318){\rule{1pt}{1pt}}
\put(1436,318){\rule{1pt}{1pt}}
\put(342,372){\rule{1pt}{1pt}}
\put(372,350){\rule{1pt}{1pt}}
\put(402,340){\rule{1pt}{1pt}}
\put(433,334){\rule{1pt}{1pt}}
\put(463,330){\rule{1pt}{1pt}}
\put(494,328){\rule{1pt}{1pt}}
\put(524,326){\rule{1pt}{1pt}}
\put(554,324){\rule{1pt}{1pt}}
\put(585,323){\rule{1pt}{1pt}}
\put(615,323){\rule{1pt}{1pt}}
\put(646,322){\rule{1pt}{1pt}}
\put(676,322){\rule{1pt}{1pt}}
\put(706,321){\rule{1pt}{1pt}}
\put(737,321){\rule{1pt}{1pt}}
\put(767,321){\rule{1pt}{1pt}}
\put(798,320){\rule{1pt}{1pt}}
\put(828,320){\rule{1pt}{1pt}}
\put(858,320){\rule{1pt}{1pt}}
\put(889,320){\rule{1pt}{1pt}}
\put(919,320){\rule{1pt}{1pt}}
\put(950,320){\rule{1pt}{1pt}}
\put(980,320){\rule{1pt}{1pt}}
\put(1010,319){\rule{1pt}{1pt}}
\put(1041,319){\rule{1pt}{1pt}}
\put(1071,319){\rule{1pt}{1pt}}
\put(1102,319){\rule{1pt}{1pt}}
\put(1132,319){\rule{1pt}{1pt}}
\put(1162,319){\rule{1pt}{1pt}}
\put(1193,319){\rule{1pt}{1pt}}
\put(1223,319){\rule{1pt}{1pt}}
\put(1254,319){\rule{1pt}{1pt}}
\put(1284,319){\rule{1pt}{1pt}}
\put(1314,319){\rule{1pt}{1pt}}
\put(1345,319){\rule{1pt}{1pt}}
\put(1375,319){\rule{1pt}{1pt}}
\put(1406,319){\rule{1pt}{1pt}}
\put(1436,319){\rule{1pt}{1pt}}
\put(244,550){\usebox{\plotpoint}}
\put(244.0,550.0){\rule[-0.200pt]{281.371pt}{0.400pt}}
\put(244,440){\usebox{\plotpoint}}
\put(244.0,440.0){\rule[-0.200pt]{281.371pt}{0.400pt}}
\end{picture}

%% file: fignp.tex
% GNUPLOT: LaTeX picture
\setlength{\unitlength}{0.240900pt}
\ifx\plotpoint\undefined\newsavebox{\plotpoint}\fi
\sbox{\plotpoint}{\rule[-0.200pt]{0.400pt}{0.400pt}}%
\begin{picture}(1500,900)(0,0)
\font\gnuplot=cmr10 at 10pt
\gnuplot
\sbox{\plotpoint}{\rule[-0.200pt]{0.400pt}{0.400pt}}%
\put(220.0,750.0){\rule[-0.200pt]{292.934pt}{0.400pt}}
\put(220.0,113.0){\rule[-0.200pt]{0.400pt}{184.048pt}}
\put(220.0,113.0){\rule[-0.200pt]{4.818pt}{0.400pt}}
\put(198,113){\makebox(0,0)[r]{-2.5}}
\put(1416.0,113.0){\rule[-0.200pt]{4.818pt}{0.400pt}}
\put(220.0,240.0){\rule[-0.200pt]{4.818pt}{0.400pt}}
\put(198,240){\makebox(0,0)[r]{-2}}
\put(1416.0,240.0){\rule[-0.200pt]{4.818pt}{0.400pt}}
\put(220.0,368.0){\rule[-0.200pt]{4.818pt}{0.400pt}}
\put(198,368){\makebox(0,0)[r]{-1.5}}
\put(1416.0,368.0){\rule[-0.200pt]{4.818pt}{0.400pt}}
\put(220.0,495.0){\rule[-0.200pt]{4.818pt}{0.400pt}}
\put(198,495){\makebox(0,0)[r]{-1}}
\put(1416.0,495.0){\rule[-0.200pt]{4.818pt}{0.400pt}}
\put(220.0,622.0){\rule[-0.200pt]{4.818pt}{0.400pt}}
\put(198,622){\makebox(0,0)[r]{-0.5}}
\put(1416.0,622.0){\rule[-0.200pt]{4.818pt}{0.400pt}}
\put(220.0,750.0){\rule[-0.200pt]{4.818pt}{0.400pt}}
\put(198,750){\makebox(0,0)[r]{0}}
\put(1416.0,750.0){\rule[-0.200pt]{4.818pt}{0.400pt}}
\put(220.0,877.0){\rule[-0.200pt]{4.818pt}{0.400pt}}
\put(198,877){\makebox(0,0)[r]{0.5}}
\put(1416.0,877.0){\rule[-0.200pt]{4.818pt}{0.400pt}}
\put(220.0,113.0){\rule[-0.200pt]{0.400pt}{4.818pt}}
\put(220,68){\makebox(0,0){0}}
\put(220.0,857.0){\rule[-0.200pt]{0.400pt}{4.818pt}}
\put(355.0,113.0){\rule[-0.200pt]{0.400pt}{4.818pt}}
\put(355,68){\makebox(0,0){0.1}}
\put(355.0,857.0){\rule[-0.200pt]{0.400pt}{4.818pt}}
\put(490.0,113.0){\rule[-0.200pt]{0.400pt}{4.818pt}}
\put(490,68){\makebox(0,0){0.2}}
\put(490.0,857.0){\rule[-0.200pt]{0.400pt}{4.818pt}}
\put(625.0,113.0){\rule[-0.200pt]{0.400pt}{4.818pt}}
\put(625,68){\makebox(0,0){0.3}}
\put(625.0,857.0){\rule[-0.200pt]{0.400pt}{4.818pt}}
\put(760.0,113.0){\rule[-0.200pt]{0.400pt}{4.818pt}}
\put(760,68){\makebox(0,0){0.4}}
\put(760.0,857.0){\rule[-0.200pt]{0.400pt}{4.818pt}}
\put(896.0,113.0){\rule[-0.200pt]{0.400pt}{4.818pt}}
\put(896,68){\makebox(0,0){0.5}}
\put(896.0,857.0){\rule[-0.200pt]{0.400pt}{4.818pt}}
\put(1031.0,113.0){\rule[-0.200pt]{0.400pt}{4.818pt}}
\put(1031,68){\makebox(0,0){0.6}}
\put(1031.0,857.0){\rule[-0.200pt]{0.400pt}{4.818pt}}
\put(1166.0,113.0){\rule[-0.200pt]{0.400pt}{4.818pt}}
\put(1166,68){\makebox(0,0){0.7}}
\put(1166.0,857.0){\rule[-0.200pt]{0.400pt}{4.818pt}}
\put(1301.0,113.0){\rule[-0.200pt]{0.400pt}{4.818pt}}
\put(1301,68){\makebox(0,0){0.8}}
\put(1301.0,857.0){\rule[-0.200pt]{0.400pt}{4.818pt}}
\put(1436.0,113.0){\rule[-0.200pt]{0.400pt}{4.818pt}}
\put(1436,68){\makebox(0,0){0.9}}
\put(1436.0,857.0){\rule[-0.200pt]{0.400pt}{4.818pt}}
\put(220.0,113.0){\rule[-0.200pt]{292.934pt}{0.400pt}}
\put(1436.0,113.0){\rule[-0.200pt]{0.400pt}{184.048pt}}
\put(220.0,877.0){\rule[-0.200pt]{292.934pt}{0.400pt}}
\put(45,495)
{\makebox(0,0){$\displaystyle{\frac{B_{\pm}(M_X)}{M_{1/2}^{(0)}}}$}}
\put(828,23){\makebox(0,0){$\cos \theta$}}
\put(220.0,113.0){\rule[-0.200pt]{0.400pt}{184.048pt}}
\put(220,853){\rule{1pt}{1pt}}
\put(247,853){\rule{1pt}{1pt}}
\put(274,853){\rule{1pt}{1pt}}
\put(301,853){\rule{1pt}{1pt}}
\put(328,852){\rule{1pt}{1pt}}
\put(355,852){\rule{1pt}{1pt}}
\put(382,852){\rule{1pt}{1pt}}
\put(409,852){\rule{1pt}{1pt}}
\put(436,852){\rule{1pt}{1pt}}
\put(463,852){\rule{1pt}{1pt}}
\put(490,852){\rule{1pt}{1pt}}
\put(517,851){\rule{1pt}{1pt}}
\put(544,851){\rule{1pt}{1pt}}
\put(571,851){\rule{1pt}{1pt}}
\put(598,851){\rule{1pt}{1pt}}
\put(625,850){\rule{1pt}{1pt}}
\put(652,850){\rule{1pt}{1pt}}
\put(679,850){\rule{1pt}{1pt}}
\put(706,849){\rule{1pt}{1pt}}
\put(733,849){\rule{1pt}{1pt}}
\put(760,848){\rule{1pt}{1pt}}
\put(787,848){\rule{1pt}{1pt}}
\put(814,847){\rule{1pt}{1pt}}
\put(842,847){\rule{1pt}{1pt}}
\put(869,846){\rule{1pt}{1pt}}
\put(896,845){\rule{1pt}{1pt}}
\put(923,844){\rule{1pt}{1pt}}
\put(950,843){\rule{1pt}{1pt}}
\put(977,842){\rule{1pt}{1pt}}
\put(1004,841){\rule{1pt}{1pt}}
\put(1031,840){\rule{1pt}{1pt}}
\put(1058,839){\rule{1pt}{1pt}}
\put(1085,837){\rule{1pt}{1pt}}
\put(1112,835){\rule{1pt}{1pt}}
\put(1139,833){\rule{1pt}{1pt}}
\put(1166,831){\rule{1pt}{1pt}}
\put(1193,828){\rule{1pt}{1pt}}
\put(1220,825){\rule{1pt}{1pt}}
\put(1247,822){\rule{1pt}{1pt}}
\put(1274,818){\rule{1pt}{1pt}}
\put(1301,812){\rule{1pt}{1pt}}
\put(1328,806){\rule{1pt}{1pt}}
\put(1355,798){\rule{1pt}{1pt}}
\put(1382,788){\rule{1pt}{1pt}}
\put(1409,773){\rule{1pt}{1pt}}
\put(1436,753){\rule{1pt}{1pt}}
\put(220,199){\rule{1pt}{1pt}}
\put(247,199){\rule{1pt}{1pt}}
\put(274,199){\rule{1pt}{1pt}}
\put(301,199){\rule{1pt}{1pt}}
\put(328,200){\rule{1pt}{1pt}}
\put(355,200){\rule{1pt}{1pt}}
\put(382,200){\rule{1pt}{1pt}}
\put(409,200){\rule{1pt}{1pt}}
\put(436,200){\rule{1pt}{1pt}}
\put(463,200){\rule{1pt}{1pt}}
\put(490,200){\rule{1pt}{1pt}}
\put(517,201){\rule{1pt}{1pt}}
\put(544,201){\rule{1pt}{1pt}}
\put(571,201){\rule{1pt}{1pt}}
\put(598,201){\rule{1pt}{1pt}}
\put(625,202){\rule{1pt}{1pt}}
\put(652,202){\rule{1pt}{1pt}}
\put(679,202){\rule{1pt}{1pt}}
\put(706,203){\rule{1pt}{1pt}}
\put(733,203){\rule{1pt}{1pt}}
\put(760,204){\rule{1pt}{1pt}}
\put(787,204){\rule{1pt}{1pt}}
\put(814,205){\rule{1pt}{1pt}}
\put(842,205){\rule{1pt}{1pt}}
\put(869,206){\rule{1pt}{1pt}}
\put(896,207){\rule{1pt}{1pt}}
\put(923,208){\rule{1pt}{1pt}}
\put(950,209){\rule{1pt}{1pt}}
\put(977,210){\rule{1pt}{1pt}}
\put(1004,211){\rule{1pt}{1pt}}
\put(1031,212){\rule{1pt}{1pt}}
\put(1058,213){\rule{1pt}{1pt}}
\put(1085,215){\rule{1pt}{1pt}}
\put(1112,217){\rule{1pt}{1pt}}
\put(1139,219){\rule{1pt}{1pt}}
\put(1166,221){\rule{1pt}{1pt}}
\put(1193,224){\rule{1pt}{1pt}}
\put(1220,227){\rule{1pt}{1pt}}
\put(1247,230){\rule{1pt}{1pt}}
\put(1274,234){\rule{1pt}{1pt}}
\put(1301,240){\rule{1pt}{1pt}}
\put(1328,246){\rule{1pt}{1pt}}
\put(1355,254){\rule{1pt}{1pt}}
\put(1382,264){\rule{1pt}{1pt}}
\put(1409,279){\rule{1pt}{1pt}}
\put(1436,299){\rule{1pt}{1pt}}
\put(220,468){\circle*{12}}
\put(247,468){\circle*{12}}
\put(274,468){\circle*{12}}
\put(301,468){\circle*{12}}
\put(328,468){\circle*{12}}
\put(355,468){\circle*{12}}
\put(382,468){\circle*{12}}
\put(409,468){\circle*{12}}
\put(436,468){\circle*{12}}
\put(463,468){\circle*{12}}
\put(490,469){\circle*{12}}
\put(517,469){\circle*{12}}
\put(544,469){\circle*{12}}
\put(571,469){\circle*{12}}
\put(598,469){\circle*{12}}
\put(625,469){\circle*{12}}
\put(652,469){\circle*{12}}
\put(679,469){\circle*{12}}
\put(706,470){\circle*{12}}
\put(733,470){\circle*{12}}
\put(760,470){\circle*{12}}
\put(787,470){\circle*{12}}
\put(814,470){\circle*{12}}
\put(842,471){\circle*{12}}
\put(869,471){\circle*{12}}
\put(896,471){\circle*{12}}
\put(923,472){\circle*{12}}
\put(950,472){\circle*{12}}
\put(977,472){\circle*{12}}
\put(1004,473){\circle*{12}}
\put(1031,473){\circle*{12}}
\put(1058,474){\circle*{12}}
\put(1085,475){\circle*{12}}
\put(1112,475){\circle*{12}}
\put(1139,476){\circle*{12}}
\put(1166,477){\circle*{12}}
\put(1193,478){\circle*{12}}
\put(1220,479){\circle*{12}}
\put(1247,481){\circle*{12}}
\put(1274,483){\circle*{12}}
\put(1301,485){\circle*{12}}
\put(1328,487){\circle*{12}}
\put(1355,491){\circle*{12}}
\put(1382,495){\circle*{12}}
\put(1409,500){\circle*{12}}
\put(1436,508){\circle*{12}}
\put(220,703){\circle*{12}}
\put(247,703){\circle*{12}}
\put(274,703){\circle*{12}}
\put(301,703){\circle*{12}}
\put(328,703){\circle*{12}}
\put(355,703){\circle*{12}}
\put(382,703){\circle*{12}}
\put(409,703){\circle*{12}}
\put(436,703){\circle*{12}}
\put(463,703){\circle*{12}}
\put(490,703){\circle*{12}}
\put(517,703){\circle*{12}}
\put(544,703){\circle*{12}}
\put(571,702){\circle*{12}}
\put(598,702){\circle*{12}}
\put(625,702){\circle*{12}}
\put(652,702){\circle*{12}}
\put(679,702){\circle*{12}}
\put(706,702){\circle*{12}}
\put(733,702){\circle*{12}}
\put(760,701){\circle*{12}}
\put(787,701){\circle*{12}}
\put(814,701){\circle*{12}}
\put(842,701){\circle*{12}}
\put(869,700){\circle*{12}}
\put(896,700){\circle*{12}}
\put(923,700){\circle*{12}}
\put(950,699){\circle*{12}}
\put(977,699){\circle*{12}}
\put(1004,698){\circle*{12}}
\put(1031,698){\circle*{12}}
\put(1058,697){\circle*{12}}
\put(1085,697){\circle*{12}}
\put(1112,696){\circle*{12}}
\put(1139,695){\circle*{12}}
\put(1166,694){\circle*{12}}
\put(1193,693){\circle*{12}}
\put(1220,692){\circle*{12}}
\put(1247,690){\circle*{12}}
\put(1274,689){\circle*{12}}
\put(1301,686){\circle*{12}}
\put(1328,684){\circle*{12}}
\put(1355,681){\circle*{12}}
\put(1382,676){\circle*{12}}
\put(1409,671){\circle*{12}}
\put(1436,663){\circle*{12}}
\put(220,652){\circle{18}}
\put(247,652){\circle{18}}
\put(274,652){\circle{18}}
\put(301,652){\circle{18}}
\put(328,652){\circle{18}}
\put(355,652){\circle{18}}
\put(382,652){\circle{18}}
\put(409,652){\circle{18}}
\put(436,652){\circle{18}}
\put(463,652){\circle{18}}
\put(490,652){\circle{18}}
\put(517,652){\circle{18}}
\put(544,652){\circle{18}}
\put(571,652){\circle{18}}
\put(598,652){\circle{18}}
\put(625,652){\circle{18}}
\put(652,652){\circle{18}}
\put(679,652){\circle{18}}
\put(706,651){\circle{18}}
\put(733,651){\circle{18}}
\put(760,651){\circle{18}}
\put(787,651){\circle{18}}
\put(814,651){\circle{18}}
\put(842,651){\circle{18}}
\put(869,651){\circle{18}}
\put(896,651){\circle{18}}
\put(923,650){\circle{18}}
\put(950,650){\circle{18}}
\put(977,650){\circle{18}}
\put(1004,650){\circle{18}}
\put(1031,649){\circle{18}}
\put(1058,649){\circle{18}}
\put(1085,649){\circle{18}}
\put(1112,648){\circle{18}}
\put(1139,648){\circle{18}}
\put(1166,648){\circle{18}}
\put(1193,647){\circle{18}}
\put(1220,646){\circle{18}}
\put(1247,646){\circle{18}}
\put(1274,645){\circle{18}}
\put(1301,644){\circle{18}}
\put(1328,642){\circle{18}}
\put(1355,641){\circle{18}}
\put(1382,639){\circle{18}}
\put(1409,636){\circle{18}}
\put(1436,632){\circle{18}}
\put(220,536){\circle{18}}
\put(247,536){\circle{18}}
\put(274,536){\circle{18}}
\put(301,536){\circle{18}}
\put(328,536){\circle{18}}
\put(355,536){\circle{18}}
\put(382,536){\circle{18}}
\put(409,536){\circle{18}}
\put(436,536){\circle{18}}
\put(463,536){\circle{18}}
\put(490,537){\circle{18}}
\put(517,537){\circle{18}}
\put(544,537){\circle{18}}
\put(571,537){\circle{18}}
\put(598,537){\circle{18}}
\put(625,537){\circle{18}}
\put(652,537){\circle{18}}
\put(679,537){\circle{18}}
\put(706,537){\circle{18}}
\put(733,537){\circle{18}}
\put(760,537){\circle{18}}
\put(787,537){\circle{18}}
\put(814,537){\circle{18}}
\put(842,538){\circle{18}}
\put(869,538){\circle{18}}
\put(896,538){\circle{18}}
\put(923,538){\circle{18}}
\put(950,538){\circle{18}}
\put(977,538){\circle{18}}
\put(1004,539){\circle{18}}
\put(1031,539){\circle{18}}
\put(1058,539){\circle{18}}
\put(1085,540){\circle{18}}
\put(1112,540){\circle{18}}
\put(1139,540){\circle{18}}
\put(1166,541){\circle{18}}
\put(1193,541){\circle{18}}
\put(1220,542){\circle{18}}
\put(1247,543){\circle{18}}
\put(1274,544){\circle{18}}
\put(1301,545){\circle{18}}
\put(1328,546){\circle{18}}
\put(1355,548){\circle{18}}
\put(1382,550){\circle{18}}
\put(1409,553){\circle{18}}
\put(1436,557){\circle{18}}
\end{picture}

%% file: Bstring22.tex
% GNUPLOT: LaTeX picture
\setlength{\unitlength}{0.240900pt}
\ifx\plotpoint\undefined\newsavebox{\plotpoint}\fi
\sbox{\plotpoint}{\rule[-0.200pt]{0.400pt}{0.400pt}}%
\begin{picture}(1500,900)(0,0)
\font\gnuplot=cmr10 at 10pt
\gnuplot
\sbox{\plotpoint}{\rule[-0.200pt]{0.400pt}{0.400pt}}%
\put(220.0,495.0){\rule[-0.200pt]{292.934pt}{0.400pt}}
\put(828.0,113.0){\rule[-0.200pt]{0.400pt}{184.048pt}}
\put(220.0,113.0){\rule[-0.200pt]{4.818pt}{0.400pt}}
\put(198,113){\makebox(0,0)[r]{-3}}
\put(1416.0,113.0){\rule[-0.200pt]{4.818pt}{0.400pt}}
\put(220.0,240.0){\rule[-0.200pt]{4.818pt}{0.400pt}}
\put(198,240){\makebox(0,0)[r]{-2}}
\put(1416.0,240.0){\rule[-0.200pt]{4.818pt}{0.400pt}}
\put(220.0,368.0){\rule[-0.200pt]{4.818pt}{0.400pt}}
\put(198,368){\makebox(0,0)[r]{-1}}
\put(1416.0,368.0){\rule[-0.200pt]{4.818pt}{0.400pt}}
\put(220.0,495.0){\rule[-0.200pt]{4.818pt}{0.400pt}}
\put(198,495){\makebox(0,0)[r]{0}}
\put(1416.0,495.0){\rule[-0.200pt]{4.818pt}{0.400pt}}
\put(220.0,622.0){\rule[-0.200pt]{4.818pt}{0.400pt}}
\put(198,622){\makebox(0,0)[r]{1}}
\put(1416.0,622.0){\rule[-0.200pt]{4.818pt}{0.400pt}}
\put(220.0,750.0){\rule[-0.200pt]{4.818pt}{0.400pt}}
\put(198,750){\makebox(0,0)[r]{2}}
\put(1416.0,750.0){\rule[-0.200pt]{4.818pt}{0.400pt}}
\put(220.0,877.0){\rule[-0.200pt]{4.818pt}{0.400pt}}
\put(198,877){\makebox(0,0)[r]{3}}
\put(1416.0,877.0){\rule[-0.200pt]{4.818pt}{0.400pt}}
\put(288.0,113.0){\rule[-0.200pt]{0.400pt}{4.818pt}}
\put(288,68){\makebox(0,0){-0.8}}
\put(288.0,857.0){\rule[-0.200pt]{0.400pt}{4.818pt}}
\put(423.0,113.0){\rule[-0.200pt]{0.400pt}{4.818pt}}
\put(423,68){\makebox(0,0){-0.6}}
\put(423.0,857.0){\rule[-0.200pt]{0.400pt}{4.818pt}}
\put(558.0,113.0){\rule[-0.200pt]{0.400pt}{4.818pt}}
\put(558,68){\makebox(0,0){-0.4}}
\put(558.0,857.0){\rule[-0.200pt]{0.400pt}{4.818pt}}
\put(693.0,113.0){\rule[-0.200pt]{0.400pt}{4.818pt}}
\put(693,68){\makebox(0,0){-0.2}}
\put(693.0,857.0){\rule[-0.200pt]{0.400pt}{4.818pt}}
\put(828.0,113.0){\rule[-0.200pt]{0.400pt}{4.818pt}}
\put(828,68){\makebox(0,0){0}}
\put(828.0,857.0){\rule[-0.200pt]{0.400pt}{4.818pt}}
\put(963.0,113.0){\rule[-0.200pt]{0.400pt}{4.818pt}}
\put(963,68){\makebox(0,0){0.2}}
\put(963.0,857.0){\rule[-0.200pt]{0.400pt}{4.818pt}}
\put(1098.0,113.0){\rule[-0.200pt]{0.400pt}{4.818pt}}
\put(1098,68){\makebox(0,0){0.4}}
\put(1098.0,857.0){\rule[-0.200pt]{0.400pt}{4.818pt}}
\put(1233.0,113.0){\rule[-0.200pt]{0.400pt}{4.818pt}}
\put(1233,68){\makebox(0,0){0.6}}
\put(1233.0,857.0){\rule[-0.200pt]{0.400pt}{4.818pt}}
\put(1368.0,113.0){\rule[-0.200pt]{0.400pt}{4.818pt}}
\put(1368,68){\makebox(0,0){0.8}}
\put(1368.0,857.0){\rule[-0.200pt]{0.400pt}{4.818pt}}
\put(220.0,113.0){\rule[-0.200pt]{292.934pt}{0.400pt}}
\put(1436.0,113.0){\rule[-0.200pt]{0.400pt}{184.048pt}}
\put(220.0,877.0){\rule[-0.200pt]{292.934pt}{0.400pt}}
\put(45,495){\makebox(0,0)
{$\displaystyle{\frac{B^{(0)}}{M_{1/2}^{(0)}}}$}}
\put(828,23){\makebox(0,0){$\cos \theta$}}
\put(423,686){\makebox(0,0)[r]{$B_Z^{(0)}$}}
\put(423,304){\makebox(0,0)[r]{$B_Z^{(0)}$}}
\put(389,559){\makebox(0,0)[r]{$B_\lambda^{(0)}$ }}
\put(693,495){\makebox(0,0)[r]{$B_\mu^{(0)}$ }}
\put(693,240){\makebox(0,0)[r]{$B_\mu^{(0)}$ }}
\put(220.0,113.0){\rule[-0.200pt]{0.400pt}{184.048pt}}
\sbox{\plotpoint}{\rule[-0.400pt]{0.800pt}{0.800pt}}%
\put(220,681){\usebox{\plotpoint}}
\multiput(220.00,679.08)(0.491,-0.511){17}{\rule{1.000pt}{0.123pt}}
\multiput(220.00,679.34)(9.924,-12.000){2}{\rule{0.500pt}{0.800pt}}
\multiput(232.00,667.08)(0.847,-0.520){9}{\rule{1.500pt}{0.125pt}}
\multiput(232.00,667.34)(9.887,-8.000){2}{\rule{0.750pt}{0.800pt}}
\multiput(245.00,659.08)(0.913,-0.526){7}{\rule{1.571pt}{0.127pt}}
\multiput(245.00,659.34)(8.738,-7.000){2}{\rule{0.786pt}{0.800pt}}
\multiput(257.00,652.07)(1.132,-0.536){5}{\rule{1.800pt}{0.129pt}}
\multiput(257.00,652.34)(8.264,-6.000){2}{\rule{0.900pt}{0.800pt}}
\put(269,644.34){\rule{2.600pt}{0.800pt}}
\multiput(269.00,646.34)(6.604,-4.000){2}{\rule{1.300pt}{0.800pt}}
\put(281,640.34){\rule{2.800pt}{0.800pt}}
\multiput(281.00,642.34)(7.188,-4.000){2}{\rule{1.400pt}{0.800pt}}
\put(294,636.84){\rule{2.891pt}{0.800pt}}
\multiput(294.00,638.34)(6.000,-3.000){2}{\rule{1.445pt}{0.800pt}}
\put(306,634.34){\rule{2.891pt}{0.800pt}}
\multiput(306.00,635.34)(6.000,-2.000){2}{\rule{1.445pt}{0.800pt}}
\put(318,631.84){\rule{3.132pt}{0.800pt}}
\multiput(318.00,633.34)(6.500,-3.000){2}{\rule{1.566pt}{0.800pt}}
\put(331,629.34){\rule{2.891pt}{0.800pt}}
\multiput(331.00,630.34)(6.000,-2.000){2}{\rule{1.445pt}{0.800pt}}
\put(343,627.84){\rule{2.891pt}{0.800pt}}
\multiput(343.00,628.34)(6.000,-1.000){2}{\rule{1.445pt}{0.800pt}}
\put(355,626.34){\rule{2.891pt}{0.800pt}}
\multiput(355.00,627.34)(6.000,-2.000){2}{\rule{1.445pt}{0.800pt}}
\put(367,624.84){\rule{3.132pt}{0.800pt}}
\multiput(367.00,625.34)(6.500,-1.000){2}{\rule{1.566pt}{0.800pt}}
\put(380,623.84){\rule{2.891pt}{0.800pt}}
\multiput(380.00,624.34)(6.000,-1.000){2}{\rule{1.445pt}{0.800pt}}
\put(404,622.84){\rule{3.132pt}{0.800pt}}
\multiput(404.00,623.34)(6.500,-1.000){2}{\rule{1.566pt}{0.800pt}}
\put(417,621.84){\rule{2.891pt}{0.800pt}}
\multiput(417.00,622.34)(6.000,-1.000){2}{\rule{1.445pt}{0.800pt}}
\put(392.0,625.0){\rule[-0.400pt]{2.891pt}{0.800pt}}
\put(453,620.84){\rule{3.132pt}{0.800pt}}
\multiput(453.00,621.34)(6.500,-1.000){2}{\rule{1.566pt}{0.800pt}}
\put(429.0,623.0){\rule[-0.400pt]{5.782pt}{0.800pt}}
\put(515,620.84){\rule{2.891pt}{0.800pt}}
\multiput(515.00,620.34)(6.000,1.000){2}{\rule{1.445pt}{0.800pt}}
\put(466.0,622.0){\rule[-0.400pt]{11.804pt}{0.800pt}}
\put(552,621.84){\rule{2.891pt}{0.800pt}}
\multiput(552.00,621.34)(6.000,1.000){2}{\rule{1.445pt}{0.800pt}}
\put(527.0,623.0){\rule[-0.400pt]{6.022pt}{0.800pt}}
\put(588,622.84){\rule{3.132pt}{0.800pt}}
\multiput(588.00,622.34)(6.500,1.000){2}{\rule{1.566pt}{0.800pt}}
\put(564.0,624.0){\rule[-0.400pt]{5.782pt}{0.800pt}}
\put(613,623.84){\rule{2.891pt}{0.800pt}}
\multiput(613.00,623.34)(6.000,1.000){2}{\rule{1.445pt}{0.800pt}}
\put(625,624.84){\rule{3.132pt}{0.800pt}}
\multiput(625.00,624.34)(6.500,1.000){2}{\rule{1.566pt}{0.800pt}}
\put(601.0,625.0){\rule[-0.400pt]{2.891pt}{0.800pt}}
\put(650,625.84){\rule{2.891pt}{0.800pt}}
\multiput(650.00,625.34)(6.000,1.000){2}{\rule{1.445pt}{0.800pt}}
\put(662,626.84){\rule{2.891pt}{0.800pt}}
\multiput(662.00,626.34)(6.000,1.000){2}{\rule{1.445pt}{0.800pt}}
\put(674,627.84){\rule{3.132pt}{0.800pt}}
\multiput(674.00,627.34)(6.500,1.000){2}{\rule{1.566pt}{0.800pt}}
\put(638.0,627.0){\rule[-0.400pt]{2.891pt}{0.800pt}}
\put(699,628.84){\rule{2.891pt}{0.800pt}}
\multiput(699.00,628.34)(6.000,1.000){2}{\rule{1.445pt}{0.800pt}}
\put(711,629.84){\rule{3.132pt}{0.800pt}}
\multiput(711.00,629.34)(6.500,1.000){2}{\rule{1.566pt}{0.800pt}}
\put(724,630.84){\rule{2.891pt}{0.800pt}}
\multiput(724.00,630.34)(6.000,1.000){2}{\rule{1.445pt}{0.800pt}}
\put(736,631.84){\rule{2.891pt}{0.800pt}}
\multiput(736.00,631.34)(6.000,1.000){2}{\rule{1.445pt}{0.800pt}}
\put(748,632.84){\rule{2.891pt}{0.800pt}}
\multiput(748.00,632.34)(6.000,1.000){2}{\rule{1.445pt}{0.800pt}}
\put(760,633.84){\rule{3.132pt}{0.800pt}}
\multiput(760.00,633.34)(6.500,1.000){2}{\rule{1.566pt}{0.800pt}}
\put(773,635.34){\rule{2.891pt}{0.800pt}}
\multiput(773.00,634.34)(6.000,2.000){2}{\rule{1.445pt}{0.800pt}}
\put(785,636.84){\rule{2.891pt}{0.800pt}}
\multiput(785.00,636.34)(6.000,1.000){2}{\rule{1.445pt}{0.800pt}}
\put(797,637.84){\rule{3.132pt}{0.800pt}}
\multiput(797.00,637.34)(6.500,1.000){2}{\rule{1.566pt}{0.800pt}}
\put(810,638.84){\rule{2.891pt}{0.800pt}}
\multiput(810.00,638.34)(6.000,1.000){2}{\rule{1.445pt}{0.800pt}}
\put(822,640.34){\rule{2.891pt}{0.800pt}}
\multiput(822.00,639.34)(6.000,2.000){2}{\rule{1.445pt}{0.800pt}}
\put(834,641.84){\rule{2.891pt}{0.800pt}}
\multiput(834.00,641.34)(6.000,1.000){2}{\rule{1.445pt}{0.800pt}}
\put(846,643.34){\rule{3.132pt}{0.800pt}}
\multiput(846.00,642.34)(6.500,2.000){2}{\rule{1.566pt}{0.800pt}}
\put(859,644.84){\rule{2.891pt}{0.800pt}}
\multiput(859.00,644.34)(6.000,1.000){2}{\rule{1.445pt}{0.800pt}}
\put(871,646.34){\rule{2.891pt}{0.800pt}}
\multiput(871.00,645.34)(6.000,2.000){2}{\rule{1.445pt}{0.800pt}}
\put(883,647.84){\rule{3.132pt}{0.800pt}}
\multiput(883.00,647.34)(6.500,1.000){2}{\rule{1.566pt}{0.800pt}}
\put(896,649.34){\rule{2.891pt}{0.800pt}}
\multiput(896.00,648.34)(6.000,2.000){2}{\rule{1.445pt}{0.800pt}}
\put(908,651.34){\rule{2.891pt}{0.800pt}}
\multiput(908.00,650.34)(6.000,2.000){2}{\rule{1.445pt}{0.800pt}}
\put(920,652.84){\rule{2.891pt}{0.800pt}}
\multiput(920.00,652.34)(6.000,1.000){2}{\rule{1.445pt}{0.800pt}}
\put(932,654.34){\rule{3.132pt}{0.800pt}}
\multiput(932.00,653.34)(6.500,2.000){2}{\rule{1.566pt}{0.800pt}}
\put(945,656.34){\rule{2.891pt}{0.800pt}}
\multiput(945.00,655.34)(6.000,2.000){2}{\rule{1.445pt}{0.800pt}}
\put(957,658.34){\rule{2.891pt}{0.800pt}}
\multiput(957.00,657.34)(6.000,2.000){2}{\rule{1.445pt}{0.800pt}}
\put(969,660.34){\rule{3.132pt}{0.800pt}}
\multiput(969.00,659.34)(6.500,2.000){2}{\rule{1.566pt}{0.800pt}}
\put(982,662.34){\rule{2.891pt}{0.800pt}}
\multiput(982.00,661.34)(6.000,2.000){2}{\rule{1.445pt}{0.800pt}}
\put(994,664.84){\rule{2.891pt}{0.800pt}}
\multiput(994.00,663.34)(6.000,3.000){2}{\rule{1.445pt}{0.800pt}}
\put(1006,667.34){\rule{2.891pt}{0.800pt}}
\multiput(1006.00,666.34)(6.000,2.000){2}{\rule{1.445pt}{0.800pt}}
\put(1018,669.34){\rule{3.132pt}{0.800pt}}
\multiput(1018.00,668.34)(6.500,2.000){2}{\rule{1.566pt}{0.800pt}}
\put(1031,671.84){\rule{2.891pt}{0.800pt}}
\multiput(1031.00,670.34)(6.000,3.000){2}{\rule{1.445pt}{0.800pt}}
\put(1043,674.34){\rule{2.891pt}{0.800pt}}
\multiput(1043.00,673.34)(6.000,2.000){2}{\rule{1.445pt}{0.800pt}}
\put(1055,676.84){\rule{3.132pt}{0.800pt}}
\multiput(1055.00,675.34)(6.500,3.000){2}{\rule{1.566pt}{0.800pt}}
\put(1068,679.84){\rule{2.891pt}{0.800pt}}
\multiput(1068.00,678.34)(6.000,3.000){2}{\rule{1.445pt}{0.800pt}}
\put(1080,682.84){\rule{2.891pt}{0.800pt}}
\multiput(1080.00,681.34)(6.000,3.000){2}{\rule{1.445pt}{0.800pt}}
\put(1092,685.84){\rule{2.891pt}{0.800pt}}
\multiput(1092.00,684.34)(6.000,3.000){2}{\rule{1.445pt}{0.800pt}}
\put(1104,688.84){\rule{3.132pt}{0.800pt}}
\multiput(1104.00,687.34)(6.500,3.000){2}{\rule{1.566pt}{0.800pt}}
\put(1117,692.34){\rule{2.600pt}{0.800pt}}
\multiput(1117.00,690.34)(6.604,4.000){2}{\rule{1.300pt}{0.800pt}}
\put(1129,695.84){\rule{2.891pt}{0.800pt}}
\multiput(1129.00,694.34)(6.000,3.000){2}{\rule{1.445pt}{0.800pt}}
\put(1141,699.34){\rule{2.600pt}{0.800pt}}
\multiput(1141.00,697.34)(6.604,4.000){2}{\rule{1.300pt}{0.800pt}}
\put(1153,703.34){\rule{2.800pt}{0.800pt}}
\multiput(1153.00,701.34)(7.188,4.000){2}{\rule{1.400pt}{0.800pt}}
\put(1166,707.34){\rule{2.600pt}{0.800pt}}
\multiput(1166.00,705.34)(6.604,4.000){2}{\rule{1.300pt}{0.800pt}}
\multiput(1178.00,712.38)(1.600,0.560){3}{\rule{2.120pt}{0.135pt}}
\multiput(1178.00,709.34)(7.600,5.000){2}{\rule{1.060pt}{0.800pt}}
\multiput(1190.00,717.38)(1.768,0.560){3}{\rule{2.280pt}{0.135pt}}
\multiput(1190.00,714.34)(8.268,5.000){2}{\rule{1.140pt}{0.800pt}}
\multiput(1203.00,722.38)(1.600,0.560){3}{\rule{2.120pt}{0.135pt}}
\multiput(1203.00,719.34)(7.600,5.000){2}{\rule{1.060pt}{0.800pt}}
\multiput(1215.00,727.38)(1.600,0.560){3}{\rule{2.120pt}{0.135pt}}
\multiput(1215.00,724.34)(7.600,5.000){2}{\rule{1.060pt}{0.800pt}}
\multiput(1227.00,732.39)(1.132,0.536){5}{\rule{1.800pt}{0.129pt}}
\multiput(1227.00,729.34)(8.264,6.000){2}{\rule{0.900pt}{0.800pt}}
\multiput(1239.00,738.39)(1.244,0.536){5}{\rule{1.933pt}{0.129pt}}
\multiput(1239.00,735.34)(8.987,6.000){2}{\rule{0.967pt}{0.800pt}}
\multiput(1252.00,744.40)(0.913,0.526){7}{\rule{1.571pt}{0.127pt}}
\multiput(1252.00,741.34)(8.738,7.000){2}{\rule{0.786pt}{0.800pt}}
\multiput(1264.00,751.40)(0.913,0.526){7}{\rule{1.571pt}{0.127pt}}
\multiput(1264.00,748.34)(8.738,7.000){2}{\rule{0.786pt}{0.800pt}}
\multiput(1276.00,758.40)(0.847,0.520){9}{\rule{1.500pt}{0.125pt}}
\multiput(1276.00,755.34)(9.887,8.000){2}{\rule{0.750pt}{0.800pt}}
\multiput(1289.00,766.40)(0.774,0.520){9}{\rule{1.400pt}{0.125pt}}
\multiput(1289.00,763.34)(9.094,8.000){2}{\rule{0.700pt}{0.800pt}}
\multiput(1301.00,774.40)(0.674,0.516){11}{\rule{1.267pt}{0.124pt}}
\multiput(1301.00,771.34)(9.371,9.000){2}{\rule{0.633pt}{0.800pt}}
\multiput(1313.00,783.40)(0.599,0.514){13}{\rule{1.160pt}{0.124pt}}
\multiput(1313.00,780.34)(9.592,10.000){2}{\rule{0.580pt}{0.800pt}}
\multiput(1325.00,793.41)(0.536,0.511){17}{\rule{1.067pt}{0.123pt}}
\multiput(1325.00,790.34)(10.786,12.000){2}{\rule{0.533pt}{0.800pt}}
\multiput(1338.00,805.41)(0.491,0.511){17}{\rule{1.000pt}{0.123pt}}
\multiput(1338.00,802.34)(9.924,12.000){2}{\rule{0.500pt}{0.800pt}}
\multiput(1351.41,816.00)(0.511,0.581){17}{\rule{0.123pt}{1.133pt}}
\multiput(1348.34,816.00)(12.000,11.648){2}{\rule{0.800pt}{0.567pt}}
\multiput(1363.41,830.00)(0.509,0.616){19}{\rule{0.123pt}{1.185pt}}
\multiput(1360.34,830.00)(13.000,13.541){2}{\rule{0.800pt}{0.592pt}}
\multiput(1376.41,846.00)(0.511,0.807){17}{\rule{0.123pt}{1.467pt}}
\multiput(1373.34,846.00)(12.000,15.956){2}{\rule{0.800pt}{0.733pt}}
\multiput(1388.40,865.00)(0.526,0.913){7}{\rule{0.127pt}{1.571pt}}
\multiput(1385.34,865.00)(7.000,8.738){2}{\rule{0.800pt}{0.786pt}}
\put(687.0,630.0){\rule[-0.400pt]{2.891pt}{0.800pt}}
\put(220,309){\usebox{\plotpoint}}
\multiput(220.00,310.41)(0.491,0.511){17}{\rule{1.000pt}{0.123pt}}
\multiput(220.00,307.34)(9.924,12.000){2}{\rule{0.500pt}{0.800pt}}
\multiput(232.00,322.40)(0.847,0.520){9}{\rule{1.500pt}{0.125pt}}
\multiput(232.00,319.34)(9.887,8.000){2}{\rule{0.750pt}{0.800pt}}
\multiput(245.00,330.40)(0.913,0.526){7}{\rule{1.571pt}{0.127pt}}
\multiput(245.00,327.34)(8.738,7.000){2}{\rule{0.786pt}{0.800pt}}
\multiput(257.00,337.39)(1.132,0.536){5}{\rule{1.800pt}{0.129pt}}
\multiput(257.00,334.34)(8.264,6.000){2}{\rule{0.900pt}{0.800pt}}
\put(269,342.34){\rule{2.600pt}{0.800pt}}
\multiput(269.00,340.34)(6.604,4.000){2}{\rule{1.300pt}{0.800pt}}
\put(281,346.34){\rule{2.800pt}{0.800pt}}
\multiput(281.00,344.34)(7.188,4.000){2}{\rule{1.400pt}{0.800pt}}
\put(294,349.84){\rule{2.891pt}{0.800pt}}
\multiput(294.00,348.34)(6.000,3.000){2}{\rule{1.445pt}{0.800pt}}
\put(306,352.34){\rule{2.891pt}{0.800pt}}
\multiput(306.00,351.34)(6.000,2.000){2}{\rule{1.445pt}{0.800pt}}
\put(318,354.84){\rule{3.132pt}{0.800pt}}
\multiput(318.00,353.34)(6.500,3.000){2}{\rule{1.566pt}{0.800pt}}
\put(331,357.34){\rule{2.891pt}{0.800pt}}
\multiput(331.00,356.34)(6.000,2.000){2}{\rule{1.445pt}{0.800pt}}
\put(343,358.84){\rule{2.891pt}{0.800pt}}
\multiput(343.00,358.34)(6.000,1.000){2}{\rule{1.445pt}{0.800pt}}
\put(355,360.34){\rule{2.891pt}{0.800pt}}
\multiput(355.00,359.34)(6.000,2.000){2}{\rule{1.445pt}{0.800pt}}
\put(367,361.84){\rule{3.132pt}{0.800pt}}
\multiput(367.00,361.34)(6.500,1.000){2}{\rule{1.566pt}{0.800pt}}
\put(380,362.84){\rule{2.891pt}{0.800pt}}
\multiput(380.00,362.34)(6.000,1.000){2}{\rule{1.445pt}{0.800pt}}
\put(404,363.84){\rule{3.132pt}{0.800pt}}
\multiput(404.00,363.34)(6.500,1.000){2}{\rule{1.566pt}{0.800pt}}
\put(417,364.84){\rule{2.891pt}{0.800pt}}
\multiput(417.00,364.34)(6.000,1.000){2}{\rule{1.445pt}{0.800pt}}
\put(392.0,365.0){\rule[-0.400pt]{2.891pt}{0.800pt}}
\put(453,365.84){\rule{3.132pt}{0.800pt}}
\multiput(453.00,365.34)(6.500,1.000){2}{\rule{1.566pt}{0.800pt}}
\put(429.0,367.0){\rule[-0.400pt]{5.782pt}{0.800pt}}
\put(515,365.84){\rule{2.891pt}{0.800pt}}
\multiput(515.00,366.34)(6.000,-1.000){2}{\rule{1.445pt}{0.800pt}}
\put(466.0,368.0){\rule[-0.400pt]{11.804pt}{0.800pt}}
\put(552,364.84){\rule{2.891pt}{0.800pt}}
\multiput(552.00,365.34)(6.000,-1.000){2}{\rule{1.445pt}{0.800pt}}
\put(527.0,367.0){\rule[-0.400pt]{6.022pt}{0.800pt}}
\put(588,363.84){\rule{3.132pt}{0.800pt}}
\multiput(588.00,364.34)(6.500,-1.000){2}{\rule{1.566pt}{0.800pt}}
\put(564.0,366.0){\rule[-0.400pt]{5.782pt}{0.800pt}}
\put(613,362.84){\rule{2.891pt}{0.800pt}}
\multiput(613.00,363.34)(6.000,-1.000){2}{\rule{1.445pt}{0.800pt}}
\put(625,361.84){\rule{3.132pt}{0.800pt}}
\multiput(625.00,362.34)(6.500,-1.000){2}{\rule{1.566pt}{0.800pt}}
\put(601.0,365.0){\rule[-0.400pt]{2.891pt}{0.800pt}}
\put(650,360.84){\rule{2.891pt}{0.800pt}}
\multiput(650.00,361.34)(6.000,-1.000){2}{\rule{1.445pt}{0.800pt}}
\put(662,359.84){\rule{2.891pt}{0.800pt}}
\multiput(662.00,360.34)(6.000,-1.000){2}{\rule{1.445pt}{0.800pt}}
\put(674,358.84){\rule{3.132pt}{0.800pt}}
\multiput(674.00,359.34)(6.500,-1.000){2}{\rule{1.566pt}{0.800pt}}
\put(638.0,363.0){\rule[-0.400pt]{2.891pt}{0.800pt}}
\put(699,357.84){\rule{2.891pt}{0.800pt}}
\multiput(699.00,358.34)(6.000,-1.000){2}{\rule{1.445pt}{0.800pt}}
\put(711,356.84){\rule{3.132pt}{0.800pt}}
\multiput(711.00,357.34)(6.500,-1.000){2}{\rule{1.566pt}{0.800pt}}
\put(724,355.84){\rule{2.891pt}{0.800pt}}
\multiput(724.00,356.34)(6.000,-1.000){2}{\rule{1.445pt}{0.800pt}}
\put(736,354.84){\rule{2.891pt}{0.800pt}}
\multiput(736.00,355.34)(6.000,-1.000){2}{\rule{1.445pt}{0.800pt}}
\put(748,353.84){\rule{2.891pt}{0.800pt}}
\multiput(748.00,354.34)(6.000,-1.000){2}{\rule{1.445pt}{0.800pt}}
\put(760,352.84){\rule{3.132pt}{0.800pt}}
\multiput(760.00,353.34)(6.500,-1.000){2}{\rule{1.566pt}{0.800pt}}
\put(773,351.34){\rule{2.891pt}{0.800pt}}
\multiput(773.00,352.34)(6.000,-2.000){2}{\rule{1.445pt}{0.800pt}}
\put(785,349.84){\rule{2.891pt}{0.800pt}}
\multiput(785.00,350.34)(6.000,-1.000){2}{\rule{1.445pt}{0.800pt}}
\put(797,348.84){\rule{3.132pt}{0.800pt}}
\multiput(797.00,349.34)(6.500,-1.000){2}{\rule{1.566pt}{0.800pt}}
\put(810,347.84){\rule{2.891pt}{0.800pt}}
\multiput(810.00,348.34)(6.000,-1.000){2}{\rule{1.445pt}{0.800pt}}
\put(822,346.34){\rule{2.891pt}{0.800pt}}
\multiput(822.00,347.34)(6.000,-2.000){2}{\rule{1.445pt}{0.800pt}}
\put(834,344.84){\rule{2.891pt}{0.800pt}}
\multiput(834.00,345.34)(6.000,-1.000){2}{\rule{1.445pt}{0.800pt}}
\put(846,343.34){\rule{3.132pt}{0.800pt}}
\multiput(846.00,344.34)(6.500,-2.000){2}{\rule{1.566pt}{0.800pt}}
\put(859,341.84){\rule{2.891pt}{0.800pt}}
\multiput(859.00,342.34)(6.000,-1.000){2}{\rule{1.445pt}{0.800pt}}
\put(871,340.34){\rule{2.891pt}{0.800pt}}
\multiput(871.00,341.34)(6.000,-2.000){2}{\rule{1.445pt}{0.800pt}}
\put(883,338.84){\rule{3.132pt}{0.800pt}}
\multiput(883.00,339.34)(6.500,-1.000){2}{\rule{1.566pt}{0.800pt}}
\put(896,337.34){\rule{2.891pt}{0.800pt}}
\multiput(896.00,338.34)(6.000,-2.000){2}{\rule{1.445pt}{0.800pt}}
\put(908,335.34){\rule{2.891pt}{0.800pt}}
\multiput(908.00,336.34)(6.000,-2.000){2}{\rule{1.445pt}{0.800pt}}
\put(920,333.84){\rule{2.891pt}{0.800pt}}
\multiput(920.00,334.34)(6.000,-1.000){2}{\rule{1.445pt}{0.800pt}}
\put(932,332.34){\rule{3.132pt}{0.800pt}}
\multiput(932.00,333.34)(6.500,-2.000){2}{\rule{1.566pt}{0.800pt}}
\put(945,330.34){\rule{2.891pt}{0.800pt}}
\multiput(945.00,331.34)(6.000,-2.000){2}{\rule{1.445pt}{0.800pt}}
\put(957,328.34){\rule{2.891pt}{0.800pt}}
\multiput(957.00,329.34)(6.000,-2.000){2}{\rule{1.445pt}{0.800pt}}
\put(969,326.34){\rule{3.132pt}{0.800pt}}
\multiput(969.00,327.34)(6.500,-2.000){2}{\rule{1.566pt}{0.800pt}}
\put(982,324.34){\rule{2.891pt}{0.800pt}}
\multiput(982.00,325.34)(6.000,-2.000){2}{\rule{1.445pt}{0.800pt}}
\put(994,321.84){\rule{2.891pt}{0.800pt}}
\multiput(994.00,323.34)(6.000,-3.000){2}{\rule{1.445pt}{0.800pt}}
\put(1006,319.34){\rule{2.891pt}{0.800pt}}
\multiput(1006.00,320.34)(6.000,-2.000){2}{\rule{1.445pt}{0.800pt}}
\put(1018,317.34){\rule{3.132pt}{0.800pt}}
\multiput(1018.00,318.34)(6.500,-2.000){2}{\rule{1.566pt}{0.800pt}}
\put(1031,314.84){\rule{2.891pt}{0.800pt}}
\multiput(1031.00,316.34)(6.000,-3.000){2}{\rule{1.445pt}{0.800pt}}
\put(1043,312.34){\rule{2.891pt}{0.800pt}}
\multiput(1043.00,313.34)(6.000,-2.000){2}{\rule{1.445pt}{0.800pt}}
\put(1055,309.84){\rule{3.132pt}{0.800pt}}
\multiput(1055.00,311.34)(6.500,-3.000){2}{\rule{1.566pt}{0.800pt}}
\put(1068,306.84){\rule{2.891pt}{0.800pt}}
\multiput(1068.00,308.34)(6.000,-3.000){2}{\rule{1.445pt}{0.800pt}}
\put(1080,303.84){\rule{2.891pt}{0.800pt}}
\multiput(1080.00,305.34)(6.000,-3.000){2}{\rule{1.445pt}{0.800pt}}
\put(1092,300.84){\rule{2.891pt}{0.800pt}}
\multiput(1092.00,302.34)(6.000,-3.000){2}{\rule{1.445pt}{0.800pt}}
\put(1104,297.84){\rule{3.132pt}{0.800pt}}
\multiput(1104.00,299.34)(6.500,-3.000){2}{\rule{1.566pt}{0.800pt}}
\put(1117,294.34){\rule{2.600pt}{0.800pt}}
\multiput(1117.00,296.34)(6.604,-4.000){2}{\rule{1.300pt}{0.800pt}}
\put(1129,290.84){\rule{2.891pt}{0.800pt}}
\multiput(1129.00,292.34)(6.000,-3.000){2}{\rule{1.445pt}{0.800pt}}
\put(1141,287.34){\rule{2.600pt}{0.800pt}}
\multiput(1141.00,289.34)(6.604,-4.000){2}{\rule{1.300pt}{0.800pt}}
\put(1153,283.34){\rule{2.800pt}{0.800pt}}
\multiput(1153.00,285.34)(7.188,-4.000){2}{\rule{1.400pt}{0.800pt}}
\put(1166,279.34){\rule{2.600pt}{0.800pt}}
\multiput(1166.00,281.34)(6.604,-4.000){2}{\rule{1.300pt}{0.800pt}}
\multiput(1178.00,277.06)(1.600,-0.560){3}{\rule{2.120pt}{0.135pt}}
\multiput(1178.00,277.34)(7.600,-5.000){2}{\rule{1.060pt}{0.800pt}}
\multiput(1190.00,272.06)(1.768,-0.560){3}{\rule{2.280pt}{0.135pt}}
\multiput(1190.00,272.34)(8.268,-5.000){2}{\rule{1.140pt}{0.800pt}}
\multiput(1203.00,267.06)(1.600,-0.560){3}{\rule{2.120pt}{0.135pt}}
\multiput(1203.00,267.34)(7.600,-5.000){2}{\rule{1.060pt}{0.800pt}}
\multiput(1215.00,262.06)(1.600,-0.560){3}{\rule{2.120pt}{0.135pt}}
\multiput(1215.00,262.34)(7.600,-5.000){2}{\rule{1.060pt}{0.800pt}}
\multiput(1227.00,257.07)(1.132,-0.536){5}{\rule{1.800pt}{0.129pt}}
\multiput(1227.00,257.34)(8.264,-6.000){2}{\rule{0.900pt}{0.800pt}}
\multiput(1239.00,251.07)(1.244,-0.536){5}{\rule{1.933pt}{0.129pt}}
\multiput(1239.00,251.34)(8.987,-6.000){2}{\rule{0.967pt}{0.800pt}}
\multiput(1252.00,245.08)(0.913,-0.526){7}{\rule{1.571pt}{0.127pt}}
\multiput(1252.00,245.34)(8.738,-7.000){2}{\rule{0.786pt}{0.800pt}}
\multiput(1264.00,238.08)(0.913,-0.526){7}{\rule{1.571pt}{0.127pt}}
\multiput(1264.00,238.34)(8.738,-7.000){2}{\rule{0.786pt}{0.800pt}}
\multiput(1276.00,231.08)(0.847,-0.520){9}{\rule{1.500pt}{0.125pt}}
\multiput(1276.00,231.34)(9.887,-8.000){2}{\rule{0.750pt}{0.800pt}}
\multiput(1289.00,223.08)(0.774,-0.520){9}{\rule{1.400pt}{0.125pt}}
\multiput(1289.00,223.34)(9.094,-8.000){2}{\rule{0.700pt}{0.800pt}}
\multiput(1301.00,215.08)(0.674,-0.516){11}{\rule{1.267pt}{0.124pt}}
\multiput(1301.00,215.34)(9.371,-9.000){2}{\rule{0.633pt}{0.800pt}}
\multiput(1313.00,206.08)(0.599,-0.514){13}{\rule{1.160pt}{0.124pt}}
\multiput(1313.00,206.34)(9.592,-10.000){2}{\rule{0.580pt}{0.800pt}}
\multiput(1325.00,196.08)(0.536,-0.511){17}{\rule{1.067pt}{0.123pt}}
\multiput(1325.00,196.34)(10.786,-12.000){2}{\rule{0.533pt}{0.800pt}}
\multiput(1338.00,184.08)(0.491,-0.511){17}{\rule{1.000pt}{0.123pt}}
\multiput(1338.00,184.34)(9.924,-12.000){2}{\rule{0.500pt}{0.800pt}}
\multiput(1351.41,169.30)(0.511,-0.581){17}{\rule{0.123pt}{1.133pt}}
\multiput(1348.34,171.65)(12.000,-11.648){2}{\rule{0.800pt}{0.567pt}}
\multiput(1363.41,155.08)(0.509,-0.616){19}{\rule{0.123pt}{1.185pt}}
\multiput(1360.34,157.54)(13.000,-13.541){2}{\rule{0.800pt}{0.592pt}}
\multiput(1376.41,137.91)(0.511,-0.807){17}{\rule{0.123pt}{1.467pt}}
\multiput(1373.34,140.96)(12.000,-15.956){2}{\rule{0.800pt}{0.733pt}}
\multiput(1388.40,118.48)(0.526,-0.913){7}{\rule{0.127pt}{1.571pt}}
\multiput(1385.34,121.74)(7.000,-8.738){2}{\rule{0.800pt}{0.786pt}}
\put(687.0,360.0){\rule[-0.400pt]{2.891pt}{0.800pt}}
\sbox{\plotpoint}{\rule[-0.500pt]{1.000pt}{1.000pt}}%
\put(220,377){\usebox{\plotpoint}}
\multiput(220,377)(11.969,16.957){2}{\usebox{\plotpoint}}
\multiput(232,394)(13.593,15.685){0}{\usebox{\plotpoint}}
\put(245.60,409.60){\usebox{\plotpoint}}
\put(260.42,424.13){\usebox{\plotpoint}}
\put(276.00,437.83){\usebox{\plotpoint}}
\put(293.14,449.47){\usebox{\plotpoint}}
\multiput(294,450)(17.270,11.513){0}{\usebox{\plotpoint}}
\put(310.59,460.68){\usebox{\plotpoint}}
\put(328.73,470.78){\usebox{\plotpoint}}
\multiput(331,472)(17.928,10.458){0}{\usebox{\plotpoint}}
\put(346.83,480.91){\usebox{\plotpoint}}
\put(365.73,489.47){\usebox{\plotpoint}}
\multiput(367,490)(18.845,8.698){0}{\usebox{\plotpoint}}
\put(384.67,497.94){\usebox{\plotpoint}}
\put(403.83,505.93){\usebox{\plotpoint}}
\multiput(404,506)(19.372,7.451){0}{\usebox{\plotpoint}}
\put(423.13,513.55){\usebox{\plotpoint}}
\multiput(429,516)(19.690,6.563){0}{\usebox{\plotpoint}}
\put(442.61,520.67){\usebox{\plotpoint}}
\put(462.08,527.79){\usebox{\plotpoint}}
\multiput(466,529)(19.690,6.563){0}{\usebox{\plotpoint}}
\put(481.80,534.27){\usebox{\plotpoint}}
\put(501.31,541.35){\usebox{\plotpoint}}
\multiput(503,542)(19.690,6.563){0}{\usebox{\plotpoint}}
\put(521.10,547.53){\usebox{\plotpoint}}
\multiput(527,549)(19.690,6.563){0}{\usebox{\plotpoint}}
\put(540.94,553.60){\usebox{\plotpoint}}
\put(560.71,559.90){\usebox{\plotpoint}}
\multiput(564,561)(19.690,6.563){0}{\usebox{\plotpoint}}
\put(580.40,566.47){\usebox{\plotpoint}}
\put(600.42,571.87){\usebox{\plotpoint}}
\multiput(601,572)(19.690,6.563){0}{\usebox{\plotpoint}}
\put(620.12,578.37){\usebox{\plotpoint}}
\multiput(625,580)(20.224,4.667){0}{\usebox{\plotpoint}}
\put(640.16,583.72){\usebox{\plotpoint}}
\put(659.85,590.28){\usebox{\plotpoint}}
\multiput(662,591)(19.690,6.563){0}{\usebox{\plotpoint}}
\put(679.69,596.31){\usebox{\plotpoint}}
\multiput(687,598)(19.690,6.563){0}{\usebox{\plotpoint}}
\put(699.57,602.19){\usebox{\plotpoint}}
\put(719.49,607.96){\usebox{\plotpoint}}
\multiput(724,609)(19.690,6.563){0}{\usebox{\plotpoint}}
\put(739.30,614.10){\usebox{\plotpoint}}
\put(758.99,620.66){\usebox{\plotpoint}}
\multiput(760,621)(20.224,4.667){0}{\usebox{\plotpoint}}
\put(779.02,626.01){\usebox{\plotpoint}}
\multiput(785,628)(19.690,6.563){0}{\usebox{\plotpoint}}
\put(798.72,632.53){\usebox{\plotpoint}}
\put(818.50,638.83){\usebox{\plotpoint}}
\multiput(822,640)(19.690,6.563){0}{\usebox{\plotpoint}}
\put(838.19,645.40){\usebox{\plotpoint}}
\put(857.97,651.68){\usebox{\plotpoint}}
\multiput(859,652)(19.690,6.563){0}{\usebox{\plotpoint}}
\put(877.48,658.70){\usebox{\plotpoint}}
\multiput(883,661)(19.838,6.104){0}{\usebox{\plotpoint}}
\put(897.12,665.37){\usebox{\plotpoint}}
\put(916.57,672.57){\usebox{\plotpoint}}
\multiput(920,674)(19.690,6.563){0}{\usebox{\plotpoint}}
\put(936.10,679.58){\usebox{\plotpoint}}
\put(955.36,687.32){\usebox{\plotpoint}}
\multiput(957,688)(19.159,7.983){0}{\usebox{\plotpoint}}
\put(974.71,694.76){\usebox{\plotpoint}}
\put(993.74,702.87){\usebox{\plotpoint}}
\multiput(994,703)(19.159,7.983){0}{\usebox{\plotpoint}}
\put(1012.89,710.87){\usebox{\plotpoint}}
\multiput(1018,713)(19.372,7.451){0}{\usebox{\plotpoint}}
\put(1032.16,718.58){\usebox{\plotpoint}}
\put(1050.72,727.86){\usebox{\plotpoint}}
\multiput(1055,730)(18.845,8.698){0}{\usebox{\plotpoint}}
\put(1069.48,736.74){\usebox{\plotpoint}}
\put(1088.04,746.02){\usebox{\plotpoint}}
\multiput(1092,748)(17.928,10.458){0}{\usebox{\plotpoint}}
\put(1106.15,756.16){\usebox{\plotpoint}}
\put(1124.28,766.25){\usebox{\plotpoint}}
\multiput(1129,769)(17.928,10.458){0}{\usebox{\plotpoint}}
\put(1142.17,776.78){\usebox{\plotpoint}}
\put(1159.59,788.05){\usebox{\plotpoint}}
\put(1176.58,799.94){\usebox{\plotpoint}}
\multiput(1178,801)(17.270,11.513){0}{\usebox{\plotpoint}}
\put(1193.61,811.78){\usebox{\plotpoint}}
\put(1210.13,824.35){\usebox{\plotpoint}}
\put(1225.81,837.91){\usebox{\plotpoint}}
\multiput(1227,839)(15.300,14.025){0}{\usebox{\plotpoint}}
\put(1241.19,851.85){\usebox{\plotpoint}}
\put(1256.47,865.85){\usebox{\plotpoint}}
\multiput(1264,874)(14.676,14.676){0}{\usebox{\plotpoint}}
\put(1267,877){\usebox{\plotpoint}}
\put(220,613){\usebox{\plotpoint}}
\multiput(220,613)(11.969,-16.957){2}{\usebox{\plotpoint}}
\multiput(232,596)(13.593,-15.685){0}{\usebox{\plotpoint}}
\put(245.60,580.40){\usebox{\plotpoint}}
\put(260.42,565.87){\usebox{\plotpoint}}
\put(276.00,552.17){\usebox{\plotpoint}}
\put(293.14,540.53){\usebox{\plotpoint}}
\multiput(294,540)(17.270,-11.513){0}{\usebox{\plotpoint}}
\put(310.59,529.32){\usebox{\plotpoint}}
\put(328.73,519.22){\usebox{\plotpoint}}
\multiput(331,518)(17.928,-10.458){0}{\usebox{\plotpoint}}
\put(346.83,509.09){\usebox{\plotpoint}}
\put(365.73,500.53){\usebox{\plotpoint}}
\multiput(367,500)(18.845,-8.698){0}{\usebox{\plotpoint}}
\put(384.67,492.06){\usebox{\plotpoint}}
\put(403.83,484.07){\usebox{\plotpoint}}
\multiput(404,484)(19.372,-7.451){0}{\usebox{\plotpoint}}
\put(423.13,476.45){\usebox{\plotpoint}}
\multiput(429,474)(19.690,-6.563){0}{\usebox{\plotpoint}}
\put(442.61,469.33){\usebox{\plotpoint}}
\put(462.08,462.21){\usebox{\plotpoint}}
\multiput(466,461)(19.690,-6.563){0}{\usebox{\plotpoint}}
\put(481.80,455.73){\usebox{\plotpoint}}
\put(501.31,448.65){\usebox{\plotpoint}}
\multiput(503,448)(19.690,-6.563){0}{\usebox{\plotpoint}}
\put(521.10,442.47){\usebox{\plotpoint}}
\multiput(527,441)(19.690,-6.563){0}{\usebox{\plotpoint}}
\put(540.94,436.40){\usebox{\plotpoint}}
\put(560.71,430.10){\usebox{\plotpoint}}
\multiput(564,429)(19.690,-6.563){0}{\usebox{\plotpoint}}
\put(580.40,423.53){\usebox{\plotpoint}}
\put(600.42,418.13){\usebox{\plotpoint}}
\multiput(601,418)(19.690,-6.563){0}{\usebox{\plotpoint}}
\put(620.12,411.63){\usebox{\plotpoint}}
\multiput(625,410)(20.224,-4.667){0}{\usebox{\plotpoint}}
\put(640.16,406.28){\usebox{\plotpoint}}
\put(659.85,399.72){\usebox{\plotpoint}}
\multiput(662,399)(19.690,-6.563){0}{\usebox{\plotpoint}}
\put(679.69,393.69){\usebox{\plotpoint}}
\multiput(687,392)(19.690,-6.563){0}{\usebox{\plotpoint}}
\put(699.57,387.81){\usebox{\plotpoint}}
\put(719.49,382.04){\usebox{\plotpoint}}
\multiput(724,381)(19.690,-6.563){0}{\usebox{\plotpoint}}
\put(739.30,375.90){\usebox{\plotpoint}}
\put(758.99,369.34){\usebox{\plotpoint}}
\multiput(760,369)(20.224,-4.667){0}{\usebox{\plotpoint}}
\put(779.02,363.99){\usebox{\plotpoint}}
\multiput(785,362)(19.690,-6.563){0}{\usebox{\plotpoint}}
\put(798.72,357.47){\usebox{\plotpoint}}
\put(818.50,351.17){\usebox{\plotpoint}}
\multiput(822,350)(19.690,-6.563){0}{\usebox{\plotpoint}}
\put(838.19,344.60){\usebox{\plotpoint}}
\put(857.97,338.32){\usebox{\plotpoint}}
\multiput(859,338)(19.690,-6.563){0}{\usebox{\plotpoint}}
\put(877.48,331.30){\usebox{\plotpoint}}
\multiput(883,329)(19.838,-6.104){0}{\usebox{\plotpoint}}
\put(897.12,324.63){\usebox{\plotpoint}}
\put(916.57,317.43){\usebox{\plotpoint}}
\multiput(920,316)(19.690,-6.563){0}{\usebox{\plotpoint}}
\put(936.10,310.42){\usebox{\plotpoint}}
\put(955.36,302.68){\usebox{\plotpoint}}
\multiput(957,302)(19.159,-7.983){0}{\usebox{\plotpoint}}
\put(974.71,295.24){\usebox{\plotpoint}}
\put(993.74,287.13){\usebox{\plotpoint}}
\multiput(994,287)(19.159,-7.983){0}{\usebox{\plotpoint}}
\put(1012.89,279.13){\usebox{\plotpoint}}
\multiput(1018,277)(19.372,-7.451){0}{\usebox{\plotpoint}}
\put(1032.16,271.42){\usebox{\plotpoint}}
\put(1050.72,262.14){\usebox{\plotpoint}}
\multiput(1055,260)(18.845,-8.698){0}{\usebox{\plotpoint}}
\put(1069.48,253.26){\usebox{\plotpoint}}
\put(1088.04,243.98){\usebox{\plotpoint}}
\multiput(1092,242)(17.928,-10.458){0}{\usebox{\plotpoint}}
\put(1106.15,233.84){\usebox{\plotpoint}}
\put(1124.28,223.75){\usebox{\plotpoint}}
\multiput(1129,221)(17.928,-10.458){0}{\usebox{\plotpoint}}
\put(1142.17,213.22){\usebox{\plotpoint}}
\put(1159.59,201.95){\usebox{\plotpoint}}
\put(1176.58,190.06){\usebox{\plotpoint}}
\multiput(1178,189)(17.270,-11.513){0}{\usebox{\plotpoint}}
\put(1193.61,178.22){\usebox{\plotpoint}}
\put(1210.13,165.65){\usebox{\plotpoint}}
\put(1225.81,152.09){\usebox{\plotpoint}}
\multiput(1227,151)(15.300,-14.025){0}{\usebox{\plotpoint}}
\put(1241.19,138.15){\usebox{\plotpoint}}
\put(1256.47,124.15){\usebox{\plotpoint}}
\multiput(1264,116)(14.676,-14.676){0}{\usebox{\plotpoint}}
\put(1267,113){\usebox{\plotpoint}}
\sbox{\plotpoint}{\rule[-0.600pt]{1.200pt}{1.200pt}}%
\put(220,199){\usebox{\plotpoint}}
\multiput(222.24,199.00)(0.501,0.489){14}{\rule{0.121pt}{1.600pt}}
\multiput(217.51,199.00)(12.000,9.679){2}{\rule{1.200pt}{0.800pt}}
\multiput(232.00,214.24)(0.587,0.502){10}{\rule{1.860pt}{0.121pt}}
\multiput(232.00,209.51)(9.139,10.000){2}{\rule{0.930pt}{1.200pt}}
\multiput(245.00,224.24)(0.657,0.503){6}{\rule{2.100pt}{0.121pt}}
\multiput(245.00,219.51)(7.641,8.000){2}{\rule{1.050pt}{1.200pt}}
\multiput(257.00,232.24)(0.738,0.505){4}{\rule{2.357pt}{0.122pt}}
\multiput(257.00,227.51)(7.108,7.000){2}{\rule{1.179pt}{1.200pt}}
\multiput(269.00,239.24)(0.792,0.509){2}{\rule{2.700pt}{0.123pt}}
\multiput(269.00,234.51)(6.396,6.000){2}{\rule{1.350pt}{1.200pt}}
\put(281,243.01){\rule{3.132pt}{1.200pt}}
\multiput(281.00,240.51)(6.500,5.000){2}{\rule{1.566pt}{1.200pt}}
\put(294,247.51){\rule{2.891pt}{1.200pt}}
\multiput(294.00,245.51)(6.000,4.000){2}{\rule{1.445pt}{1.200pt}}
\put(306,251.51){\rule{2.891pt}{1.200pt}}
\multiput(306.00,249.51)(6.000,4.000){2}{\rule{1.445pt}{1.200pt}}
\put(318,255.01){\rule{3.132pt}{1.200pt}}
\multiput(318.00,253.51)(6.500,3.000){2}{\rule{1.566pt}{1.200pt}}
\put(331,258.01){\rule{2.891pt}{1.200pt}}
\multiput(331.00,256.51)(6.000,3.000){2}{\rule{1.445pt}{1.200pt}}
\put(343,261.01){\rule{2.891pt}{1.200pt}}
\multiput(343.00,259.51)(6.000,3.000){2}{\rule{1.445pt}{1.200pt}}
\put(355,263.51){\rule{2.891pt}{1.200pt}}
\multiput(355.00,262.51)(6.000,2.000){2}{\rule{1.445pt}{1.200pt}}
\put(367,265.51){\rule{3.132pt}{1.200pt}}
\multiput(367.00,264.51)(6.500,2.000){2}{\rule{1.566pt}{1.200pt}}
\put(380,267.51){\rule{2.891pt}{1.200pt}}
\multiput(380.00,266.51)(6.000,2.000){2}{\rule{1.445pt}{1.200pt}}
\put(392,269.51){\rule{2.891pt}{1.200pt}}
\multiput(392.00,268.51)(6.000,2.000){2}{\rule{1.445pt}{1.200pt}}
\put(404,271.51){\rule{3.132pt}{1.200pt}}
\multiput(404.00,270.51)(6.500,2.000){2}{\rule{1.566pt}{1.200pt}}
\put(417,273.51){\rule{2.891pt}{1.200pt}}
\multiput(417.00,272.51)(6.000,2.000){2}{\rule{1.445pt}{1.200pt}}
\put(429,275.01){\rule{2.891pt}{1.200pt}}
\multiput(429.00,274.51)(6.000,1.000){2}{\rule{1.445pt}{1.200pt}}
\put(441,276.01){\rule{2.891pt}{1.200pt}}
\multiput(441.00,275.51)(6.000,1.000){2}{\rule{1.445pt}{1.200pt}}
\put(453,277.51){\rule{3.132pt}{1.200pt}}
\multiput(453.00,276.51)(6.500,2.000){2}{\rule{1.566pt}{1.200pt}}
\put(466,279.01){\rule{2.891pt}{1.200pt}}
\multiput(466.00,278.51)(6.000,1.000){2}{\rule{1.445pt}{1.200pt}}
\put(478,280.01){\rule{2.891pt}{1.200pt}}
\multiput(478.00,279.51)(6.000,1.000){2}{\rule{1.445pt}{1.200pt}}
\put(490,281.01){\rule{3.132pt}{1.200pt}}
\multiput(490.00,280.51)(6.500,1.000){2}{\rule{1.566pt}{1.200pt}}
\put(503,282.01){\rule{2.891pt}{1.200pt}}
\multiput(503.00,281.51)(6.000,1.000){2}{\rule{1.445pt}{1.200pt}}
\put(515,283.01){\rule{2.891pt}{1.200pt}}
\multiput(515.00,282.51)(6.000,1.000){2}{\rule{1.445pt}{1.200pt}}
\put(539,284.01){\rule{3.132pt}{1.200pt}}
\multiput(539.00,283.51)(6.500,1.000){2}{\rule{1.566pt}{1.200pt}}
\put(552,285.01){\rule{2.891pt}{1.200pt}}
\multiput(552.00,284.51)(6.000,1.000){2}{\rule{1.445pt}{1.200pt}}
\put(527.0,286.0){\rule[-0.600pt]{2.891pt}{1.200pt}}
\put(576,286.01){\rule{2.891pt}{1.200pt}}
\multiput(576.00,285.51)(6.000,1.000){2}{\rule{1.445pt}{1.200pt}}
\put(588,287.01){\rule{3.132pt}{1.200pt}}
\multiput(588.00,286.51)(6.500,1.000){2}{\rule{1.566pt}{1.200pt}}
\put(564.0,288.0){\rule[-0.600pt]{2.891pt}{1.200pt}}
\put(613,288.01){\rule{2.891pt}{1.200pt}}
\multiput(613.00,287.51)(6.000,1.000){2}{\rule{1.445pt}{1.200pt}}
\put(601.0,290.0){\rule[-0.600pt]{2.891pt}{1.200pt}}
\put(650,289.01){\rule{2.891pt}{1.200pt}}
\multiput(650.00,288.51)(6.000,1.000){2}{\rule{1.445pt}{1.200pt}}
\put(625.0,291.0){\rule[-0.600pt]{6.022pt}{1.200pt}}
\put(687,290.01){\rule{2.891pt}{1.200pt}}
\multiput(687.00,289.51)(6.000,1.000){2}{\rule{1.445pt}{1.200pt}}
\put(662.0,292.0){\rule[-0.600pt]{6.022pt}{1.200pt}}
\put(736,291.01){\rule{2.891pt}{1.200pt}}
\multiput(736.00,290.51)(6.000,1.000){2}{\rule{1.445pt}{1.200pt}}
\put(699.0,293.0){\rule[-0.600pt]{8.913pt}{1.200pt}}
\put(908,291.01){\rule{2.891pt}{1.200pt}}
\multiput(908.00,291.51)(6.000,-1.000){2}{\rule{1.445pt}{1.200pt}}
\put(748.0,294.0){\rule[-0.600pt]{38.544pt}{1.200pt}}
\put(957,290.01){\rule{2.891pt}{1.200pt}}
\multiput(957.00,290.51)(6.000,-1.000){2}{\rule{1.445pt}{1.200pt}}
\put(920.0,293.0){\rule[-0.600pt]{8.913pt}{1.200pt}}
\put(994,289.01){\rule{2.891pt}{1.200pt}}
\multiput(994.00,289.51)(6.000,-1.000){2}{\rule{1.445pt}{1.200pt}}
\put(969.0,292.0){\rule[-0.600pt]{6.022pt}{1.200pt}}
\put(1031,288.01){\rule{2.891pt}{1.200pt}}
\multiput(1031.00,288.51)(6.000,-1.000){2}{\rule{1.445pt}{1.200pt}}
\put(1006.0,291.0){\rule[-0.600pt]{6.022pt}{1.200pt}}
\put(1055,287.01){\rule{3.132pt}{1.200pt}}
\multiput(1055.00,287.51)(6.500,-1.000){2}{\rule{1.566pt}{1.200pt}}
\put(1068,286.01){\rule{2.891pt}{1.200pt}}
\multiput(1068.00,286.51)(6.000,-1.000){2}{\rule{1.445pt}{1.200pt}}
\put(1043.0,290.0){\rule[-0.600pt]{2.891pt}{1.200pt}}
\put(1092,285.01){\rule{2.891pt}{1.200pt}}
\multiput(1092.00,285.51)(6.000,-1.000){2}{\rule{1.445pt}{1.200pt}}
\put(1104,284.01){\rule{3.132pt}{1.200pt}}
\multiput(1104.00,284.51)(6.500,-1.000){2}{\rule{1.566pt}{1.200pt}}
\put(1080.0,288.0){\rule[-0.600pt]{2.891pt}{1.200pt}}
\put(1129,283.01){\rule{2.891pt}{1.200pt}}
\multiput(1129.00,283.51)(6.000,-1.000){2}{\rule{1.445pt}{1.200pt}}
\put(1141,282.01){\rule{2.891pt}{1.200pt}}
\multiput(1141.00,282.51)(6.000,-1.000){2}{\rule{1.445pt}{1.200pt}}
\put(1153,281.01){\rule{3.132pt}{1.200pt}}
\multiput(1153.00,281.51)(6.500,-1.000){2}{\rule{1.566pt}{1.200pt}}
\put(1166,280.01){\rule{2.891pt}{1.200pt}}
\multiput(1166.00,280.51)(6.000,-1.000){2}{\rule{1.445pt}{1.200pt}}
\put(1178,279.01){\rule{2.891pt}{1.200pt}}
\multiput(1178.00,279.51)(6.000,-1.000){2}{\rule{1.445pt}{1.200pt}}
\put(1190,277.51){\rule{3.132pt}{1.200pt}}
\multiput(1190.00,278.51)(6.500,-2.000){2}{\rule{1.566pt}{1.200pt}}
\put(1203,276.01){\rule{2.891pt}{1.200pt}}
\multiput(1203.00,276.51)(6.000,-1.000){2}{\rule{1.445pt}{1.200pt}}
\put(1215,275.01){\rule{2.891pt}{1.200pt}}
\multiput(1215.00,275.51)(6.000,-1.000){2}{\rule{1.445pt}{1.200pt}}
\put(1227,273.51){\rule{2.891pt}{1.200pt}}
\multiput(1227.00,274.51)(6.000,-2.000){2}{\rule{1.445pt}{1.200pt}}
\put(1239,271.51){\rule{3.132pt}{1.200pt}}
\multiput(1239.00,272.51)(6.500,-2.000){2}{\rule{1.566pt}{1.200pt}}
\put(1252,269.51){\rule{2.891pt}{1.200pt}}
\multiput(1252.00,270.51)(6.000,-2.000){2}{\rule{1.445pt}{1.200pt}}
\put(1264,267.51){\rule{2.891pt}{1.200pt}}
\multiput(1264.00,268.51)(6.000,-2.000){2}{\rule{1.445pt}{1.200pt}}
\put(1276,265.51){\rule{3.132pt}{1.200pt}}
\multiput(1276.00,266.51)(6.500,-2.000){2}{\rule{1.566pt}{1.200pt}}
\put(1289,263.51){\rule{2.891pt}{1.200pt}}
\multiput(1289.00,264.51)(6.000,-2.000){2}{\rule{1.445pt}{1.200pt}}
\put(1301,261.01){\rule{2.891pt}{1.200pt}}
\multiput(1301.00,262.51)(6.000,-3.000){2}{\rule{1.445pt}{1.200pt}}
\put(1313,258.01){\rule{2.891pt}{1.200pt}}
\multiput(1313.00,259.51)(6.000,-3.000){2}{\rule{1.445pt}{1.200pt}}
\put(1325,255.01){\rule{3.132pt}{1.200pt}}
\multiput(1325.00,256.51)(6.500,-3.000){2}{\rule{1.566pt}{1.200pt}}
\put(1338,251.51){\rule{2.891pt}{1.200pt}}
\multiput(1338.00,253.51)(6.000,-4.000){2}{\rule{1.445pt}{1.200pt}}
\put(1350,247.51){\rule{2.891pt}{1.200pt}}
\multiput(1350.00,249.51)(6.000,-4.000){2}{\rule{1.445pt}{1.200pt}}
\put(1362,243.01){\rule{3.132pt}{1.200pt}}
\multiput(1362.00,245.51)(6.500,-5.000){2}{\rule{1.566pt}{1.200pt}}
\multiput(1375.00,240.25)(0.792,-0.509){2}{\rule{2.700pt}{0.123pt}}
\multiput(1375.00,240.51)(6.396,-6.000){2}{\rule{1.350pt}{1.200pt}}
\multiput(1387.00,234.26)(0.738,-0.505){4}{\rule{2.357pt}{0.122pt}}
\multiput(1387.00,234.51)(7.108,-7.000){2}{\rule{1.179pt}{1.200pt}}
\multiput(1399.00,227.26)(0.657,-0.503){6}{\rule{2.100pt}{0.121pt}}
\multiput(1399.00,227.51)(7.641,-8.000){2}{\rule{1.050pt}{1.200pt}}
\multiput(1411.00,219.26)(0.587,-0.502){10}{\rule{1.860pt}{0.121pt}}
\multiput(1411.00,219.51)(9.139,-10.000){2}{\rule{0.930pt}{1.200pt}}
\multiput(1426.24,205.36)(0.501,-0.489){14}{\rule{0.121pt}{1.600pt}}
\multiput(1421.51,208.68)(12.000,-9.679){2}{\rule{1.200pt}{0.800pt}}
\put(1117.0,286.0){\rule[-0.600pt]{2.891pt}{1.200pt}}
\put(220,536){\usebox{\plotpoint}}
\multiput(220.00,533.26)(0.444,-0.501){14}{\rule{1.500pt}{0.121pt}}
\multiput(220.00,533.51)(8.887,-12.000){2}{\rule{0.750pt}{1.200pt}}
\multiput(232.00,521.26)(0.533,-0.502){12}{\rule{1.718pt}{0.121pt}}
\multiput(232.00,521.51)(9.434,-11.000){2}{\rule{0.859pt}{1.200pt}}
\multiput(245.00,510.26)(0.657,-0.503){6}{\rule{2.100pt}{0.121pt}}
\multiput(245.00,510.51)(7.641,-8.000){2}{\rule{1.050pt}{1.200pt}}
\multiput(257.00,502.25)(0.792,-0.509){2}{\rule{2.700pt}{0.123pt}}
\multiput(257.00,502.51)(6.396,-6.000){2}{\rule{1.350pt}{1.200pt}}
\multiput(269.00,496.25)(0.792,-0.509){2}{\rule{2.700pt}{0.123pt}}
\multiput(269.00,496.51)(6.396,-6.000){2}{\rule{1.350pt}{1.200pt}}
\put(281,488.01){\rule{3.132pt}{1.200pt}}
\multiput(281.00,490.51)(6.500,-5.000){2}{\rule{1.566pt}{1.200pt}}
\put(294,483.01){\rule{2.891pt}{1.200pt}}
\multiput(294.00,485.51)(6.000,-5.000){2}{\rule{1.445pt}{1.200pt}}
\put(306,479.01){\rule{2.891pt}{1.200pt}}
\multiput(306.00,480.51)(6.000,-3.000){2}{\rule{1.445pt}{1.200pt}}
\put(318,475.51){\rule{3.132pt}{1.200pt}}
\multiput(318.00,477.51)(6.500,-4.000){2}{\rule{1.566pt}{1.200pt}}
\put(331,472.01){\rule{2.891pt}{1.200pt}}
\multiput(331.00,473.51)(6.000,-3.000){2}{\rule{1.445pt}{1.200pt}}
\put(343,469.51){\rule{2.891pt}{1.200pt}}
\multiput(343.00,470.51)(6.000,-2.000){2}{\rule{1.445pt}{1.200pt}}
\put(355,467.01){\rule{2.891pt}{1.200pt}}
\multiput(355.00,468.51)(6.000,-3.000){2}{\rule{1.445pt}{1.200pt}}
\put(367,464.51){\rule{3.132pt}{1.200pt}}
\multiput(367.00,465.51)(6.500,-2.000){2}{\rule{1.566pt}{1.200pt}}
\put(380,462.51){\rule{2.891pt}{1.200pt}}
\multiput(380.00,463.51)(6.000,-2.000){2}{\rule{1.445pt}{1.200pt}}
\put(392,460.51){\rule{2.891pt}{1.200pt}}
\multiput(392.00,461.51)(6.000,-2.000){2}{\rule{1.445pt}{1.200pt}}
\put(404,458.51){\rule{3.132pt}{1.200pt}}
\multiput(404.00,459.51)(6.500,-2.000){2}{\rule{1.566pt}{1.200pt}}
\put(417,457.01){\rule{2.891pt}{1.200pt}}
\multiput(417.00,457.51)(6.000,-1.000){2}{\rule{1.445pt}{1.200pt}}
\put(429,455.51){\rule{2.891pt}{1.200pt}}
\multiput(429.00,456.51)(6.000,-2.000){2}{\rule{1.445pt}{1.200pt}}
\put(441,454.01){\rule{2.891pt}{1.200pt}}
\multiput(441.00,454.51)(6.000,-1.000){2}{\rule{1.445pt}{1.200pt}}
\put(453,453.01){\rule{3.132pt}{1.200pt}}
\multiput(453.00,453.51)(6.500,-1.000){2}{\rule{1.566pt}{1.200pt}}
\put(466,452.01){\rule{2.891pt}{1.200pt}}
\multiput(466.00,452.51)(6.000,-1.000){2}{\rule{1.445pt}{1.200pt}}
\put(478,451.01){\rule{2.891pt}{1.200pt}}
\multiput(478.00,451.51)(6.000,-1.000){2}{\rule{1.445pt}{1.200pt}}
\put(490,450.01){\rule{3.132pt}{1.200pt}}
\multiput(490.00,450.51)(6.500,-1.000){2}{\rule{1.566pt}{1.200pt}}
\put(503,449.01){\rule{2.891pt}{1.200pt}}
\multiput(503.00,449.51)(6.000,-1.000){2}{\rule{1.445pt}{1.200pt}}
\put(515,448.01){\rule{2.891pt}{1.200pt}}
\multiput(515.00,448.51)(6.000,-1.000){2}{\rule{1.445pt}{1.200pt}}
\put(527,447.01){\rule{2.891pt}{1.200pt}}
\multiput(527.00,447.51)(6.000,-1.000){2}{\rule{1.445pt}{1.200pt}}
\put(539,446.01){\rule{3.132pt}{1.200pt}}
\multiput(539.00,446.51)(6.500,-1.000){2}{\rule{1.566pt}{1.200pt}}
\put(564,445.01){\rule{2.891pt}{1.200pt}}
\multiput(564.00,445.51)(6.000,-1.000){2}{\rule{1.445pt}{1.200pt}}
\put(576,444.01){\rule{2.891pt}{1.200pt}}
\multiput(576.00,444.51)(6.000,-1.000){2}{\rule{1.445pt}{1.200pt}}
\put(552.0,448.0){\rule[-0.600pt]{2.891pt}{1.200pt}}
\put(601,443.01){\rule{2.891pt}{1.200pt}}
\multiput(601.00,443.51)(6.000,-1.000){2}{\rule{1.445pt}{1.200pt}}
\put(588.0,446.0){\rule[-0.600pt]{3.132pt}{1.200pt}}
\put(625,442.01){\rule{3.132pt}{1.200pt}}
\multiput(625.00,442.51)(6.500,-1.000){2}{\rule{1.566pt}{1.200pt}}
\put(613.0,445.0){\rule[-0.600pt]{2.891pt}{1.200pt}}
\put(662,441.01){\rule{2.891pt}{1.200pt}}
\multiput(662.00,441.51)(6.000,-1.000){2}{\rule{1.445pt}{1.200pt}}
\put(638.0,444.0){\rule[-0.600pt]{5.782pt}{1.200pt}}
\put(699,440.01){\rule{2.891pt}{1.200pt}}
\multiput(699.00,440.51)(6.000,-1.000){2}{\rule{1.445pt}{1.200pt}}
\put(674.0,443.0){\rule[-0.600pt]{6.022pt}{1.200pt}}
\put(760,439.01){\rule{3.132pt}{1.200pt}}
\multiput(760.00,439.51)(6.500,-1.000){2}{\rule{1.566pt}{1.200pt}}
\put(711.0,442.0){\rule[-0.600pt]{11.804pt}{1.200pt}}
\put(883,439.01){\rule{3.132pt}{1.200pt}}
\multiput(883.00,438.51)(6.500,1.000){2}{\rule{1.566pt}{1.200pt}}
\put(773.0,441.0){\rule[-0.600pt]{26.499pt}{1.200pt}}
\put(945,440.01){\rule{2.891pt}{1.200pt}}
\multiput(945.00,439.51)(6.000,1.000){2}{\rule{1.445pt}{1.200pt}}
\put(896.0,442.0){\rule[-0.600pt]{11.804pt}{1.200pt}}
\put(982,441.01){\rule{2.891pt}{1.200pt}}
\multiput(982.00,440.51)(6.000,1.000){2}{\rule{1.445pt}{1.200pt}}
\put(957.0,443.0){\rule[-0.600pt]{6.022pt}{1.200pt}}
\put(1018,442.01){\rule{3.132pt}{1.200pt}}
\multiput(1018.00,441.51)(6.500,1.000){2}{\rule{1.566pt}{1.200pt}}
\put(994.0,444.0){\rule[-0.600pt]{5.782pt}{1.200pt}}
\put(1043,443.01){\rule{2.891pt}{1.200pt}}
\multiput(1043.00,442.51)(6.000,1.000){2}{\rule{1.445pt}{1.200pt}}
\put(1031.0,445.0){\rule[-0.600pt]{2.891pt}{1.200pt}}
\put(1068,444.01){\rule{2.891pt}{1.200pt}}
\multiput(1068.00,443.51)(6.000,1.000){2}{\rule{1.445pt}{1.200pt}}
\put(1080,445.01){\rule{2.891pt}{1.200pt}}
\multiput(1080.00,444.51)(6.000,1.000){2}{\rule{1.445pt}{1.200pt}}
\put(1055.0,446.0){\rule[-0.600pt]{3.132pt}{1.200pt}}
\put(1104,446.01){\rule{3.132pt}{1.200pt}}
\multiput(1104.00,445.51)(6.500,1.000){2}{\rule{1.566pt}{1.200pt}}
\put(1117,447.01){\rule{2.891pt}{1.200pt}}
\multiput(1117.00,446.51)(6.000,1.000){2}{\rule{1.445pt}{1.200pt}}
\put(1129,448.01){\rule{2.891pt}{1.200pt}}
\multiput(1129.00,447.51)(6.000,1.000){2}{\rule{1.445pt}{1.200pt}}
\put(1141,449.01){\rule{2.891pt}{1.200pt}}
\multiput(1141.00,448.51)(6.000,1.000){2}{\rule{1.445pt}{1.200pt}}
\put(1153,450.01){\rule{3.132pt}{1.200pt}}
\multiput(1153.00,449.51)(6.500,1.000){2}{\rule{1.566pt}{1.200pt}}
\put(1166,451.01){\rule{2.891pt}{1.200pt}}
\multiput(1166.00,450.51)(6.000,1.000){2}{\rule{1.445pt}{1.200pt}}
\put(1178,452.01){\rule{2.891pt}{1.200pt}}
\multiput(1178.00,451.51)(6.000,1.000){2}{\rule{1.445pt}{1.200pt}}
\put(1190,453.01){\rule{3.132pt}{1.200pt}}
\multiput(1190.00,452.51)(6.500,1.000){2}{\rule{1.566pt}{1.200pt}}
\put(1203,454.01){\rule{2.891pt}{1.200pt}}
\multiput(1203.00,453.51)(6.000,1.000){2}{\rule{1.445pt}{1.200pt}}
\put(1215,455.51){\rule{2.891pt}{1.200pt}}
\multiput(1215.00,454.51)(6.000,2.000){2}{\rule{1.445pt}{1.200pt}}
\put(1227,457.01){\rule{2.891pt}{1.200pt}}
\multiput(1227.00,456.51)(6.000,1.000){2}{\rule{1.445pt}{1.200pt}}
\put(1239,458.51){\rule{3.132pt}{1.200pt}}
\multiput(1239.00,457.51)(6.500,2.000){2}{\rule{1.566pt}{1.200pt}}
\put(1252,460.51){\rule{2.891pt}{1.200pt}}
\multiput(1252.00,459.51)(6.000,2.000){2}{\rule{1.445pt}{1.200pt}}
\put(1264,462.51){\rule{2.891pt}{1.200pt}}
\multiput(1264.00,461.51)(6.000,2.000){2}{\rule{1.445pt}{1.200pt}}
\put(1276,464.51){\rule{3.132pt}{1.200pt}}
\multiput(1276.00,463.51)(6.500,2.000){2}{\rule{1.566pt}{1.200pt}}
\put(1289,467.01){\rule{2.891pt}{1.200pt}}
\multiput(1289.00,465.51)(6.000,3.000){2}{\rule{1.445pt}{1.200pt}}
\put(1301,469.51){\rule{2.891pt}{1.200pt}}
\multiput(1301.00,468.51)(6.000,2.000){2}{\rule{1.445pt}{1.200pt}}
\put(1313,472.01){\rule{2.891pt}{1.200pt}}
\multiput(1313.00,470.51)(6.000,3.000){2}{\rule{1.445pt}{1.200pt}}
\put(1325,475.51){\rule{3.132pt}{1.200pt}}
\multiput(1325.00,473.51)(6.500,4.000){2}{\rule{1.566pt}{1.200pt}}
\put(1338,479.01){\rule{2.891pt}{1.200pt}}
\multiput(1338.00,477.51)(6.000,3.000){2}{\rule{1.445pt}{1.200pt}}
\put(1350,483.01){\rule{2.891pt}{1.200pt}}
\multiput(1350.00,480.51)(6.000,5.000){2}{\rule{1.445pt}{1.200pt}}
\put(1362,488.01){\rule{3.132pt}{1.200pt}}
\multiput(1362.00,485.51)(6.500,5.000){2}{\rule{1.566pt}{1.200pt}}
\multiput(1375.00,495.24)(0.792,0.509){2}{\rule{2.700pt}{0.123pt}}
\multiput(1375.00,490.51)(6.396,6.000){2}{\rule{1.350pt}{1.200pt}}
\multiput(1387.00,501.24)(0.792,0.509){2}{\rule{2.700pt}{0.123pt}}
\multiput(1387.00,496.51)(6.396,6.000){2}{\rule{1.350pt}{1.200pt}}
\multiput(1399.00,507.24)(0.657,0.503){6}{\rule{2.100pt}{0.121pt}}
\multiput(1399.00,502.51)(7.641,8.000){2}{\rule{1.050pt}{1.200pt}}
\multiput(1411.00,515.24)(0.533,0.502){12}{\rule{1.718pt}{0.121pt}}
\multiput(1411.00,510.51)(9.434,11.000){2}{\rule{0.859pt}{1.200pt}}
\multiput(1424.00,526.24)(0.444,0.501){14}{\rule{1.500pt}{0.121pt}}
\multiput(1424.00,521.51)(8.887,12.000){2}{\rule{0.750pt}{1.200pt}}
\put(1092.0,448.0){\rule[-0.600pt]{2.891pt}{1.200pt}}
\end{picture}

%% file: tanbb.tex
% GNUPLOT: LaTeX picture
\setlength{\unitlength}{0.240900pt}
\ifx\plotpoint\undefined\newsavebox{\plotpoint}\fi
\sbox{\plotpoint}{\rule[-0.200pt]{0.400pt}{0.400pt}}%
\begin{picture}(1500,900)(0,0)
\font\gnuplot=cmr10 at 10pt
\gnuplot
\sbox{\plotpoint}{\rule[-0.200pt]{0.400pt}{0.400pt}}%
\put(220.0,113.0){\rule[-0.200pt]{4.818pt}{0.400pt}}
\put(198,113){\makebox(0,0)[r]{-2}}
\put(1416.0,113.0){\rule[-0.200pt]{4.818pt}{0.400pt}}
\put(220.0,189.0){\rule[-0.200pt]{4.818pt}{0.400pt}}
\put(198,189){\makebox(0,0)[r]{-1.8}}
\put(1416.0,189.0){\rule[-0.200pt]{4.818pt}{0.400pt}}
\put(220.0,266.0){\rule[-0.200pt]{4.818pt}{0.400pt}}
\put(198,266){\makebox(0,0)[r]{-1.6}}
\put(1416.0,266.0){\rule[-0.200pt]{4.818pt}{0.400pt}}
\put(220.0,342.0){\rule[-0.200pt]{4.818pt}{0.400pt}}
\put(198,342){\makebox(0,0)[r]{-1.4}}
\put(1416.0,342.0){\rule[-0.200pt]{4.818pt}{0.400pt}}
\put(220.0,419.0){\rule[-0.200pt]{4.818pt}{0.400pt}}
\put(198,419){\makebox(0,0)[r]{-1.2}}
\put(1416.0,419.0){\rule[-0.200pt]{4.818pt}{0.400pt}}
\put(220.0,495.0){\rule[-0.200pt]{4.818pt}{0.400pt}}
\put(198,495){\makebox(0,0)[r]{-1}}
\put(1416.0,495.0){\rule[-0.200pt]{4.818pt}{0.400pt}}
\put(220.0,571.0){\rule[-0.200pt]{4.818pt}{0.400pt}}
\put(198,571){\makebox(0,0)[r]{-0.8}}
\put(1416.0,571.0){\rule[-0.200pt]{4.818pt}{0.400pt}}
\put(220.0,648.0){\rule[-0.200pt]{4.818pt}{0.400pt}}
\put(198,648){\makebox(0,0)[r]{-0.6}}
\put(1416.0,648.0){\rule[-0.200pt]{4.818pt}{0.400pt}}
\put(220.0,724.0){\rule[-0.200pt]{4.818pt}{0.400pt}}
\put(198,724){\makebox(0,0)[r]{-0.4}}
\put(1416.0,724.0){\rule[-0.200pt]{4.818pt}{0.400pt}}
\put(220.0,801.0){\rule[-0.200pt]{4.818pt}{0.400pt}}
\put(198,801){\makebox(0,0)[r]{-0.2}}
\put(1416.0,801.0){\rule[-0.200pt]{4.818pt}{0.400pt}}
\put(220.0,877.0){\rule[-0.200pt]{4.818pt}{0.400pt}}
\put(198,877){\makebox(0,0)[r]{0}}
\put(1416.0,877.0){\rule[-0.200pt]{4.818pt}{0.400pt}}
\put(220.0,113.0){\rule[-0.200pt]{0.400pt}{4.818pt}}
\put(220,68){\makebox(0,0){20}}
\put(220.0,857.0){\rule[-0.200pt]{0.400pt}{4.818pt}}
\put(372.0,113.0){\rule[-0.200pt]{0.400pt}{4.818pt}}
\put(372,68){\makebox(0,0){25}}
\put(372.0,857.0){\rule[-0.200pt]{0.400pt}{4.818pt}}
\put(524.0,113.0){\rule[-0.200pt]{0.400pt}{4.818pt}}
\put(524,68){\makebox(0,0){30}}
\put(524.0,857.0){\rule[-0.200pt]{0.400pt}{4.818pt}}
\put(676.0,113.0){\rule[-0.200pt]{0.400pt}{4.818pt}}
\put(676,68){\makebox(0,0){35}}
\put(676.0,857.0){\rule[-0.200pt]{0.400pt}{4.818pt}}
\put(828.0,113.0){\rule[-0.200pt]{0.400pt}{4.818pt}}
\put(828,68){\makebox(0,0){40}}
\put(828.0,857.0){\rule[-0.200pt]{0.400pt}{4.818pt}}
\put(980.0,113.0){\rule[-0.200pt]{0.400pt}{4.818pt}}
\put(980,68){\makebox(0,0){45}}
\put(980.0,857.0){\rule[-0.200pt]{0.400pt}{4.818pt}}
\put(1132.0,113.0){\rule[-0.200pt]{0.400pt}{4.818pt}}
\put(1132,68){\makebox(0,0){50}}
\put(1132.0,857.0){\rule[-0.200pt]{0.400pt}{4.818pt}}
\put(1284.0,113.0){\rule[-0.200pt]{0.400pt}{4.818pt}}
\put(1284,68){\makebox(0,0){55}}
\put(1284.0,857.0){\rule[-0.200pt]{0.400pt}{4.818pt}}
\put(1436.0,113.0){\rule[-0.200pt]{0.400pt}{4.818pt}}
\put(1436,68){\makebox(0,0){60}}
\put(1436.0,857.0){\rule[-0.200pt]{0.400pt}{4.818pt}}
\put(220.0,113.0){\rule[-0.200pt]{292.934pt}{0.400pt}}
\put(1436.0,113.0){\rule[-0.200pt]{0.400pt}{184.048pt}}
\put(220.0,877.0){\rule[-0.200pt]{292.934pt}{0.400pt}}
\put(45,495){\makebox(0,0){$\displaystyle{\frac{\Delta B}{M^{(0)}_{1/2}}}$}}
\put(828,23){\makebox(0,0){$\tan \beta$}}
\put(220.0,113.0){\rule[-0.200pt]{0.400pt}{184.048pt}}
\put(220,571){\usebox{\plotpoint}}
\multiput(220.00,569.92)(1.423,-0.499){211}{\rule{1.236pt}{0.120pt}}
\multiput(220.00,570.17)(301.434,-107.000){2}{\rule{0.618pt}{0.400pt}}
\multiput(524.00,462.92)(1.539,-0.499){195}{\rule{1.328pt}{0.120pt}}
\multiput(524.00,463.17)(301.243,-99.000){2}{\rule{0.664pt}{0.400pt}}
\multiput(828.00,363.92)(1.338,-0.499){111}{\rule{1.167pt}{0.120pt}}
\multiput(828.00,364.17)(149.579,-57.000){2}{\rule{0.583pt}{0.400pt}}
\multiput(980.00,306.92)(1.338,-0.499){111}{\rule{1.167pt}{0.120pt}}
\multiput(980.00,307.17)(149.579,-57.000){2}{\rule{0.583pt}{0.400pt}}
\multiput(1132.00,249.92)(1.542,-0.496){37}{\rule{1.320pt}{0.119pt}}
\multiput(1132.00,250.17)(58.260,-20.000){2}{\rule{0.660pt}{0.400pt}}
\multiput(1193.00,229.92)(1.400,-0.496){41}{\rule{1.209pt}{0.120pt}}
\multiput(1193.00,230.17)(58.490,-22.000){2}{\rule{0.605pt}{0.400pt}}
\multiput(1254.00,207.93)(1.947,-0.488){13}{\rule{1.600pt}{0.117pt}}
\multiput(1254.00,208.17)(26.679,-8.000){2}{\rule{0.800pt}{0.400pt}}
\multiput(1284.00,199.92)(1.638,-0.497){53}{\rule{1.400pt}{0.120pt}}
\multiput(1284.00,200.17)(88.094,-28.000){2}{\rule{0.700pt}{0.400pt}}
\multiput(1375.00,171.92)(2.223,-0.494){25}{\rule{1.843pt}{0.119pt}}
\multiput(1375.00,172.17)(57.175,-14.000){2}{\rule{0.921pt}{0.400pt}}
\end{picture}

%% file: muMZ.tex
% GNUPLOT: LaTeX picture
\setlength{\unitlength}{0.240900pt}
\ifx\plotpoint\undefined\newsavebox{\plotpoint}\fi
\sbox{\plotpoint}{\rule[-0.200pt]{0.400pt}{0.400pt}}%
\begin{picture}(1500,900)(0,0)
\font\gnuplot=cmr10 at 10pt
\gnuplot
\sbox{\plotpoint}{\rule[-0.200pt]{0.400pt}{0.400pt}}%
\put(220.0,113.0){\rule[-0.200pt]{292.934pt}{0.400pt}}
\put(220.0,113.0){\rule[-0.200pt]{0.400pt}{184.048pt}}
\put(220.0,113.0){\rule[-0.200pt]{4.818pt}{0.400pt}}
\put(198,113){\makebox(0,0)[r]{0}}
\put(1416.0,113.0){\rule[-0.200pt]{4.818pt}{0.400pt}}
\put(220.0,266.0){\rule[-0.200pt]{4.818pt}{0.400pt}}
\put(198,266){\makebox(0,0)[r]{1}}
\put(1416.0,266.0){\rule[-0.200pt]{4.818pt}{0.400pt}}
\put(220.0,419.0){\rule[-0.200pt]{4.818pt}{0.400pt}}
\put(198,419){\makebox(0,0)[r]{2}}
\put(1416.0,419.0){\rule[-0.200pt]{4.818pt}{0.400pt}}
\put(220.0,571.0){\rule[-0.200pt]{4.818pt}{0.400pt}}
\put(198,571){\makebox(0,0)[r]{3}}
\put(1416.0,571.0){\rule[-0.200pt]{4.818pt}{0.400pt}}
\put(220.0,724.0){\rule[-0.200pt]{4.818pt}{0.400pt}}
\put(198,724){\makebox(0,0)[r]{4}}
\put(1416.0,724.0){\rule[-0.200pt]{4.818pt}{0.400pt}}
\put(220.0,877.0){\rule[-0.200pt]{4.818pt}{0.400pt}}
\put(198,877){\makebox(0,0)[r]{5}}
\put(1416.0,877.0){\rule[-0.200pt]{4.818pt}{0.400pt}}
\put(220.0,113.0){\rule[-0.200pt]{0.400pt}{4.818pt}}
\put(220,68){\makebox(0,0){0}}
\put(220.0,857.0){\rule[-0.200pt]{0.400pt}{4.818pt}}
\put(342.0,113.0){\rule[-0.200pt]{0.400pt}{4.818pt}}
\put(342,68){\makebox(0,0){50}}
\put(342.0,857.0){\rule[-0.200pt]{0.400pt}{4.818pt}}
\put(463.0,113.0){\rule[-0.200pt]{0.400pt}{4.818pt}}
\put(463,68){\makebox(0,0){100}}
\put(463.0,857.0){\rule[-0.200pt]{0.400pt}{4.818pt}}
\put(585.0,113.0){\rule[-0.200pt]{0.400pt}{4.818pt}}
\put(585,68){\makebox(0,0){150}}
\put(585.0,857.0){\rule[-0.200pt]{0.400pt}{4.818pt}}
\put(706.0,113.0){\rule[-0.200pt]{0.400pt}{4.818pt}}
\put(706,68){\makebox(0,0){200}}
\put(706.0,857.0){\rule[-0.200pt]{0.400pt}{4.818pt}}
\put(828.0,113.0){\rule[-0.200pt]{0.400pt}{4.818pt}}
\put(828,68){\makebox(0,0){250}}
\put(828.0,857.0){\rule[-0.200pt]{0.400pt}{4.818pt}}
\put(950.0,113.0){\rule[-0.200pt]{0.400pt}{4.818pt}}
\put(950,68){\makebox(0,0){300}}
\put(950.0,857.0){\rule[-0.200pt]{0.400pt}{4.818pt}}
\put(1071.0,113.0){\rule[-0.200pt]{0.400pt}{4.818pt}}
\put(1071,68){\makebox(0,0){350}}
\put(1071.0,857.0){\rule[-0.200pt]{0.400pt}{4.818pt}}
\put(1193.0,113.0){\rule[-0.200pt]{0.400pt}{4.818pt}}
\put(1193,68){\makebox(0,0){400}}
\put(1193.0,857.0){\rule[-0.200pt]{0.400pt}{4.818pt}}
\put(1314.0,113.0){\rule[-0.200pt]{0.400pt}{4.818pt}}
\put(1314,68){\makebox(0,0){450}}
\put(1314.0,857.0){\rule[-0.200pt]{0.400pt}{4.818pt}}
\put(1436.0,113.0){\rule[-0.200pt]{0.400pt}{4.818pt}}
\put(1436,68){\makebox(0,0){500}}
\put(1436.0,857.0){\rule[-0.200pt]{0.400pt}{4.818pt}}
\put(220.0,113.0){\rule[-0.200pt]{292.934pt}{0.400pt}}
\put(1436.0,113.0){\rule[-0.200pt]{0.400pt}{184.048pt}}
\put(220.0,877.0){\rule[-0.200pt]{292.934pt}{0.400pt}}
\put(45,495){\makebox(0,0)
{$\displaystyle{\frac{|\mu_{\pm}(M_X)|}{M_{1/2}^{(0)}}}$}}
\put(828,23){\makebox(0,0){\shortstack{\\ \\ \\ $M_{1/2}^{(0)}$}}}
\put(1314,342){\makebox(0,0)[r]{$\tan \beta =20$}}
\put(1314,694){\makebox(0,0)[r]{$\tan \beta =60$}}
\put(220.0,113.0){\rule[-0.200pt]{0.400pt}{184.048pt}}
\put(294,256){\usebox{\plotpoint}}
\multiput(294.58,256.00)(0.492,2.047){21}{\rule{0.119pt}{1.700pt}}
\multiput(293.17,256.00)(12.000,44.472){2}{\rule{0.400pt}{0.850pt}}
\multiput(306.58,304.00)(0.492,1.056){21}{\rule{0.119pt}{0.933pt}}
\multiput(305.17,304.00)(12.000,23.063){2}{\rule{0.400pt}{0.467pt}}
\multiput(318.58,329.00)(0.493,0.616){23}{\rule{0.119pt}{0.592pt}}
\multiput(317.17,329.00)(13.000,14.771){2}{\rule{0.400pt}{0.296pt}}
\multiput(331.00,345.58)(0.543,0.492){19}{\rule{0.536pt}{0.118pt}}
\multiput(331.00,344.17)(10.887,11.000){2}{\rule{0.268pt}{0.400pt}}
\multiput(343.00,356.59)(0.758,0.488){13}{\rule{0.700pt}{0.117pt}}
\multiput(343.00,355.17)(10.547,8.000){2}{\rule{0.350pt}{0.400pt}}
\multiput(355.00,364.59)(1.267,0.477){7}{\rule{1.060pt}{0.115pt}}
\multiput(355.00,363.17)(9.800,5.000){2}{\rule{0.530pt}{0.400pt}}
\multiput(367.00,369.59)(1.378,0.477){7}{\rule{1.140pt}{0.115pt}}
\multiput(367.00,368.17)(10.634,5.000){2}{\rule{0.570pt}{0.400pt}}
\multiput(380.00,374.61)(2.472,0.447){3}{\rule{1.700pt}{0.108pt}}
\multiput(380.00,373.17)(8.472,3.000){2}{\rule{0.850pt}{0.400pt}}
\multiput(392.00,377.61)(2.472,0.447){3}{\rule{1.700pt}{0.108pt}}
\multiput(392.00,376.17)(8.472,3.000){2}{\rule{0.850pt}{0.400pt}}
\put(404,380.17){\rule{2.700pt}{0.400pt}}
\multiput(404.00,379.17)(7.396,2.000){2}{\rule{1.350pt}{0.400pt}}
\put(417,382.17){\rule{2.500pt}{0.400pt}}
\multiput(417.00,381.17)(6.811,2.000){2}{\rule{1.250pt}{0.400pt}}
\put(429,383.67){\rule{2.891pt}{0.400pt}}
\multiput(429.00,383.17)(6.000,1.000){2}{\rule{1.445pt}{0.400pt}}
\put(441,384.67){\rule{2.891pt}{0.400pt}}
\multiput(441.00,384.17)(6.000,1.000){2}{\rule{1.445pt}{0.400pt}}
\put(453,386.17){\rule{2.700pt}{0.400pt}}
\multiput(453.00,385.17)(7.396,2.000){2}{\rule{1.350pt}{0.400pt}}
\put(478,387.67){\rule{2.891pt}{0.400pt}}
\multiput(478.00,387.17)(6.000,1.000){2}{\rule{1.445pt}{0.400pt}}
\put(490,388.67){\rule{3.132pt}{0.400pt}}
\multiput(490.00,388.17)(6.500,1.000){2}{\rule{1.566pt}{0.400pt}}
\put(503,389.67){\rule{2.891pt}{0.400pt}}
\multiput(503.00,389.17)(6.000,1.000){2}{\rule{1.445pt}{0.400pt}}
\put(466.0,388.0){\rule[-0.200pt]{2.891pt}{0.400pt}}
\put(527,390.67){\rule{2.891pt}{0.400pt}}
\multiput(527.00,390.17)(6.000,1.000){2}{\rule{1.445pt}{0.400pt}}
\put(515.0,391.0){\rule[-0.200pt]{2.891pt}{0.400pt}}
\put(564,391.67){\rule{2.891pt}{0.400pt}}
\multiput(564.00,391.17)(6.000,1.000){2}{\rule{1.445pt}{0.400pt}}
\put(539.0,392.0){\rule[-0.200pt]{6.022pt}{0.400pt}}
\put(613,392.67){\rule{2.891pt}{0.400pt}}
\multiput(613.00,392.17)(6.000,1.000){2}{\rule{1.445pt}{0.400pt}}
\put(576.0,393.0){\rule[-0.200pt]{8.913pt}{0.400pt}}
\put(674,393.67){\rule{3.132pt}{0.400pt}}
\multiput(674.00,393.17)(6.500,1.000){2}{\rule{1.566pt}{0.400pt}}
\put(625.0,394.0){\rule[-0.200pt]{11.804pt}{0.400pt}}
\put(785,394.67){\rule{2.891pt}{0.400pt}}
\multiput(785.00,394.17)(6.000,1.000){2}{\rule{1.445pt}{0.400pt}}
\put(687.0,395.0){\rule[-0.200pt]{23.608pt}{0.400pt}}
\put(1092,395.67){\rule{2.891pt}{0.400pt}}
\multiput(1092.00,395.17)(6.000,1.000){2}{\rule{1.445pt}{0.400pt}}
\put(797.0,396.0){\rule[-0.200pt]{71.065pt}{0.400pt}}
\put(1104.0,397.0){\rule[-0.200pt]{79.979pt}{0.400pt}}
\put(294,257){\usebox{\plotpoint}}
\multiput(294,257)(4.503,20.261){3}{\usebox{\plotpoint}}
\multiput(306,311)(8.176,19.077){2}{\usebox{\plotpoint}}
\put(328.54,353.59){\usebox{\plotpoint}}
\put(342.70,368.70){\usebox{\plotpoint}}
\multiput(343,369)(17.270,11.513){0}{\usebox{\plotpoint}}
\put(360.29,379.64){\usebox{\plotpoint}}
\put(379.37,387.76){\usebox{\plotpoint}}
\multiput(380,388)(19.690,6.563){0}{\usebox{\plotpoint}}
\put(399.21,393.80){\usebox{\plotpoint}}
\multiput(404,395)(20.514,3.156){0}{\usebox{\plotpoint}}
\put(419.63,397.44){\usebox{\plotpoint}}
\put(440.10,400.85){\usebox{\plotpoint}}
\multiput(441,401)(20.684,1.724){0}{\usebox{\plotpoint}}
\put(460.78,402.60){\usebox{\plotpoint}}
\multiput(466,403)(20.684,1.724){0}{\usebox{\plotpoint}}
\put(481.46,404.29){\usebox{\plotpoint}}
\put(502.15,405.93){\usebox{\plotpoint}}
\multiput(503,406)(20.756,0.000){0}{\usebox{\plotpoint}}
\put(522.88,406.66){\usebox{\plotpoint}}
\multiput(527,407)(20.756,0.000){0}{\usebox{\plotpoint}}
\put(543.61,407.35){\usebox{\plotpoint}}
\multiput(552,408)(20.756,0.000){0}{\usebox{\plotpoint}}
\put(564.34,408.03){\usebox{\plotpoint}}
\put(585.05,409.00){\usebox{\plotpoint}}
\multiput(588,409)(20.756,0.000){0}{\usebox{\plotpoint}}
\put(605.79,409.40){\usebox{\plotpoint}}
\multiput(613,410)(20.756,0.000){0}{\usebox{\plotpoint}}
\put(626.52,410.00){\usebox{\plotpoint}}
\put(647.28,410.00){\usebox{\plotpoint}}
\multiput(650,410)(20.756,0.000){0}{\usebox{\plotpoint}}
\put(668.01,410.50){\usebox{\plotpoint}}
\multiput(674,411)(20.756,0.000){0}{\usebox{\plotpoint}}
\put(688.75,411.00){\usebox{\plotpoint}}
\put(709.50,411.00){\usebox{\plotpoint}}
\multiput(711,411)(20.756,0.000){0}{\usebox{\plotpoint}}
\put(730.26,411.00){\usebox{\plotpoint}}
\multiput(736,411)(20.756,0.000){0}{\usebox{\plotpoint}}
\put(751.00,411.25){\usebox{\plotpoint}}
\put(771.73,412.00){\usebox{\plotpoint}}
\multiput(773,412)(20.756,0.000){0}{\usebox{\plotpoint}}
\put(792.48,412.00){\usebox{\plotpoint}}
\multiput(797,412)(20.756,0.000){0}{\usebox{\plotpoint}}
\put(813.24,412.00){\usebox{\plotpoint}}
\put(833.99,412.00){\usebox{\plotpoint}}
\multiput(834,412)(20.756,0.000){0}{\usebox{\plotpoint}}
\put(854.75,412.00){\usebox{\plotpoint}}
\multiput(859,412)(20.756,0.000){0}{\usebox{\plotpoint}}
\put(875.50,412.00){\usebox{\plotpoint}}
\multiput(883,412)(20.756,0.000){0}{\usebox{\plotpoint}}
\put(896.26,412.00){\usebox{\plotpoint}}
\put(917.02,412.00){\usebox{\plotpoint}}
\multiput(920,412)(20.756,0.000){0}{\usebox{\plotpoint}}
\put(937.77,412.00){\usebox{\plotpoint}}
\multiput(945,412)(20.684,1.724){0}{\usebox{\plotpoint}}
\put(958.48,413.00){\usebox{\plotpoint}}
\put(979.24,413.00){\usebox{\plotpoint}}
\multiput(982,413)(20.756,0.000){0}{\usebox{\plotpoint}}
\put(1000.00,413.00){\usebox{\plotpoint}}
\multiput(1006,413)(20.756,0.000){0}{\usebox{\plotpoint}}
\put(1020.75,413.00){\usebox{\plotpoint}}
\put(1041.51,413.00){\usebox{\plotpoint}}
\multiput(1043,413)(20.756,0.000){0}{\usebox{\plotpoint}}
\put(1062.26,413.00){\usebox{\plotpoint}}
\multiput(1068,413)(20.756,0.000){0}{\usebox{\plotpoint}}
\put(1083.02,413.00){\usebox{\plotpoint}}
\put(1103.77,413.00){\usebox{\plotpoint}}
\multiput(1104,413)(20.756,0.000){0}{\usebox{\plotpoint}}
\put(1124.53,413.00){\usebox{\plotpoint}}
\multiput(1129,413)(20.756,0.000){0}{\usebox{\plotpoint}}
\put(1145.28,413.00){\usebox{\plotpoint}}
\multiput(1153,413)(20.756,0.000){0}{\usebox{\plotpoint}}
\put(1166.04,413.00){\usebox{\plotpoint}}
\put(1186.80,413.00){\usebox{\plotpoint}}
\multiput(1190,413)(20.756,0.000){0}{\usebox{\plotpoint}}
\put(1207.55,413.00){\usebox{\plotpoint}}
\multiput(1215,413)(20.756,0.000){0}{\usebox{\plotpoint}}
\put(1228.31,413.00){\usebox{\plotpoint}}
\put(1249.06,413.00){\usebox{\plotpoint}}
\multiput(1252,413)(20.756,0.000){0}{\usebox{\plotpoint}}
\put(1269.82,413.00){\usebox{\plotpoint}}
\multiput(1276,413)(20.756,0.000){0}{\usebox{\plotpoint}}
\put(1290.57,413.00){\usebox{\plotpoint}}
\put(1311.33,413.00){\usebox{\plotpoint}}
\multiput(1313,413)(20.756,0.000){0}{\usebox{\plotpoint}}
\put(1332.08,413.00){\usebox{\plotpoint}}
\multiput(1338,413)(20.756,0.000){0}{\usebox{\plotpoint}}
\put(1352.84,413.00){\usebox{\plotpoint}}
\put(1373.59,413.00){\usebox{\plotpoint}}
\multiput(1375,413)(20.756,0.000){0}{\usebox{\plotpoint}}
\put(1394.35,413.00){\usebox{\plotpoint}}
\multiput(1399,413)(20.756,0.000){0}{\usebox{\plotpoint}}
\put(1415.11,413.00){\usebox{\plotpoint}}
\put(1435.86,413.00){\usebox{\plotpoint}}
\put(1436,413){\usebox{\plotpoint}}
\sbox{\plotpoint}{\rule[-0.400pt]{0.800pt}{0.800pt}}%
\put(294,258){\usebox{\plotpoint}}
\multiput(295.41,258.00)(0.511,2.932){17}{\rule{0.123pt}{4.600pt}}
\multiput(292.34,258.00)(12.000,56.452){2}{\rule{0.800pt}{2.300pt}}
\multiput(307.41,324.00)(0.511,1.440){17}{\rule{0.123pt}{2.400pt}}
\multiput(304.34,324.00)(12.000,28.019){2}{\rule{0.800pt}{1.200pt}}
\multiput(319.41,357.00)(0.509,0.781){19}{\rule{0.123pt}{1.431pt}}
\multiput(316.34,357.00)(13.000,17.030){2}{\rule{0.800pt}{0.715pt}}
\multiput(332.41,377.00)(0.511,0.581){17}{\rule{0.123pt}{1.133pt}}
\multiput(329.34,377.00)(12.000,11.648){2}{\rule{0.800pt}{0.567pt}}
\multiput(343.00,392.40)(0.599,0.514){13}{\rule{1.160pt}{0.124pt}}
\multiput(343.00,389.34)(9.592,10.000){2}{\rule{0.580pt}{0.800pt}}
\multiput(355.00,402.40)(0.913,0.526){7}{\rule{1.571pt}{0.127pt}}
\multiput(355.00,399.34)(8.738,7.000){2}{\rule{0.786pt}{0.800pt}}
\multiput(367.00,409.38)(1.768,0.560){3}{\rule{2.280pt}{0.135pt}}
\multiput(367.00,406.34)(8.268,5.000){2}{\rule{1.140pt}{0.800pt}}
\multiput(380.00,414.38)(1.600,0.560){3}{\rule{2.120pt}{0.135pt}}
\multiput(380.00,411.34)(7.600,5.000){2}{\rule{1.060pt}{0.800pt}}
\put(392,417.84){\rule{2.891pt}{0.800pt}}
\multiput(392.00,416.34)(6.000,3.000){2}{\rule{1.445pt}{0.800pt}}
\put(404,420.84){\rule{3.132pt}{0.800pt}}
\multiput(404.00,419.34)(6.500,3.000){2}{\rule{1.566pt}{0.800pt}}
\put(417,423.34){\rule{2.891pt}{0.800pt}}
\multiput(417.00,422.34)(6.000,2.000){2}{\rule{1.445pt}{0.800pt}}
\put(429,425.34){\rule{2.891pt}{0.800pt}}
\multiput(429.00,424.34)(6.000,2.000){2}{\rule{1.445pt}{0.800pt}}
\put(441,427.34){\rule{2.891pt}{0.800pt}}
\multiput(441.00,426.34)(6.000,2.000){2}{\rule{1.445pt}{0.800pt}}
\put(453,428.84){\rule{3.132pt}{0.800pt}}
\multiput(453.00,428.34)(6.500,1.000){2}{\rule{1.566pt}{0.800pt}}
\put(466,429.84){\rule{2.891pt}{0.800pt}}
\multiput(466.00,429.34)(6.000,1.000){2}{\rule{1.445pt}{0.800pt}}
\put(478,430.84){\rule{2.891pt}{0.800pt}}
\multiput(478.00,430.34)(6.000,1.000){2}{\rule{1.445pt}{0.800pt}}
\put(490,431.84){\rule{3.132pt}{0.800pt}}
\multiput(490.00,431.34)(6.500,1.000){2}{\rule{1.566pt}{0.800pt}}
\put(503,432.84){\rule{2.891pt}{0.800pt}}
\multiput(503.00,432.34)(6.000,1.000){2}{\rule{1.445pt}{0.800pt}}
\put(527,433.84){\rule{2.891pt}{0.800pt}}
\multiput(527.00,433.34)(6.000,1.000){2}{\rule{1.445pt}{0.800pt}}
\put(539,434.84){\rule{3.132pt}{0.800pt}}
\multiput(539.00,434.34)(6.500,1.000){2}{\rule{1.566pt}{0.800pt}}
\put(515.0,435.0){\rule[-0.400pt]{2.891pt}{0.800pt}}
\put(576,435.84){\rule{2.891pt}{0.800pt}}
\multiput(576.00,435.34)(6.000,1.000){2}{\rule{1.445pt}{0.800pt}}
\put(552.0,437.0){\rule[-0.400pt]{5.782pt}{0.800pt}}
\put(613,436.84){\rule{2.891pt}{0.800pt}}
\multiput(613.00,436.34)(6.000,1.000){2}{\rule{1.445pt}{0.800pt}}
\put(588.0,438.0){\rule[-0.400pt]{6.022pt}{0.800pt}}
\put(662,437.84){\rule{2.891pt}{0.800pt}}
\multiput(662.00,437.34)(6.000,1.000){2}{\rule{1.445pt}{0.800pt}}
\put(625.0,439.0){\rule[-0.400pt]{8.913pt}{0.800pt}}
\put(736,438.84){\rule{2.891pt}{0.800pt}}
\multiput(736.00,438.34)(6.000,1.000){2}{\rule{1.445pt}{0.800pt}}
\put(674.0,440.0){\rule[-0.400pt]{14.936pt}{0.800pt}}
\put(871,439.84){\rule{2.891pt}{0.800pt}}
\multiput(871.00,439.34)(6.000,1.000){2}{\rule{1.445pt}{0.800pt}}
\put(748.0,441.0){\rule[-0.400pt]{29.631pt}{0.800pt}}
\put(1239,440.84){\rule{3.132pt}{0.800pt}}
\multiput(1239.00,440.34)(6.500,1.000){2}{\rule{1.566pt}{0.800pt}}
\put(883.0,442.0){\rule[-0.400pt]{85.760pt}{0.800pt}}
\put(1252.0,443.0){\rule[-0.400pt]{44.326pt}{0.800pt}}
\sbox{\plotpoint}{\rule[-0.500pt]{1.000pt}{1.000pt}}%
\put(294,270){\usebox{\plotpoint}}
\multiput(294,270)(2.804,20.565){5}{\usebox{\plotpoint}}
\multiput(306,358)(5.579,19.992){2}{\usebox{\plotpoint}}
\put(323.13,411.66){\usebox{\plotpoint}}
\put(332.51,430.14){\usebox{\plotpoint}}
\put(344.82,446.82){\usebox{\plotpoint}}
\put(359.88,461.07){\usebox{\plotpoint}}
\put(377.11,472.45){\usebox{\plotpoint}}
\multiput(380,474)(19.159,7.983){0}{\usebox{\plotpoint}}
\put(396.25,480.42){\usebox{\plotpoint}}
\put(416.03,486.70){\usebox{\plotpoint}}
\multiput(417,487)(20.136,5.034){0}{\usebox{\plotpoint}}
\put(436.27,491.21){\usebox{\plotpoint}}
\multiput(441,492)(20.473,3.412){0}{\usebox{\plotpoint}}
\put(456.75,494.58){\usebox{\plotpoint}}
\put(477.36,496.95){\usebox{\plotpoint}}
\multiput(478,497)(20.684,1.724){0}{\usebox{\plotpoint}}
\put(497.97,499.23){\usebox{\plotpoint}}
\multiput(503,500)(20.684,1.724){0}{\usebox{\plotpoint}}
\put(518.63,501.00){\usebox{\plotpoint}}
\multiput(527,501)(20.684,1.724){0}{\usebox{\plotpoint}}
\put(539.34,502.03){\usebox{\plotpoint}}
\put(560.06,503.00){\usebox{\plotpoint}}
\multiput(564,503)(20.684,1.724){0}{\usebox{\plotpoint}}
\put(580.77,504.00){\usebox{\plotpoint}}
\multiput(588,504)(20.694,1.592){0}{\usebox{\plotpoint}}
\put(601.49,505.00){\usebox{\plotpoint}}
\put(622.21,505.77){\usebox{\plotpoint}}
\multiput(625,506)(20.756,0.000){0}{\usebox{\plotpoint}}
\put(642.96,506.00){\usebox{\plotpoint}}
\multiput(650,506)(20.756,0.000){0}{\usebox{\plotpoint}}
\put(663.71,506.14){\usebox{\plotpoint}}
\put(684.43,507.00){\usebox{\plotpoint}}
\multiput(687,507)(20.756,0.000){0}{\usebox{\plotpoint}}
\put(705.19,507.00){\usebox{\plotpoint}}
\multiput(711,507)(20.694,1.592){0}{\usebox{\plotpoint}}
\put(725.90,508.00){\usebox{\plotpoint}}
\put(746.66,508.00){\usebox{\plotpoint}}
\multiput(748,508)(20.756,0.000){0}{\usebox{\plotpoint}}
\put(767.41,508.00){\usebox{\plotpoint}}
\multiput(773,508)(20.756,0.000){0}{\usebox{\plotpoint}}
\put(788.17,508.00){\usebox{\plotpoint}}
\put(808.89,508.91){\usebox{\plotpoint}}
\multiput(810,509)(20.756,0.000){0}{\usebox{\plotpoint}}
\put(829.64,509.00){\usebox{\plotpoint}}
\multiput(834,509)(20.756,0.000){0}{\usebox{\plotpoint}}
\put(850.40,509.00){\usebox{\plotpoint}}
\multiput(859,509)(20.756,0.000){0}{\usebox{\plotpoint}}
\put(871.15,509.00){\usebox{\plotpoint}}
\put(891.91,509.00){\usebox{\plotpoint}}
\multiput(896,509)(20.756,0.000){0}{\usebox{\plotpoint}}
\put(912.66,509.00){\usebox{\plotpoint}}
\multiput(920,509)(20.756,0.000){0}{\usebox{\plotpoint}}
\put(933.42,509.00){\usebox{\plotpoint}}
\put(954.17,509.00){\usebox{\plotpoint}}
\multiput(957,509)(20.684,1.724){0}{\usebox{\plotpoint}}
\put(974.89,510.00){\usebox{\plotpoint}}
\multiput(982,510)(20.756,0.000){0}{\usebox{\plotpoint}}
\put(995.64,510.00){\usebox{\plotpoint}}
\put(1016.40,510.00){\usebox{\plotpoint}}
\multiput(1018,510)(20.756,0.000){0}{\usebox{\plotpoint}}
\put(1037.16,510.00){\usebox{\plotpoint}}
\multiput(1043,510)(20.756,0.000){0}{\usebox{\plotpoint}}
\put(1057.91,510.00){\usebox{\plotpoint}}
\put(1078.67,510.00){\usebox{\plotpoint}}
\multiput(1080,510)(20.756,0.000){0}{\usebox{\plotpoint}}
\put(1099.42,510.00){\usebox{\plotpoint}}
\multiput(1104,510)(20.756,0.000){0}{\usebox{\plotpoint}}
\put(1120.18,510.00){\usebox{\plotpoint}}
\put(1140.93,510.00){\usebox{\plotpoint}}
\multiput(1141,510)(20.756,0.000){0}{\usebox{\plotpoint}}
\put(1161.69,510.00){\usebox{\plotpoint}}
\multiput(1166,510)(20.756,0.000){0}{\usebox{\plotpoint}}
\put(1182.44,510.00){\usebox{\plotpoint}}
\multiput(1190,510)(20.756,0.000){0}{\usebox{\plotpoint}}
\put(1203.20,510.00){\usebox{\plotpoint}}
\put(1223.95,510.00){\usebox{\plotpoint}}
\multiput(1227,510)(20.756,0.000){0}{\usebox{\plotpoint}}
\put(1244.71,510.00){\usebox{\plotpoint}}
\multiput(1252,510)(20.756,0.000){0}{\usebox{\plotpoint}}
\put(1265.47,510.00){\usebox{\plotpoint}}
\put(1286.22,510.00){\usebox{\plotpoint}}
\multiput(1289,510)(20.756,0.000){0}{\usebox{\plotpoint}}
\put(1306.98,510.00){\usebox{\plotpoint}}
\multiput(1313,510)(20.756,0.000){0}{\usebox{\plotpoint}}
\put(1327.73,510.00){\usebox{\plotpoint}}
\put(1348.49,510.00){\usebox{\plotpoint}}
\multiput(1350,510)(20.756,0.000){0}{\usebox{\plotpoint}}
\put(1369.24,510.00){\usebox{\plotpoint}}
\multiput(1375,510)(20.756,0.000){0}{\usebox{\plotpoint}}
\put(1389.99,510.25){\usebox{\plotpoint}}
\put(1410.71,511.00){\usebox{\plotpoint}}
\multiput(1411,511)(20.756,0.000){0}{\usebox{\plotpoint}}
\put(1431.47,511.00){\usebox{\plotpoint}}
\put(1436,511){\usebox{\plotpoint}}
\sbox{\plotpoint}{\rule[-0.600pt]{1.200pt}{1.200pt}}%
\put(294,322){\usebox{\plotpoint}}
\multiput(296.24,322.00)(0.501,7.463){14}{\rule{0.121pt}{17.100pt}}
\multiput(291.51,322.00)(12.000,132.508){2}{\rule{1.200pt}{8.550pt}}
\multiput(308.24,490.00)(0.501,3.369){14}{\rule{0.121pt}{8.000pt}}
\multiput(303.51,490.00)(12.000,60.396){2}{\rule{1.200pt}{4.000pt}}
\multiput(320.24,567.00)(0.501,1.809){16}{\rule{0.121pt}{4.546pt}}
\multiput(315.51,567.00)(13.000,36.564){2}{\rule{1.200pt}{2.273pt}}
\multiput(333.24,613.00)(0.501,1.254){14}{\rule{0.121pt}{3.300pt}}
\multiput(328.51,613.00)(12.000,23.151){2}{\rule{1.200pt}{1.650pt}}
\multiput(345.24,643.00)(0.501,0.894){14}{\rule{0.121pt}{2.500pt}}
\multiput(340.51,643.00)(12.000,16.811){2}{\rule{1.200pt}{1.250pt}}
\multiput(357.24,665.00)(0.501,0.624){14}{\rule{0.121pt}{1.900pt}}
\multiput(352.51,665.00)(12.000,12.056){2}{\rule{1.200pt}{0.950pt}}
\multiput(367.00,683.24)(0.489,0.501){14}{\rule{1.600pt}{0.121pt}}
\multiput(367.00,678.51)(9.679,12.000){2}{\rule{0.800pt}{1.200pt}}
\multiput(380.00,695.24)(0.588,0.502){8}{\rule{1.900pt}{0.121pt}}
\multiput(380.00,690.51)(8.056,9.000){2}{\rule{0.950pt}{1.200pt}}
\multiput(392.00,704.24)(0.738,0.505){4}{\rule{2.357pt}{0.122pt}}
\multiput(392.00,699.51)(7.108,7.000){2}{\rule{1.179pt}{1.200pt}}
\multiput(404.00,711.24)(0.962,0.509){2}{\rule{2.900pt}{0.123pt}}
\multiput(404.00,706.51)(6.981,6.000){2}{\rule{1.450pt}{1.200pt}}
\put(417,715.01){\rule{2.891pt}{1.200pt}}
\multiput(417.00,712.51)(6.000,5.000){2}{\rule{1.445pt}{1.200pt}}
\put(429,719.51){\rule{2.891pt}{1.200pt}}
\multiput(429.00,717.51)(6.000,4.000){2}{\rule{1.445pt}{1.200pt}}
\put(441,723.51){\rule{2.891pt}{1.200pt}}
\multiput(441.00,721.51)(6.000,4.000){2}{\rule{1.445pt}{1.200pt}}
\put(453,727.01){\rule{3.132pt}{1.200pt}}
\multiput(453.00,725.51)(6.500,3.000){2}{\rule{1.566pt}{1.200pt}}
\put(466,729.51){\rule{2.891pt}{1.200pt}}
\multiput(466.00,728.51)(6.000,2.000){2}{\rule{1.445pt}{1.200pt}}
\put(478,731.51){\rule{2.891pt}{1.200pt}}
\multiput(478.00,730.51)(6.000,2.000){2}{\rule{1.445pt}{1.200pt}}
\put(490,733.51){\rule{3.132pt}{1.200pt}}
\multiput(490.00,732.51)(6.500,2.000){2}{\rule{1.566pt}{1.200pt}}
\put(503,735.51){\rule{2.891pt}{1.200pt}}
\multiput(503.00,734.51)(6.000,2.000){2}{\rule{1.445pt}{1.200pt}}
\put(515,737.01){\rule{2.891pt}{1.200pt}}
\multiput(515.00,736.51)(6.000,1.000){2}{\rule{1.445pt}{1.200pt}}
\put(527,738.51){\rule{2.891pt}{1.200pt}}
\multiput(527.00,737.51)(6.000,2.000){2}{\rule{1.445pt}{1.200pt}}
\put(539,740.01){\rule{3.132pt}{1.200pt}}
\multiput(539.00,739.51)(6.500,1.000){2}{\rule{1.566pt}{1.200pt}}
\put(552,741.01){\rule{2.891pt}{1.200pt}}
\multiput(552.00,740.51)(6.000,1.000){2}{\rule{1.445pt}{1.200pt}}
\put(564,742.01){\rule{2.891pt}{1.200pt}}
\multiput(564.00,741.51)(6.000,1.000){2}{\rule{1.445pt}{1.200pt}}
\put(576,743.01){\rule{2.891pt}{1.200pt}}
\multiput(576.00,742.51)(6.000,1.000){2}{\rule{1.445pt}{1.200pt}}
\put(601,744.01){\rule{2.891pt}{1.200pt}}
\multiput(601.00,743.51)(6.000,1.000){2}{\rule{1.445pt}{1.200pt}}
\put(613,745.01){\rule{2.891pt}{1.200pt}}
\multiput(613.00,744.51)(6.000,1.000){2}{\rule{1.445pt}{1.200pt}}
\put(588.0,746.0){\rule[-0.600pt]{3.132pt}{1.200pt}}
\put(638,746.01){\rule{2.891pt}{1.200pt}}
\multiput(638.00,745.51)(6.000,1.000){2}{\rule{1.445pt}{1.200pt}}
\put(625.0,748.0){\rule[-0.600pt]{3.132pt}{1.200pt}}
\put(662,747.01){\rule{2.891pt}{1.200pt}}
\multiput(662.00,746.51)(6.000,1.000){2}{\rule{1.445pt}{1.200pt}}
\put(650.0,749.0){\rule[-0.600pt]{2.891pt}{1.200pt}}
\put(699,748.01){\rule{2.891pt}{1.200pt}}
\multiput(699.00,747.51)(6.000,1.000){2}{\rule{1.445pt}{1.200pt}}
\put(674.0,750.0){\rule[-0.600pt]{6.022pt}{1.200pt}}
\put(736,749.01){\rule{2.891pt}{1.200pt}}
\multiput(736.00,748.51)(6.000,1.000){2}{\rule{1.445pt}{1.200pt}}
\put(711.0,751.0){\rule[-0.600pt]{6.022pt}{1.200pt}}
\put(785,750.01){\rule{2.891pt}{1.200pt}}
\multiput(785.00,749.51)(6.000,1.000){2}{\rule{1.445pt}{1.200pt}}
\put(748.0,752.0){\rule[-0.600pt]{8.913pt}{1.200pt}}
\put(859,751.01){\rule{2.891pt}{1.200pt}}
\multiput(859.00,750.51)(6.000,1.000){2}{\rule{1.445pt}{1.200pt}}
\put(797.0,753.0){\rule[-0.600pt]{14.936pt}{1.200pt}}
\put(969,752.01){\rule{3.132pt}{1.200pt}}
\multiput(969.00,751.51)(6.500,1.000){2}{\rule{1.566pt}{1.200pt}}
\put(871.0,754.0){\rule[-0.600pt]{23.608pt}{1.200pt}}
\put(1166,753.01){\rule{2.891pt}{1.200pt}}
\multiput(1166.00,752.51)(6.000,1.000){2}{\rule{1.445pt}{1.200pt}}
\put(982.0,755.0){\rule[-0.600pt]{44.326pt}{1.200pt}}
\put(1178.0,756.0){\rule[-0.600pt]{62.152pt}{1.200pt}}
\end{picture}

%% file: tanbsin2.tex
% GNUPLOT: LaTeX picture
\setlength{\unitlength}{0.240900pt}
\ifx\plotpoint\undefined\newsavebox{\plotpoint}\fi
\sbox{\plotpoint}{\rule[-0.200pt]{0.400pt}{0.400pt}}%
\begin{picture}(1500,900)(0,0)
\font\gnuplot=cmr10 at 10pt
\gnuplot
\sbox{\plotpoint}{\rule[-0.200pt]{0.400pt}{0.400pt}}%
\put(220.0,453.0){\rule[-0.200pt]{292.934pt}{0.400pt}}
\put(220.0,113.0){\rule[-0.200pt]{4.818pt}{0.400pt}}
\put(198,113){\makebox(0,0)[r]{-0.8}}
\put(1416.0,113.0){\rule[-0.200pt]{4.818pt}{0.400pt}}
\put(220.0,198.0){\rule[-0.200pt]{4.818pt}{0.400pt}}
\put(198,198){\makebox(0,0)[r]{-0.6}}
\put(1416.0,198.0){\rule[-0.200pt]{4.818pt}{0.400pt}}
\put(220.0,283.0){\rule[-0.200pt]{4.818pt}{0.400pt}}
\put(198,283){\makebox(0,0)[r]{-0.4}}
\put(1416.0,283.0){\rule[-0.200pt]{4.818pt}{0.400pt}}
\put(220.0,368.0){\rule[-0.200pt]{4.818pt}{0.400pt}}
\put(198,368){\makebox(0,0)[r]{-0.2}}
\put(1416.0,368.0){\rule[-0.200pt]{4.818pt}{0.400pt}}
\put(220.0,453.0){\rule[-0.200pt]{4.818pt}{0.400pt}}
\put(198,453){\makebox(0,0)[r]{0}}
\put(1416.0,453.0){\rule[-0.200pt]{4.818pt}{0.400pt}}
\put(220.0,537.0){\rule[-0.200pt]{4.818pt}{0.400pt}}
\put(198,537){\makebox(0,0)[r]{0.2}}
\put(1416.0,537.0){\rule[-0.200pt]{4.818pt}{0.400pt}}
\put(220.0,622.0){\rule[-0.200pt]{4.818pt}{0.400pt}}
\put(198,622){\makebox(0,0)[r]{0.4}}
\put(1416.0,622.0){\rule[-0.200pt]{4.818pt}{0.400pt}}
\put(220.0,707.0){\rule[-0.200pt]{4.818pt}{0.400pt}}
\put(198,707){\makebox(0,0)[r]{0.6}}
\put(1416.0,707.0){\rule[-0.200pt]{4.818pt}{0.400pt}}
\put(220.0,792.0){\rule[-0.200pt]{4.818pt}{0.400pt}}
\put(198,792){\makebox(0,0)[r]{0.8}}
\put(1416.0,792.0){\rule[-0.200pt]{4.818pt}{0.400pt}}
\put(220.0,877.0){\rule[-0.200pt]{4.818pt}{0.400pt}}
\put(198,877){\makebox(0,0)[r]{1}}
\put(1416.0,877.0){\rule[-0.200pt]{4.818pt}{0.400pt}}
\put(220.0,113.0){\rule[-0.200pt]{0.400pt}{4.818pt}}
\put(220,68){\makebox(0,0){20}}
\put(220.0,857.0){\rule[-0.200pt]{0.400pt}{4.818pt}}
\put(372.0,113.0){\rule[-0.200pt]{0.400pt}{4.818pt}}
\put(372,68){\makebox(0,0){25}}
\put(372.0,857.0){\rule[-0.200pt]{0.400pt}{4.818pt}}
\put(524.0,113.0){\rule[-0.200pt]{0.400pt}{4.818pt}}
\put(524,68){\makebox(0,0){30}}
\put(524.0,857.0){\rule[-0.200pt]{0.400pt}{4.818pt}}
\put(676.0,113.0){\rule[-0.200pt]{0.400pt}{4.818pt}}
\put(676,68){\makebox(0,0){35}}
\put(676.0,857.0){\rule[-0.200pt]{0.400pt}{4.818pt}}
\put(828.0,113.0){\rule[-0.200pt]{0.400pt}{4.818pt}}
\put(828,68){\makebox(0,0){40}}
\put(828.0,857.0){\rule[-0.200pt]{0.400pt}{4.818pt}}
\put(980.0,113.0){\rule[-0.200pt]{0.400pt}{4.818pt}}
\put(980,68){\makebox(0,0){45}}
\put(980.0,857.0){\rule[-0.200pt]{0.400pt}{4.818pt}}
\put(1132.0,113.0){\rule[-0.200pt]{0.400pt}{4.818pt}}
\put(1132,68){\makebox(0,0){50}}
\put(1132.0,857.0){\rule[-0.200pt]{0.400pt}{4.818pt}}
\put(1284.0,113.0){\rule[-0.200pt]{0.400pt}{4.818pt}}
\put(1284,68){\makebox(0,0){55}}
\put(1284.0,857.0){\rule[-0.200pt]{0.400pt}{4.818pt}}
\put(1436.0,113.0){\rule[-0.200pt]{0.400pt}{4.818pt}}
\put(1436,68){\makebox(0,0){60}}
\put(1436.0,857.0){\rule[-0.200pt]{0.400pt}{4.818pt}}
\put(220.0,113.0){\rule[-0.200pt]{292.934pt}{0.400pt}}
\put(1436.0,113.0){\rule[-0.200pt]{0.400pt}{184.048pt}}
\put(220.0,877.0){\rule[-0.200pt]{292.934pt}{0.400pt}}
\put(45,495){\makebox(0,0){$\sin \theta$}}
\put(828,23){\makebox(0,0){$\tan \beta$}}
\put(220.0,113.0){\rule[-0.200pt]{0.400pt}{184.048pt}}
\put(220,190){\circle*{18}}
\put(250,214){\circle*{18}}
\put(281,240){\circle*{18}}
\put(311,268){\circle*{18}}
\put(342,297){\circle*{18}}
\put(372,327){\circle*{18}}
\put(402,374){\circle*{18}}
\put(433,423){\circle*{18}}
\put(463,472){\circle*{18}}
\put(494,521){\circle*{18}}
\put(524,568){\circle*{18}}
\put(554,598){\circle*{18}}
\put(585,626){\circle*{18}}
\put(615,653){\circle*{18}}
\put(646,679){\circle*{18}}
\put(676,703){\circle*{18}}
\put(706,734){\circle*{18}}
\put(737,761){\circle*{18}}
\put(767,785){\circle*{18}}
\put(798,806){\circle*{18}}
\put(828,824){\circle*{18}}
\put(858,838){\circle*{18}}
\put(889,850){\circle*{18}}
\put(919,859){\circle*{18}}
\put(950,866){\circle*{18}}
\put(980,871){\circle*{18}}
\put(1132,875){\circle*{18}}
\put(1193,870){\circle*{18}}
\put(1254,863){\circle*{18}}
\put(1284,860){\circle*{18}}
\put(1375,848){\circle*{18}}
\put(1436,842){\circle*{18}}
\end{picture}

%% file: Sub1.tex
% GNUPLOT: LaTeX picture
\setlength{\unitlength}{0.240900pt}
\ifx\plotpoint\undefined\newsavebox{\plotpoint}\fi
\begin{picture}(1500,900)(0,0)
\font\gnuplot=cmr10 at 10pt
\gnuplot
\sbox{\plotpoint}{\rule[-0.200pt]{0.400pt}{0.400pt}}%
\put(220.0,113.0){\rule[-0.200pt]{292.934pt}{0.400pt}}
\put(220.0,113.0){\rule[-0.200pt]{4.818pt}{0.400pt}}
\put(198,113){\makebox(0,0)[r]{0}}
\put(1416.0,113.0){\rule[-0.200pt]{4.818pt}{0.400pt}}
\put(220.0,198.0){\rule[-0.200pt]{4.818pt}{0.400pt}}
\put(198,198){\makebox(0,0)[r]{0.1}}
\put(1416.0,198.0){\rule[-0.200pt]{4.818pt}{0.400pt}}
\put(220.0,283.0){\rule[-0.200pt]{4.818pt}{0.400pt}}
\put(198,283){\makebox(0,0)[r]{0.2}}
\put(1416.0,283.0){\rule[-0.200pt]{4.818pt}{0.400pt}}
\put(220.0,368.0){\rule[-0.200pt]{4.818pt}{0.400pt}}
\put(198,368){\makebox(0,0)[r]{0.3}}
\put(1416.0,368.0){\rule[-0.200pt]{4.818pt}{0.400pt}}
\put(220.0,453.0){\rule[-0.200pt]{4.818pt}{0.400pt}}
\put(198,453){\makebox(0,0)[r]{0.4}}
\put(1416.0,453.0){\rule[-0.200pt]{4.818pt}{0.400pt}}
\put(220.0,537.0){\rule[-0.200pt]{4.818pt}{0.400pt}}
\put(198,537){\makebox(0,0)[r]{0.5}}
\put(1416.0,537.0){\rule[-0.200pt]{4.818pt}{0.400pt}}
\put(220.0,622.0){\rule[-0.200pt]{4.818pt}{0.400pt}}
\put(198,622){\makebox(0,0)[r]{0.6}}
\put(1416.0,622.0){\rule[-0.200pt]{4.818pt}{0.400pt}}
\put(220.0,707.0){\rule[-0.200pt]{4.818pt}{0.400pt}}
\put(198,707){\makebox(0,0)[r]{0.7}}
\put(1416.0,707.0){\rule[-0.200pt]{4.818pt}{0.400pt}}
\put(220.0,792.0){\rule[-0.200pt]{4.818pt}{0.400pt}}
\put(198,792){\makebox(0,0)[r]{0.8}}
\put(1416.0,792.0){\rule[-0.200pt]{4.818pt}{0.400pt}}
\put(220.0,877.0){\rule[-0.200pt]{4.818pt}{0.400pt}}
\put(198,877){\makebox(0,0)[r]{0.9}}
\put(1416.0,877.0){\rule[-0.200pt]{4.818pt}{0.400pt}}
\put(220.0,113.0){\rule[-0.200pt]{0.400pt}{4.818pt}}
\put(220,68){\makebox(0,0){2}}
\put(220.0,857.0){\rule[-0.200pt]{0.400pt}{4.818pt}}
\put(372.0,113.0){\rule[-0.200pt]{0.400pt}{4.818pt}}
\put(372,68){\makebox(0,0){3}}
\put(372.0,857.0){\rule[-0.200pt]{0.400pt}{4.818pt}}
\put(524.0,113.0){\rule[-0.200pt]{0.400pt}{4.818pt}}
\put(524,68){\makebox(0,0){4}}
\put(524.0,857.0){\rule[-0.200pt]{0.400pt}{4.818pt}}
\put(676.0,113.0){\rule[-0.200pt]{0.400pt}{4.818pt}}
\put(676,68){\makebox(0,0){5}}
\put(676.0,857.0){\rule[-0.200pt]{0.400pt}{4.818pt}}
\put(828.0,113.0){\rule[-0.200pt]{0.400pt}{4.818pt}}
\put(828,68){\makebox(0,0){6}}
\put(828.0,857.0){\rule[-0.200pt]{0.400pt}{4.818pt}}
\put(980.0,113.0){\rule[-0.200pt]{0.400pt}{4.818pt}}
\put(980,68){\makebox(0,0){7}}
\put(980.0,857.0){\rule[-0.200pt]{0.400pt}{4.818pt}}
\put(1132.0,113.0){\rule[-0.200pt]{0.400pt}{4.818pt}}
\put(1132,68){\makebox(0,0){8}}
\put(1132.0,857.0){\rule[-0.200pt]{0.400pt}{4.818pt}}
\put(1284.0,113.0){\rule[-0.200pt]{0.400pt}{4.818pt}}
\put(1284,68){\makebox(0,0){9}}
\put(1284.0,857.0){\rule[-0.200pt]{0.400pt}{4.818pt}}
\put(1436.0,113.0){\rule[-0.200pt]{0.400pt}{4.818pt}}
\put(1436,68){\makebox(0,0){10}}
\put(1436.0,857.0){\rule[-0.200pt]{0.400pt}{4.818pt}}
\put(220.0,113.0){\rule[-0.200pt]{292.934pt}{0.400pt}}
\put(1436.0,113.0){\rule[-0.200pt]{0.400pt}{184.048pt}}
\put(220.0,877.0){\rule[-0.200pt]{292.934pt}{0.400pt}}
\put(45,495){\makebox(0,0){$h$}}
\put(828,23){\makebox(0,0){$\tan \beta$}}
\put(220.0,113.0){\rule[-0.200pt]{0.400pt}{184.048pt}}
\put(220,141){\rule{1pt}{1pt}}
\put(235,148){\rule{1pt}{1pt}}
\put(250,155){\rule{1pt}{1pt}}
\put(266,161){\rule{1pt}{1pt}}
\put(281,166){\rule{1pt}{1pt}}
\put(296,171){\rule{1pt}{1pt}}
\put(311,176){\rule{1pt}{1pt}}
\put(326,181){\rule{1pt}{1pt}}
\put(342,185){\rule{1pt}{1pt}}
\put(357,189){\rule{1pt}{1pt}}
\put(372,193){\rule{1pt}{1pt}}
\put(387,197){\rule{1pt}{1pt}}
\put(402,201){\rule{1pt}{1pt}}
\put(418,204){\rule{1pt}{1pt}}
\put(433,208){\rule{1pt}{1pt}}
\put(448,211){\rule{1pt}{1pt}}
\put(463,214){\rule{1pt}{1pt}}
\put(478,217){\rule{1pt}{1pt}}
\put(494,220){\rule{1pt}{1pt}}
\put(509,222){\rule{1pt}{1pt}}
\put(524,225){\rule{1pt}{1pt}}
\put(531,737){\rule{1pt}{1pt}}
\put(531,738){\rule{1pt}{1pt}}
\put(531,735){\rule{1pt}{1pt}}
\put(531,740){\rule{1pt}{1pt}}
\put(532,729){\rule{1pt}{1pt}}
\put(532,745){\rule{1pt}{1pt}}
\put(533,722){\rule{1pt}{1pt}}
\put(533,751){\rule{1pt}{1pt}}
\put(535,717){\rule{1pt}{1pt}}
\put(535,755){\rule{1pt}{1pt}}
\put(536,713){\rule{1pt}{1pt}}
\put(536,758){\rule{1pt}{1pt}}
\put(538,710){\rule{1pt}{1pt}}
\put(538,761){\rule{1pt}{1pt}}
\put(539,227){\rule{1pt}{1pt}}
\put(539,707){\rule{1pt}{1pt}}
\put(539,763){\rule{1pt}{1pt}}
\put(541,704){\rule{1pt}{1pt}}
\put(541,765){\rule{1pt}{1pt}}
\put(542,701){\rule{1pt}{1pt}}
\put(542,766){\rule{1pt}{1pt}}
\put(544,699){\rule{1pt}{1pt}}
\put(544,768){\rule{1pt}{1pt}}
\put(545,696){\rule{1pt}{1pt}}
\put(547,694){\rule{1pt}{1pt}}
\put(548,692){\rule{1pt}{1pt}}
\put(550,690){\rule{1pt}{1pt}}
\put(551,688){\rule{1pt}{1pt}}
\put(554,230){\rule{1pt}{1pt}}
\put(554,684){\rule{1pt}{1pt}}
\put(554,776){\rule{1pt}{1pt}}
\put(562,675){\rule{1pt}{1pt}}
\put(562,781){\rule{1pt}{1pt}}
\put(570,232){\rule{1pt}{1pt}}
\put(570,667){\rule{1pt}{1pt}}
\put(570,784){\rule{1pt}{1pt}}
\put(577,660){\rule{1pt}{1pt}}
\put(585,234){\rule{1pt}{1pt}}
\put(585,653){\rule{1pt}{1pt}}
\put(585,790){\rule{1pt}{1pt}}
\put(592,647){\rule{1pt}{1pt}}
\put(600,237){\rule{1pt}{1pt}}
\put(600,641){\rule{1pt}{1pt}}
\put(600,794){\rule{1pt}{1pt}}
\put(608,636){\rule{1pt}{1pt}}
\put(615,239){\rule{1pt}{1pt}}
\put(615,630){\rule{1pt}{1pt}}
\put(615,797){\rule{1pt}{1pt}}
\put(623,625){\rule{1pt}{1pt}}
\put(630,241){\rule{1pt}{1pt}}
\put(630,621){\rule{1pt}{1pt}}
\put(630,799){\rule{1pt}{1pt}}
\put(638,616){\rule{1pt}{1pt}}
\put(646,243){\rule{1pt}{1pt}}
\put(646,612){\rule{1pt}{1pt}}
\put(646,802){\rule{1pt}{1pt}}
\put(661,245){\rule{1pt}{1pt}}
\put(661,604){\rule{1pt}{1pt}}
\put(661,803){\rule{1pt}{1pt}}
\put(676,247){\rule{1pt}{1pt}}
\put(676,597){\rule{1pt}{1pt}}
\put(676,805){\rule{1pt}{1pt}}
\put(691,248){\rule{1pt}{1pt}}
\put(691,590){\rule{1pt}{1pt}}
\put(691,806){\rule{1pt}{1pt}}
\put(706,250){\rule{1pt}{1pt}}
\put(706,583){\rule{1pt}{1pt}}
\put(706,807){\rule{1pt}{1pt}}
\put(722,252){\rule{1pt}{1pt}}
\put(722,577){\rule{1pt}{1pt}}
\put(722,808){\rule{1pt}{1pt}}
\put(737,253){\rule{1pt}{1pt}}
\put(737,571){\rule{1pt}{1pt}}
\put(737,809){\rule{1pt}{1pt}}
\put(752,255){\rule{1pt}{1pt}}
\put(752,566){\rule{1pt}{1pt}}
\put(752,809){\rule{1pt}{1pt}}
\put(767,257){\rule{1pt}{1pt}}
\put(767,561){\rule{1pt}{1pt}}
\put(767,810){\rule{1pt}{1pt}}
\put(782,258){\rule{1pt}{1pt}}
\put(782,556){\rule{1pt}{1pt}}
\put(782,810){\rule{1pt}{1pt}}
\put(798,260){\rule{1pt}{1pt}}
\put(798,552){\rule{1pt}{1pt}}
\put(798,811){\rule{1pt}{1pt}}
\put(813,261){\rule{1pt}{1pt}}
\put(813,547){\rule{1pt}{1pt}}
\put(813,811){\rule{1pt}{1pt}}
\put(828,263){\rule{1pt}{1pt}}
\put(828,543){\rule{1pt}{1pt}}
\put(828,812){\rule{1pt}{1pt}}
\put(843,264){\rule{1pt}{1pt}}
\put(843,539){\rule{1pt}{1pt}}
\put(843,812){\rule{1pt}{1pt}}
\put(858,265){\rule{1pt}{1pt}}
\put(858,535){\rule{1pt}{1pt}}
\put(858,812){\rule{1pt}{1pt}}
\put(874,267){\rule{1pt}{1pt}}
\put(874,532){\rule{1pt}{1pt}}
\put(874,813){\rule{1pt}{1pt}}
\put(889,268){\rule{1pt}{1pt}}
\put(889,528){\rule{1pt}{1pt}}
\put(889,813){\rule{1pt}{1pt}}
\put(904,269){\rule{1pt}{1pt}}
\put(904,525){\rule{1pt}{1pt}}
\put(904,813){\rule{1pt}{1pt}}
\put(919,270){\rule{1pt}{1pt}}
\put(919,522){\rule{1pt}{1pt}}
\put(919,813){\rule{1pt}{1pt}}
\put(934,271){\rule{1pt}{1pt}}
\put(934,519){\rule{1pt}{1pt}}
\put(934,813){\rule{1pt}{1pt}}
\put(950,273){\rule{1pt}{1pt}}
\put(950,516){\rule{1pt}{1pt}}
\put(950,813){\rule{1pt}{1pt}}
\put(965,274){\rule{1pt}{1pt}}
\put(965,513){\rule{1pt}{1pt}}
\put(965,814){\rule{1pt}{1pt}}
\put(980,275){\rule{1pt}{1pt}}
\put(980,511){\rule{1pt}{1pt}}
\put(980,814){\rule{1pt}{1pt}}
\put(995,276){\rule{1pt}{1pt}}
\put(995,508){\rule{1pt}{1pt}}
\put(995,814){\rule{1pt}{1pt}}
\put(1010,277){\rule{1pt}{1pt}}
\put(1010,506){\rule{1pt}{1pt}}
\put(1010,814){\rule{1pt}{1pt}}
\put(1026,278){\rule{1pt}{1pt}}
\put(1026,503){\rule{1pt}{1pt}}
\put(1026,814){\rule{1pt}{1pt}}
\put(1041,279){\rule{1pt}{1pt}}
\put(1041,279){\rule{1pt}{1pt}}
\put(1041,814){\rule{1pt}{1pt}}
\put(1056,280){\rule{1pt}{1pt}}
\put(1056,499){\rule{1pt}{1pt}}
\put(1056,814){\rule{1pt}{1pt}}
\put(1071,281){\rule{1pt}{1pt}}
\put(1071,496){\rule{1pt}{1pt}}
\put(1071,814){\rule{1pt}{1pt}}
\put(1086,282){\rule{1pt}{1pt}}
\put(1086,494){\rule{1pt}{1pt}}
\put(1086,814){\rule{1pt}{1pt}}
\put(1102,283){\rule{1pt}{1pt}}
\put(1102,492){\rule{1pt}{1pt}}
\put(1102,814){\rule{1pt}{1pt}}
\put(1117,284){\rule{1pt}{1pt}}
\put(1117,490){\rule{1pt}{1pt}}
\put(1117,814){\rule{1pt}{1pt}}
\put(1132,284){\rule{1pt}{1pt}}
\put(1132,488){\rule{1pt}{1pt}}
\put(1132,814){\rule{1pt}{1pt}}
\put(1147,285){\rule{1pt}{1pt}}
\put(1147,487){\rule{1pt}{1pt}}
\put(1147,814){\rule{1pt}{1pt}}
\put(1162,286){\rule{1pt}{1pt}}
\put(1162,485){\rule{1pt}{1pt}}
\put(1162,815){\rule{1pt}{1pt}}
\put(1178,287){\rule{1pt}{1pt}}
\put(1178,483){\rule{1pt}{1pt}}
\put(1178,815){\rule{1pt}{1pt}}
\put(1193,288){\rule{1pt}{1pt}}
\put(1193,481){\rule{1pt}{1pt}}
\put(1193,815){\rule{1pt}{1pt}}
\put(1208,289){\rule{1pt}{1pt}}
\put(1208,480){\rule{1pt}{1pt}}
\put(1208,815){\rule{1pt}{1pt}}
\put(1223,289){\rule{1pt}{1pt}}
\put(1223,478){\rule{1pt}{1pt}}
\put(1223,815){\rule{1pt}{1pt}}
\put(1238,290){\rule{1pt}{1pt}}
\put(1238,477){\rule{1pt}{1pt}}
\put(1238,815){\rule{1pt}{1pt}}
\put(1254,291){\rule{1pt}{1pt}}
\put(1254,475){\rule{1pt}{1pt}}
\put(1254,815){\rule{1pt}{1pt}}
\put(1269,292){\rule{1pt}{1pt}}
\put(1269,474){\rule{1pt}{1pt}}
\put(1269,815){\rule{1pt}{1pt}}
\put(1284,292){\rule{1pt}{1pt}}
\put(1284,472){\rule{1pt}{1pt}}
\put(1284,815){\rule{1pt}{1pt}}
\put(1299,293){\rule{1pt}{1pt}}
\put(1299,471){\rule{1pt}{1pt}}
\put(1299,815){\rule{1pt}{1pt}}
\put(1314,294){\rule{1pt}{1pt}}
\put(1314,469){\rule{1pt}{1pt}}
\put(1314,815){\rule{1pt}{1pt}}
\put(1330,294){\rule{1pt}{1pt}}
\put(1330,468){\rule{1pt}{1pt}}
\put(1330,815){\rule{1pt}{1pt}}
\put(1345,295){\rule{1pt}{1pt}}
\put(1345,467){\rule{1pt}{1pt}}
\put(1345,815){\rule{1pt}{1pt}}
\put(1360,296){\rule{1pt}{1pt}}
\put(1360,466){\rule{1pt}{1pt}}
\put(1360,815){\rule{1pt}{1pt}}
\put(1375,296){\rule{1pt}{1pt}}
\put(1375,464){\rule{1pt}{1pt}}
\put(1375,815){\rule{1pt}{1pt}}
\put(1390,297){\rule{1pt}{1pt}}
\put(1390,463){\rule{1pt}{1pt}}
\put(1390,815){\rule{1pt}{1pt}}
\put(1406,298){\rule{1pt}{1pt}}
\put(1406,462){\rule{1pt}{1pt}}
\put(1406,815){\rule{1pt}{1pt}}
\put(1421,298){\rule{1pt}{1pt}}
\put(1421,461){\rule{1pt}{1pt}}
\put(1421,815){\rule{1pt}{1pt}}
\put(1436,299){\rule{1pt}{1pt}}
\put(1436,460){\rule{1pt}{1pt}}
\put(1436,815){\rule{1pt}{1pt}}
\end{picture}

%% file: Sub2.tex
% GNUPLOT: LaTeX picture
\setlength{\unitlength}{0.240900pt}
\ifx\plotpoint\undefined\newsavebox{\plotpoint}\fi
\begin{picture}(1500,900)(0,0)
\font\gnuplot=cmr10 at 10pt
\gnuplot
\sbox{\plotpoint}{\rule[-0.200pt]{0.400pt}{0.400pt}}%
\put(220.0,648.0){\rule[-0.200pt]{292.934pt}{0.400pt}}
\put(220.0,113.0){\rule[-0.200pt]{0.400pt}{184.048pt}}
\put(220.0,113.0){\rule[-0.200pt]{4.818pt}{0.400pt}}
\put(198,113){\makebox(0,0)[r]{-7}}
\put(1416.0,113.0){\rule[-0.200pt]{4.818pt}{0.400pt}}
\put(220.0,189.0){\rule[-0.200pt]{4.818pt}{0.400pt}}
\put(198,189){\makebox(0,0)[r]{-6}}
\put(1416.0,189.0){\rule[-0.200pt]{4.818pt}{0.400pt}}
\put(220.0,266.0){\rule[-0.200pt]{4.818pt}{0.400pt}}
\put(198,266){\makebox(0,0)[r]{-5}}
\put(1416.0,266.0){\rule[-0.200pt]{4.818pt}{0.400pt}}
\put(220.0,342.0){\rule[-0.200pt]{4.818pt}{0.400pt}}
\put(198,342){\makebox(0,0)[r]{-4}}
\put(1416.0,342.0){\rule[-0.200pt]{4.818pt}{0.400pt}}
\put(220.0,419.0){\rule[-0.200pt]{4.818pt}{0.400pt}}
\put(198,419){\makebox(0,0)[r]{-3}}
\put(1416.0,419.0){\rule[-0.200pt]{4.818pt}{0.400pt}}
\put(220.0,495.0){\rule[-0.200pt]{4.818pt}{0.400pt}}
\put(198,495){\makebox(0,0)[r]{-2}}
\put(1416.0,495.0){\rule[-0.200pt]{4.818pt}{0.400pt}}
\put(220.0,571.0){\rule[-0.200pt]{4.818pt}{0.400pt}}
\put(198,571){\makebox(0,0)[r]{-1}}
\put(1416.0,571.0){\rule[-0.200pt]{4.818pt}{0.400pt}}
\put(220.0,648.0){\rule[-0.200pt]{4.818pt}{0.400pt}}
\put(198,648){\makebox(0,0)[r]{0}}
\put(1416.0,648.0){\rule[-0.200pt]{4.818pt}{0.400pt}}
\put(220.0,724.0){\rule[-0.200pt]{4.818pt}{0.400pt}}
\put(198,724){\makebox(0,0)[r]{1}}
\put(1416.0,724.0){\rule[-0.200pt]{4.818pt}{0.400pt}}
\put(220.0,801.0){\rule[-0.200pt]{4.818pt}{0.400pt}}
\put(198,801){\makebox(0,0)[r]{2}}
\put(1416.0,801.0){\rule[-0.200pt]{4.818pt}{0.400pt}}
\put(220.0,877.0){\rule[-0.200pt]{4.818pt}{0.400pt}}
\put(198,877){\makebox(0,0)[r]{3}}
\put(1416.0,877.0){\rule[-0.200pt]{4.818pt}{0.400pt}}
\put(220.0,113.0){\rule[-0.200pt]{0.400pt}{4.818pt}}
\put(220,68){\makebox(0,0){0}}
\put(220.0,857.0){\rule[-0.200pt]{0.400pt}{4.818pt}}
\put(463.0,113.0){\rule[-0.200pt]{0.400pt}{4.818pt}}
\put(463,68){\makebox(0,0){1}}
\put(463.0,857.0){\rule[-0.200pt]{0.400pt}{4.818pt}}
\put(706.0,113.0){\rule[-0.200pt]{0.400pt}{4.818pt}}
\put(706,68){\makebox(0,0){2}}
\put(706.0,857.0){\rule[-0.200pt]{0.400pt}{4.818pt}}
\put(950.0,113.0){\rule[-0.200pt]{0.400pt}{4.818pt}}
\put(950,68){\makebox(0,0){3}}
\put(950.0,857.0){\rule[-0.200pt]{0.400pt}{4.818pt}}
\put(1193.0,113.0){\rule[-0.200pt]{0.400pt}{4.818pt}}
\put(1193,68){\makebox(0,0){4}}
\put(1193.0,857.0){\rule[-0.200pt]{0.400pt}{4.818pt}}
\put(1436.0,113.0){\rule[-0.200pt]{0.400pt}{4.818pt}}
\put(1436,68){\makebox(0,0){5}}
\put(1436.0,857.0){\rule[-0.200pt]{0.400pt}{4.818pt}}
\put(220.0,113.0){\rule[-0.200pt]{292.934pt}{0.400pt}}
\put(1436.0,113.0){\rule[-0.200pt]{0.400pt}{184.048pt}}
\put(220.0,877.0){\rule[-0.200pt]{292.934pt}{0.400pt}}
\put(45,495){\makebox(0,0){$h_{H_{2}}$}}
\put(828,23){\makebox(0,0){$h_{H_{1}}$}}
\put(220.0,113.0){\rule[-0.200pt]{0.400pt}{184.048pt}}
\put(220,655){\rule{1pt}{1pt}}
\put(244,641){\rule{1pt}{1pt}}
\put(269,628){\rule{1pt}{1pt}}
\put(293,616){\rule{1pt}{1pt}}
\put(317,604){\rule{1pt}{1pt}}
\put(342,592){\rule{1pt}{1pt}}
\put(366,580){\rule{1pt}{1pt}}
\put(390,569){\rule{1pt}{1pt}}
\put(415,557){\rule{1pt}{1pt}}
\put(439,546){\rule{1pt}{1pt}}
\put(463,535){\rule{1pt}{1pt}}
\put(488,524){\rule{1pt}{1pt}}
\put(512,514){\rule{1pt}{1pt}}
\put(536,503){\rule{1pt}{1pt}}
\put(560,492){\rule{1pt}{1pt}}
\put(585,482){\rule{1pt}{1pt}}
\put(609,471){\rule{1pt}{1pt}}
\put(633,461){\rule{1pt}{1pt}}
\put(658,451){\rule{1pt}{1pt}}
\put(658,707){\rule{1pt}{1pt}}
\put(658,731){\rule{1pt}{1pt}}
\put(682,440){\rule{1pt}{1pt}}
\put(682,698){\rule{1pt}{1pt}}
\put(682,736){\rule{1pt}{1pt}}
\put(706,430){\rule{1pt}{1pt}}
\put(706,691){\rule{1pt}{1pt}}
\put(706,740){\rule{1pt}{1pt}}
\put(731,420){\rule{1pt}{1pt}}
\put(731,684){\rule{1pt}{1pt}}
\put(731,744){\rule{1pt}{1pt}}
\put(755,410){\rule{1pt}{1pt}}
\put(755,678){\rule{1pt}{1pt}}
\put(755,747){\rule{1pt}{1pt}}
\put(779,400){\rule{1pt}{1pt}}
\put(779,672){\rule{1pt}{1pt}}
\put(779,750){\rule{1pt}{1pt}}
\put(804,390){\rule{1pt}{1pt}}
\put(804,666){\rule{1pt}{1pt}}
\put(804,752){\rule{1pt}{1pt}}
\put(828,380){\rule{1pt}{1pt}}
\put(828,660){\rule{1pt}{1pt}}
\put(828,755){\rule{1pt}{1pt}}
\put(852,370){\rule{1pt}{1pt}}
\put(852,654){\rule{1pt}{1pt}}
\put(852,757){\rule{1pt}{1pt}}
\put(877,360){\rule{1pt}{1pt}}
\put(877,648){\rule{1pt}{1pt}}
\put(877,760){\rule{1pt}{1pt}}
\put(901,350){\rule{1pt}{1pt}}
\put(901,642){\rule{1pt}{1pt}}
\put(901,762){\rule{1pt}{1pt}}
\put(925,340){\rule{1pt}{1pt}}
\put(925,636){\rule{1pt}{1pt}}
\put(925,764){\rule{1pt}{1pt}}
\put(950,331){\rule{1pt}{1pt}}
\put(950,631){\rule{1pt}{1pt}}
\put(950,767){\rule{1pt}{1pt}}
\put(974,321){\rule{1pt}{1pt}}
\put(974,625){\rule{1pt}{1pt}}
\put(974,769){\rule{1pt}{1pt}}
\put(998,311){\rule{1pt}{1pt}}
\put(998,619){\rule{1pt}{1pt}}
\put(998,771){\rule{1pt}{1pt}}
\put(1023,301){\rule{1pt}{1pt}}
\put(1023,613){\rule{1pt}{1pt}}
\put(1023,773){\rule{1pt}{1pt}}
\put(1047,292){\rule{1pt}{1pt}}
\put(1047,607){\rule{1pt}{1pt}}
\put(1047,775){\rule{1pt}{1pt}}
\put(1071,282){\rule{1pt}{1pt}}
\put(1071,601){\rule{1pt}{1pt}}
\put(1071,777){\rule{1pt}{1pt}}
\put(1096,272){\rule{1pt}{1pt}}
\put(1096,596){\rule{1pt}{1pt}}
\put(1096,779){\rule{1pt}{1pt}}
\put(1120,263){\rule{1pt}{1pt}}
\put(1120,590){\rule{1pt}{1pt}}
\put(1120,781){\rule{1pt}{1pt}}
\put(1144,253){\rule{1pt}{1pt}}
\put(1144,584){\rule{1pt}{1pt}}
\put(1144,784){\rule{1pt}{1pt}}
\put(1168,244){\rule{1pt}{1pt}}
\put(1168,578){\rule{1pt}{1pt}}
\put(1168,786){\rule{1pt}{1pt}}
\put(1193,234){\rule{1pt}{1pt}}
\put(1193,572){\rule{1pt}{1pt}}
\put(1193,788){\rule{1pt}{1pt}}
\put(1217,225){\rule{1pt}{1pt}}
\put(1217,566){\rule{1pt}{1pt}}
\put(1217,790){\rule{1pt}{1pt}}
\put(1241,215){\rule{1pt}{1pt}}
\put(1241,560){\rule{1pt}{1pt}}
\put(1241,792){\rule{1pt}{1pt}}
\put(1266,206){\rule{1pt}{1pt}}
\put(1266,554){\rule{1pt}{1pt}}
\put(1266,794){\rule{1pt}{1pt}}
\put(1290,196){\rule{1pt}{1pt}}
\put(1290,548){\rule{1pt}{1pt}}
\put(1290,796){\rule{1pt}{1pt}}
\put(1314,187){\rule{1pt}{1pt}}
\put(1314,542){\rule{1pt}{1pt}}
\put(1314,798){\rule{1pt}{1pt}}
\put(1339,178){\rule{1pt}{1pt}}
\put(1339,536){\rule{1pt}{1pt}}
\put(1339,800){\rule{1pt}{1pt}}
\put(1363,168){\rule{1pt}{1pt}}
\put(1363,530){\rule{1pt}{1pt}}
\put(1363,802){\rule{1pt}{1pt}}
\put(1387,159){\rule{1pt}{1pt}}
\put(1387,524){\rule{1pt}{1pt}}
\put(1387,804){\rule{1pt}{1pt}}
\put(1412,149){\rule{1pt}{1pt}}
\put(1412,518){\rule{1pt}{1pt}}
\put(1412,806){\rule{1pt}{1pt}}
\put(1436,140){\rule{1pt}{1pt}}
\put(1436,512){\rule{1pt}{1pt}}
\put(1436,808){\rule{1pt}{1pt}}
\put(220,675){\circle*{12}}
\put(244,665){\circle*{12}}
\put(269,656){\circle*{12}}
\put(293,646){\circle*{12}}
\put(317,637){\circle*{12}}
\put(342,628){\circle*{12}}
\put(342,701){\circle*{12}}
\put(342,701){\circle*{12}}
\put(366,619){\circle*{12}}
\put(366,689){\circle*{12}}
\put(366,707){\circle*{12}}
\put(390,610){\circle*{12}}
\put(390,681){\circle*{12}}
\put(390,709){\circle*{12}}
\put(415,601){\circle*{12}}
\put(415,674){\circle*{12}}
\put(415,710){\circle*{12}}
\put(439,592){\circle*{12}}
\put(439,667){\circle*{12}}
\put(439,711){\circle*{12}}
\put(463,583){\circle*{12}}
\put(463,661){\circle*{12}}
\put(463,711){\circle*{12}}
\put(488,574){\circle*{12}}
\put(488,654){\circle*{12}}
\put(488,712){\circle*{12}}
\put(512,565){\circle*{12}}
\put(512,647){\circle*{12}}
\put(512,713){\circle*{12}}
\put(536,556){\circle*{12}}
\put(536,640){\circle*{12}}
\put(536,713){\circle*{12}}
\put(560,548){\circle*{12}}
\put(560,634){\circle*{12}}
\put(560,714){\circle*{12}}
\put(585,539){\circle*{12}}
\put(585,627){\circle*{12}}
\put(585,714){\circle*{12}}
\put(609,530){\circle*{12}}
\put(609,620){\circle*{12}}
\put(609,714){\circle*{12}}
\put(633,521){\circle*{12}}
\put(633,614){\circle*{12}}
\put(633,715){\circle*{12}}
\put(658,513){\circle*{12}}
\put(658,607){\circle*{12}}
\put(658,715){\circle*{12}}
\put(682,504){\circle*{12}}
\put(682,600){\circle*{12}}
\put(682,716){\circle*{12}}
\put(706,495){\circle*{12}}
\put(706,594){\circle*{12}}
\put(706,716){\circle*{12}}
\put(731,487){\circle*{12}}
\put(731,587){\circle*{12}}
\put(731,716){\circle*{12}}
\put(755,478){\circle*{12}}
\put(755,580){\circle*{12}}
\put(755,717){\circle*{12}}
\put(779,470){\circle*{12}}
\put(779,573){\circle*{12}}
\put(779,717){\circle*{12}}
\put(804,461){\circle*{12}}
\put(804,567){\circle*{12}}
\put(804,717){\circle*{12}}
\put(828,453){\circle*{12}}
\put(828,560){\circle*{12}}
\put(828,718){\circle*{12}}
\put(852,444){\circle*{12}}
\put(852,553){\circle*{12}}
\put(852,718){\circle*{12}}
\put(877,436){\circle*{12}}
\put(877,546){\circle*{12}}
\put(877,718){\circle*{12}}
\put(901,427){\circle*{12}}
\put(901,539){\circle*{12}}
\put(901,719){\circle*{12}}
\put(925,419){\circle*{12}}
\put(925,533){\circle*{12}}
\put(925,719){\circle*{12}}
\put(950,410){\circle*{12}}
\put(950,526){\circle*{12}}
\put(950,719){\circle*{12}}
\put(974,402){\circle*{12}}
\put(974,519){\circle*{12}}
\put(974,720){\circle*{12}}
\put(998,393){\circle*{12}}
\put(998,512){\circle*{12}}
\put(998,720){\circle*{12}}
\put(1023,385){\circle*{12}}
\put(1023,505){\circle*{12}}
\put(1023,720){\circle*{12}}
\put(1047,376){\circle*{12}}
\put(1047,498){\circle*{12}}
\put(1047,721){\circle*{12}}
\put(1071,368){\circle*{12}}
\put(1071,491){\circle*{12}}
\put(1071,721){\circle*{12}}
\put(1096,360){\circle*{12}}
\put(1096,484){\circle*{12}}
\put(1096,721){\circle*{12}}
\put(1120,351){\circle*{12}}
\put(1120,478){\circle*{12}}
\put(1120,722){\circle*{12}}
\put(1144,343){\circle*{12}}
\put(1144,471){\circle*{12}}
\put(1144,722){\circle*{12}}
\put(1168,335){\circle*{12}}
\put(1168,464){\circle*{12}}
\put(1168,722){\circle*{12}}
\put(1193,326){\circle*{12}}
\put(1193,457){\circle*{12}}
\put(1193,723){\circle*{12}}
\put(1217,318){\circle*{12}}
\put(1217,450){\circle*{12}}
\put(1217,723){\circle*{12}}
\put(1241,309){\circle*{12}}
\put(1241,443){\circle*{12}}
\put(1241,723){\circle*{12}}
\put(1266,301){\circle*{12}}
\put(1266,436){\circle*{12}}
\put(1266,724){\circle*{12}}
\put(1290,293){\circle*{12}}
\put(1290,429){\circle*{12}}
\put(1290,724){\circle*{12}}
\put(1314,284){\circle*{12}}
\put(1314,422){\circle*{12}}
\put(1314,724){\circle*{12}}
\put(1339,276){\circle*{12}}
\put(1339,415){\circle*{12}}
\put(1339,725){\circle*{12}}
\put(1363,268){\circle*{12}}
\put(1363,408){\circle*{12}}
\put(1363,725){\circle*{12}}
\put(1387,260){\circle*{12}}
\put(1387,401){\circle*{12}}
\put(1387,725){\circle*{12}}
\put(1412,251){\circle*{12}}
\put(1412,394){\circle*{12}}
\put(1412,725){\circle*{12}}
\put(1436,243){\circle*{12}}
\put(1436,387){\circle*{12}}
\put(1436,726){\circle*{12}}
\put(220,683){\circle{12}}
\put(244,675){\circle{12}}
\put(244,705){\circle{12}}
\put(244,705){\circle{12}}
\put(269,666){\circle{12}}
\put(269,694){\circle{12}}
\put(269,709){\circle{12}}
\put(293,658){\circle{12}}
\put(293,687){\circle{12}}
\put(293,710){\circle{12}}
\put(317,649){\circle{12}}
\put(317,680){\circle{12}}
\put(317,710){\circle{12}}
\put(342,641){\circle{12}}
\put(342,673){\circle{12}}
\put(342,710){\circle{12}}
\put(366,632){\circle{12}}
\put(366,665){\circle{12}}
\put(366,711){\circle{12}}
\put(390,624){\circle{12}}
\put(390,658){\circle{12}}
\put(390,711){\circle{12}}
\put(415,616){\circle{12}}
\put(415,651){\circle{12}}
\put(415,711){\circle{12}}
\put(439,607){\circle{12}}
\put(439,644){\circle{12}}
\put(439,711){\circle{12}}
\put(463,599){\circle{12}}
\put(463,637){\circle{12}}
\put(463,711){\circle{12}}
\put(488,591){\circle{12}}
\put(488,630){\circle{12}}
\put(488,711){\circle{12}}
\put(512,583){\circle{12}}
\put(512,623){\circle{12}}
\put(512,711){\circle{12}}
\put(536,574){\circle{12}}
\put(536,616){\circle{12}}
\put(536,712){\circle{12}}
\put(560,566){\circle{12}}
\put(560,609){\circle{12}}
\put(560,712){\circle{12}}
\put(585,558){\circle{12}}
\put(585,602){\circle{12}}
\put(585,712){\circle{12}}
\put(609,550){\circle{12}}
\put(609,595){\circle{12}}
\put(609,712){\circle{12}}
\put(633,542){\circle{12}}
\put(633,588){\circle{12}}
\put(633,712){\circle{12}}
\put(658,534){\circle{12}}
\put(658,580){\circle{12}}
\put(658,712){\circle{12}}
\put(682,526){\circle{12}}
\put(682,573){\circle{12}}
\put(682,712){\circle{12}}
\put(706,517){\circle{12}}
\put(706,566){\circle{12}}
\put(706,712){\circle{12}}
\put(731,509){\circle{12}}
\put(731,559){\circle{12}}
\put(731,712){\circle{12}}
\put(755,501){\circle{12}}
\put(755,552){\circle{12}}
\put(755,712){\circle{12}}
\put(779,493){\circle{12}}
\put(779,545){\circle{12}}
\put(779,712){\circle{12}}
\put(804,485){\circle{12}}
\put(804,537){\circle{12}}
\put(804,713){\circle{12}}
\put(828,477){\circle{12}}
\put(828,530){\circle{12}}
\put(828,713){\circle{12}}
\put(852,469){\circle{12}}
\put(852,523){\circle{12}}
\put(852,713){\circle{12}}
\put(877,461){\circle{12}}
\put(877,516){\circle{12}}
\put(877,713){\circle{12}}
\put(901,453){\circle{12}}
\put(901,508){\circle{12}}
\put(901,713){\circle{12}}
\put(925,445){\circle{12}}
\put(925,501){\circle{12}}
\put(925,713){\circle{12}}
\put(950,437){\circle{12}}
\put(950,494){\circle{12}}
\put(950,713){\circle{12}}
\put(974,429){\circle{12}}
\put(974,487){\circle{12}}
\put(974,713){\circle{12}}
\put(998,420){\circle{12}}
\put(998,480){\circle{12}}
\put(998,713){\circle{12}}
\put(1023,412){\circle{12}}
\put(1023,472){\circle{12}}
\put(1023,713){\circle{12}}
\put(1047,404){\circle{12}}
\put(1047,465){\circle{12}}
\put(1047,713){\circle{12}}
\put(1071,396){\circle{12}}
\put(1071,458){\circle{12}}
\put(1071,713){\circle{12}}
\put(1096,388){\circle{12}}
\put(1096,451){\circle{12}}
\put(1096,714){\circle{12}}
\put(1120,380){\circle{12}}
\put(1120,443){\circle{12}}
\put(1120,714){\circle{12}}
\put(1144,372){\circle{12}}
\put(1144,436){\circle{12}}
\put(1144,714){\circle{12}}
\put(1168,364){\circle{12}}
\put(1168,429){\circle{12}}
\put(1168,714){\circle{12}}
\put(1193,356){\circle{12}}
\put(1193,421){\circle{12}}
\put(1193,714){\circle{12}}
\put(1217,348){\circle{12}}
\put(1217,414){\circle{12}}
\put(1217,714){\circle{12}}
\put(1241,340){\circle{12}}
\put(1241,407){\circle{12}}
\put(1241,714){\circle{12}}
\put(1266,332){\circle{12}}
\put(1266,400){\circle{12}}
\put(1266,714){\circle{12}}
\put(1290,324){\circle{12}}
\put(1290,392){\circle{12}}
\put(1290,714){\circle{12}}
\put(1314,316){\circle{12}}
\put(1314,385){\circle{12}}
\put(1314,714){\circle{12}}
\put(1339,308){\circle{12}}
\put(1339,378){\circle{12}}
\put(1339,714){\circle{12}}
\put(1363,300){\circle{12}}
\put(1363,370){\circle{12}}
\put(1363,714){\circle{12}}
\put(1387,292){\circle{12}}
\put(1387,363){\circle{12}}
\put(1387,714){\circle{12}}
\put(1412,284){\circle{12}}
\put(1412,356){\circle{12}}
\put(1412,715){\circle{12}}
\put(1436,277){\circle{12}}
\put(1436,348){\circle{12}}
\put(1436,715){\circle{12}}
\end{picture}

%% file: newfig3.tex
% GNUPLOT: LaTeX picture
\setlength{\unitlength}{0.240900pt}
\ifx\plotpoint\undefined\newsavebox{\plotpoint}\fi
\begin{picture}(1500,900)(0,0)
\font\gnuplot=cmr10 at 10pt
\gnuplot
\sbox{\plotpoint}{\rule[-0.200pt]{0.400pt}{0.400pt}}%
\put(220.0,622.0){\rule[-0.200pt]{292.934pt}{0.400pt}}
\put(220.0,113.0){\rule[-0.200pt]{4.818pt}{0.400pt}}
\put(198,113){\makebox(0,0)[r]{-2}}
\put(1416.0,113.0){\rule[-0.200pt]{4.818pt}{0.400pt}}
\put(220.0,240.0){\rule[-0.200pt]{4.818pt}{0.400pt}}
\put(198,240){\makebox(0,0)[r]{-1.5}}
\put(1416.0,240.0){\rule[-0.200pt]{4.818pt}{0.400pt}}
\put(220.0,368.0){\rule[-0.200pt]{4.818pt}{0.400pt}}
\put(198,368){\makebox(0,0)[r]{-1}}
\put(1416.0,368.0){\rule[-0.200pt]{4.818pt}{0.400pt}}
\put(220.0,495.0){\rule[-0.200pt]{4.818pt}{0.400pt}}
\put(198,495){\makebox(0,0)[r]{-0.5}}
\put(1416.0,495.0){\rule[-0.200pt]{4.818pt}{0.400pt}}
\put(220.0,622.0){\rule[-0.200pt]{4.818pt}{0.400pt}}
\put(198,622){\makebox(0,0)[r]{0}}
\put(1416.0,622.0){\rule[-0.200pt]{4.818pt}{0.400pt}}
\put(220.0,750.0){\rule[-0.200pt]{4.818pt}{0.400pt}}
\put(198,750){\makebox(0,0)[r]{0.5}}
\put(1416.0,750.0){\rule[-0.200pt]{4.818pt}{0.400pt}}
\put(220.0,877.0){\rule[-0.200pt]{4.818pt}{0.400pt}}
\put(198,877){\makebox(0,0)[r]{1}}
\put(1416.0,877.0){\rule[-0.200pt]{4.818pt}{0.400pt}}
\put(220.0,113.0){\rule[-0.200pt]{0.400pt}{4.818pt}}
\put(220,68){\makebox(0,0){2}}
\put(220.0,857.0){\rule[-0.200pt]{0.400pt}{4.818pt}}
\put(372.0,113.0){\rule[-0.200pt]{0.400pt}{4.818pt}}
\put(372,68){\makebox(0,0){3}}
\put(372.0,857.0){\rule[-0.200pt]{0.400pt}{4.818pt}}
\put(524.0,113.0){\rule[-0.200pt]{0.400pt}{4.818pt}}
\put(524,68){\makebox(0,0){4}}
\put(524.0,857.0){\rule[-0.200pt]{0.400pt}{4.818pt}}
\put(676.0,113.0){\rule[-0.200pt]{0.400pt}{4.818pt}}
\put(676,68){\makebox(0,0){5}}
\put(676.0,857.0){\rule[-0.200pt]{0.400pt}{4.818pt}}
\put(828.0,113.0){\rule[-0.200pt]{0.400pt}{4.818pt}}
\put(828,68){\makebox(0,0){6}}
\put(828.0,857.0){\rule[-0.200pt]{0.400pt}{4.818pt}}
\put(980.0,113.0){\rule[-0.200pt]{0.400pt}{4.818pt}}
\put(980,68){\makebox(0,0){7}}
\put(980.0,857.0){\rule[-0.200pt]{0.400pt}{4.818pt}}
\put(1132.0,113.0){\rule[-0.200pt]{0.400pt}{4.818pt}}
\put(1132,68){\makebox(0,0){8}}
\put(1132.0,857.0){\rule[-0.200pt]{0.400pt}{4.818pt}}
\put(1284.0,113.0){\rule[-0.200pt]{0.400pt}{4.818pt}}
\put(1284,68){\makebox(0,0){9}}
\put(1284.0,857.0){\rule[-0.200pt]{0.400pt}{4.818pt}}
\put(1436.0,113.0){\rule[-0.200pt]{0.400pt}{4.818pt}}
\put(1436,68){\makebox(0,0){10}}
\put(1436.0,857.0){\rule[-0.200pt]{0.400pt}{4.818pt}}
\put(220.0,113.0){\rule[-0.200pt]{292.934pt}{0.400pt}}
\put(1436.0,113.0){\rule[-0.200pt]{0.400pt}{184.048pt}}
\put(220.0,877.0){\rule[-0.200pt]{292.934pt}{0.400pt}}
\put(45,495)
{\makebox(0,0){$\displaystyle{\frac{B_{\pm}(M_X)}{M_{1/2}^{(0)}}}$}}
\put(828,23){\makebox(0,0){$\tan \beta$}}
\put(220.0,113.0){\rule[-0.200pt]{0.400pt}{184.048pt}}
\put(220,811){\rule{1pt}{1pt}}
\put(235,797){\rule{1pt}{1pt}}
\put(250,784){\rule{1pt}{1pt}}
\put(266,773){\rule{1pt}{1pt}}
\put(281,764){\rule{1pt}{1pt}}
\put(296,755){\rule{1pt}{1pt}}
\put(311,747){\rule{1pt}{1pt}}
\put(326,740){\rule{1pt}{1pt}}
\put(342,733){\rule{1pt}{1pt}}
\put(357,727){\rule{1pt}{1pt}}
\put(372,722){\rule{1pt}{1pt}}
\put(387,717){\rule{1pt}{1pt}}
\put(402,712){\rule{1pt}{1pt}}
\put(418,707){\rule{1pt}{1pt}}
\put(433,703){\rule{1pt}{1pt}}
\put(448,699){\rule{1pt}{1pt}}
\put(463,696){\rule{1pt}{1pt}}
\put(478,692){\rule{1pt}{1pt}}
\put(494,689){\rule{1pt}{1pt}}
\put(509,686){\rule{1pt}{1pt}}
\put(524,683){\rule{1pt}{1pt}}
\put(539,680){\rule{1pt}{1pt}}
\put(554,677){\rule{1pt}{1pt}}
\put(570,675){\rule{1pt}{1pt}}
\put(585,672){\rule{1pt}{1pt}}
\put(600,670){\rule{1pt}{1pt}}
\put(615,668){\rule{1pt}{1pt}}
\put(630,666){\rule{1pt}{1pt}}
\put(646,664){\rule{1pt}{1pt}}
\put(661,662){\rule{1pt}{1pt}}
\put(676,660){\rule{1pt}{1pt}}
\put(691,658){\rule{1pt}{1pt}}
\put(706,656){\rule{1pt}{1pt}}
\put(722,655){\rule{1pt}{1pt}}
\put(737,653){\rule{1pt}{1pt}}
\put(752,651){\rule{1pt}{1pt}}
\put(767,650){\rule{1pt}{1pt}}
\put(782,649){\rule{1pt}{1pt}}
\put(798,647){\rule{1pt}{1pt}}
\put(813,646){\rule{1pt}{1pt}}
\put(828,645){\rule{1pt}{1pt}}
\put(843,643){\rule{1pt}{1pt}}
\put(858,642){\rule{1pt}{1pt}}
\put(874,641){\rule{1pt}{1pt}}
\put(889,640){\rule{1pt}{1pt}}
\put(904,639){\rule{1pt}{1pt}}
\put(919,638){\rule{1pt}{1pt}}
\put(934,637){\rule{1pt}{1pt}}
\put(950,636){\rule{1pt}{1pt}}
\put(965,635){\rule{1pt}{1pt}}
\put(980,634){\rule{1pt}{1pt}}
\put(995,633){\rule{1pt}{1pt}}
\put(1010,632){\rule{1pt}{1pt}}
\put(1026,631){\rule{1pt}{1pt}}
\put(1041,630){\rule{1pt}{1pt}}
\put(1056,629){\rule{1pt}{1pt}}
\put(1071,628){\rule{1pt}{1pt}}
\put(1086,628){\rule{1pt}{1pt}}
\put(1102,627){\rule{1pt}{1pt}}
\put(1117,626){\rule{1pt}{1pt}}
\put(1132,625){\rule{1pt}{1pt}}
\put(1147,625){\rule{1pt}{1pt}}
\put(1162,624){\rule{1pt}{1pt}}
\put(1178,623){\rule{1pt}{1pt}}
\put(1193,623){\rule{1pt}{1pt}}
\put(1208,622){\rule{1pt}{1pt}}
\put(1223,621){\rule{1pt}{1pt}}
\put(1238,621){\rule{1pt}{1pt}}
\put(1254,620){\rule{1pt}{1pt}}
\put(1269,620){\rule{1pt}{1pt}}
\put(1284,619){\rule{1pt}{1pt}}
\put(1299,618){\rule{1pt}{1pt}}
\put(1314,618){\rule{1pt}{1pt}}
\put(1330,617){\rule{1pt}{1pt}}
\put(1345,617){\rule{1pt}{1pt}}
\put(1360,616){\rule{1pt}{1pt}}
\put(1375,616){\rule{1pt}{1pt}}
\put(1390,615){\rule{1pt}{1pt}}
\put(1406,615){\rule{1pt}{1pt}}
\put(1421,614){\rule{1pt}{1pt}}
\put(1436,614){\rule{1pt}{1pt}}
\put(220,224){\rule{1pt}{1pt}}
\put(235,247){\rule{1pt}{1pt}}
\put(250,267){\rule{1pt}{1pt}}
\put(266,285){\rule{1pt}{1pt}}
\put(281,300){\rule{1pt}{1pt}}
\put(296,314){\rule{1pt}{1pt}}
\put(311,327){\rule{1pt}{1pt}}
\put(326,338){\rule{1pt}{1pt}}
\put(342,348){\rule{1pt}{1pt}}
\put(357,358){\rule{1pt}{1pt}}
\put(372,366){\rule{1pt}{1pt}}
\put(387,374){\rule{1pt}{1pt}}
\put(402,382){\rule{1pt}{1pt}}
\put(418,388){\rule{1pt}{1pt}}
\put(433,395){\rule{1pt}{1pt}}
\put(448,400){\rule{1pt}{1pt}}
\put(463,406){\rule{1pt}{1pt}}
\put(478,411){\rule{1pt}{1pt}}
\put(494,416){\rule{1pt}{1pt}}
\put(509,420){\rule{1pt}{1pt}}
\put(524,424){\rule{1pt}{1pt}}
\put(539,428){\rule{1pt}{1pt}}
\put(554,432){\rule{1pt}{1pt}}
\put(570,436){\rule{1pt}{1pt}}
\put(585,439){\rule{1pt}{1pt}}
\put(600,442){\rule{1pt}{1pt}}
\put(615,445){\rule{1pt}{1pt}}
\put(630,448){\rule{1pt}{1pt}}
\put(646,451){\rule{1pt}{1pt}}
\put(661,454){\rule{1pt}{1pt}}
\put(676,456){\rule{1pt}{1pt}}
\put(691,459){\rule{1pt}{1pt}}
\put(706,461){\rule{1pt}{1pt}}
\put(722,463){\rule{1pt}{1pt}}
\put(737,465){\rule{1pt}{1pt}}
\put(752,467){\rule{1pt}{1pt}}
\put(767,469){\rule{1pt}{1pt}}
\put(782,471){\rule{1pt}{1pt}}
\put(798,473){\rule{1pt}{1pt}}
\put(813,475){\rule{1pt}{1pt}}
\put(828,476){\rule{1pt}{1pt}}
\put(843,478){\rule{1pt}{1pt}}
\put(858,479){\rule{1pt}{1pt}}
\put(874,481){\rule{1pt}{1pt}}
\put(889,482){\rule{1pt}{1pt}}
\put(904,484){\rule{1pt}{1pt}}
\put(919,485){\rule{1pt}{1pt}}
\put(934,486){\rule{1pt}{1pt}}
\put(950,488){\rule{1pt}{1pt}}
\put(965,489){\rule{1pt}{1pt}}
\put(980,490){\rule{1pt}{1pt}}
\put(995,491){\rule{1pt}{1pt}}
\put(1010,492){\rule{1pt}{1pt}}
\put(1026,493){\rule{1pt}{1pt}}
\put(1041,494){\rule{1pt}{1pt}}
\put(1056,495){\rule{1pt}{1pt}}
\put(1071,496){\rule{1pt}{1pt}}
\put(1086,497){\rule{1pt}{1pt}}
\put(1102,498){\rule{1pt}{1pt}}
\put(1117,499){\rule{1pt}{1pt}}
\put(1132,500){\rule{1pt}{1pt}}
\put(1147,501){\rule{1pt}{1pt}}
\put(1162,502){\rule{1pt}{1pt}}
\put(1178,503){\rule{1pt}{1pt}}
\put(1193,503){\rule{1pt}{1pt}}
\put(1208,504){\rule{1pt}{1pt}}
\put(1223,505){\rule{1pt}{1pt}}
\put(1238,506){\rule{1pt}{1pt}}
\put(1254,506){\rule{1pt}{1pt}}
\put(1269,507){\rule{1pt}{1pt}}
\put(1284,508){\rule{1pt}{1pt}}
\put(1299,508){\rule{1pt}{1pt}}
\put(1314,509){\rule{1pt}{1pt}}
\put(1330,510){\rule{1pt}{1pt}}
\put(1345,510){\rule{1pt}{1pt}}
\put(1360,511){\rule{1pt}{1pt}}
\put(1375,512){\rule{1pt}{1pt}}
\put(1390,512){\rule{1pt}{1pt}}
\put(1406,513){\rule{1pt}{1pt}}
\put(1421,513){\rule{1pt}{1pt}}
\put(1436,514){\rule{1pt}{1pt}}
\end{picture}